\pdfoutput=1
\documentclass[10pt, twoside]{scrbook}	

\usepackage[english]{babel} 						
\usepackage[protrusion=true,expansion=true]{microtype}			
\usepackage{amsmath,amsfonts,amsthm}					
\usepackage[pdftex]{graphicx}						
\usepackage[svgnames]{xcolor}						
\usepackage[hang, small,labelfont=bf,up,textfont=it,up]{caption}	
\usepackage{epstopdf}							           
\usepackage{subfigure}							
\usepackage{booktabs}							
\usepackage{fix-cm}	
\usepackage[final]{pdfpages}
\usepackage{amssymb}
\usepackage{rotating}
\usepackage{multirow}
\usepackage{paralist}
\usepackage{amstext}
\usepackage{CJKutf8}
\usepackage[colorlinks, citecolor=blue, linkcolor=black, urlcolor=blue]{hyperref}

\usepackage{dsfont}
\usepackage[normalem]{ulem}
\usepackage[square,numbers,comma,sort&compress]{natbib}
\usepackage{bm}
\usepackage{amssymb}
\usepackage{enumitem}
\usepackage[utf8]{inputenc} 

\usepackage[paperwidth=170mm, paperheight=244mm]{geometry}

\usepackage{setspace}
\usepackage{fancyhdr}

\renewcommand{\chaptermark}[1]{\markboth{\textit{Chapter \thechapter. #1}}{}}

\fancypagestyle{plain}{ 
     \fancyhead{} 
      
     } 
 
\fancypagestyle{chapter}{ 
     \fancyhead{} 

     \fancyfoot[LE, RO]{\footnotesize  \thepage}} 

\usepackage[explicit]{titlesec}
\usepackage{sectsty}				

\usepackage{lmodern}

\newlength\chapnumb
\titleformat{\chapter}[block]
{\doublespacing\normalfont\sffamily}{}{0pt}
{\parbox[b]{\chapnumb}{%
   \fontsize{75}{75}\selectfont\thechapter}%
  \parbox[b]{\dimexpr\textwidth-\chapnumb\relax}{%
    \raggedleft%
    \hfill{\doublespacing\normalfont\sffamily\huge#1}\\
    \vspace{-15pt}\rule{\dimexpr\textwidth-\chapnumb\relax}{0.4pt}}}  

\titleformat{name=\chapter,numberless}[block]
{\doublespacing\normalfont\sffamily}{}{0pt}
{\parbox[b]{\chapnumb}{%
   \mbox{}}%
  \parbox[b]{\dimexpr\textwidth-\chapnumb\relax}{%
    \raggedleft%
    \hfill{\doublespacing\normalfont\sffamily\huge#1}\\
    \vspace{-15pt}\rule{\dimexpr\textwidth-\chapnumb\relax}{0.4pt}}}

\titleclass{\part}{top} 
\titleformat{\part}
[display]
{\vspace{15pc}\centering\normalfont\Huge\bfseries}
{\MakeUppercase{\partname} \thepart}
{0pt}
{\vspace{1pc}\huge\MakeUppercase}

\titlespacing*{\part}{0pt}{0pt}{20pt}

\usepackage{lettrine}

\newfont{\gwpfont}{cmssq8 scaled 1000}

\begin{document}

\cleardoublepage

\begin{titlepage}
\thispagestyle{empty} 
\begin{figure}[h]
{\centering
{\includegraphics[width=0.45\textwidth]{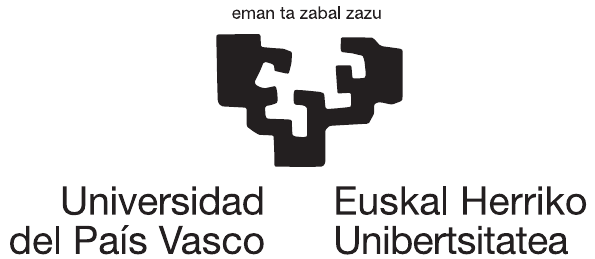}}\par}
\end{figure}

\begin{center}
{\Large Department of Physical Chemistry}\\ 

\vspace{1.5cm} 

{\LARGE\textbf{Design of Light-Matter Interactions}}\\
\vspace{0.3cm}
{\LARGE \textbf{for Quantum Technologies}}\\
\vspace{2.5cm}

{\Large \textbf{I\~nigo Arrazola}}\\

\vfill

{\Large Ph.D. Thesis}\\
\vspace{0.2cm}
{\Large Leioa 2020}\\
\end{center}

\end{titlepage}

\thispagestyle{empty} 

\begin{center}
  Department of Physical Chemistry\break
  University of the Basque Country (UPV/EHU)\break
  Postal Box 644, 48080 Bilbao, Spain

  
  \vspace{10pt}
  
\end{center}

\vfill

\begin{flushleft}

\vspace{5pt}

\noindent
This document was generated with the 2020 \LaTeX~distribution.\break
The plots and figures of this thesis were generated with MATLAB and Apple's Keynote. \break
The cover painting was done by \href{https://www.anderetxaniz.com/}{Ander Etxaniz}.

\vspace{5pt}

\noindent
This work was funded by the Basque Government grant PRE-2015-1-0394

\vspace{15pt}

\noindent
\includegraphics[height=20pt]{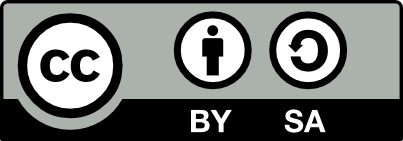}\break
2016-2020 I\~nigo Arrazola.
This work is licensed under the Creative Commons Attribution-ShareAlike 4.0 International License.
To view a copy of this license, visit\break
\href{http://creativecommons.org/licenses/by-sa/4.0/deed.en_US}
{http://creativecommons.org/licenses/by-sa/4.0/deed.en\_US}
\end{flushleft}

\clearpage

\begin{titlepage}
\thispagestyle{empty} 
\begin{figure}[h]
{\centering
{\includegraphics[width=0.45\textwidth]{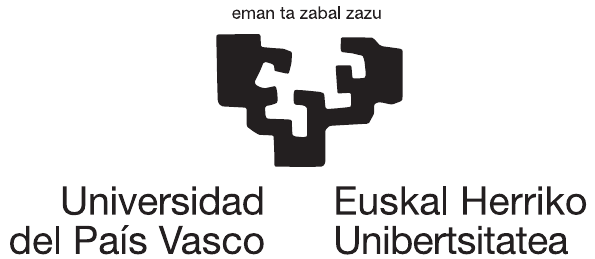}}\par}
\end{figure}

\begin{center}
{\Large Department of Physical Chemistry}\\ 

\vspace{1.5cm} 

{\LARGE\textbf{Design of Light-Matter Interactions}}\\
\vspace{0.3cm}
{\LARGE \textbf{for Quantum Technologies}}\\
\end{center}

\vspace{3.75cm}

\begin{center}
\large{\textit{Supervisors}:\\
\vspace{0.1cm}
Prof. Enrique Solano \\
\vspace{0.1cm}
Dr. Jorge Casanova\\}
\end{center}

\vfill

\begin{flushright}
\large{ Submitted by I\~nigo Arrazola\\
for the degree of Doctor of Physics}
\end{flushright}

\end{titlepage}

\newpage\null\thispagestyle{empty}\newpage

\thispagestyle{empty}

\vspace*{2cm}
\begin{flushright}
{\emph{{
Neure etxekueri \\
(to my family)
}}}
\end{flushright}
\newpage\null\thispagestyle{empty}\newpage

\thispagestyle{empty}

\vspace*{2cm}
{{\emph{ 
Urek ere ezin zehatz esan, erreka nondik nora doan \\
(Nor can the waters say exactly from where the river flows)
}}}
\begin{flushright}
\textbf{Ruper Ordorika} \\
\textit{Bihotz begiekin, Hurrengo goizean (Metak CD), 2001}
\end{flushright}

\newpage\null\thispagestyle{empty}\newpage


{\pagestyle{plain}
\cleardoublepage
\pagenumbering{gobble}
\tableofcontents
\cleardoublepage}


\frontmatter

{\pagestyle{plain}

\chapter*{Abstract}
\addcontentsline{toc}{section}{Abstract}

Quantum mechanics, is at the heart of many of the technological and scientific milestones of the last century such as the laser, the integrated circuit, or the magnetic resonance imaging scanner. However, only few decades have passed since we have the possibility to coherently manipulate the quantum states encoded in physical registers of specific quantum platforms. Understanding the light-matter interaction mechanisms that govern the dynamics of these systems is crucial for manipulating these registers and scaling up their quantum volume, which is a must for building a full-fledged quantum computer with vast implications in physics, chemistry, economics, and beyond. In 2019, the group of Prof.~John Martinis at Google achieved a technological milestone when they performed an experiment with 53 superconducting quantum bits (qubits). Although this can still be considered a small quantum system, the Google group claimed that, to solve the problem run by their quantum setup, the best classical computers would need thousands of years, and in consequence, they had achieved quantum supremacy. This is a matter of debate, as, short after, IBM argued that their classical computers could run the algorithm in a few days. Nevertheless, it is believed that noisy intermediate-scale quantum devices are capable of outperforming the best classical computers in specific tasks such as calculating properties of many-body quantum systems. In this regard, quantum simulators are specific purpose quantum computers that are expected to boost our understanding of, e.g., high-temperature superconductors or light-matter interactions beyond perturbative regimes. A different application of quantum technologies is that of quantum sensors. These quantum devices, that can be manipulated with suited radiation patterns, enable the measurement of physical quantities with unprecedented spatial resolution. This procedure is known as quantum sensing, and, among other things, it is a field that promises a better understanding of biological systems. In light of the above, it is clear that the endeavour of investigating light-matter interactions and finding optimal scenarios for their manipulation is of major importance for quantum technology and its emerging applications.

In this Thesis, we develop novel proposals for efficient quantum information processing and quantum sensing, quantum simulation of generalized light-matter interactions beyond the strong-coupling regime, and quantum supremacy experiments with neutral atoms. In particular, we propose two different methods to generate quantum logic gates with trapped ions driven by microwave radiation. One is aimed to be applied in current setups, while the other one assumes experimental parameters reachable in the near term. We demonstrate that both methods are robust against the main sources of decoherence in these systems. Moreover, our quantum gates work without laser radiation which is an advantage for scaling up trapped-ion quantum processors, as optical tables are substituted by microwave antennas that can be easily integrated in microtrap arrays. We also study different models of light-matter interaction, more specifically, the Rabi-Stark and the nonlinear quantum Rabi models, and propose a method for their implementation using laser-driven trapped ions, where quantum simulations of light-matter interaction have already been realized. In these models, we discover interesting properties such as the appearance of selective multi-photon interactions or the blockade of population distribution in the Hilbert space. Furthermore, we propose how ultracold atoms in optical lattices can be controlled with microwave and laser radiation in order to realize the boson sampling problem, a model of computation capable of showing quantum supremacy with tens of particles.  Taking into account experimental error sources, such as particle losses, we estimate that, using neutral atoms in spin-dependent optical lattices and, within realistic conditions, quantum supremacy could be achieved with tens of atoms. Finally, we develop a method to achieve selective interactions between a nitrogen-vacancy center and nearby carbon-13 atoms. These interactions are obtained by suitably designed microwave pulse sequences, and can be used to perform nuclear magnetic resonance at the nanoscale, with applications in biological sciences. Compared to other methods, ours is energy efficient, and thus, less invasive and suitable to be applied in biological samples.

All in all, in this Thesis we design radiation patterns capable of creating effective light-matter interactions suited to applications in quantum computing, quantum simulation and quantum sensing. In this manner, the results presented here significantly expand our knowledge on the control of light-matter interactions, and provide optimal scenarios for current quantum devices to generate the next-generation of quantum applications. Moreover, we introduce novel methods to simulate generalized light-matter interactions beyond perturbative regimes. Thereby, we believe our results will boost the construction of better quantum sensors, quantum simulators and trapped-ion quantum processors, as well as the first experimental realization of quantum supremacy using neutral atoms.

\cleardoublepage
\chapter*{Acknowledgements}
\addcontentsline{toc}{section}{Acknowledgements}

During the development of this thesis I have been educated in the exciting field of quantum technologies, and been able to make a scientific contribution to it. Along the way, I have met great people and visit amazing cities like Boston, Munich or Shanghai. All this would not have been possible without QUTIS and its leader, Prof.~Enrique Solano, who from the very beginning has supported and guided me. His eagerness for avoiding comfort zones and keep evolving, his determination to give the best every day, and his sporadic musical recommendations have been always inspiring. I consider him a true brilliant minded rebel. If this thesis has another father, this is Dr.~Jorge Casanova. Jorge was leaving when I entered in QUTIS; fortunately he taught me all I needed to know in order to carry out my bachelor thesis. He put me on the right track, and occasionally come back from Ulm to give me a good push. For that I am deeply grateful. I still have to figure out where those ideas come from! 

In QUTIS I have always felt valued and appreciated, from the more senior members to the younger ones. I had the opportunity to work under the supervision of Prof.~Lucas Lamata and Prof.~Enrique Rico, and I would like to thank them for their work and guidance. I would also like to thank Prof.~I\~nigo Egusquiza, who, lucky for us, has always been around whenever we needed an oracle. Of course, the younger group mates have also contributed to the positive work environment, and all that I have learned from long conversations with Dr.~Laura Garc\'ia-\'Alvarez, Dr.~Urtzi Las Heras and Dr.~Unai \'Alvarez-Rodr\'iguez, I subconsciously keep in my mind. I want to mention my office mates and travel companions Adrian Parra and Dr.~Julen S.~Pedernales, who have suffered and enjoyed my concerns and reflexions on a daily basis. Thank you guys.

\begin{CJK}{UTF8}{gbsn}
Throughout this thesis, I had the opportunity to visit other top-level research groups that have received me warmly. I want to thank Prof.~Xi Chen for inviting me, more than once, to Shanghai University. In such a different country, the unprecedented hospitality that I received was key to make my trip to China one of the best experiences in my life.  From Shanghai, I want to especially thank Dr.~Xiao-Hang Cheng, Dr.~Lei Cong and Lijuan Dong, my closest collaborators and good friends. 谢谢. From the beginning, I have maintained a long-standing collaboration with the group of Prof.~Daniel Rodr\'iguez in Granada. I want to thank Prof.~Rodr\'iguez, Manu, and Fran for that, and for giving me the opportunity to see your lab evolving, while always answering my questions patiently. Much\'isimas gracias a todos. I wish to see the 7 Tesla Penning trap at its best soon! For this thesis, the collaboration with the group of Dr.~Andrea Alberti and Prof.~Dieter Meschede at the University of Bonn was very important. I want to thank Dr.~Alberti for bringing the topic of boson sampling to my table, for hosting me in his group and for helping me with the math. I also want to thank Dr.~Carsten Robens for his hospitality (both in Bonn and Boston), his guidance and for being always supportive. Danke Sch\"on. In addition, I would like to thank Prof.~Kihwan Kim for hosting me at Tsinghua University in Beijing, Prof.~Gerhard Kirchmair for inviting me to the IQOQI in Innsbruck, and Prof.~Tobias Sch\"atz for receiving me at the University of Freiburg, as well as their group members, that made me enjoy the stay. 
\end{CJK}

Last but not least, I want to thank GNT for turning lunchtime into an exciting religious ceremony. I declare myself a devote GNTer with a single golden rule: 12:50 sharp. I am thankful to Dr.~Pablo Jimeno for the \LaTeX \ template and Dr.~Sof\'ia Mart\'inez-Garaot for the style of the references. Finally, I want to thank the natural mechanism that has made me a terrible soccer player, as this has allowed me to focus on important things such as mathematics, music, and cinema, and, of course, my deepest appreciation to my friends and family for all the unconditional support.

\cleardoublepage
\chapter*{List of publications}
\addcontentsline{toc}{section}{List of publications}

This thesis is based in the following publications:
\\

\textbf{ Chapter 2: \nameref{chapter:chapter_1}}

\begin{enumerate}

\item {I. Arrazola, J. Casanova, J. S. Pedernales, Z.-Y. Wang, E. Solano, and M. B. Plenio, \\
\textit{ Pulsed dynamical decoupling for fast and robust two-qubit gates on trapped ions},\\
\href{https://journals.aps.org/pra/abstract/10.1103/PhysRevA.97.052312}{Physical Review A \textbf{97}, 052312 (2018).}}

\item {I. Arrazola, M. B. Plenio, E. Solano, and J. Casanova \\
\textit{ Hybrid Microwave-Radiation Patterns for High-Fidelity Quantum Gates with Trapped Ions},\\
\href{https://journals.aps.org/prapplied/abstract/10.1103/PhysRevApplied.13.024068}{Physical Review Applied \textbf{13}, 024068 (2020).}}

\end{enumerate}

\textbf{ Chapter 3: \nameref{chapter:chapter_2}}

\begin{enumerate}[resume]

\item { X.-H. Cheng, I. Arrazola, J. S. Pedernales, L. Lamata, X. Chen, and E. Solano\\
\textit{ Nonlinear quantum Rabi model in trapped ions},\\
\href{https://journals.aps.org/pra/abstract/10.1103/PhysRevA.97.023624}{Physical Review A \textbf{97}, 023624 (2018).}}

\item { L. Cong, S. Felicetti, J. Casanova, L. Lamata, E. Solano, and I. Arrazola \\
\textit{ Selective interactions in the quantum Rabi model},\\
\href{https://journals.aps.org/pra/abstract/10.1103/PhysRevA.101.032350}{Physical Review A \textbf{101}, 032350 (2020).}}

\end{enumerate}

\textbf{ Chapter 5: \nameref{chapter:chapter_4}}

\begin{enumerate}[resume]

\item { I. Arrazola, E. Solano, and J. Casanova \\
\textit{ Selective hybrid spin interactions with low radiation power},\\
\href{https://journals.aps.org/prb/abstract/10.1103/PhysRevB.99.245405}{Physical Review B \textbf{99}, 245405 (2019).}}

\end{enumerate}

Other articles published in the course of this thesis yet not included in it are:

\begin{enumerate}[resume]

\item { I. Arrazola, J. S. Pedernales, L. Lamata, and E. Solano \\
\textit{ Digital-Analog Quantum Simulation of Spin Models in Trapped Ions},\\
\href{https://www.nature.com/articles/srep30534}{Scientific Reports \textbf{6}, 30534 (2016).}}

\item { X.-H. Cheng, I. Arrazola, J. S. Pedernales, L. Lamata, X. Chen, and E. Solano \\
\textit{ Switchable particle statistics with an embedding quantum simulator},\\
\href{https://journals.aps.org/pra/abstract/10.1103/PhysRevA.95.022305}{Physical Review A \textbf{95}, 022305 (2017).}}

\item { F. Dom\'{i}nguez, I. Arrazola, J. Dom\'{e}nech, J. S. Pedernales, L. Lamata, E. Solano, and D. Rodr\'{i}guez \\
\textit{ A Single-Ion Reservoir as a High-Sensitive Sensor of Electric Signals},\\
\href{https://www.nature.com/articles/s41598-017-08782-5}{Scientific Reports \textbf{ 7}, 8336 (2017).}}

\item {F. Dom\'{i}nguez, M. J. Guti\'{e}rrez, I. Arrazola, J. Berrocal, J. M. Cornejo, J. J. Del Pozo, R. A. Rica, S. Schmidt, E. Solano, and D. Rodr\'{i}guez \\
\textit{ Motional studies of one and two laser-cooled trapped ions for electric-field sensing applications},\\
\href{https://www.tandfonline.com/doi/abs/10.1080/09500340.2017.1406157}{Journal of Modern Optics \textbf{65}, 613 (2018).}}

\item {M. J Guti\'errez, J. Berrocal, J. M Cornejo, F. Dom\'{i}nguez, J. J. Del Pozo, I. Arrazola, J. Ba\~nuelos, P. Escobedo, O. Kaleja, L. Lamata, R. A. Rica, S. Schmidt, M. Block, E. Solano, and D. Rodr\'{i}guez \\ 
\textit{ The TRAPSENSOR facility: an open-ring 7 tesla Penning trap for laser-based precision experiments}, \\
\href{https://iopscience.iop.org/article/10.1088/1367-2630/aafa45}{New Journal of Physics \textbf{21} 023023 (2019).}}

\item {R. Puebla, G. Zicari, I. Arrazola, E. Solano, M. Paternostro, and J. Casanova \\ 
\textit{ Spin-boson model as a simulator of non-Markovian multiphoton Jaynes-Cummings models}, \\
\href{https://www.mdpi.com/2073-8994/11/5/695}{Symmetry \textbf{11}, 695 (2019).}}

\item {M. J. Guti\'{e}rrez, J. Berrocal, F. Dom\'{i}nguez, I. Arrazola, M. Block, E. Solano, and D. Rodr\'{i}guez \\ 
\textit{ Dynamics of an unbalanced two-ion crystal in a Penning trap for application in optical mass spectrometry}, \\
\href{https://journals.aps.org/pra/abstract/10.1103/PhysRevA.100.063415}{Physical Review A \textbf{100}, 063415 (2019).}}

\item {T. Xin, S. Wei, J. Cui, J. Xiao, I. Arrazola, L. Lamata, X. Kong, D. Lu, E. Solano, and G. Long \\ 
\textit{ Quantum algorithm for solving linear differential equations: Theory and experiment}, \\
\href{https://journals.aps.org/pra/abstract/10.1103/PhysRevA.101.032307}{Physical Review A \textbf{101}, 032307 (2020).}}

\item {C. Munuera-Javaloy, I. Arrazola, E. Solano, and J. Casanova \\ 
\textit{ Double quantum magnetometry at large static magnetic fields}, \\
\href{https://journals.aps.org/prb/abstract/10.1103/PhysRevB.101.104411}{Physical Review B \textbf{101}, 104411 (2020).}}

\item {J.-N. Zhang, I. Arrazola, J. Casanova, L. Lamata, K. Kim, and E. Solano \\ 
\textit{ Probabilistic eigensolver with a trapped-ion quantum processor}, \\
\href{https://journals.aps.org/pra/abstract/10.1103/PhysRevA.101.052333}{Physical Review A \textbf{101}, 052333 (2020).}}

\end{enumerate}

\cleardoublepage
\chapter*{List of abbreviations}
\addcontentsline{toc}{section}{List of abbreviations}

\begin{enumerate}[leftmargin=5cm]
\item [\textbf{ BSPD}]{Boson-Sampling Probability Distribution}
\item [\textbf{ DSC}]{Deep-Strong Coupling}
\item [\textbf{DD}]{Dynamical Decoupling}
\item [\textbf{JC}]{Jaynes-Cummings}
\item [\textbf{ LD}]{Lamb-Dicke}
\item [\textbf{ MW}]{Microwave}
\item [\textbf{ NV}]{Nitrogen-Vacancy}
\item [\textbf{ NISQ}]{Noisy Intermediate-Scale Quantum}
\item [\textbf{NQRM}]{Nonlinear Quantum Rabi Model}
\item [\textbf{ NMR}]{Nuclear Magnetic Resonance}
\item [\textbf{ OU}]{Ornstein-Uhlenbeck}
\item [\textbf{QED}]{Quantum Electrodynamics}
\item [\textbf{QRM}]{Quantum Rabi Model}
\item [\textbf{RWA}]{Rotating-Wave Approximation}
\item [\textbf{SC}]{Strong Coupling}
\item [\textbf{USC}]{Ultrastrong Coupling}
\end{enumerate}

\cleardoublepage
\chapter*{List of symbols}
\addcontentsline{toc}{section}{List of symbols}

\vspace{0.1cm}

\textbf{  \ \ \ Physical constants:}

\begin{tabular}{p{0.185\textwidth}p{0.7\textwidth}ll}
  $\hbar$ & Reduced Planck constant $1.054571817\times10^{-34}$ m$^2$ kg / s \\
    $e$ & Electric charge of the electron $1.602176634\times10^{-19}$ C \\
   $c$ & Speed of light $299792458$ m/s \\ 
    $\varepsilon_0$ & Vacuum permittivity  $8.8541878128(13) \times10^{-12}$ F/m \\
    $\mu_0$ & Vacuum permeability  $4\pi\times10^{-7}$  T m/A 
\end{tabular}

\vspace{0.5cm}

\textbf{ Chapter 1: \nameref{chapter:chapter_0}}

\begin{tabular}{p{0.185\textwidth}p{0.70\textwidth}ll}
  $B_0, B_z, \vec{B}(t)$ & Intensity of static magnetic field, time-varying magnetic field \\
  $\vec{\mu}$ & Magnetic dipole moment \\
  $\gamma$ & Gyromagnetic ratio \\
  $\sigma_{x,y,z}, \sigma_{+} (\sigma_{-})$ & Pauli matrices, creation (annihilation) operator for a two-level system  \\
  $\omega_0, \omega_0^\textrm{ R}, \omega^\textrm{ R}$ & Larmor frequency, frequency of electronic transition, frequency of MW or light field  \\
  $\Omega,\tilde{\Omega}$ &  Rabi frequency \\
  $\omega,\tilde{\omega}$ & Frequency of $\vec{B}(t)$ field \\
  $\Delta,\tilde{\Delta}$ & Detuning between field and qubit frequency \\
  $|\!\!\uparrow\rangle,|\!\!\downarrow\rangle$ & ``Up" and ``down" states of the magnetic dipole \\
  $T_1, T_2$ & Depolarization time, dephasing time\\
  $\delta$ & Unknown shift of the Larmor frequency \\
  $g$ & Dipole-field coupling strengh (in units of angular frequency) \\
  $a^\dagger , a$ & Bosonic creation and annihilation operators \\
  $ |g\rangle, |e\rangle$ & ``Ground" and ``excited" electronic states \\
  $\Gamma$ & Electronic relaxation (depolarization) rate\\
  $\kappa$ & Cavity photon-loss rate \\
  $|n\rangle$ & Fock state number $n$ \\
  $P_{e,n}, P_{g,n}$ & Population of $|e,n\rangle$ and $|g,n\rangle$ states \\
  $J_{i,j}$ & Coupling between spins $i$ and $j$ (in units of angular frequency)\\
  $k$ & Wavenumber 
\end{tabular}

\begin{tabular}{p{0.185\textwidth}p{0.7\textwidth}ll}
 $x$ & Position of the atom \\
  $V(x)$ & Potential energy \\
  $S_{1/2}, P_{1/2}, D_{5/2}$ & Electronic subspaces of the atom \\
  $D$ & Zero-field splitting \\
  $^3A_2, ^3\!\!E, ^1\!\!A_1, ^1\!\!E$ & Electronic subspaces of the NV 
\end{tabular}

\vspace{0.5cm}

\textbf{ Chapter 2: \nameref{chapter:chapter_1}}

\begin{tabular}{p{0.185\textwidth}p{0.7\textwidth}ll}
  $\eta, \eta_m$ &  Effective LD parameter $\eta=\frac{\gamma_eg_B}{8\nu}\sqrt{\frac{\hbar}{M\nu}}$, $\eta_m= \frac{\gamma_e g_B}{8\nu_m} \sqrt{\frac{\hbar}{M \nu_m}}$ \\
  $\omega_e,\omega_g, \omega_j$ &  Energy of excited and ground states (in units of angular frequency), $j$-th qubit frequency  \\
   $z^0_j, \Delta z$ &  Equilibrium position of ion $j$, distance between equilibrium positions \\
   $B(z)$ &  Intensity of the magnetic field as a function of the position $z$ \\
  $\gamma_e$ &  Electronic gyromagnetic ratio \\
   $M$ & Mass of the ion \\
  $a^\dagger, c^\dagger (a,c)$ & Creation (annihilation) operators of the center-of-mass and breathing modes \\
  $\nu, \nu_1 ,\nu_2$ & Frequencies of the trap, center-of-mass and breathing modes \\
  $g_B$ & Magnetic field gradient \\
  $b^\dagger (b)$ & Redefined creation (annihilation) operator of center-of-mass mode \\
    $\Omega_j, \phi$ & Rabi frequency and phase of the MW driving with frequency $\omega_j$ \\
    $f_j$ &  Modulation function representing the effect of $\pi$ pulses on the $j$-th qubit \\
    $G_{jm}$  & Function quantifying the displacement of ion $j$ in the phase space \\
    $U_s, U_c$ & Time-evolution operators of spin-force and two-qubit gate \\
    $\varphi, \theta_n$ & Accumulated gate phase \\
    $\varphi_m$ & Gate phase accumulated by mode $m$ \\
     $\tilde{\varphi}, \tilde{\varphi}_m$ &  Rescaled $\varphi, \varphi_m$  \\  
    $T_\textrm{ G}$  &  Gate final time \\
    $\vec{\phi^x},\vec{\phi^y}$ & List of phases in X and Y pulse blocks \\
    $\tau, \tau_a (\tau_b)$ & Duration of the pulse block, time of execution of the first (second) pulse of the block \\
    $n_\textrm{ B}, n_\textrm{ RT}, n_\textrm{ PF}$ & Number of blocks applied, phase-space round trips, phase flips \\
$r$ & Number of periods $2\pi/\nu_1$ in $\tau$ \\
$\delta_1,\delta_2$ & Difference between qubit frequencies $\delta_2 = - \delta_1 = \omega_2 - \omega_1$ \\
$t_\pi$ & $\pi$-pulse time \\
    $\bar{N}_b,\bar{N}_c $ & Average number of phonons for thermal states of the center of mass and breathing modes \\
    $\Delta t$ & Time between MW pulse in ion 1 and ion 2 
       \end{tabular}

              
       \begin{tabular}{p{0.185\textwidth}p{0.7\textwidth}ll}
        $U^{(1)}_\textrm{XY}, U^{(2)}_\textrm{XY}$ & Time-evolution operator of an XY block acting on the first (second) ion subspace \\
       $\rho$ & Density matrix \\  
       $\Gamma_b,\Gamma_c $ & Heating rates of center-of-mass and breathing modes  \\
    $T $ & Temperature of the trap electrodes \\    
    $\delta$ & Detuning with respect $\omega_j$ \\
    $S_{\alpha}$ & Collective spin-$1/2$ operator, e.g. $S_{\alpha}=\sigma_1^{\alpha}+\sigma_2^{\alpha}$ \\
    $\Omega, \Omega_\textrm{ DD}, \tilde{\Omega}_\textrm{DD}$ & Rabi frequency of bichromatic field, Rabi frequency of DD field, reescaled $\Omega_\textrm{DD}$ \\
      $J_n(z)$ & Bessel function of the first kind \\
    $\xi$ & Detuning with respect sideband frequency $\delta=\nu+\xi$\\
    $t_n$ & Time after $n_\textrm{ RT}$ phase-space round trips \\ 
    $\epsilon_j(t)$ & Energy fluctuation in the $j$-th qubit (in units of angular frequency) \\
    $\phi(t), \dot{\phi},\phi_\textrm{ DD} $ &  Time-varying phase of bichromatic driving, time-derivative of $\phi(t)$, phase of DD field \\
    $g_{\tilde{\Omega}}, g_{\nu}$ & Coupling strengths of effective second-order terms \\
   $\bar{n}$ & Average number of phonons of initial state \\
    $\dot{\bar{n}}$ & Reescaled heating rate of center-of-mass mode $\dot{\bar{n}}=\Gamma_b\bar{N}_b$ \\
   $\tau_{B}, T_2$ & Correlation time of magnetic field fluctuations, dephasing time induced by magnetic field fluctuations \\ 
      $\tau_{\Omega}, \delta_{\Omega}$ & Correlation time of MW field fluctuations, relative amplitude of MW field fluctuations   
   \end{tabular}

{In the corresponding appendices:}

\begin{tabular}{p{0.185\textwidth}p{0.7\textwidth}ll}
$|0\rangle, |1\rangle , |2\rangle , |3\rangle $ & States of the hyperfine subspace \\
$E_0, E_1, E_2, E_3 $ & Energies of hyperfine states \\
$X(t)$ & Change of the qubit frequency due to magnetic field fluctuations \\
$\epsilon_\perp$ & Relative amplitude (in terms of $\Omega(t)$) of the MW field inducing transitions outside the qubit subspace \\
$c_d$ & Diffusion constant of the OU process \\
 $\nu_r, \Delta_r$ & Radial trapping frequency, radial coupling strength of qubit-mode interaction \\
 $d^\dagger (d)$ & Creation (annihilation) operators of a collective radial mode \\
 $\beta$ & Ratio between axial and radial qubit-mode coupling strengths \\
$q_j$ & $j$-th ion's displacement around the equilibrium position \\
$Q_j$ & Normal mode coordinates \\
$a^\dagger_m (a_m)$ & Normal mode creation (annihilation) operator where $a_1=b$ and $a_2=c$ \\
$\Omega_k^\textrm{ M}$ & $k$-th order of the Magnus expansion \\
$\tilde{\tau}_a, \tilde{\tau}_b, x$ & $\tau_a, \tau_b, t$ in terms of $\tau$ \\
 $\gamma, \hat{n}_{0}, \varphi_0$ & Frequency, unit vector and phase characterizing crosstalk defined before Eq.~(\ref{unitcross}), in Eq.~(\ref{unitcross}) and after Eq.~(\ref{unitcross}) respectively 
 \end{tabular}

\begin{tabular}{p{0.185\textwidth}p{0.7\textwidth}ll} 
 $\dot{n}_\textrm{com}^\textrm{ref}, \dot{n}_\textrm{bre}^\textrm{ref} $ & Reference center-of-mass and breathing mode heating rates \\
$\nu^\textrm{ref}_1, \nu^\textrm{ref}_2$ & Reference center-of-mass and breathing mode frequencies \\
$T^\textrm{ref}, d_\textrm{ i-e}^\textrm{ref} $ & Reference temperature, reference ion-electrode distance \\
$d_\textrm{ i-e}$ & Ion-electrode distance \\
$N-1$ & Number of trap periods in $2\pi/\xi$ \\
$m$ & Number of $2\pi/\tilde{\Omega}_\textrm{ DD}$ periods in $t_n$ \\
$\tilde{\Omega}$ & Equivalent to $\tilde{\Omega}_\textrm{ DD}$ \\
$\tilde{S}_{\pm}$ & Redefined spin operators $\tilde{S}_{\pm}=\frac{1}{2}(S_z\pm i S_x)$
 \end{tabular}

\vspace{0.5cm}

\textbf{ Chapter 3: \nameref{chapter:chapter_2}}

\begin{tabular}{p{0.185\textwidth}p{0.7\textwidth}ll}
    $\omega_0, \omega, \omega_n^0, \omega_I$ & Frequency of qubit, frequency of bosonic mode, $\omega_n^0=\omega_0+\gamma(2 n +1)$, ion-qubit frequency \\
    $ a^\dagger (a)$ & Creation (annihilation) operator of a light mode in the QRM and, also, of a vibrational mode of a single trapped ion \\ 
    $g, \gamma$ & Coupling strengths of Rabi and Stark terms (in units of angular frequency) \\
    $\Omega_n, \Omega_{n,n+1}$ & Rabi frequency of the QRM \\
    $\delta_n^-, \delta_n^+$ & Frequency of rotating and counter-rotating terms  \\
    $\Delta_n^{\textrm e}, \Delta_n^{\textrm g}$ & Second-order energy shift associated with the $|e\rangle$ and $|g\rangle$ state \\
    $\Omega^{(3)}_{n-} (\Omega^{(3)}_{n+})$ & Rabi frequency of third-order JC-like (anti-JC-like) interaction \\
      $\delta^{(3)}_{n-}, \delta^{(3)}_{n+}$ & Frequency of third-order rotating and counter-rotating terms \\
       $\tilde{\delta}^{(3)}_{n-}, \tilde{\delta}^{(3)}_{n+}$ & Frequency of third-order rotating and counter-rotating terms, corrected up to second-order shifts \\    
      $\Omega^{(k)}_{n-} (\Omega^{(k)}_{n+})$ & Rabi frequency of $k$-order JC-like (anti-JC-like) interaction \\
      $\delta^{(k)}_{n-}, \delta^{(k)}_{n+}$ & Frequency of $k$-order rotating and counter-rotating terms \\
      $\omega^c_0$ & Approximate value of resonance frequency \\
    $\Omega_{\textrm{ S},r,b}, \omega_{\textrm{ S},r,b}, \phi_{\textrm{ S},r,b}$ & Rabi frequency, frequency, and phase of carrier, red-detuned and blue-detuned drivings \\
    $g_{r}, g_{b},\hat{g}_\textrm{ S}$ & Coupling strength of first red sideband, first blue sideband and carrier interactions \\
    $\Omega_0$ & $\Omega_0\equiv \Omega_\textrm{ S}(1-\eta^2/2)$ \\
    $\omega_0^\textrm{ R}, \omega^\textrm{ R}$ & Frequency of simulated qubit and simulated bosonic mode  \\
    $g^\textrm{ R}, \gamma^\textrm{ R}$ & Coupling strengths of simulated qubit-boson interaction and simulated Stark interaction  \\
     $\Omega, \omega_\textrm{ L}, \phi, \delta$ & Rabi frequency, frequency, phase and detuning of a generic laser driving \\
        $\hat{f}_1(\hat{n}), f_1(\hat{n})$ & Nonlinear operator, definition in Eq.~(\ref{NLfunc}), nonlinear operator evaluated in Fock state $|n\rangle$  \\
        $\tilde{\Omega}_{n,n+1}$ & Rabi frequency of the NQRM, $|f_1(n)|\Omega_{n,n+1}$  \\   
          $\langle n\rangle$ & Average number of phonons of initial state \\
         $\alpha$ & Complex number characterizing a coherent state, where $|\alpha|$ and ${\arg}(\alpha)$ are called ``amplitude" and ``phase" respectively \\   
   \end{tabular} 
   
     In the corresponding appendix :
       
   \begin{tabular}{p{0.185\textwidth}p{0.7\textwidth}ll}
  $\omega_n^e, \omega_n^g$ & $\omega_n^e=(\omega  +\gamma )n+\omega_0/2, \omega_n^g=(\omega  -\gamma )n-\omega_0/2$ \\ 
  $S_{n}(t)$ & $S_{n}(t)\equiv\sigma_+ e^{i\delta^+_{n} t}+\sigma_- e^{i\delta^-_{n} t}$ \\
  $g_{r,b}^{(1)}$ & Coupling strength of red and blue sideband (in units of angular frequency)  \\
  $g_{r,b}^{(2)}$ & Second-order coupling strength of red and blue sideband (in units of angular frequency)  \\
  $\tilde{\sigma}_{\pm}$ & $\tilde{\sigma}_{\pm}\equiv(\sigma_y\pm i\sigma_z)/2$ \\
  $g_\textrm{ JC}, g_\textrm{ aJC}$ & Coupling strength of JC and anti-JC terms (in units of angular frequency) 
   \end{tabular} 
   
  \vspace{0.5cm}
  
\textbf{ Chapter 4: \nameref{chapter:chapter_3}}

\begin{tabular}{p{0.185\textwidth}p{0.7\textwidth}ll}
 $N, M$  & Number of bosons, number of modes \\
 $t, \tau$ & Natural number describing discrete time steps, actual duration of a time step (spin-addressing operation) \\
 $a^\dagger_m (a_m), \hat{n}_m, n_m$  & Creation (annihilation) operator of bosonic mode $m$, number operator at mode $m$, number of particles in mode $m$ \\  
 $\hat{N}$ & Total number operator $\hat{N}=\sum_m\hat{n}_m$ \\
 $U, U_{ij}$ & Haar random unitary matrix, matrix element ($i$-th row and $j$-th column) \\
 $P_{BS}$ & Boson sampling probability distribution \\
 $|{\uparrow}\rangle, |{\downarrow}\rangle$ & Atomic hyperfine states \\
 $\lambda_L$ & Optical lattice wavelength \\
 $V_{\uparrow,\downarrow}(x)$ & Potential energy due to optical trapping \\ 
 $x_{\uparrow,\downarrow}(t)$ & ``Position" of $|\!\!\uparrow\rangle$ and $|\!\!\downarrow\rangle$ lattices \\
 $T(s,t)$ & $2\times 2$ unitary matrix, building block of $U$ \\
  $H_{2\times 2}, A(\theta), A(\phi)$ & $2\times 2$ unitary matrices, building blocks of $T(s,t)$ \\
  $|\psi_0\rangle, |0\rangle, |n_m\rangle$ & Initial state, vacuum state of all $M$ modes, Fock state of $m$-th mode\\
  $|\psi_u\rangle$ & Uniform initial state \\
  $P(n_1,n_2,..., n_{M})$ & Probability of generic final configuration $n_1,n_2,..., n_{M}$ \\ 
  $\hat{U}$ & Boson sampling time-evolution operator \\
  $t_\textrm{ in}, t_\textrm{ op}, t_{det}, t_\textrm{ pr}$ & Time required for initial state preparation, interference operation, measurement, and the whole process \\
  $R_\textrm{ pr} (R_0), R$ & Generation-rate of atomic (photonic) experimental samples, generation-rate of valid experimental samples \\
  $\eta_\textrm{ d}, \eta, \eta_\textrm{f}, \eta_\textrm{c} $ & Atomic detection efficiency, single-photon survival probability, single-photon fixed survival probability, single-photon survival probability per unit of length of circuit  \\
   $\tau_\textrm{ bg}, \tau_\textrm{ tb}$ &  One-body loss lifetime, two-body loss lifetime 
    \end{tabular} 
    
 \begin{tabular}{p{0.185\textwidth}p{0.7\textwidth}ll}
    $P_\textrm{surv, step, pair}(k)$ & Total survival probability of $N$-body sample, survival probability per time step $\tau$, probability of finding $k$ particle pairs (a pair of particles in the same lattice site) \\
  $d$ & Length of photonic circuit \\
 $ k_l$ & Number of lost photons \\
 $\tilde{a}$ & Real positive value quantifying the speed of classical computer \\
 $\omega_{\uparrow}, \omega_{\downarrow}$ & Energy of $|\!\!\uparrow\rangle$ and $|\!\!\downarrow\rangle$ states (in units of angular frequency) \\
  $\Omega_0 (\varphi_0), \Omega_s, $ & Rabi frequency (phase) of MW driving, Rabi frequency of light field at site $s$ \\
   $\hat{U}, \hat{U}_t, \hat{U}_{t,1} $ & Boson sampling time-evolution operator, time-evolution operator corresponding to step $t$, time-evolution operator from time step $1$ to $t$, i.e. $\hat{U}_{t,1}=\prod_{j=1}^{t}\hat{U}_j$  \\
  $\theta_s^t, \phi_s^t$ & Phases characterizing $T(s,t)$ transformation \\
   $\vec{\theta}_t, \vec{\phi}_t$ & Lists of phases at time step $t$ \\  
  $\mathcal{L}_b(\rho), \Gamma_b, F_b$ & Generic Lindblad superoperator, decay rate, and jump operator \\  
  $\Gamma_\textrm{ bg}, \Gamma_\textrm{ tb}$ & One-body-loss rate, two-body-loss rate \\
   $F$ & Fidelity with respect the boson sampling final state \\
  $p$ & Survival probability after all interference operations \\
  $V_t$ & Two-body-loss Hamiltonian at time step $t$, in an interaction picture with respect the boson-sampling Hamiltonian, i.e. $\hat{U}^\dagger_{t-1,1}V\hat{U}_{t-1,1}$ or $\hat{U}^\dagger_{t-1,1}V'\hat{U}_{t-1,1}$, depending if $j$ is odd or even \\
  $\epsilon$ & Fluctuation in energy difference $\omega_\uparrow-\omega_\downarrow$, in terms of $\Omega_0$
   \end{tabular}

     In the corresponding appendix :
   
   \begin{tabular}{p{0.185\textwidth}p{0.7\textwidth}ll}
   $k_2, k_3, k_4$ & Number of pairs, trios, and quartets \\
   $P(k_2,k_3,k_4)$ & Probability of having $k_2$ pairs, $k_3$ trios, and $k_4$ quartets \\
   $c$ & Ratio between $M$ and $N^2$ \\
   $D, d, |d\rangle$ & Total number of configurations, index for a possible configuration, state of a possible configuration \\
   $p_k, p_{k,k'}$ & Ratio between number of configurations where $k$ atoms are at mode $m$ and $D$, ratio between number of configurations where $k$ atoms are at mode $m$ while $k'$ atoms at mode $m'$ ($m\neq m'$), and $D$ \\
   $a_j$ & $a_j\equiv\prod_{i=1}^j\frac{M-i}{M+N-i}$ \\
   $\lambda_j$ & $\lambda_j\equiv\frac{N}{M+N-j}$ 
      \end{tabular} 
   
  \newpage
   \textbf{ Chapter 5: \nameref{chapter:chapter_4}}

\begin{tabular}{p{0.185\textwidth}p{0.7\textwidth}ll}
  $S_{x,y,z}$ & Spin operators for spin-$1$  \\
  $|0\rangle (|1\rangle), |+\rangle$ & Ground (excited) state of the NV qubit, superposition state $|+\rangle=\frac{1}{\sqrt{2}}(|0\rangle+|1\rangle)$ \\
  $\omega_\textrm{ n}, \omega_{j} $ & Nuclear Larmor frequency, Larmor frequency of $j$-th nucleus accounting for a shift due to the coupling with the NV  \\
   $I_j^{\alpha}$ & $j$-th nucleus spin-$1/2$ operator, $I_j^{\alpha}=1/2\sigma_j^{\alpha}$ with $\alpha=x,y,z,+,-$  \\
   $\gamma_\textrm{ n}$ & Nuclear gyromagnetic ratio \\
   $\vec{A}_j$  & Vector characterizing the NV-nucleus coupling, definition in Eq.~(\ref{hypervec}) \\
      $\vec{r}_j$ & Position vector of $j$-th nucleus, taking the vacancy site as the origin \\
   $\omega(t), \omega, \phi$ & Rabi frequency, frequency and phase of MW field \\
   $\hat{\omega}_j$ & Unit vector representing the new precession axis of nucleus $j$ \\
   $F(t)$ & Modulation function representing the effect of $\pi$ pulses\\
   $T$ & Period of $F(t)$ \\
   $n, l$ & Number of the harmonic in the Fourier series, number of the harmonic chosen such that $l \omega_\textrm{ M} \approx \omega_k$ \\
   $\omega_k$ & Larmor frequency of the nucleus in which we want to induce resonance \\
   $f_n, f^\textrm{ m}_n$ and $f^\textrm{ th}_n$ & Fourier coefficient corresponding to the $n$-th harmonic, Fourier coefficient with modulated  and top-hat pulses defined in Eq.~(\ref{modulatedf}) and Eq.~(\ref{tophatcoeff}) \\
   $\omega_\textrm{ M}$ & Angular frequency associated with period $T$ \\
   $\hat{x}_j, \hat{y}_j, \hat{z}_j$ & Unit vector of new Cartesian basis defined below Eq.~(\ref{modulated}) \\
   $t_m, t_p$ & The instant we star applying the $m$th pulse, central point of the $m$th pulse $t_p=t_m+t_\pi/2$ \\
   $\alpha_q(t)$ & Real, positive, time-varying function, $q$ being a natural number \\
   $a_1, c_1$ & Free parameters of Gaussian function $\alpha_1(t)$ \\
   $\gamma_{^{13}\textrm{ C}}, \gamma_\textrm{ H}$ & ${^{13}\textrm{ C}}$ nuclear gyromagnetic ratio, $^{1}\textrm{ H}$ nuclear (proton) gyromagnetic ratio \\
   $t_f$ & Final time of the sequence \\
   $E^\textrm{ th}$ and $E^\textrm{ ext}$ & Energy delivered by top-hat and extended (modulated) $\pi$ pulses
   \end{tabular} 
   
   In the corresponding appendix :
   
   \begin{tabular}{p{0.185\textwidth}p{0.7\textwidth}ll}
     $\tau_m, \tau_\pi$ & $t_m$ and $t_\pi$ in terms of $T/2$ \\
     $x, y$ & Integrating variables \\
     $\vec{P}, \vec{E}, \vec{B}$ & Poynting vector, electric field and magnetic field of MW driving \\
     $\vec{k}$ & Wavevector of MW driving \\
     $\vec{x}$ & Position of the NV center \\
     $B_0(t)$ & Time-varying amplitude of magnetic field \\
     
      \end{tabular} 
\cleardoublepage}

\mainmatter
\pagestyle{fancy}

\chapter{Introduction}
\label{chapter:chapter_0}
\thispagestyle{chapter}

Quantum mechanics, which describes the natural processes that take place at the atomic scale, is one of the most successful theories in physics. From the very beginning, light-matter interaction, or the interaction between atoms and electromagnetic fields, has played a major role in the development of the quantum theory. The quantisation of energy was first proposed by Max Plack in 1900 to describe the electromagnetic spectral distribution produced by a thermal source~\cite{Planck00}. In 1905, Einstein explained the photoelectric effect~\cite{Einstein05}, and, by 1917, he had developed a model for light-matter interaction that accounted for the absorption and stimulated or spontaneous emission of light by atoms~\cite{Einstein17}. The quantum theory was further formalised by Erwin Schr\"{o}dinger in 1926, when he proposed a wave equation that correctly predicted the spectral lines of the hydrogen atom~\cite{Schrodinger26}. By the same years, Paul Dirac made the first attempt to quantise the electromagnetic field~\cite{Dirac27}. However, it was not until the late 1940's that a theory for the interaction between quantised light and matter became available~\cite{Feynman85}, namely quantum electrodynamics (QED), accurately predicting the Lamb shift measured in the hydrogen microwave (MW) spectrum~\cite{Lamb47}. 

The rapid growth of MW and radio-frequency technology for telecommunications in the first half of the twentieth century led to the invention of the maser in 1953~\cite{Gordon55}. The maser produces coherent MW radiation through amplification by stimulated emission. This, along with other discoveries, such as optical pumping~\cite{Kastler50}, paved the way to the construction of the first laser in 1960~\cite{Taylor00}. The laser helped to formalise the theory of optical coherence~\cite{Glauber63a,Glauber63b}, and enabled coherently controlled interactions between light and matter. In the mid 70's, the ability to trap charged atoms~\cite{Paul90} was combined with lasers, leading to the first laser cooling protocol~\cite{Wineland75,Hansch75}. During the 80's laser cooling and trapping techniques for neutral atoms were developed~\cite{Phillips98}, quantum jumps were observed~\cite{Nagourney86,Sauter86,Bergquist86} and ground-state cooling of a single trapped ion was achieved~\cite{Diedrich89}. Lasers also allowed to study Rydberg atoms in optical or MW cavity resonators, a research field called cavity QED~\cite{Haroche89}. The interaction between these atoms with highly-excited electronic states and cavity modes leads to changes in the atomic properties, e.g. the enhancement or suppression of the spontaneous emission rate\footnote{The enhancement of the spontaneous emission rate is called the Purcell effect, and was previously discovered by Edward M. Purcell in the context of nuclear magnetic resonance~\cite{Purcell46}.}~\cite{Hulet85,Jhe87}. In 1992, the so-called strong-coupling (SC) regime between light and matter was achieved for a single atom~\cite{Thompson92}, allowing to study the interaction between atoms and photons at the level of single quanta~\cite{Brune96}. Light-matter interaction at the SC regime allows the formation of hybrid modes called polaritons, which share properties of both light and matter. These play a role, for example, in the localisation of light at small volumes via surface plasmons for subwavelength imaging~\cite{Vasa19}.

In a parallel effort, in the 1930's, Alan Turing set the grounds of the theory of computation when he showed that a machine with enough memory and following a very limited set of instructions, the so-called Turing machine, could efficiently perform any algorithmic process~\cite{Turing37}. This idea, along with the discovery of the first transistor\footnote{The Nobel Prize in Physics 1956 was awarded jointly to William B. Shockley, John Bardeen and Walter H. Brattain ``for their researches on semiconductors and their discovery of the transistor effect".}, and later, the MOSFET~\cite{Lojek07} in 1959, led to the exponential development of modern digital computers~\cite{Moore65}. Algorithms started to be classified in terms of the amount of resources, time and memory needed in order to find a solution using a Turing machine~\cite{Arora09}. For some problems, like multiplication, this amount of resources increased polynomially with the size of the input number, making them easy to solve. Other problems were hard to solve, for example, factoring integer numbers, for which no algorithm is known that scales polynomially with the input size. Another important issue was that, in the model of computation described by Turing, logical operations are irreversible. In 1961, Landauer used Shannon's information theory~\cite{Shannon48} to state that each of these irreversible operations would increase entropy, setting a fundamental energy cost to each operation~\cite{Landauer61}. Landauer's principle motivated the search for reversible computation models~\cite{Bennett73}, and, in the early 80's, the idea of a quantum Turing machine was proposed and developed by Benioff~\cite{Benioff80}, Manin~\cite{Manin80} and Deutsch~\cite{Deutsch85}. This quantum computer would store the information in two-level quantum systems, called quantum bits or qubits. Due to the linearity of the Schr\"odinger equation, this model of computation would be reversible. Furthermore, Feynman suggested that the quantum computer would be a natural testbed for simulating quantum physics~\cite{Feynman82}, while this would typically require an exponential amount of resources in a classical Turing machine~\cite{Poplavskii75}. In this way, quantum simulation appeared as the first practical application of a quantum computer. During the same years, the first ideas for secure communication using quantum mechanical variables were introduced~\cite{Wiesner83,Bennett14}, originating a research field that is known today as quantum cryptography~\cite{Bennett92}. In 1994, Peter Shor proposed a quantum algorithm for efficiently factoring integer numbers~\cite{Shor94}, which boosted what has been later called ``the second quantum revolution" by Dowling and Milburn~\cite{Dowling03}.

By that time, the control of individual quantum systems was already possible\footnote{In 2012, Serge Haroche and David J. Wineland received the Nobel Prize ``for ground-breaking experimental methods that enable measuring and manipulation of individual quantum systems".}, and the race for building a quantum computer started. Few months after the announcement of Shor's algorithm, Cirac and Zoller proposed a method to implement the controlled-NOT gate using two trapped ions~\cite{Cirac95}. Based on that proposal, the first two-qubit gate was realized by the group led by Wineland~\cite{Monroe95}. In the years that followed, numerous proposals for the physical realization of qubits and gates using photons~\cite{Chuang95,Knill01}, ions~\cite{Sorensen99,Sorensen00,Solano99,Jonathan00}, atoms in cavity QED~\cite{Pellizzari95} or optical lattices~\cite{Brennen99}, anyons~\cite{Kitaev03}, semiconductors~\cite{Loss98}, nuclear magnetic resonance (NMR)~\cite{Cory97,Gershenfeld97}, crystallographic defects in diamond~\cite{Cappellaro09}, and superconducting circuits~\cite{Nakamura99} were introduced. Besides, new quantum algorithms were developed~\cite{Grover96,Harrow09}, and the first protocols for quantum error correction were introduced~\cite{Shor95,Steane96}. Ideally, a quantum computer ought to be built by qubits that interact strongly among them, yet are isolated from the environment to avoid any source of decoherence. On the other hand, we should be able to control and measure the system from the exterior, which requires the aforementioned isolation to be highly selective. In this regard, quantum error correction~\cite{Gottesman09} provides pathways to build fault-tolerant quantum computers with qubit-errors below an acceptable given threshold~\cite{Campbell17}. In exchange, one needs to encode the quantum information of a logical qubit in many physical qubits.

Despite the impressive progress made increasing both the number of qubits and their coherence properties~\cite{Monz11,Barends14,Kaufman15,Ebert15,Lekitsch17,Popkin16,Wang16c,Barredo18,Zeng17,Bernien17,Kandala17,Wang20,Bermudez17,Zajac17}, the present-day quantum technology, sometimes referred to as noisy intermediate-scale quantum (NISQ) technology~\cite{Preskill18}, remains insufficient for a physical realisation of a fault-tolerant universal quantum computer~\cite{Wilczek16}. This has stimulated the development of alternative models of quantum computation adapted to NISQ devices, such as digital-analog quantum computation~\cite{Parra20}, variational quantum eigensolvers~\cite{McClean16}, constant-depth quantum circuits \cite{Terhal04}, temporally unstructured quantum circuits~\cite{Shepherd09}, quantum circuits with identical noninteracting bosons (also known as boson-sampling devices)~\cite{Aaronson11}, random circuits of coupled qubits~\cite{Boixo18} or quantum annealers \cite{Das08}. Because of their reduced experimental complexity, all these models of quantum computation are promising candidates to show a speedup over classical algorithms in NISQ architectures, a feat often named as quantum supremacy or quantum advantage~\cite{Preskill13,Papageorgiou13,Harrow17}. In fact, last year, quantum supremacy was claimed for the first time by the group led by Martinis using random circuits of superconducting qubits~\cite{Arute19}. 

Arguably the most important application of quantum computation is quantum simulation~\cite{Cirac12,BlochDalibard12,Blatt12,AspuruGuzik12,Houck12,VanHoucke12,Georgescu14,Gross17}, which consists in the reproduction of relevant quantum models using controllable quantum systems. The idea was first proposed by Manin~\cite{Manin80} and Feynman~\cite{Feynman82} in the early 80's, and was further formalised by Lloyd~\cite{Lloyd96}, who theoretically proved that quantum computers can efficiently simulate any local quantum system. Because the dimension of the Hilbert space grows exponentially with the number of constituents of the system, the classical simulation of quantum many-body phenomena is, in general, extremely inefficient, and, often, impossible~\cite{Laughlin00}.
On the contrary, quantum simulators can efficiently simulate models in quantum field theory~\cite{Byrnes06,Cirac10,Casanova11,Mazza12,Jordan12,Hauke13}, quantum chemistry~\cite{Kassal11,Yung14,ArguelloLuengo19}, condensed matter~\cite{Jotzu14,Yan20,Muniz20,Tang20}, or even quantum gravity~\cite{Keren19}. The universal simulator described by Lloyd was later called digital quantum simulator, and it is different from analog~\cite{Britton12,Parsons16,Zhang17} or digital-analog~\cite{Arrazola16,Lamata18} quantum simulators which are non-universal platform-dependent models of computation designed to mimic the behaviour of specific quantum systems. These, however, are better adapted to NISQ systems, and have already been used to realise physical predictions beyond the reach of classical methods~\cite{Trotzky12}. Quantum computers and simulators are better than classical computers at the task of simulating nature~\cite{Bernstein97}, and, thus, are expected to become an important tool for scientific discovery~\cite{Alexeev19}.

The same extraordinary sensitivity to external agents that makes the construction of a quantum computer challenging, can be exploited to build quantum sensors. Quantum sensing is the use of quantum objects or quantum coherence to measure a physical quantity~\cite{Degen17}. It also refers to the use of entanglement to improve the precision of a measurement beyond classical limits~\cite{Giovannetti11}. Examples of quantum sensors include atomic clocks~\cite{Brewer19}, which serve as the best time and frequency standards, superconducting quantum interference devices and the thermal vapour of atoms, which make the most precise magnetometers~\cite{Dang10}, nitrogen-vacancy (NV) centers in diamond as magnetic sensors at the nanoscale~\cite{Maletinsky12,Rondin12}, trapped ions as electric-field and force sensors~\cite{Maiwald09,Biercuk10}, or squeezed states of light for gravitational wave detection~\cite{Abadie11}. Among quantum technologies, quantum sensors have the greatest potential for practical applications in the near term. Quantum correlations among photons can be exploited to achieve target detection in unfavourable scenarios with bright background noise and a low-reflectivity target. Extending this sensing scheme, called quantum illumination~\cite{Tan08,Lloyd08,Zhang13}, to the MW regime is believed to be the way forward in the construction of the first quantum radar~\cite{Barzanjeh15}. Another example of practical applications is given by atomic clocks. Clocks in relative motion or at different gravitational potentials experience time differently. Quantifying this time difference can be then useful to determine the structure of massive objects in geophysics and hydrology~\cite{Chou10,McGrew18}. Furthermore, quantum sensing may also play a crucial role in scientific discovery. For example, creating quantum superpositions with massive particles~\cite{Bose99,Chang10,RomeroIsart11b,Scala13,Pedernales20,Hall98,Fein19} is useful to test collapse models, which state that the Schr\"odinger equation is an approximation that breaks down with large enough masses delocalised above a critical distance~\cite{RomeroIsart11a,Bassi13}, or investigate the quantum nature of gravity, for instance, observing gravity-mediated entanglement~\cite{Bose17,Marletto17}.

Quantum science and technology is a burgeoning research field with promising applications that will have a tremendous impact in society. Light-matter interactions are at the heart of quantum platforms, and are essential to exploit the quantum behaviour of these systems. As an example, MW radiation or laser fields can be applied to quantum systems based on trapped atoms for the sake of high-fidelity quantum information processing. Existing models of light-matter interaction provide us with the theoretical framework needed to explore new forms of manipulating quantum states, and, helped by numerical simulations, search for methods that achieve, e.g., energy-efficient quantum sensing or robust quantum logic operations. Conversely, controlled quantum systems can be used to study models of light-matter interaction without conventional approximations, in regimes where classical numerical methods breakdown. In summary, this thesis focuses on the design of electromagnetic radiation patterns capable of tailoring light-matter interactions in quantum systems to achieve fast and robust  quantum information processing, energy-efficient quantum sensing or the simulation of quantum models whose dynamics can be engineered.


\section{What you will find in this thesis}

In this thesis, we design new forms to control light-matter interactions for specific applications in quantum computing, quantum simulation and quantum sensing with trapped ions, ultracold atoms in optical lattices, and NV centers. For that, \textbf{ in chapter~\ref{chapter:chapter_0}} we first review central models of light-matter interaction, namely the semiclassical and quantum Rabi models. We also explain basic concepts of dynamical decoupling (DD), such as the spin echo or the dressed state approach. Finally, we present the quantum platforms studied in the thesis, and review their importance in the current ecosystem of quantum technologies.

\textbf{ In chapter~\ref{chapter:chapter_1}}, we propose two different methods to realise high-fidelity entangling gates with trapped ions using MW fields. The first uses pulsed MW radiation to generate fast gates with experimental parameters achievable in the near term. The second method applies to current experimental regimes and uses continuous radiation patterns to drive the gate. As opposed to lasers, MW sources are easy to control and can be incorporated to scalable trap designs, which make MW-driven trapped ions a leading approach to scale up trapped-ion quantum processors. Our presented gate-designs account for the main sources of decoherence present in those processors, and, using pulsed and continuous DD techniques, we show how to minimise their effect reaching fidelities above $99.9\%$ in realistic experimental scenarios. 

\textbf{ In chapter~\ref{chapter:chapter_2}}, we explore two models of light-matter interaction and study their dynamics in different parameter regimes. In the case of the Rabi-Stark model, we find selective multiphoton interactions in the SC and USC regimes. Using time-dependent perturbation theory, we develop an analytical framework to explain these interactions. In the case of the nonlinear quantum Rabi model, we find its dynamical behaviour is limited to certain regions in the Fock space, diving the latter into different sections. Combining this property with dissipation, we design a method to generate large-$n$ Fock states with trapped ions. We also provide methods to simulate both the Rabi-Stark and the nonlinear quantum Rabi models with a laser-driven trapped ion.  

\textbf{ In chapter~\ref{chapter:chapter_3}}, we introduce a method to realise boson sampling with ultracold atoms in optical lattices. The control of such system is achieved using both MW and laser fields, and combining pulsed and continuous radiation patterns. Boson sampling is a model of quantum computation with potential to demonstrate quantum supremacy in the near future. Using simple error-scaling models, we estimate how the experimental errors should scale with the number of bosons (atoms, in this case) in order to show advantage with respect to the best classical algorithms. We benchmark our error model with exact numerical simulations of non-Hermitian Hamiltonian models that include particle loss.

\textbf{ In chapter~\ref{chapter:chapter_4}}, we present a design of amplitude modulated MW pulses able to achieve selective NV-nuclei interactions at strong magnetic fields. Working with strong magnetic fields can provide an enhancement on the resolution of NMR spectra, however, if the MW power is not accordingly increased, it leads to a decay of the NMR signal. On the other hand, working with a high MW power may not be convenient, especially when dealing with biological samples that could be damaged when exposed to a large amount of radiation. Our presented method circumvents these issues by using amplitude modulated pulses that enhances the resolution of the measured signals.

\section{Models for light-matter interaction}\label{sec:Intro}

\subsection{The Rabi model}\label{subsec:RabiModel}

\begin{figure}[t!]
\centering
\includegraphics[width=1\textwidth]{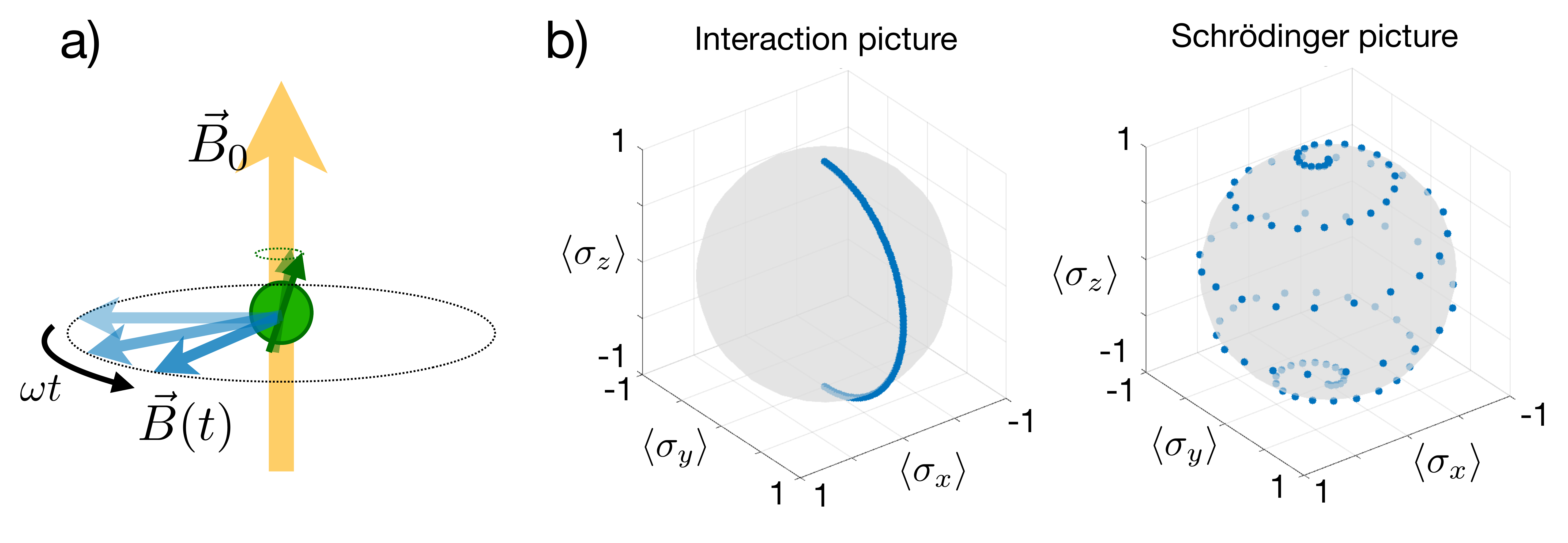}
\caption{a) A particle with spin precessing under a static magnetic field $\vec{B}_0$ and an oscillating transverse field $\vec{B}(t)$. b) A $\pi$ pulse in the Bloch sphere. Evolution of initial state $|\!\!\downarrow\rangle$, according to the resonant Rabi model and during a time $\pi/\Omega$, is shown, both in the interaction and Schr\"odinger pictures, with $\Omega=\omega_0/10$.}\label{fig:IntSch}
\end{figure}

The Rabi model or the semiclassical Rabi model was introduced by Isaac Rabi in 1937 to describe the effect of oscillating magnetic fields onto atoms with nuclear spin~\cite{Rabi37}. Despite being originally introduced in the context of NMR, the Rabi model is also the most simple model that describes the interaction between a two-level atom and a classical electromagnetic field. Think of a spin-$1/2$ particle under the influence of a static magnetic field $\vec{B}_0=B_0\hat{z}$, where $\hat{z}$ is the unit vector in the $z$ direction. The energy of such a particle would be described by the following Hamiltonian 
\begin{equation}\label{Rabi1}
H_0=-\vec{\mu}\cdot \vec{B}_0=\frac{\hbar\omega_0}{2}\sigma_z,
\end{equation}
where $\vec{\mu}=-\gamma\frac{\hbar}{2} \vec{\sigma}$, $\gamma$ being the particle's gyromagnetic ratio, $\hbar$ the reduced Planck constant, and $\sigma_{x,y,z}$ the Pauli matrices. In NMR, $\omega_0=-\gamma B_0$ is known as the Larmor frequency, and, according to Eq.~(\ref{Rabi1}) any spin state will rotate around the $z$ axis with a period $2\pi/\omega_0$. When a orthogonal gyrating magnetic field $\vec{B}(t)=B[\cos{(\omega t)} \hat{x}+\sin{(\omega t)}\hat{y}]$ is applied, we obtain the Rabi model
\begin{equation}\label{Rabi2}
H_\textrm{ R}=\frac{\omega_0}{2}\sigma_z + \frac{\Omega}{2}(\sigma_+e^{-i\omega t}+ \sigma_-e^{i\omega t}),
\end{equation}
where $\sigma_\pm=(\sigma_x\pm i\sigma_y)/2$ and $\Omega=\gamma B/2$. In Eq.~(\ref{Rabi2}) and from now on, all Hamiltonians will be redefined as $H \rightarrow H/\hbar$, thus, will be given in units of angular frequency. To solve the dynamics given by Eq.~(\ref{Rabi2}) we move to an interaction picture~\cite{Sakurai94} with respect to $H=\frac{\omega}{2}\sigma_z$, $H_\textrm{ R}^I=e^{iHt}H_\textrm{ R}e^{-iHt}-H$, obtaining 
\begin{equation}\label{Rabi3}
H_\textrm{ R}^I=\frac{\Delta}{2}\sigma_z + \frac{\Omega}{2}\sigma_x,
\end{equation}
where $\Delta=\omega_0-\omega$ is the detuning with respect to the Larmor frequency. At resonance, Eq.~(\ref{Rabi3}) describes the rotation of the spin around the $\hat{x}$ axis at a frequency $\Omega$, called the Rabi frequency. In Fig.~\ref{fig:IntSch}(b), the evolution of state $|\!\!\downarrow\rangle$ ($\sigma_z|\!\!\downarrow\rangle=-|\!\!\downarrow\rangle$) is shown, in terms of the average values of $\sigma_{x,y,z}$\footnote{Quantum mechanics is a probabilistic theory and it predicts average values of observables. The measurement of these average values requires several experimental runs to collect statistical data.} for a time $\pi/\Omega$. This operation, where the population of state $|\!\!\downarrow\rangle$ is transferred to state $|\!\!\uparrow\rangle$, is known as a $\pi$ pulse.

The Rabi model describes the evolution of a two-level system under an oscillating magnetic field, however, it does not account for environmental effects that cause decoherence. In 1946, Felix Bloch introduced phenomenological equations that described the dynamics of the spin including those effects~\cite{Bloch46}, characterised by two coherence times $T_1$ and $T_2$. The former, called relaxation time, is related with population decay from the $|\!\!\uparrow\rangle$ to the $|\!\!\downarrow\rangle$ state, while the latter represents the unpredictability (after a time $T_2$) of the relative phase between $|\!\!\downarrow\rangle$ and $|\!\!\uparrow\rangle$ states. Just as the Rabi model, Bloch equations are valid to describe the evolution of a two-level atom driven by a coherent electromagnetic field and affected by decoherence, which in that case, take the name of Maxwell-Bloch or optical Bloch equations~\cite{Arecchi65}.

\subsubsection{The rotating wave approximation}

Instead of the gyrating field used in Eq.~(\ref{Rabi2}), it is more typical to consider a time-varying field of the form $\vec{B}(t)=B\cos(\omega t)\hat{x}$. In such case, Eq.~(\ref{Rabi3}) reads
\begin{equation}\label{Rabi4}
H_\textrm{ R}^I=\frac{\Delta}{2}\sigma_z + \frac{\Omega}{2}\sigma_x +\frac{\Omega}{2}(\sigma_+e^{-i2\omega t} +\sigma_-e^{i2\omega t}).
\end{equation}
If the intensity of the transverse field is small compared to the static magnetic field, then $\omega\approx\omega_0\gg\Omega$, and the last terms in Eq.~(\ref{Rabi4}), called counter-rotating terms, can be neglected under the rotating-wave approximation (RWA). A measurable effect of those counter-rotating terms is a shift of the Larmor frequency by an amount $\Omega^2/4\omega_0$, known as the Bloch-Siegert shift\footnote{This shift was first measured by Felix Bloch and Arnold J. F. Siegert in 1940~\cite{Bloch40}.}. To derive this, one has to use time-dependent perturbation theory, see for example, Eqs.~(\ref{Magnus}) and (\ref{MagnusTerms}) in appendix~\ref{app:TimeEvol}.

\subsubsection{Dressed states}
We call dressed states to electronic or nuclear-spin states whose energy has been shifted because of the interaction with the radiation field. A typical example is the so-called light shift~\cite{Foot07}, obtained when the absolute value of the detuning $|\Delta|$ is much larger than the Rabi frequency $\Omega$, yet $|\Delta|\ll\omega_0$. The effective Hamiltonian in this regime is
\begin{equation}\label{HamilLight}
H^I_\textrm{ light}\approx\frac{\Omega^2}{4\Delta}\sigma_z,
\end{equation}
which describes a change of the energy difference between states $|\!\!\downarrow\rangle$ and $|\!\!\uparrow\rangle$, positive or negative depending on the sign of the detuning $\Delta$. When the Rabi frequency $\Omega$ (which is proportional to the intensity of the radiation field) changes with the position $\vec{x}$, the shift can induce a force in the particle.

Another kind of dressed state is also described by Eq.~(\ref{Rabi3}). When $\Delta=0$, an energy difference of $\Omega$ between states $|\pm\rangle=|\!\!\uparrow\rangle\pm |\!\!\downarrow\rangle$ (up to normalisation and in the interaction picture) is produced. The introduction of a weaker field $\vec{B}(t)=~\tilde{B}(\cos{(\tilde{\omega} t)} \hat{x}+\sin{(\tilde\omega t)}\hat{y}$, represented by
\begin{equation}\label{Rabi2}
H^I_\textrm{ R}=\frac{\Omega}{2}\sigma_x + \frac{\tilde{\Omega}}{2}(\sigma_+e^{i\tilde{\Delta} t}+ \sigma_-e^{-i\tilde{\Delta} t}),
\end{equation}
where $\tilde{\Omega}=\gamma \tilde{B}/2$ and $\tilde{\Delta}=\omega_0-\tilde{\omega}$, can lead to transitions between states $|+\rangle$ and $|-\rangle$ when $\tilde{\Delta}=\pm\Omega$. In atomic physics, these two transitions at $\omega\pm\Omega$ and the central transition at $\omega_0$ can be observed and are known as the Mollow triplet~\cite{Mollow69,Schuda74}. Both the light-shift and the Mollow triplet are often referred in the literature as the alternating current (AC) Stark effect~\cite{Foot07} or Autler-Townes effect~\cite{Autler55}. 
States dressed with resonant driving fields are naturally protected from small shifts in $\omega_0$~\cite{Puebla16} and form the basis of continuous DD techniques~\cite{Bermudez12,Lemmer13}.

\subsubsection{Spin echo}
The spin echo or Hahn echo\footnote{The spin echo was first observed by Erwin L. Hahn in 1950~\cite{Hahn50}.} is a method by which a revival of the spin's coherence occurs after the application of a $\pi$ pulse. It is also the most common method to combat dephasing caused by uncontrolled shifts on $\omega_0$ due to fluctuations of the intensity of the magnetic field. For a simple analysis, let us assume we start with the spin in state $|\!\!\uparrow\rangle+|\!\!\downarrow\rangle$. According to Eq.~(\ref{Rabi1}) the state should evolve as $e^{i\omega_0t}|\!\!\uparrow\rangle+e^{-i\omega_0t}|\!\!\downarrow\rangle$, however, in a rotating frame with respect to $\frac{\omega_0}{2}\sigma_z$, the state should stay as $|\!\!\uparrow\rangle+|\!\!\downarrow\rangle$. If the Larmor frequency undergoes an unknown shift $\delta$, at a time $t$, the state would be $e^{i\delta t}|\!\!\uparrow\rangle+e^{-i\delta t}|\!\!\downarrow\rangle$. This would imply the loss of information about the relative phase between the two states, and therefore, a coherence loss\footnote{More precisely, the loss of coherence would occur if $\delta$ changes stochastically with each experimental run.}. The application of a $\pi$ pulse produces a population exchange between both states, transforming the state into $e^{-i\delta t}|\!\!\uparrow\rangle+e^{i\delta t}|\!\!\downarrow\rangle$. After a time $t$, the state would be $e^{i\delta t}e^{-i\delta t}|\!\!\uparrow\rangle+e^{-i\delta t}e^{i\delta t}|\!\!\downarrow\rangle=|\!\!\uparrow\rangle+|\!\!\downarrow\rangle$, and the relative phase produced by the unknown shift disappears, recovering coherence. It is important to remark that spin echo is used as the basis of pulsed DD techniques~\cite{Carr54,Meiboom58,Souza12}. 

\subsection{The quantum Rabi model}

\begin{figure}[t!]
\centering
\includegraphics[width=0.8\textwidth]{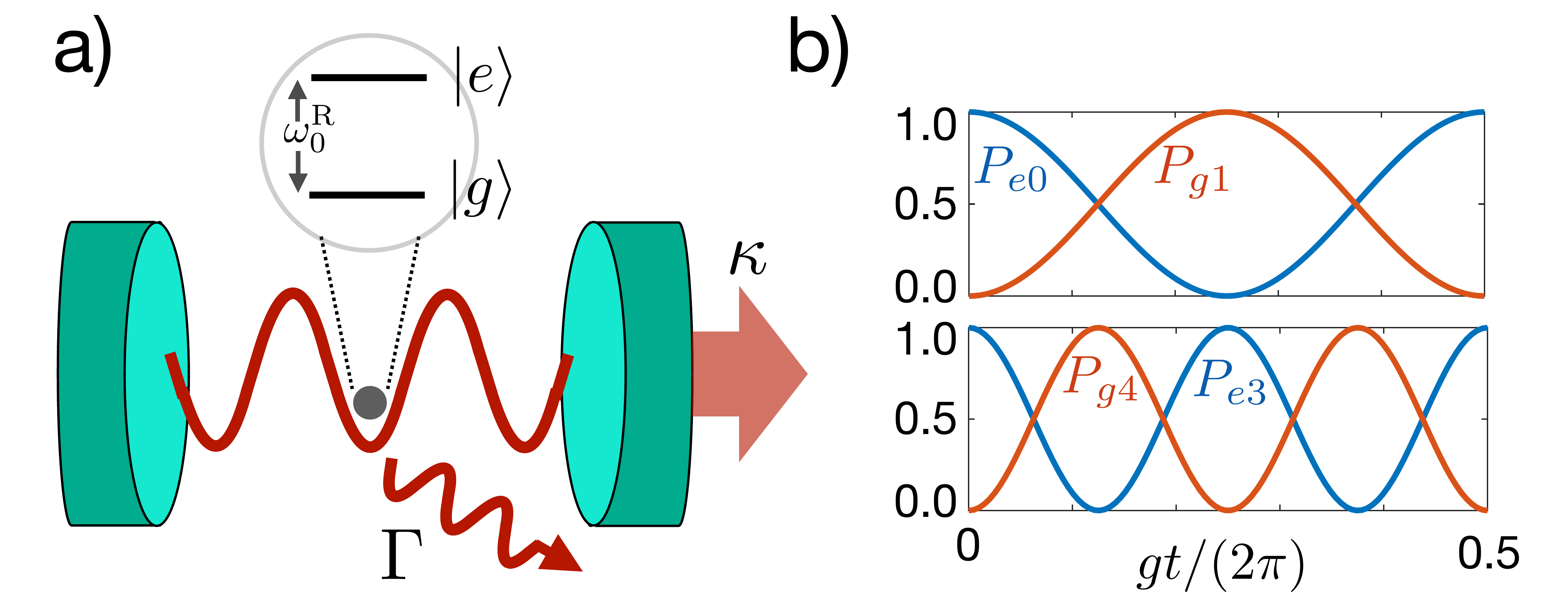}
\caption{a) Sketch of an atom inside a cavity, where $\Gamma$ and $\kappa$ are the relaxation rates of the atom and the cavity, respectively. b) Time evolution of state populations for $g/\omega^\textrm{ R}=0.01$, $\omega_0^\textrm{ R}=\omega^\textrm{ R}$ according to the QRM. For initial states $|e,0\rangle$ (up) and $|e,3\rangle$ (down), periodic population exchange is shown, with states $|g,1\rangle$ and $|g,4\rangle$ and at rates $g$ and $2g$ respectively.}\label{fig:QRM}
\end{figure}

The quantum Rabi model (QRM) describes the interaction between a two-level atom and a quantised mode of the electromagnetic field. The model was first introduced by Jaynes and Cummings in 1963~\cite{Jaynes63}, however, its predictions were not measured until the late 80's in the context of cavity QED~\cite{Rempe87,Haroche89,Raimond01}. Besides, the QRM describes the basic interactions that take place in trapped ions~\cite{Leibfried03}, superconducting circuits~\cite{Clarke08} or semiconducting quantum dots~\cite{Bera10}. The Hamiltonian of the model is 
\begin{equation}\label{QRM}
H_\textrm{ QRM}=\frac{\omega^\textrm{ R}_0}{2}\sigma_z+\omega^\textrm{ R} a^\dagger a + g (a+a^\dagger)(\sigma_++\sigma_-)
\end{equation}
where $\omega_0^\textrm{ R}$ is the natural frequency of the two-level atom, $\omega^\textrm{ R}$ and $a^\dagger (a)$ are the frequency and the creation (annihilation) operators of the electromagnetic mode, and $g$ measures how strong is the coupling between the atom's dipole moment and the bosonic field mode. See Fig.~\ref{fig:QRM} for a simple sketch of the system. Near resonance, that is, when $\omega^\textrm{ R}\approx\omega_0^\textrm{ R}$, and if $g\ll\omega^\textrm{ R}$, the terms $a\sigma_-$ and $a^\dagger\sigma_+$ can be eliminated by the RWA, retrieving the so-called Jaynes-Cummings (JC) model
\begin{equation}\label{JCM}
H_\textrm{ JC}=\frac{\omega_0^\textrm{ R}}{2}\sigma_z+\omega^\textrm{ R} a^\dagger a + g (a\sigma_++a^\dagger\sigma_-).
\end{equation}
The JC model is analytically solvable and, when $\omega^\textrm{ R}=\omega_0^\textrm{ R}$, it describes a periodic population exchange between states $|e,n\rangle$ and $|g,n+1\rangle$ at a rate given by $\Omega=g\sqrt{n+1}$~\cite{Jaynes63,Paul63}, as shown in Fig.~\ref{fig:QRM}(b). Here, $|g\rangle$ and $|e\rangle$ represent the ground and excited state of the atom, and $|n\rangle$ represents the $n$-th Fock state of the bosonic mode. Notice that the frequency of these oscillations $\Omega$, also called Rabi oscillations, depends on the number of photons $n$ in the cavity. At $n=0$, an excited atom may emit a photon to an empty cavity mode, and absorb it afterwards\footnote{These oscillations are known as vacuum Rabi oscillations~\cite{Loudon00}.}. The JC model also predicts the collapses and revivals of atomic state populations when the field mode is in a coherent state\footnote{These collapses and revivals were first observed by Gerhard Rempe, Herbert Walther and Norbert Klein in 1987~\cite{Rempe87}.}~\cite{Gerry04}. 

The experimental observation of Rabi oscillations requires $\Omega$ to be larger than the relaxation rates ($1/T_1$) of both the atom and the cavity, known as the SC regime~\cite{Haroche85,Gallas85}. In contrast to the weak-coupling regime, where substantial changes in the atomic properties (e.g. the Purcell effect) can already be observed, the SC regime allows the formation of hybrid dressed states, different from the ones introduced in section~\ref{subsec:RabiModel}, called polaritonic states or simply polaritons, which share both light and matter character~\cite{Vasa19}. In the case of the resonant JC model, these are represented by $|g,n+1\rangle\pm|e,n\rangle$ eigenstates, up to normalisation.

Based on previous work by Dicke~\cite{Dicke54}, Tavis and Cummings studied the interaction of $N$ atoms with a single mode of the electromagnetic field, predicting that the frequency of (now collective) Rabi oscillations can increase by a factor of $\sqrt{N}$~\cite{Tavis68}. In fact, the first cavity QED experiments that achieved the SC regime were done with more than one atom~\cite{Raizen89,Zhu90,Bernardot92}. Since then, the effective coupling strength $g$ of light-matter interactions has progressively increased, reaching the ultrastrong coupling (USC)~\cite{Gunter09,Niemczyk10,Rossatto17,Kockum19,Forn19} regime ($g/\omega^\textrm{ R}\gtrsim0.1$) or 
the deep-strong coupling (DSC)~\cite{Yoshihara17,Bayer17,Mueller20} regime ($g/\omega^\textrm{ R}\gtrsim1$) with superconducting circuits, Landau polaritons or plasmon polaritons. These experimental developments have motivated the study of the full QRM in Eq.~(\ref{QRM})~\cite{Braak16}, as the RWA is not longer justified. Despite being the most simple representation of quantum light-matter interaction, an exact analytical solution for the QRM was only proposed recently, in 2011~\cite{Braak11}. Unlike the JC model, the QRM dynamics does not show clear features until it reaches the DSC regime, where periodic collapses and revivals of the qubit initial state survival probability are predicted~\cite{Casanova10}.

\section{Quantum technologies}

Quantum technologies that aim to achieve quantum computing or quantum sensing need a well defined qubit or quantum information register, and the ability to initialise and measure its state. In the case of quantum sensing, this qubit should also interact with the physical quantity of interest. For quantum computing, the design of the system should be scalable to a large number of qubits, while maintaining long coherence times and the ability to make single and two-qubit gates. In the following, we describe three of the most promising quantum platforms to develop both quantum sensing and quantum computing, which are also the ones studied in this thesis. 

\subsubsection{Trapped ions}
\begin{figure}[t!]
\centering
\includegraphics[width=1\textwidth]{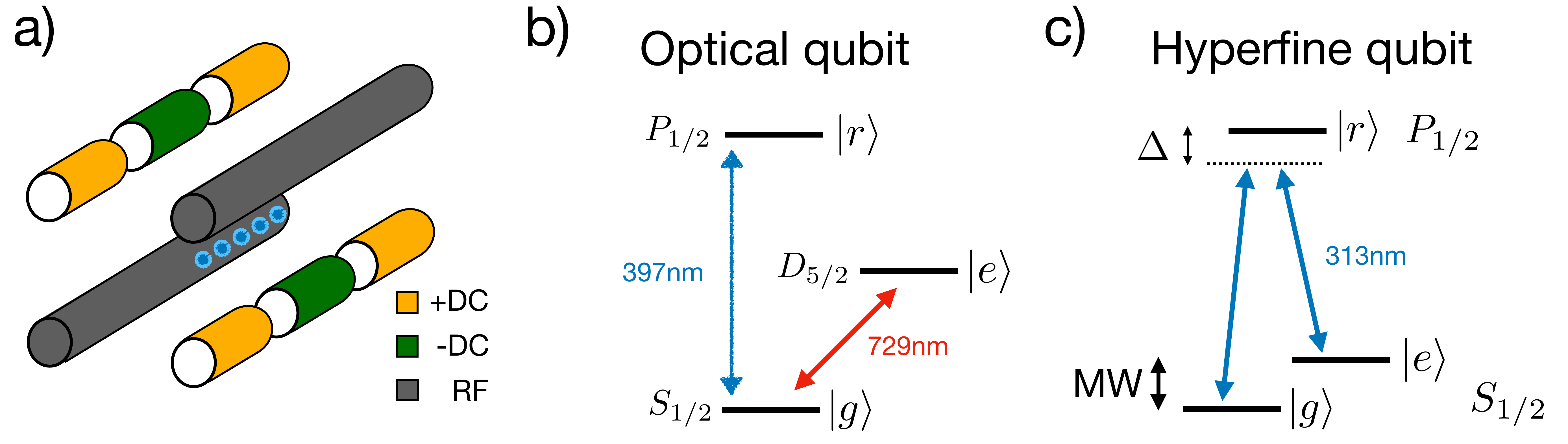}
\caption{a) Sketch of a linear radiofrequency trap similar to the one shown in Ref.~\cite{Leibfried03}. In the middle, a chain of 5 ions scattering blue light. b) Simplified level scheme of an optical qubit using a $^{40}$Ca$^+$ ion~\cite{Nagerl00}. The metastable transition $S\leftrightarrow D$ at $729$ nm it is used as a qubit with lifetime $T_1\sim1$ s. The radiative transition $S\leftrightarrow P$ at $397$ nm is used for read-out. c) Simplified level scheme of a hyperfine qubit using a $^9$Be$^+$ ion~\cite{Monroe95}. Off resonant excitation of the radiative $S\leftrightarrow P$ transition at $313$ nm is used to produce coherent population exchange between two hyperfine levels of the $S$ subspace.}\label{fig:IntroIons}
\end{figure}

Charged atomic ions are trapped and suspended in vacuum using oscillating electromagnetic fields~\cite{Brown86,Leibfried03}. The effective harmonic potential created by these fields confines the ion in all directions, with trapping frequencies on the order of few MHz. In linear traps, the ions arrange in a linear configuration in the $z$ direction, where the harmonic force is weaker than in the other directions (see Fig.~\ref{fig:IntroIons}(a) for an illustration of a linear trap). Two electronic states connected by an electric quadrupole transition, or two hyperfine states connected by Raman or MW transitions can serve as a qubit, see Figs.~\ref{fig:IntroIons}(a)~and~(b). In both cases, an electric dipole transition to a radiative state is used for cooling the ion's motion\footnote{Ground-state cooling with trapping frequencies of the order of one megahertz correspond to a temperature of tens of $\mu$K.}, initialisation and read-out of the qubit state~\cite{Wineland98}. For one of the qubit states, driving the radiative transition will induce spontaneous emission of photons that can be collected by a camera, realising the measurement of the qubit state.

Coherent operations are typically realised with lasers, these allow to manipulate the qubit state and, moreover, to entangle internal (electronic) and external (vibrational) degrees of freedom by the so-called sideband transitions. It is the control of this qubit-boson interaction which allows to realise entangling gates among different ions in few microseconds~\cite{Schafer18} and with the highest fidelity achieved so far in quantum technologies~\cite{Ballance16,Gaebler16}. Coherence times can range from milliseconds to minutes using decoherence-free subspaces and DD~\cite{Wang20}. A dispersive coupling with the collective motional modes also permits to engineer spin-spin interactions governed by the effective Hamiltonian $H=\sum_{i,j}J_{ij}\sigma_i^x\sigma_j^x$~\cite{Porras04,Schneider12,Bohnet16}, which is useful to study many-body quantum phenomena like quantum phase transitions~\cite{Zhang17,Jurcevic17,Cui20} or quantum chaos~\cite{Grass13,Sieberer19}. Moreover, trapped-ions have also proven to be an excellent testbed to simulate relativistic quantum mechanics~\cite{Lamata08,Gerritsma10,Wittemer19} or quantum field theories~\cite{Martinez16,Zhang18}. 

In this thesis, we study trapped ions as quantum simulators for generalised QRMs beyond the SC regime, and also as a scalable platform to build quantum computers. In the case of the latter, we combine laser-free entangling operations driven by MW radiation with DD techniques to achieve high-fidelity entangling gates. Finally, trapped ions can also be excellent quantum sensors of electric fields. This can be exploited to measure minute forces with a sensitivity of $1~ \textrm{ yN}/\sqrt{\textrm{Hz}}$~\cite{Maiwald09,Biercuk10}, or for high-precision mass spectrometry\footnote{Check articles 8, 9, 10, and 12 from the list of publications}.

\subsubsection{Ultracold atoms in optical lattices}
\begin{figure}[t!]
\centering
\includegraphics[width=1\textwidth]{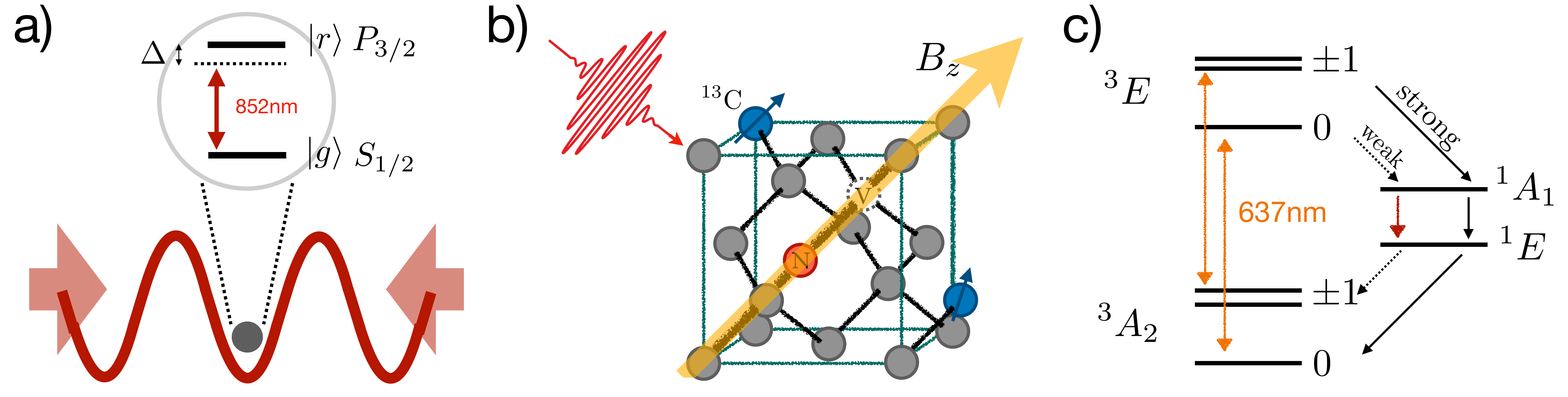}
\caption{a) A neutral particle trapped in one minimum of the optical potential $V(x)$ (in red). Also, a simplified level scheme of the $^{133}$Cs atom, where the off resonant excitation of the $S\leftrightarrow P$ transition by the standing wave is the responsible of the dipole force. b) NV center in a diamond lattice, with $^{12}$C (grey) and $^{13}$C (blue) atoms nearby. A static magnetic field is applied parallel to the NV axis, and a transverse MW field is depicted in red. c) Level scheme for the NV center, where coloured (black) arrows indicate radiative (nonradiative) transitions.}\label{fig:IntroNV}
\end{figure}

Neutral atoms can be trapped in periodic optical potentials, called optical lattices, created by the interference of two counter-propagating laser beams. The frequency of the laser fields is far detuned with respect to the frequency of an electric dipole transition of the atom. As a result, a light shift, similar to the one presented in section~\ref{subsec:RabiModel}, is induced with the Rabi frequency proportional to the intensity of the laser field. The interference between the two beams creates a periodic spatial pattern for the intensity ($\Omega\cos{(kx)}$ in one dimension), which induces a force on the atom, called optical dipole force. This force is used to trap atoms in minima of the light potential $V(x)=\Omega^2/\Delta\cos^2{(kx)}$ (see Fig.~\ref{fig:IntroNV}(a) for an illustration), while the photon scattering rate is highly reduced with a large detuning $\Delta$. Atoms trapped in optical potentials need to be cooled down to temperatures below tens of $\mu$K, achievable with laser cooling techniques\footnote{In 1997, Steven Chu, Claude Cohen-Tannoudji and William D. Phillips received the Nobel prize for their developments of methods to cool and trap atoms using laser light.}. Using different hyperfine states of the atoms, these can be initialised in predefined arrays, and similar to trapped ions, the measurement is done via resonance fluorescence. 

Ultracold atoms in optical lattices are the leading technology for analog quantum simulation of many-body physics~\cite{BlochDalibard12,Gross17}. The periodic form of the optical potentials resemble structures of real crystal lattices, and the interaction among the atoms can be controlled with light fields. Important many-body systems, such as Bose-Hubbard or Fermi-Hubbard models can be simulated~\cite{Greiner02}, mechanisms such as thermalisation~\cite{Langen13,Langen15,Kaufman16} or many-body localisation~\cite{Anderson58,Choi16} observed, or fundamental theories for high-energy and condensed-matter physics studied via synthetic gauge fields~\cite{Montvay94}.

Quantum computation can also be pursued with optical lattices. Qubits can be encoded in long-lived hyperfine states and entangling gates can be realised via the strong dipole-dipole interactions among Rydberg atoms. However, and despite the significant progress, several challenges such as the generation of a high-fidelity two-qubit gate remain unsolved~\cite{Saffman16}.

In this thesis, we base on previous experiments of quantum interference of non-interacting atoms with optical lattices, such as the realisation of Hong-Ou-Mandel interference~\cite{RobensThesis}, to propose a scalable method to realise multiparticle quantum interference experiments, and more specifically, boson sampling.

\subsubsection{Nitrogen-vacancy centers in diamond}

The NV center is a point defect in diamond with a particular set of properties that makes it suitable for the study and exploitation of quantum phenomena~\cite{Walker79}. The defect is formed when a carbon atom in the diamond lattice is substituted by a nitrogen and an adjacent lattice site presents a vacancy~\cite{Aharonovich11,Suter17}, see Fig.~\ref{fig:IntroNV}(b), and can be produced by several techniques~\cite{Doherty13}. The NV$^-$ or NV center's\footnote{Other two charge states exist, NV$^+$ and NV$^0$, whose properties differ from those of the negatively charged defect.} electronic level structure is formed by two triplet states $^3A_2$ (ground) and $^3E$ (excited) and two intermediate singlet states $^1A_1$ and $^1E$, see Fig.~\ref{fig:IntroNV}(c). A qubit is usually encoded in the $m_s=0$ and one of $m_s=\pm1$ states of the subspace $^3A_2$, separated in energy by $D=(2\pi)\times2.88$ GHz, called the zero-field splitting. Also, a magnetic field in $B_z$ the direction of the NV breaks the degeneracy between states $m_s=\pm1$ by an amount $\gamma_eB_z$. Operations within this subspace are carried out using MW radiation, while optical transitions are used for initialisation and measurement of the NV. When the electron is excited to the $^3E$ subspace through a spin-conserving radiative transition, the excited states can decay to the ground state directly or via the intermediate states. The latter occurs with a higher probability for states $|^3E,\pm1\rangle$ and the intermediate states are long-lived ($\sim250$ ns) compared to the states in $^3E$. This results in a spin-dependent fluorescence signal, which is stronger for the $|^3A_2,0\rangle$ state. Most importantly, the decay via the intermediate states changes the spin state from $m_s=\pm1$ to $m_s=0$ and vice versa, and, because of the preference of one of the pathways, this can be used for the initialisation or polarisation of the spin state.

The coherent control of the NV center is possible at room temperature, with coherence times on the order of milliseconds~\cite{Degen17} and electron spin polarisation above $90\%$~\cite{Jelezko06}. Because of this, NV centers have been proposed as quantum sensors for high-resolution scanning probe microscopy~\cite{Chernobrod05,Balasubramanian08,Degen08}, magnetometry~\cite{Maze08,Taylor08,Cole09}, thermometry~\cite{Hodges13,Kucsko13,Neumann13,Toyli13}, pressure sensing~\cite{Doherty14} or to be used as biomarkers with living organisms~\cite{Fu07}. In this thesis, we focus on the possibility of realising nanoscale NMR with NV centers. This enables the control and measurement of magnetic field emitters (as nuclear spins) with sub-$100$nm resolution, with applications for imaging of nanometric magnetic structures~\cite{Balasubramanian08,Maletinsky12,Rondin12} or optical polarisation of magnetic nuclei for high-resolution magnetic resonance imaging~\cite{Schwartz18}.

NV centers also have interesting applications for quantum information processing. At low temperatures, the fidelity of initialisation and read-out increases significantly~\cite{Robledo11}, and $T_1$ coherence times approaching to $10^3$ s have been reported~\cite{Abobeih18} at $\approx 3.7$~K. Coherence times due to dephasing $T_2$ depend mainly on the number of $^{13}$C nuclei near the NV~\cite{Maze08}, and, with a small amount of these impurities, $T_2$ can be increased up to $T_1$ with DD techniques~\cite{Souza12}. Both the NV centers and the nearby $^{13}$C nuclei can be used as quantum information registers~\cite{Neumann10,Liu18}, current experiments being able to achieve high-fidelity gates~\cite{Rong15} and registers up to $10$ qubits~\cite{Bradley19}. Still, the deterministic fabrication of these solid-state devices is challenging as requires a three dimensional precision better than $10$ nm~\cite{Scarabelli16}. Alternative approaches to scalability consider NV centers coupled to superconducting circuits~\cite{Zhu11} or photons~\cite{Togan10,Kalb17}. Using the latter, entanglement between electron spins separated by $1.3$ km has been achieved~\cite{Hensen15}.

\chapter{Quantum Logic with MW-Driven Trapped Ions}
\label{chapter:chapter_1}
\thispagestyle{chapter}

The control and manipulation of the quantum information of individual atoms via electromagnetic fields has been extremely successful with trapped ions, which has placed trapped-ion quantum technology as a leading candidate to build reliable quantum simulators and computers~\cite{Nielsen10,Haffner08,Ladd10,Blatt12,Cirac12}. In the near future, these could solve computational problems in a more efficient manner than classical devices by exploiting the quantum correlations among their atomic constituents. To this end, the systematic generation of single-qubit and two-qubit gates with high-fidelity is crucial. The latter has been achieved by using laser light that couples the internal (atomic) and external (vibrational) degrees of freedom of the ions leading to fast single-qubit, and two-qubit gates of a high fidelity~\cite{Ballance16,Gaebler16,Schafer18}. Nevertheless, to scale laser-based quantum processors while maintaining high fidelities represents a hard technological challenge, since it requires the precise and simultaneous control of multiple laser sources.

An alternative approach to laser-driven systems was proposed by Mintert and Wunderlich~\cite{Mintert01}. This involves the use of MW fields together with magnetic field gradients to create interactions among the internal states of the ions. Unlike lasers, the control of MW sources is comparatively easy, and their introduction in scalable trap designs is less demanding~\cite{Lekitsch17}. In addition, MW driven quantum gates do not use any optical transition. This avoids spontaneous emission of some atomic states that define the qubit, which is an unavoidable limiting factor for laser-driven quantum gates~\cite{Plenio97,Gaebler16,Ballance16}. After the initial proposal in~\cite{Mintert01}, the use of MW schemes has been pursued in two distinct fashions, using either static magnetic field gradients~\cite{Mintert01,Weidt15,Piltz16,Welzel19}, or MW radiation in the near-field~\cite{Ospelkaus08,Ospelkaus11,Hahn19,Zarantonello19}. In the former, a static magnetic field gradient provides spatial field variations on the size of the wave packet of the ion, coupling the atomic and vibrational degrees of freedom. This coupling is combined with MW fields in the far-field regime, which are used to modulate the interaction. In the latter, oscillating MWs in the near-field regime are used as the generators of the qubit-boson interaction. Notice that MW fields in the near field naturally provide the required spatial field variations without the need to add a static magnetic field gradient. In addition, recently a method has been proposed to couple ionic internal and external degrees of freedom, that combines oscillating magnetic field gradients with MW fields on the far-field regime~\cite{Srinivas19,Sutherland19}.

Typically, ion qubits are encoded in hyperfine atomic states which are sensitive to magnetic field fluctuations. These represent the main source of decoherence, and have to be removed to achieve quantum information processing with high fidelity. To this end, pulsed and continuous DD techniques have been introduced~\cite{Jonathan00,Szwer11,Piltz13,Casanova15,Puebla16,Puebla17,Arrazola18,Wang19}. In particular, the
creation of dressed-states has been proved useful~\cite{Timoney11,Bermudez12,Lemmer13,Mikelsons15,Cohen15,Wolk17,Webb18} and has led to the best reported gate fidelities ($>98\%$) with MW fields on the far-field regime~\cite{Weidt16}. On the other hand, the best near-field MW gates, with fidelities of $99.7\%$~\cite{Harty16,Zarantonello19}, use a driving field on the carrier transition or MW amplitude modulation to protect the gate with respect to the main sources of error. 

In this chapter we propose two different methods to achieve high-fidelity two-qubit gates using MW fields on the far-field regime combined with a static magnetic field gradient. In both cases, DD techniques are used to design gates which are robust against fluctuations of the MW and magnetic fields. The method presented in section \ref{sect:1_PDD} is specifically designed to work with a strong qubit-boson coupling and large MW power, leading to gate times of tens of microseconds. On the other hand, the gate scheme in section \ref{sect:Slow_ions} works with a lower qubit-boson coupling and MW power,  making it more accessible experimentally, and achieving gate times on the order of a millisecond.

\section{Pulsed DD for fast and robust two-qubit gates}
\label{sect:1_PDD}

Fast trapped-ion entangling gates require an effectively large qubit-boson coupling, which is usually obtained increasing either the Lamb-Dicke (LD) parameter, i.e. the original qubit-boson coupling, or the intensity of the field driving the qubit. Experiments using far-field MW with static magnetic field gradients work with a LD parameter of $\eta \sim0.01$ and less than a hundred kilohertz of MW Rabi frequency. This leads to gate times on the order of the millisecond. Reference~\cite{Cohen15}, for example, proposes to use hundreds of kilohertz of MW Rabi frequency, obtaining gate times of a few hundreds of microseconds. In this section we propose to boost the speed of two-qubit gates by increasing the magnetic field gradient and thus the LD parameter to $\eta\sim0.1$. This complicates the application of schemes where the qubit-qubit interaction is mediated by a single motional mode, as the spectroscopic discrimination of the rest is no longer possible. The use of multiple modes has also been explored theoretically~\cite{Duan04} and experimentally~\cite{Mizrahi13} for laser based systems. 

In this section, we propose a scheme leading to fast and high-fidelity two-qubit gates through a specifically designed sequence of MW $\pi$-pulses acting in the presence of a magnetic field gradient. Our method employs the two vibrational modes in the axial direction of the two-ion chain leading to gate times approaching the inverse of the trap frequency. On top of that, the sequence uses pulsed DD to protect qubits from uncontrolled noise sources. The high speed and robustness of our scheme results in two-qubit gates of high fidelity even in presence of motional heating. Our detailed numerical simulations show that state-of-the-art in MW trapped-ion technology allows for two-qubit gates sufficiently fast to pave the way for scalable quantum computers.

\subsection{System: two $^{171}$Yb$^+$ ions}\label{subsect:system}

\begin{figure}[t!]
\centering
\includegraphics[width=1\textwidth]{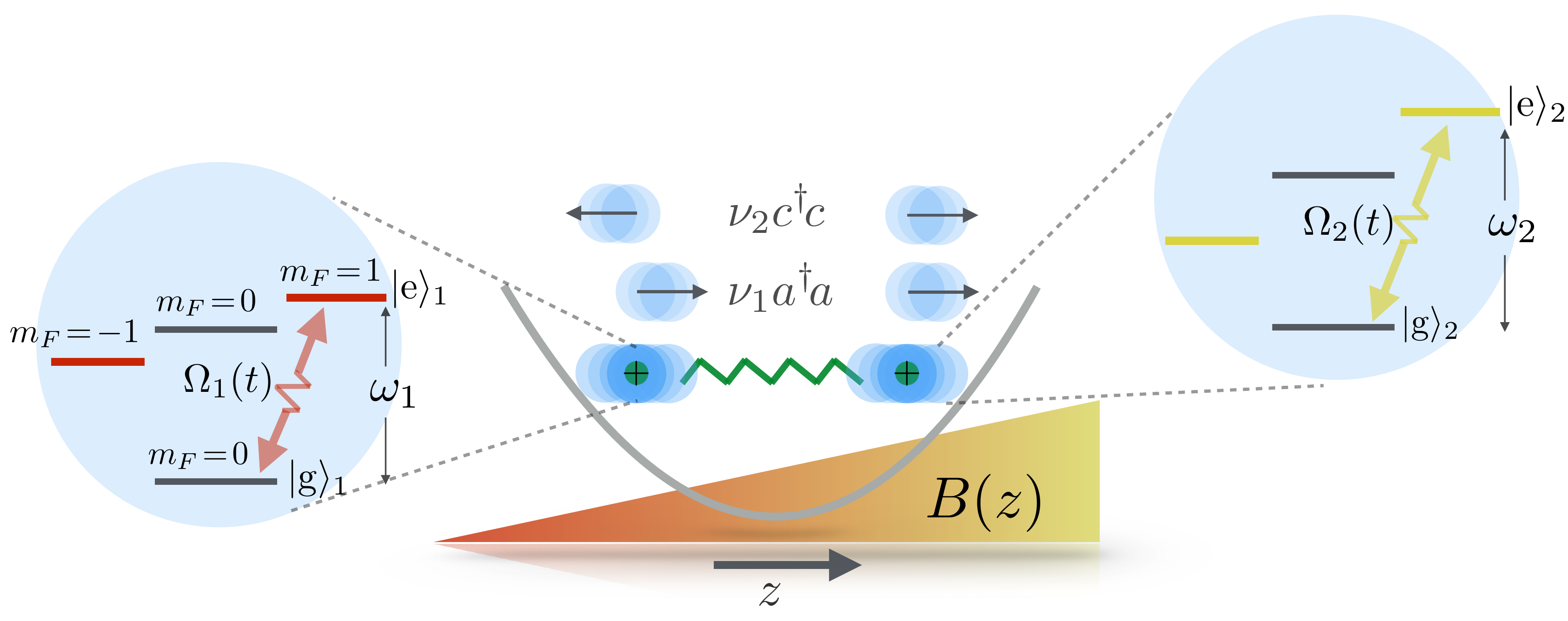}
\caption{Two trapped $^{171}$Yb$^+$ ions under a magnetic field gradient. Hyperfine levels of the two ions are shown. The magnetic field $B(z)$ removes the degeneracy of the $F=1$ manifold separating the $\{F=1, m_F=\pm1\}$ and $\{F=1, m_F=0\}$ levels of both ions by an amount of $\pm\gamma_e B(z_j)/2$ respectively. }\label{fig:Fig1}
\end{figure}

We consider a setup consisting of two $^{171}$Yb$^+$ ions in a MW quantum computer module~\cite{Lekitsch17}.  For each ion, we define our qubit between the lowest energy state ${|{\textrm g}\rangle\equiv\{F=0, m_F=0\}}$ and the first excited state with positive magnetic moment ${| {\textrm e} \rangle\equiv\{F=1, m_F=1\}}$ in the hyperfine manifold, see Fig.~\ref{fig:Fig1}. The conditions under which transitions to other hyperfine levels can be safely neglected are covered in appendix~\ref{app:InitialApp}. The motion of different ions is coupled via direct Coulomb interaction and can be collectively described by two harmonic normal modes in each of the three spatial dimensions. In the following, we will restrict our analysis to the normal modes in the axial direction $\hat{z}$, namely the center-of-mass and the breathing modes, which are independent of the radial ones. That is, we assume that the magnetic field gradient in the radial direction is negligible compared to that in  the axial direction. This configuration is described by the Hamiltonian 
\begin{eqnarray}
\label{Hamiltonianbare}
\nonumber H=   \nu_1 a^\dag a + \nu_2 c^\dag c &+& [\omega_{\textrm e} +\gamma_e B(z_1)/2] | {\textrm e} \rangle \langle {\textrm e} |_1 + \omega_{\textrm g} | {\textrm g} \rangle \langle {\textrm g} |_1 \\
 &+& [\omega_{\textrm e} + \gamma_e B(z_2)/2] | {\textrm e} \rangle \langle {\textrm e} |_2 + \omega_{\textrm g} |{\textrm g} \rangle \langle {\textrm g} |_2.
\end{eqnarray}


Here, $\gamma_e = (2\pi)\times 2.8$ MHz/Gauss is the gyromagnetic ratio of the electron, $a(a^\dag)$ and $c(c^\dag)$ are the bosonic annihilation(creation) operators of the motional modes, which have frequencies ${\nu_1=\nu}$ and ${\nu_2=\sqrt{3}\nu}$, respectively~\cite{James98}. Typical values for $\nu/(2\pi)$ range from hundreds of kilohertz to one or two megahertz, and $B(z_j)$ is the magnetic field at the position of ion $j$.  We consider a magnetic field gradient that leads to a linearly growing  $B(z_j)$ term, $\partial B/\partial z\!=\!g_B$. Then, by expressing the ion coordinates in terms of the vibrational normal modes~\cite{James98} and suitably shifting the zero-point energy of the qubits, the Hamiltonian in Eq.~(\ref{Hamiltonianbare}) can be rewritten as
\begin{eqnarray}
\label{modelHamiltonian}
\nonumber H= \nu_1 b^\dag b + \nu_2 c^\dag c &+& \frac{\omega_1}{2}\sigma_1^z + \eta_1\nu_1 (b+b^\dag) \sigma_1^z -  \eta_2\nu_2(c+c^\dag)\sigma_1^z\\
&+& \frac{\omega_2}{2}\sigma_2^z +  \eta_1\nu_1(b+b^\dag) \sigma_2^z +  \eta_2\nu_2(c+c^\dag)\sigma_2^z.
\end{eqnarray}
Here, the operators of the center-of-mass mode have been redefined as $b=a+2\eta_1$, where $M$ is the mass of each ion and $\eta_m= \frac{\gamma_e g_B}{8\nu_m} \sqrt{\frac{\hbar}{M \nu_m}}$ is the effective LD parameter which quantifies the coupling strength between the qubits and the $m$-th motional mode. The qubit energy splittings are $\omega_j= \omega_{\textrm e} - \omega_{\textrm g} - 4\eta_1^2\nu_1 + \gamma_e B_j/2$, where $B_j\equiv B(z_j^0)$ is the magnetic field on the equilibrium position of the $j$-th ion. For a detailed derivation of Hamiltonian in Eq.~(\ref{modelHamiltonian}), please refer to appendix~\ref{app:twoions}.

In presence of the magnetic field gradient $g_B$, the energy splitting of the two ion-qubits differs by $\omega_2 - \omega_1=\gamma_e g_B \Delta z/2$, where $\Delta z$ is the distance between the equilibrium positions of the ions. Later, we will see that $\omega_2- \omega_1= \frac{\gamma_e g_B}{2} \left( \frac{2e^2}{4\pi \varepsilon_0 M \nu^2}\right)^{1/3}$ is a quantity on the order of tens of megahertz for the parameters considered here. This energy difference combined with a specifically designed MW-pulse sequence to efficiently cancel crosstalk effects (see section~\ref{subsect:Tailored}) will allow us to address individually each qubit. We consider a bichromatic electromagnetic field of frequencies $\omega_j$ and phases $\phi$ as described by the Hamiltonian
\begin{eqnarray}\label{control}
\nonumber H_c(t)&=&\Omega_1(t) (\sigma_1^x + \sigma_2^x)\cos(\omega_1 t - \phi)\\ 
&+& \Omega_2(t) (\sigma_1^x + \sigma_2^x) \cos(\omega_2 t - \phi).
\end{eqnarray}
Under the influence of such MW control fields our system Hamiltonian would be given by 
\begin{eqnarray}\label{casi}
 H^{\textrm I}(t)&=& \eta_1\nu_1(b e^{-i \nu_1 t} + b^\dag e^{i\nu_1 t}) \sigma_1^z -  \eta_2\nu_2(ce^{-i\nu_2 t} + c^\dag e^{i \nu_2 t}) \sigma_1^z\\
\nonumber &+&  \eta_1\nu_1(b e^{-i \nu_1 t} + b^\dag e^{i\nu_1 t}) \sigma_2^z +  \eta_2\nu_2(ce^{-i\nu_2 t} + c^\dag e^{i \nu_2 t}) \sigma_2^z\\
\nonumber &+& \frac{\Omega_1(t)}{2}(\sigma_1^+ e^{i \phi} + \sigma_1^- e^{-i\phi}) + \frac{\Omega_2(t)}{2}(\sigma_2^+ e^{i \phi} + \sigma_2^- e^{-i\phi}). 
\end{eqnarray}
The Hamiltonian above is posed in a rotating frame with respect to $H_0= \nu_1 b^\dag b + \nu_2 c^\dag c + \frac{\omega_1}{2}\sigma^z_1 + \frac{\omega_2}{2}\sigma_2^z$. The non-resonant components of the MW driving have been eliminated under the RWA, whose validity will be later confirmed by detailed numerical simulations using the first-principles Hamiltonian. In this respect, we want to remark that terms that rotate at a rate similar to $2 \omega_j$ (which corresponds to a frequency of tens of gigahertz for the $^{171}$Yb$^{+}$ ion, see for example~\cite{Olmschenk07}) can be safely neglected by invoking the RWA, see appendix~\ref{app:InitialApp}. The slower frequencies rotating at a rate of $|\omega_2- \omega_1|$ (on the order of tens of MHz for our simulated conditions) lead to off-resonant couplings and require a specific treatment covered in section \ref{subsect:Tailored}.

Now we move to a rotating frame with respect to $\frac{\Omega_1(t)}{2}(\sigma_1^+ e^{i \phi} + \sigma_1^- e^{-i\phi}) + \frac{\Omega_2(t)}{2}(\sigma_2^+ e^{i \phi} + \sigma_2^- e^{-i\phi})$. The Rabi frequencies $\Omega_{1,2}(t)$ will be switched on and off, i.e. the driving is applied stroboscopically in the form of $\pi$-pulses, leading to
\begin{eqnarray}
\label{TheHamiltonian}
\nonumber H^\textrm{II}(t)&=& f_{1}(t) \sigma_1^z[\eta_1\nu_1 b e^{-i \nu_1 t} -  \eta_2\nu_2ce^{-i\nu_2 t} + \textrm{ H.c.}] \\
 &+& f_{2}(t)\sigma_2^z[\eta_1\nu_1b e^{-i \nu_1 t} + \eta_2\nu_2ce^{-i\nu_2 t} + \textrm{ H.c.}],
\end{eqnarray}
where the modulation functions $f_{j}(t)$ take the values $\pm 1$ depending on the number of  $\pi$-pulses applied to the $j$-th ion. More specifically, for an even (odd) number of pulses we have $f_{j} =1 (-1)$. The idealised description in Eq.~(\ref{TheHamiltonian}) assumes instantaneous $\pi$-pulses, which is a good approximation if the Rabi frequencies are much larger than any other frequency in Eq.~(\ref{TheHamiltonian}). Nevertheless, to match realistic experimental conditions, our numerical simulations will consider sequences of finite $\pi$-pulses in the form of top-hat functions of length $t_{\pi} = \pi/\Omega$.

The Schr\"odinger equation corresponding to Eq.~(\ref{TheHamiltonian}) is analytically solvable and leads to the propagator $U(t)=U_s(t) U_c(t)$ where 
\begin{equation}\label{solution1}
U_s(t)= \exp{\left[-i \sum_{j=1}^{2} \{\eta_1 G_{j1}(t) b +(-1)^j \eta_2 \ G_{j2}(t) c+ \textrm{ H.c.}\}\sigma_j^z\right]},
\end{equation}
and 
\begin{equation}\label{solutionphase}
U_c(t)=\exp \left[i \varphi(t) \sigma_1^z \sigma_2^z\right],
\end{equation}
see appendix~\ref{app:TimeEvol} for the derivation. The $G_{jm}(t)$ functions in $U_s(t)$ are
\begin{equation}\label{Gfuncs}
G_{jm}(t)=\nu_m \int_0^t dt' f_j(t') e^{-i\nu_m t'},
\end{equation}
while the achieved two-qubit phase $\varphi(t)$ in Eq.~(\ref{solutionphase}) is
\begin{eqnarray}\label{phase}
\varphi(t)=\eta_1^2[ \tilde{\varphi}_1(t)-\frac{1}{3\!\sqrt{3}} \tilde{\varphi}_2(t)]=\eta_1^2 \tilde{\varphi}(t),
\end{eqnarray}
where 
\begin{equation}\label{normphase}
\tilde{\varphi}_m(t)=\nu_m \ \! \Im \ \!\! {\int_{0}^t \!\!\! \ dt'} \big[ f_1(t')G_{2m}(t')+f_2(t')G_{1m}(t')  \big] \ e^{i\nu_m t'},
\end{equation}
and $\Im$ being the imaginary part of the subsequent integral. One can demonstrate that, at the end of the sequence,  $\tilde{\varphi}(t)$ does not depend on the values of $\eta_{1,2}$ and $\nu_{1,2}$ but on the ratio between mode frequencies $\nu_2/\nu_1=\sqrt{3}$ (appendix \ref{app:PulseP}). Hence, the study of $\tilde{\varphi}(t)$ covers all situations regardless of the value of $\eta_{1,2}$ and $\nu_{1,2}$.

From the solution $U(t)$, it is clear that a $\pi$-pulse sequence of duration $T_{\textrm G}$, satisfying conditions
\begin{eqnarray}
\label{conditions}
G_{jm}(T_{\textrm G})=0, \  \  \varphi(T_{\textrm G})\neq 0,
\end{eqnarray}
results in a phase gate between the two qubits and leaves the hyperfine levels of the ions decoupled from their motion. To accomplish these two conditions, we will design a specific MW pulse sequence that, in addition, will eliminate the dephasing noise due to magnetic field fluctuations or frequency offsets on the registers. Note that, if the latter are not averaged out, they would spoil the generation of a high-fidelity two-qubit gate.

\subsection{The AXY-$n$ MW sequence}\label{subsect:MW}


\begin{figure}[t]
\centering
\includegraphics[width=0.8\textwidth]{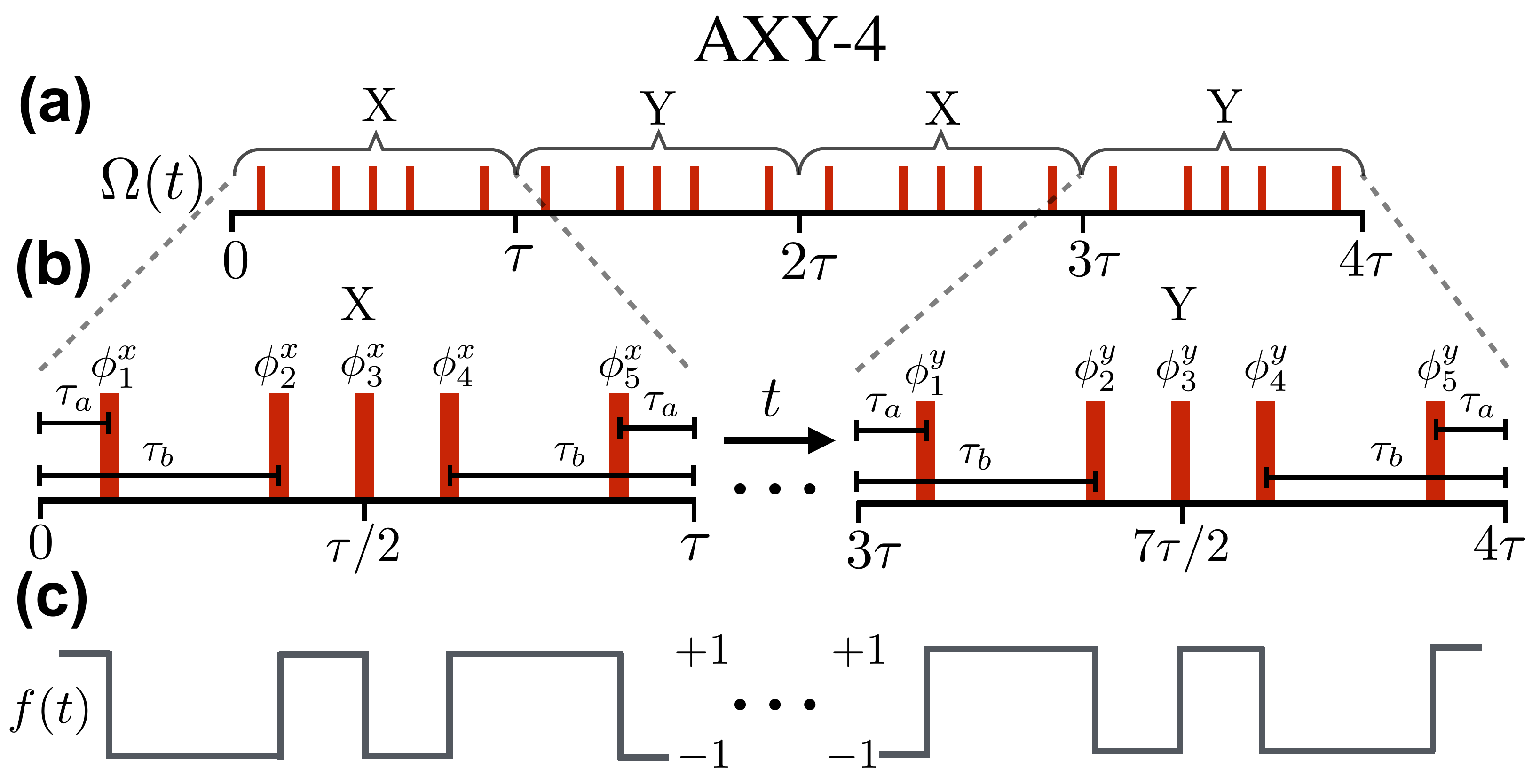}
\caption{ AXY-4 pulse sequence. (a)  Each composite pulse includes five $\pi$-pulses with tuneable distances among them. (b) Zoom on the composite X and Y pulses with the corresponding pulse-phases in $H_c(t)$. (c) Modulation function associated to the composite pulses.}\label{AXYblock}
\end{figure}

In order to satisfy Eqs.~(\ref{conditions}) we propose to use variations of the adaptive XY-$n$ (AXY-$n$) 
family of DD sequences introduced in Ref.~\cite{Casanova15} for nanoscale nuclear magnetic 
resonance~\cite{Wu16,Wang16a, Casanova16,Wang17a, Casanova17}. Unlike previously used pulsed ion-trap 
DD schemes~\cite{Piltz13}, AXY-$n_\textrm{ B}$ consists of $n_\textrm{ B}$ blocks of 5 non-equally separated $\pi$-pulses, as depicted 
in Fig.~\ref{AXYblock} for the AXY-4 case, where the inter-pulse spacing can be arbitrarily tuned while the sequence remains robust~\cite{Casanova15}. Each $\pi$-pulse is applied along an axis in the $x$-$y$ plane of the Bloch sphere of each qubit state that is rotated an angle $\phi$ with respect to the $x$ axis.

We define two blocks: the X block, made of 5 $\pi$-pulses along the axes corresponding to ${\vec\phi^x\equiv\{\phi^x_1,\phi^x_2,\phi^x_3,\phi^x_4,\phi^x_5\}=\{ \frac{\pi}{6}, \frac{\pi}{2}, 0 ,\frac{\pi}{2},\frac{\pi}{6}\}} + \zeta$, with $\zeta$ an arbitrary constant phase, and the Y block, with rotations along the same axes but shifted by a $\pi/2$ phase, i.e. ${\vec{\phi}^y=\{ \frac{\pi}{6} + \frac{\pi}{2}, \pi, \frac{\pi}{2} ,\pi,\frac{\pi}{6} + \frac{\pi}{2}\}} + \zeta$. The sequence then has $n_\textrm{ B}$ consecutive X and Y blocks with the same, tuneable, inter-pulse spacing. For example, the AXY-$4$ sequence is XYXY. As illustrated in Fig.~\ref{AXYblock}{(b)}, each block is symmetric and has a duration $\tau$. Therefore, within a five-pulse block the time of application of the first and second pulses, $\tau_a$ and $\tau_b$ where $\tau_a<\tau_b<\tau/2$, together with $\tau$ define the whole sequence.

At the end of any AXY-$n_\textrm{ B}$ sequence of length $n_\textrm{ B}\tau$, where $n_\textrm{ B}$ is an even integer, the function $G_{jm}(n\tau)$ is zero for values of $\tau$ that are a multiple of the oscillation period of mode $m$, that is for $\nu_m \tau=2\pi r$ with $r\in \mathbb{N}$. This is due to the translational symmetry of the $f_{j}(t)$ functions, for which $f_j(t'+\tau)=-f_j(t')$ and $f_j(t'+2\tau)=f_j(t')$ holds, meaning that 
\begin{eqnarray}
G_{jm}(n\tau) &=&\nu_m\int_{0}^{n\tau}dt^{\prime}f_{j}(t^{\prime})e^{-i\nu_{m}t^{\prime}}  \\
&=\sum_{p=0}^{n/2-1}&\nu_m\int_{0}^{\tau}dt' f_j(t')\Big(e^{-i\nu_m[t'+2p\tau]}-e^{-i\nu_m[t'+(2p+1)\tau]}\Big)=0\nonumber
\end{eqnarray}
if $\nu_m \tau$ is a multiple of $2\pi$, and for $n$ even.
This means that a qubit can be left in a product state with a specific motional mode $m$ regardless of the values of $\tau_a$ and $\tau_b$. Unfortunately, the two motional modes in our system have incommensurable oscillation frequencies (note that $\nu_2/\nu_1=\sqrt{3}$) which leads to the impossibility of finding a $\tau$ that, independently of $\tau_a$ and $\tau_b$,  decouples the qubits from both vibrational modes.

\begin{figure}[t]
\centering
\includegraphics[width=0.9\linewidth]{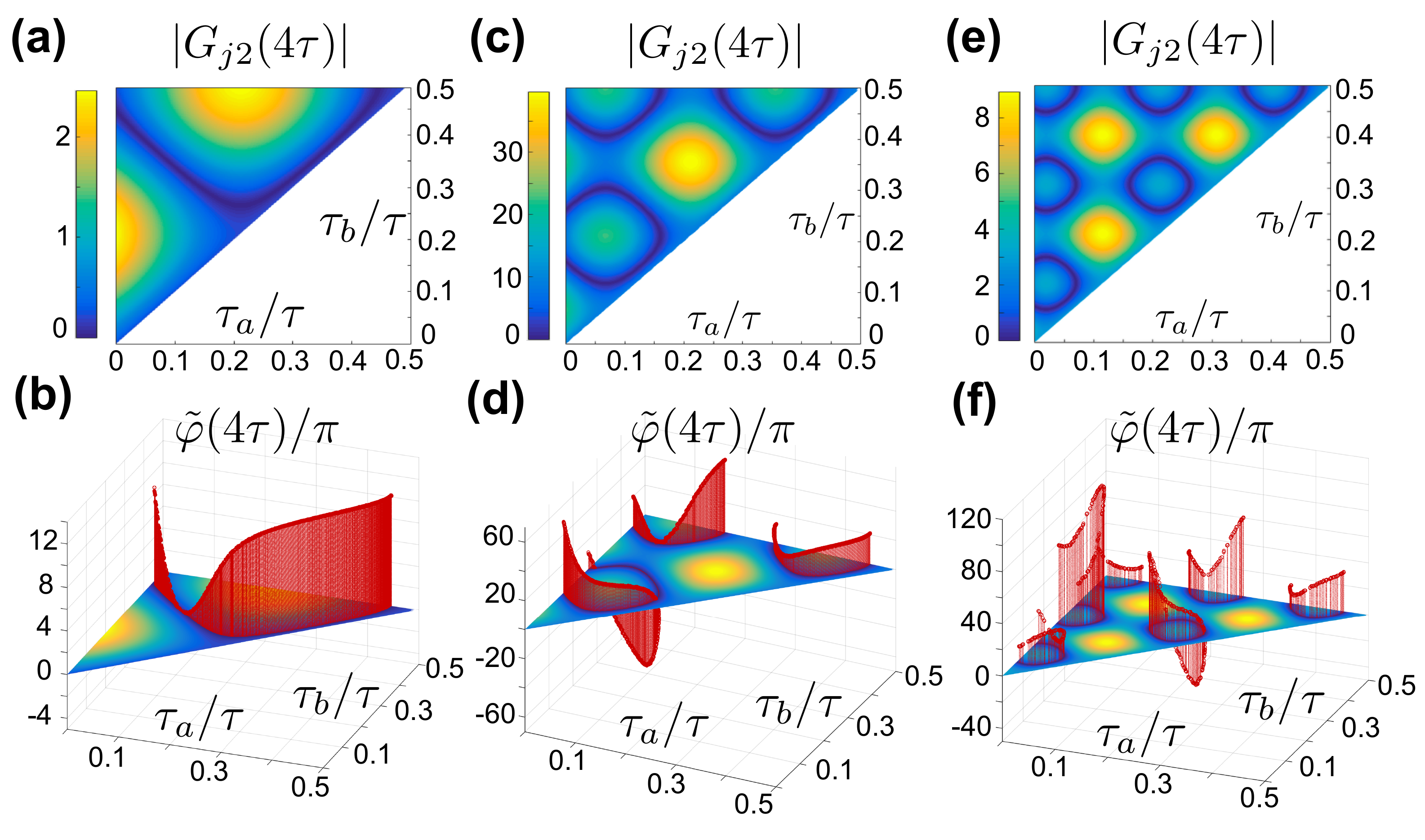}
\caption{Absolute value of $G_{j2}(t)$ after an AXY-$4$ sequence as a function of $\tau_a$ and $\tau_b$ ($\tau_a<\tau_b<\tau/2$), for (a): $\tau=1\times2\pi/\nu_1$, {(c)}: $\tau=2\times2\pi/\nu_1$, (e): $\tau=3\times2\pi/\nu_1$. The dark blue regions show the $\tau_a$ and $\tau_b$ values that correspond to a complete decoupling of the qubits with the modes at the end of the sequence. The phases $\tilde{\varphi}(t)$ are represented in {(b)}, {(d)}, {(f)} by the red panels.}
\label{Gplot}
\end{figure}

An AXY-4 sequence of a duration $4\tau$ such that $\tau=2\pi r/\nu_1$, makes $G_{j1}(4\tau)=0$ for any choice of $\tau_a$ and $\tau_b$, while we will numerically look for the values of $\tau_a$ and $\tau_b$ that minimise $|G_{j2}(4\tau)|$. For the sake of simplicity in the presentation of this part, we consider $f_1(t)=f_2(t)$, i.e. the same sequence is simultaneously applied to both qubits leading to $G_{1m}=G_{2m}$. However,  when considering real pulses, we will not use simultaneous driving in order to efficiently eliminate crosstalk effects which leads to an optimal performance of the method, see section~\ref{subsect:Tailored}. In Fig.~\ref{Gplot}{(a)} we give a contour colour plot of $|G_{j2}(4\tau)|$ with $\tau=2\pi/\nu_1$ for all combinations of $\tau_a$ and $\tau_b$. The dark blue regions represent the  values of $\tau_a$ and $\tau_b$ that minimise the $|G_{j2}(4\tau)|$ functions. Then any pair of $\tau_{a,b}$ in that region defines a valid sequence for a two-qubit phase gate. At Fig.~\ref{Gplot}{(b)}, we give the corresponding value for $\tilde{\varphi}(4\tau)$ of the resulting two-qubit gate (red panels). In Figs.~\ref{Gplot}{(c)}, \ref{Gplot}{(d)} and \ref{Gplot}{(e)}, \ref{Gplot}{(f)} the same procedure is shown for  $\tau=2\times2\pi/\nu_1$ and $\tau=3\times2\pi/\nu_1$, respectively, i.e. for values $r=2$ and $r=3$, obtaining several combinations of $\tau_a$ and $\tau_b$ that result in a phase gate. Finally, to recover the actual phase $\varphi(4\tau)$, we multiply $\tilde{\varphi}(4\tau)$ by $\eta_1^2=\frac{\hbar\gamma_e^2g_B^2}{64 M \nu^3}$, according to Eq.~(\ref{phase}), showing the dependance of the total phase $\varphi$ on $\nu$ and $g_B$.

\subsection{Tailored sequences and results}\label{subsect:Tailored}

We will benchmark the performance of our MW-pulse scheme by means of detailed numerical simulations. The total Hamiltonian governing the dynamics is $H+H_c$. In a rotating frame with respect to $H_0$ and after neglecting terms that rotate at a speed of tens of GHz (see appendices~\ref{app:InitialApp} and \ref{app:IntHamil} for more details), the effective Hamiltonian reads
\begin{eqnarray}\label{simstart}
\nonumber H^{\textrm I}(t)&=&  \eta_1\nu_1 (b e^{-i\nu_1 t}+b^\dag e^{+i\nu_1 t}) \sigma_1^z - \eta_2\nu_2(c e^{-i\nu_2 t}+c^\dag e^{+i\nu_2 t})\sigma_1^z\\
\nonumber &+&    \eta_1\nu_1(b e^{-i\nu_1 t}+b^\dag e^{+i\nu_1 t}) \sigma_2^z +\eta_2\nu_2(c e^{-i\nu_2 t}+c^\dag e^{+i\nu_2 t})\sigma_2^z\\
\nonumber &+& \frac{\Omega_1(t)}{2} \sigma_1^{\phi} + \frac{\Omega_1(t)}{2} (\sigma_2^+ e^{i\delta_2 t} e^{i\phi} + \textrm{H.c}.)\\
                  &+& \frac{\Omega_2(t)}{2} \sigma_2^{\phi} + \frac{\Omega_2(t)}{2} (\sigma_1^+ e^{i\delta_1 t} e^{i\phi} + \textrm{H.c}.).
\end{eqnarray}
Here, $\sigma_j^{\phi} = \sigma_j^+ e^{i\phi}  +  \sigma_j^- e^{-i\phi}$, and the last two lines contain both the resonant terms giving rise to the $\pi$-pulses, i.e. $ \frac{\Omega_1(t)}{2} \sigma_1^{\phi}$ and $ \frac{\Omega_2(t)}{2} \sigma_2^{\phi}$, as well as crosstalk contributions of each $\pi$-pulse on the off-resonant ion. The latter are  $\frac{\Omega_1(t)}{2} (\sigma_2^+ e^{i\delta_2t} e^{i\phi} + \textrm{H.c}.)$ and $\frac{\Omega_2(t)}{2} (\sigma_1^+ e^{i\delta_1t} e^{i\phi} + \textrm{H.c}.)$, where $\delta_2 = - \delta_1 = \omega_2 - \omega_1$. We use Eq.~(\ref{simstart}) as the starting point of our simulations without any further assumptions. In addition, our numerical simulations include motional decoherence described by a Lindblad equation accounting for an environment at a temperature of 50 K as well as static errors on $\Omega_{1,2}$, $\omega_{1,2}$, and $\nu$.

\begin{figure}[t!]
\centering
\includegraphics[width=0.9\linewidth]{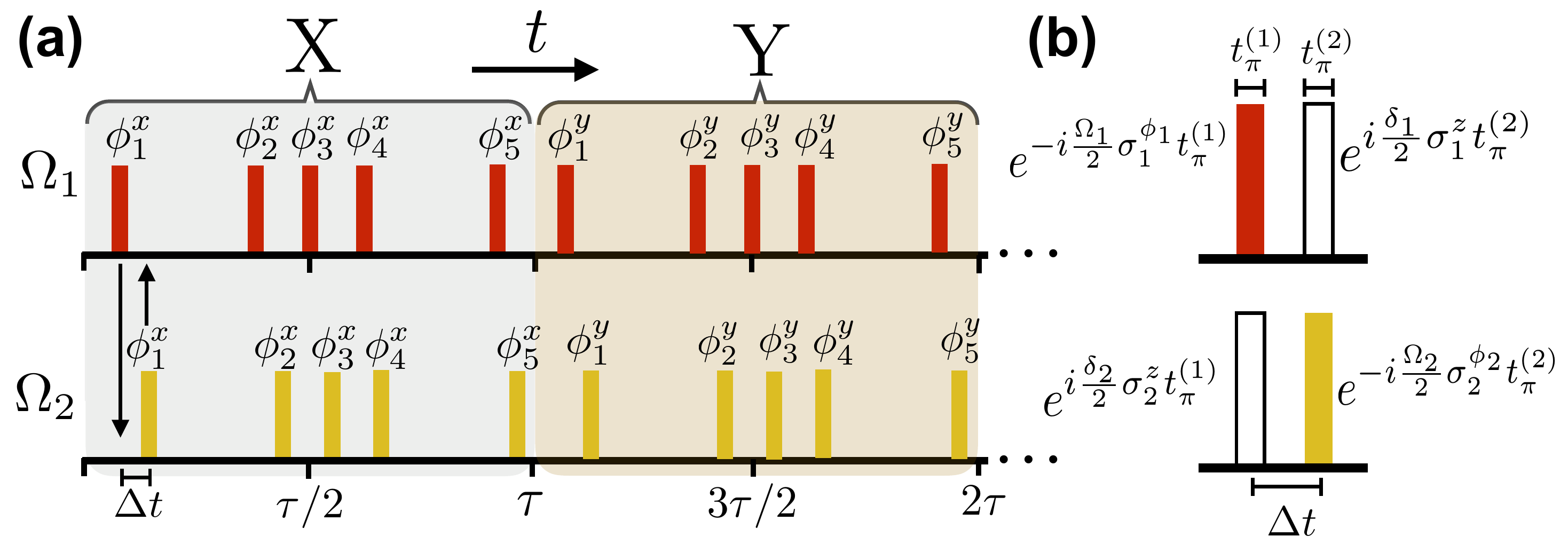}
\caption{{(a)} Pulse sequence on the first(second) ion, upper (bottom) panel. The first (second) ion is driven with an AXY-4 sequence, red (yellow) blocks represent  $\pi$-pulses.  Each pulse on the second ion is separated by $\Delta t$ from the pulses acting on the first ion.  {(b)} Zoom on two pulses. We can observe the propagators leading to $\pi$-pulses, i.e.  $\exp{[-i \frac{\Omega_{1,2}}{2} \sigma_{1,2}^{\phi} t_\pi ]}$
(red and yellow blocks), and their unwanted side-effects in the adjacent ions $\exp{[i\frac{\delta_{2,1}}{2} \sigma_{2,1}^{z} t_\pi]}$ (empty blocks). }
\label{combinedAXY}
\end{figure}

To get rid of crosstalk effects, we use a DD scheme acting non simultaneously on both ions that, at the same time, meets conditions in Eqs.~(\ref{conditions}), and gives rise to a tuneable phase gate between the ions. In this respect, one can demonstrate that a term like 
\begin{equation}
 H=\frac{\Omega_1(t)}{2} \sigma_1^{\phi} + \frac{\Omega_1(t)}{2} (\sigma_2^+ e^{i\delta_2 t} e^{i\phi} + \textrm{H.c}.),
\end{equation}
for a final time $t^{(1)}_\pi = \frac{\pi}{\Omega_1}$, i.e. the required time for a $\pi$-pulse on the first ion, has the associated propagator 
\begin{equation}
\label{crosstalk}
U_{t^{(1)}_\pi}=  e^{-i \frac{\Omega_1}{2} \sigma_1^{\phi} t_\pi } e^{i\frac{\delta_2}{2} \sigma_2^{z} t_\pi},
\end{equation}
if and only if the Rabi frequency $\Omega_1$ satisfies 
\begin{equation}
\Omega_1=\frac{\delta_2}{\sqrt{4k^2-1}}, \mbox{with} \ k\in \mathbb{N}.
\end{equation}
See appendix~\ref{app:PulseP} for a demonstration of this. In the same manner, the term $ \frac{\Omega_2(t)}{2} \sigma_2^{\phi} + \frac{\Omega_2(t)}{2} (\sigma_1^+ e^{i\delta_1 t} e^{i\phi} + \textrm{H.c}.)$ 
gives rise to $U_{t^{(2)}_\pi} =  e^{-i \frac{\Omega_2}{2} \sigma_2^{\phi} t_\pi } e^{i\frac{\delta_1}{2} \sigma_1^{z} t_\pi}$ under the  conditions  $t^{(2)}_\pi = \frac{\pi}{\Omega_2}$ and 
$\Omega_2=|\delta_1|/\sqrt{4{k}^2-1}, \mbox{with} \ k\in \mathbb{N} $. Hence, when the MW driving is applied non-simultaneously over the registers, one can clearly argue that a $\pi$-pulse on the first ion induces a dephasing-like propagator on the second ion (i.e. $e^{i\frac{\delta_2}{2} \sigma_2^{z} t_\pi}$) and vice versa. It turns out that our DD sequence successfully eliminates such undesired contribution.

Two blocks of our non simultaneous AXY-$n_\textrm{ B}$ sequence are depicted in Fig.~\ref{combinedAXY}{(a)}, where one has to select $\tau$, $\tau_{a,b}$ and $\Delta t$. While $\tau$ and $\tau_{a,b}$ define the sequence acting on the first ion, a temporal translation $\Delta t$ of each $\pi$-pulse sets the sequence on the second ion. Note that $\Delta t$ must satisfy $\Delta t>t_{\pi}$ to assure there is no pulse overlap, see Fig.~\ref{combinedAXY}{(b)}. As we said before, the construction in Fig.~\ref{combinedAXY}{(a)} eliminates the dephasing terms $e^{i\frac{\delta_{2,1}}{2} \sigma_{2,1}^{z} t_\pi}$. For example, the propagator for the first ion after a XY block $U^{(1)}_\textrm{XY}$,  upper panel in Fig.~\ref{combinedAXY}{(a)}, reads
\begin{eqnarray}\label{firstprop}
U^{(1)}_\textrm{XY}&=&\bigg[ e^{i\frac{\delta_1}{2}  \sigma_1^{z} t_\pi}\sigma_1^{\phi_5^y}\bigg]\bigg[e^{i\frac{\delta_1}{2} \sigma_1^{z} t_\pi}\sigma_1^{\phi_4^y}\bigg]\bigg[e^{i\frac{\delta_1}{2} \sigma_1^{z} t_\pi}\sigma_1^{\phi_3^y}\bigg] \bigg[ e^{i\frac{\delta_1}{2} \sigma_1^{z} t_\pi}\sigma_1^{\phi_2^y}\bigg]\bigg[ e^{i\frac{\delta_1}{2} \sigma_1^{z} t_\pi}\sigma_1^{\phi_1^y}\bigg]\nonumber\\&&
\bigg[e^{i\frac{\delta_1}{2} \sigma_1^{z} t_\pi}\sigma_1^{\phi_5^x}\bigg] \bigg[ e^{i\frac{\delta_1}{2} \sigma_1^{z} t_\pi}\sigma_1^{\phi_4^x}\bigg] \bigg[e^{i\frac{\delta_1}{2} \sigma_1^{z} t_\pi} \sigma_1^{\phi_3^x} \bigg]\bigg[e^{i\frac{\delta_1}{2} \sigma_1^{z} t_\pi} \sigma_1^{\phi_2^x} \bigg]\bigg[e^{i\frac{\delta_1}{2} \sigma_1^{z} t_\pi} \sigma_1^{\phi_1^x}\bigg]\nonumber \\
&=& \sigma_1^{\phi_5^y} \sigma_1^{\phi_4^y} \sigma_1^{\phi_3^y} \sigma_1^{\phi_2^y} \sigma_1^{\phi_1^y} \sigma_1^{\phi_5^x} \sigma_1^{\phi_4^x} \sigma_1^{\phi_3^x} \sigma_1^{\phi_2^x} \sigma_1^{\phi_1^x},
\end{eqnarray}
where the last equality can be achieved using $\{\sigma_1^z,  \sigma_1^{\phi^{x,y}}\} = 0$.  Equation~(\ref{firstprop}) describes a situation without motional degrees of freedom. However the cancelation of the dephasing terms is still valid if one includes the spin-motion coupling terms because they depend on $\sigma^z_{1,2}$, see Eq.~(\ref{Hamiltonianbare}), and the operators $e^{i\frac{\delta_{1,2}}{2} \sigma_{1,2}^z t_{\pi}}$ commute with them leading to the same cancelation.

In the same manner, one can find the propagator for the second ion $U^{(2)}_\textrm{ XY}$, see lower panel in  Fig.~\ref{combinedAXY}{(a)}. This propagator reads
\begin{eqnarray}\label{secondprop}
U^{(2)}_\textrm{XY}&=&\bigg[\sigma_2^{\phi_5^y} e^{i\frac{\delta_2}{2}  \sigma_2^{z} t_\pi}\bigg]\bigg[\sigma_2^{\phi_4^y} e^{i\frac{\delta_2}{2} \sigma_2^{z} t_\pi}\bigg]\bigg[\sigma_2^{\phi_3^y} e^{i\frac{\delta_2}{2} \sigma_2^{z} t_\pi}\bigg] \bigg[\sigma_2^{\phi_2^y} e^{i\frac{\delta_2}{2} \sigma_2^{z} t_\pi}\bigg]\bigg[ \sigma_2^{\phi_1^y} e^{i\frac{\delta_2}{2} \sigma_2^{z} t_\pi}\bigg] \nonumber\\ &&
\bigg[\sigma_2^{\phi_5^x} e^{i\frac{\delta_2}{2} \sigma_2^{z} t_\pi}\bigg] \bigg[\sigma_2^{\phi_4^x} e^{i\frac{\delta_2}{2} \sigma_2^{z} t_\pi}\bigg] \bigg[\sigma_2^{\phi_3^x} e^{i\frac{\delta_2}{2} \sigma_2^{z} t_\pi}  \bigg]\bigg[\sigma_2^{\phi_2^x} e^{i\frac{\delta_2}{2} \sigma_2^{z} t_\pi}\bigg]\bigg[ \sigma_2^{\phi_1^x}e^{i\frac{\delta_2}{2} \sigma_2^{z} t_\pi}\bigg]\nonumber \\ 
&=& \sigma_2^{\phi_5^y} \sigma_2^{\phi_4^y} \sigma_2^{\phi_3^y} \sigma_2^{\phi_2^y} \sigma_2^{\phi_1^y} \sigma_2^{\phi_5^x} \sigma_2^{\phi_4^x} \sigma_2^{\phi_3^x} \sigma_2^{\phi_2^x} \sigma_2^{\phi_1^x}.
\end{eqnarray}

We can see that after an XY block, there is no contribution of dephasing like operators, see the last lines in Eqs.~(\ref{firstprop}) and~(\ref{secondprop}). Hence, a sequence XYXY applied to both ions following the scheme in Fig.~\ref{combinedAXY}{(a)} will also share this property with the additional advantage of being robust against control errors~\cite{Casanova15}.

After simulating the application of a non-simultaneous AXY-4 sequence, we show the results (infidelities) in Table~\ref{table1}. It is noteworthy that our numerical results have been calculated including motional decoherence. More specifically, we have added to the dynamics governed by the Hamiltonian in Eq.~(\ref{simstart})  a dissipative term of the form, see for example~\cite{Brownnutt15},
\begin{eqnarray}\label{MotionalHeating}
D(\rho) &=& \frac{\Gamma_b}{2}\Big\{(\bar{N}_b+1) (2 b \rho b^{\dag} - b^{\dag} b \rho - \rho b^{\dag} b) + \bar{N}_b (2 b^\dag \rho b - b  b^{\dag} \rho - \rho b b^{\dag})\Big\}\\
&+&\frac{\Gamma_c}{2}\Big\{(\bar{N}_c+1) (2 c \rho c^{\dag} - c^{\dag} c \rho - \rho c^{\dag} c) + \bar{N}_c (2 c^\dag \rho c - c  c^{\dag} \rho - \rho c c^{\dag})\Big\}, \nonumber
\end{eqnarray}
where an estimation of the values for the heating rates $\Gamma_{b,c}$ is given in appendix~\ref{app:Heating} for each of the specific examples considered here, while $\bar{N}_{b,c} =1/(e^{\hbar \nu_{1,2}/k_{\textrm B} T}-1)$ where we have considered a temperature of $T=50$K.

\begin{table}
\centering
\caption{Infidelities (I) for two-qubit gates after the application of 20 imperfect MW pulses on each ion, according to our AXY-4 protocol, for several initial states, $\psi_j$, and different experimental conditions, see main text. We focus in $\pi/4$ and $\pi/8$ entangling phase gates, however our method is general and can achieve any phase. Initial states, up to normalisation, are $\psi_{1}=|\textrm{g}\rangle \otimes (|\textrm{g}\rangle + |\textrm{e}\rangle)$, $\psi_{2}= (|\textrm{g}\rangle + |\textrm{e}\rangle) \otimes (|\textrm{g}\rangle + |\textrm{e}\rangle)$, $\psi_{3}=|\textrm{g}\rangle \otimes (|\textrm{g}\rangle + i |\textrm{e}\rangle) + |\textrm{e}\rangle \otimes |\textrm{e}\rangle$, $\psi_{4}=|\textrm{e}\rangle \otimes (|\textrm{g}\rangle - i |\textrm{e}\rangle) + |\textrm{g}\rangle \otimes |\textrm{g}\rangle$, and $\psi_{5}=|\textrm{e}\rangle \otimes (|\textrm{g}\rangle - i |\textrm{e}\rangle) + |\textrm{g}\rangle \otimes (|\textrm{g}\rangle + i |\textrm{e}\rangle)$.}
\label{table1}
\begin{tabular}{{ |c | c | c | c | c| c|}}
\hline
I ($\times 10^{-4}$) &exp$( i\frac{\pi}{4} \sigma_1^z  \sigma_2^z)$ &exp$( i\frac{\pi}{8} \sigma_1^z  \sigma_2^z)$&exp$( i\frac{\pi}{4} \sigma_1^z  \sigma_2^z)$ &exp$( i\frac{\pi}{8} \sigma_1^z  \sigma_2^z)$\\
   &  $\eta_1 =0.069 $& $\eta_1=0.069 $  &$\eta=0.078$  &  $\eta_1=0.078$\\
 &$T_{\textrm G} = 80 \ \mu$s&$T_{\textrm G} = 80 \ \mu$s&$T_{\textrm G} = 36.3 \ \mu$s&$T_{\textrm G} = 36.3 \ \mu$s \\
\hline
$\psi_1$  & $1.172$    &$0.128$      & $2.060$   &$0.144$  \\
\hline
$\psi_2$     & $2.229 $    &$0.136 $      & $4.905$   & $0.304 $ \\
\hline
$\psi_3$  & $3.052$    &$0.116  $    & $5.899 $  & $0.371$ \\
\hline
$\psi_4$  & $ 4.631 $    &$0.172  $    & $5.946$  & $0.413$ \\
\hline
$\psi_5$  & $3.250$    &$0.110 $    & $4.635 $  & $0.293$ \\
\hline
\end{tabular}
\end{table}

We computed the gate infidelity for the following situations. Firstly, we simulated the gates exp$( i\frac{\pi}{4} \sigma_1^z  \sigma_2^z)$ and exp$( i\frac{\pi}{8} \sigma_1^z  \sigma_2^z)$, second and third columns in Table~\ref{table1}, with a gate time of $80 \ \mu$s for a magnetic field gradient of $g_B=150 \frac{T}{m} $~\cite{Weidt16}.  We designed the MW sequence such that $\tau= 3\times2\pi r/\nu_1$ leading to a gate time which is 12 times the period of the center-of-mass mode.  Other relevant parameters are $\nu_1=\nu_2/\sqrt{3} = (2\pi)\times 150$ kHz, $\pi$-pulse time of $\approx 75$ ns that implies a Rabi frequency of $\Omega_1=\Omega_2=\Omega\approx (2\pi)\times 6.63$ MHz, and $\omega_2 - \omega_1 = (2\pi)\times 25.7$ MHz, while we have chosen $\Delta t$ as 1.05 times the $\pi$-pulse time. Both bosonic modes are initially in a thermal state\footnote{A thermal state of a bosonic mode is defined as $\rho_T=\sum_{n=0}^{\infty} \frac{ \bar{n}^n}{(\bar{n}+1)^{n+1}}|n\rangle\langle n|$} with $0.2$ phonons each~\cite{Weidt15}. In addition to heating processes with rates $\Gamma_b \bar{N}_b \approx (2\pi) \times 133$ Hz and $\Gamma_c \bar{N}_c \approx (2\pi) \times 9$~Hz (appendix~\ref{app:Heating}), our simulations include a Rabi frequency mismatch of $1\%$, a trap frequency shift of $0.1\%$, and an energy shift of $(2\pi)\times 20$ kHz on both ions.

Secondly, we also target the gates exp$( i\frac{\pi}{4} \sigma_1^z  \sigma_2^z)$ and exp$( i\frac{\pi}{8} \sigma_1^z  \sigma_2^z)$, fourth and fifth columns in Table~\ref{table1}, but now with $g_B=300  \frac{T}{m}$. The gate time is $36.3 \ \mu$s, i.e. 8 times the oscillation period of the center-of-mass mode whose frequency is $\nu=\nu_1=\nu_2/\sqrt{3} = (2\pi)\times 220$ kHz. Other parameters are $\Omega\approx (2\pi)\times 10$ MHz,  $\pi$-pulse time of $\approx 49$ ns, $\omega_2 - \omega_1 = (2\pi)\times 39.8$ MHz and the energy shift upon the ions, errors on Rabi and trap frequencies,  $\Delta t$, and the initial bosonic states are the same as in the previous case. Because of the new value for $g_B$, the
heating rates had to be recalculated leading to $\Gamma_b \bar{N}_b \approx (2\pi) \times 248$ Hz  and $\Gamma_c \bar{N}_c \approx (2\pi) \times16$ Hz. 

In Table~\ref{table1} we find that, even in the presence of the errors we have included, our method leads to fast two-qubit gates with fidelities exceeding 99,9\%. Finally, we note that higher values of $g_B$ will result in faster gates.

In summary, we have demonstrated that pulsed DD schemes are efficient generators of fast and robust two-qubit gates. Our MW sequence forces the two motional modes in a certain direction to cooperate and makes the gate fast and robust against external noise sources including motional heating.

\section{Hybrid MW radiation patterns for high-fidelity quantum gates}
\label{sect:Slow_ions}

In this section, we present a method to generate two-qubit gates among trapped ions that combines pulsed and continuous MW radiation patterns in the far-field regime. As opposed to the previous method, this one is designed to be applied with a small LD parameter $\eta\sim0.01$, and it is directly applicable in chains with more than two ions as the gate is mediated by a single motional mode. Similar to the previous method, this scheme is also protected against magnetic fluctuations, errors on the delivered MW fields, and crosstalk effects caused by the use of long wavelength MW radiation. Moreover, our protocol is flexible since it runs with arbitrary values of the MW power. In particular, inspired by results in Refs.~\cite{Casanova18MW,Casanova19}, this method involves phase-modulated drivings, phase flips, and refocusing $\pi$ pulses leading to high-fidelity entangling gates within current experimental limitations. We numerically test the performance of our gates in the presence of magnetic fluctuations of different intensities, deviations on the MW Rabi frequencies, as well as under motional heating. We demonstrate the achievement of fidelities largely exceeding $99\%$ in realistic experimental scenarios, while values larger than $99.9\%$ are reachable with small improvements.

\subsection{Method: bichromatic gate with continuous DD}\label{subsect:method}
As in the previous section, we consider two $^{171}$Yb$^+$ ions sitting next to each other in the longitudinal direction $z$ of a linear harmonic trap. We define a qubit using two states of the 6s$^2S_{1/2}$ hyperfine manifold. These are $|{\textrm g}\rangle\equiv\{F=0,m_F=0\}$, and $|{\textrm e}\rangle\equiv\{F=1,m_F=1\}$. Due to the Zeeman effect, the frequency of the $j$th qubit is  $\omega_j=\omega_0 + \gamma_e B(z^0_j)/2$, where $\omega_0=(2\pi)\times12.6$ GHz, $\gamma_e=(2\pi)\times 2.8$ MHz/Gauss, see~\cite{Olmschenk07}, and $z^0_j$ is the equilibrium position of the ion. The presence of a constant magnetic field gradient $\partial B/\partial z=g_B$ in the $z$ direction results in different values of $\omega_j$ for each qubit, which allows  individual control  on each ion with MW fields~\cite{Arrazola18, Piltz14}. The Hamiltonian of the system can be written as
\begin{equation}\label{Hsys}
H = \frac{\omega_1}{2}\sigma_1^z +\frac{\omega_2}{2}\sigma_2^z +\nu a^\dagger a + \eta \nu (b+b^\dagger)S_z \, ,
\end{equation}
where $S_z=\sigma^z_1+\sigma_2^z$,  $b^\dagger$($b$) is the creation (annihilation) operator that correspond to the center-of-mass mode, $\nu$ is the trap frequency and $\eta=\frac{\gamma_eg_B}{8\nu}\sqrt{\frac{\hbar}{M\nu}}$ is the LD parameter that quantifies the strength of the qubit-boson interaction.

Bichromatic MW drivings, at detuning $\delta$, can be applied to both ions. Then, Hamiltonian~(\ref{Hsys}) in a rotating frame  with respect to $H_0=\frac{\omega_1}{2}\sigma_1^z +\frac{\omega_2}{2}\sigma_2^z +\nu b^\dagger b$ reads (see appendix~\ref{app:InteractionPic} for additional details about the involved interaction pictures)
\begin{equation}\label{HMS}
H= \eta \nu (be^{-i\nu t}+b^\dagger e^{i\nu t})S_z + \Omega\cos{(\delta t)}S_x.
\end{equation}
For the sake of clarity, we have omitted the presence of the breathing mode in Eqs.~(\ref{Hsys}) and~(\ref{HMS}), as well as the crosstalk terms in  Eq.~(\ref{HMS}). However, these will be included in our numerical simulations to demonstrate that they have a negligible impact in our scheme. Furthermore, in appendix~\ref{app:CompleteHamil} one can find a complete description of the system Hamiltonian. Now, we move to a second rotating frame with respect to $\Omega\cos{(\delta t)}S_x$ (this is known as the bichromatic interaction picture~\cite{Sutherland19,Roos08,Sutherland20}) and use the Jacobi-Anger expansion ($e^{iz\sin{(\theta)}} =  \sum_{n=-\infty}^{+\infty} J_n(z) \ e^{i n\theta}$, with $J_n(z)$ being Bessel functions of the first kind) to obtain
\begin{equation}\label{HMSI}
H= \eta \nu (be^{-i\nu t}+\textrm{H.c.})\Big\{J_0\Big(\frac{2\Omega}{\delta}\Big)S_z+2J_1\Big(\frac{2\Omega}{\delta}\Big)\sin{(\delta t)}S_y\Big\}.
\end{equation}
Note we only keep terms up to the first order of the Jacobi-Anger expansion, since higher order terms would not lead to any significant contribution if $\Omega\ll\delta$. If we choose $\delta=\nu+\xi$ with $\xi\ll \nu$, and neglect all terms that rotate with $\nu$ by invoking the RWA, we find the gate Hamiltonian
\begin{equation}\label{HMSII}
H_{\textrm G} =  i\eta\nu J_1\Big(\frac{2\Omega}{\delta}\Big) \Big\{b^\dagger e^{-i\xi t} -\textrm {H.c.}\Big\} S_y \approx  i\frac{\eta\nu\Omega}{\delta} \Big\{b^\dagger e^{-i\xi t} -\textrm {H.c.}\Big\} S_y,
\end{equation}
where we used $J_1(x)\approx x/2$ for small $x$. For evolution times $t_n=2\pi n_\textrm{ RT}/\xi$ where $n_\textrm{ RT} \in \mathbb{N} $, the time-evolution operator associated to Eq.~$(\ref{HMSII})$ is
\begin{equation}\label{HMSUnitary}
U_{\textrm G}(t_n)= \exp{(i\theta_n S_y^2)}
\end{equation}
with $\theta_n=2\pi n_\textrm{ RT} \eta^2 \nu^2J_1^2(2\Omega/\delta)/\xi^2\approx 2\pi n_\textrm{ RT} \eta^2 \Omega^2/\xi^2$~\cite{Sorensen99,Sorensen00,Solano99}. By tuning the parameters such that $\theta_n=\pi/8$, the propagator $U_{\textrm G}$ evolves the initial (separable) state $|\textrm{g,g}\rangle$ into the maximally entangled Bell state $\frac{1}{\sqrt{2}} (|\textrm{g,g}\rangle+i|\textrm{e,e}\rangle)$.

In order to protect this gate scheme from magnetic field fluctuations of the kind $\frac{\epsilon_1(t)}{2}\sigma_1^z+\frac{\epsilon_2(t)}{2}\sigma_2^z$ ($\epsilon_{1,2}(t)$ being stochastic functions) we introduce an additional MW driving that will suppress their effect.  We select a MW driving such that it enters in Eq.~(\ref{HMS}) as a carrier term of the form $\frac{\Omega_\textrm{DD}}{2}S_y$ leading to
 \begin{equation}\label{HMSDD}
H= \eta \nu (be^{-i\nu t}+b^\dagger e^{i\nu t})S_z + \Omega\cos{(\delta t)}S_x + \frac{\Omega_\textrm{DD}}{2}S_y.
\end{equation}

In the bichromatic picture, Eq.~(\ref{HMSDD}) reads (note that in the following, we adopt the convention $J_{0,1}\Big(\frac{2\Omega}{\delta}\Big) \equiv J_{0,1}$)
\begin{eqnarray}\label{HMSIDD}
H= \eta \nu (be^{-i\nu t}+b^\dagger e^{i\nu t})\Big\{J_0S_z+2J_1\sin{(\delta t)}S_y\Big\}\nonumber\\
 +\frac{\Omega_\textrm{DD}}{2}\Big\{J_0S_y-2J_1\sin{(\delta t)}S_z\Big\}.
\end{eqnarray}
The new driving $\frac{\Omega_\textrm{DD}}{2}S_y$ leads to the appearance of the second line in Eq.~(\ref{HMSIDD}). 
Here, the $\frac{\Omega_\textrm{DD}}{2}J_0 S_y$ term is the responsible of removing magnetic field fluctuations, while $J_1\Omega_\textrm{DD}\sin{(\delta t)}S_z$ interferes with the gate and has to be eliminated. This term can be neglected under a 
RWA only if $\Omega_\textrm{DD}\ll\delta$, thus, its presence limits the range of applicability of our method since larger 
values for $\Omega_\textrm{DD}$ are desirable to better remove the effect of magnetic field fluctuations. To overcome 
this problem, we introduce in all MW drivings, i.e. those leading to the terms  $\Omega\cos{(\delta t)}S_x$ and  $\frac{\Omega_\textrm{DD}}{2}S_y$ in Eq.~(\ref{HMSDD}), a time-dependent phase that will eliminate $J_1\Omega_\textrm{DD}\sin{(\delta t)}S_z$. 
This time-dependent phase follows equation
\begin{equation}\label{TP}
\phi(t)=4\frac{\Omega_\textrm{DD} J_1}{\delta J_0}\sin^2{(\delta t/2)}.
\end{equation}

\begin{figure}[t!]
\centering
\includegraphics[width=1\textwidth]{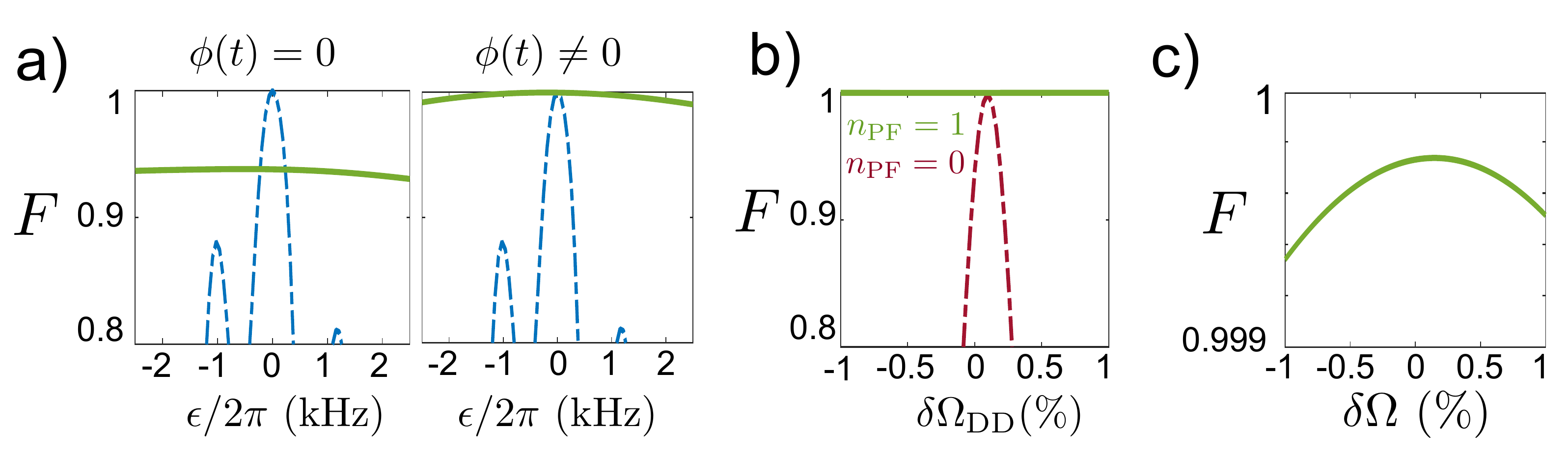}
\caption{Resilience to constant errors for $g_{B}=20.9$ T/m and $\nu=(2\pi)\times138$~kHz ($\eta=0.011$) and $\Omega=(2\pi)\times26$~kHz. (a) Bell state fidelity for $\Omega_\textrm{DD}=0$  (blue dashed curve) and $\Omega_\textrm{DD}=(2\pi)\times49$~kHz (green solid curve). Right and left panels show the cases with and without phase modulation respectively. (b) Bell state fidelity  for $n_{\textrm PF}=1$ (solid green curve) and for $n_{\textrm PF}=0$  (red dashed curve). (c) Bell state fidelity with respect to constant shifts in $\Omega(t)$.}\label{fig:StaticErrors}
\end{figure}

The presence of $\phi(t)$ changes Hamiltonian~(\ref{HMSDD}) to (see appendix~\ref{app:CompleteHamil})
\begin{equation}\label{HMSDDTP}
H= \eta \nu (be^{-i\nu t}+b^\dagger e^{i\nu t})S_z + \Omega\cos{(\delta t)}S^\parallel_\phi + \frac{\Omega_\textrm{DD}}{2}S^{\bot}_\phi,
\end{equation}
with $S^\parallel_\phi \equiv S^+e^{i\phi(t)}+\textrm{H.c.}$ and $S^\bot_\phi \equiv -iS^+e^{i\phi(t)}+\textrm{H.c.}$ In a rotating frame with respect to $-\frac{\dot{\phi}(t)}{2}S_z$ we find
\begin{equation}\label{HMSDDTPbt}
H=\Big\{ \eta \nu (be^{-i\nu t}+\textrm{H.c.})+\frac{\dot{\phi}(t)}{2}\Big\}S_z + \Omega\cos{(\delta t)}S_x+ \frac{\Omega_\textrm{DD}}{2}S_y.
\end{equation}
Now, in the bichromatic interaction picture, the previous Hamiltonian transforms as
\begin{eqnarray}\label{HMSIDDI}
\tilde{H}&=&\eta\nu (be^{-i\nu t}+\textrm{H.c.})\Big\{J_0S_z+2J_1\sin{(\delta t)}S_y\Big\} \\
&+&\frac{\tilde{\Omega}_\textrm{DD}}{2}S_y -\Omega_\textrm{DD}J_1^2/J_0\cos{(2\delta t)}S_y, \nonumber
\end{eqnarray}
where $\tilde{\Omega}_\textrm{DD}=J_0\Omega_\textrm{DD}(1+2J_1^2/J_0^2)$. Here we can see that, due to the action of $\phi(t)$, the interfering  $J_1\Omega_\textrm{DD}\sin{(\delta t)}S_z$ term is removed. Instead, in Eq.~(\ref{HMSIDDI}) we find the term $\Omega_\textrm{DD}J_1^2/J_0\cos{(2\delta t)}S_y$, which has a small coupling constant ($\Omega_\textrm{DD}J_1^2/J_0$) and that commutes with the gate Hamiltonian~(\ref{HMSII}). In Fig.~\ref{fig:StaticErrors}(a) (left panel) we show the obtained Bell state fidelity without phase modulation, i.e. by using Hamiltonian~(\ref{HMSDD}), for different values of a constant energy deviation $\epsilon$ in the qubit resonance frequencies. The blue dashed curve corresponds to the case $\Omega_\textrm{DD}=0$, where the scheme does not offer protection against $\epsilon$, while the green solid curve incorporates the driving leading to the carrier $\frac{\Omega_\textrm{DD}}{2}S_y$. Fig.~\ref{fig:StaticErrors}(a) (right panel) shows the case with phase modulation in Eq.~(\ref{HMSDDTP}) that achieves larger fidelities.

\begin{figure}[t!]
\centering
\includegraphics[width=0.8\textwidth]{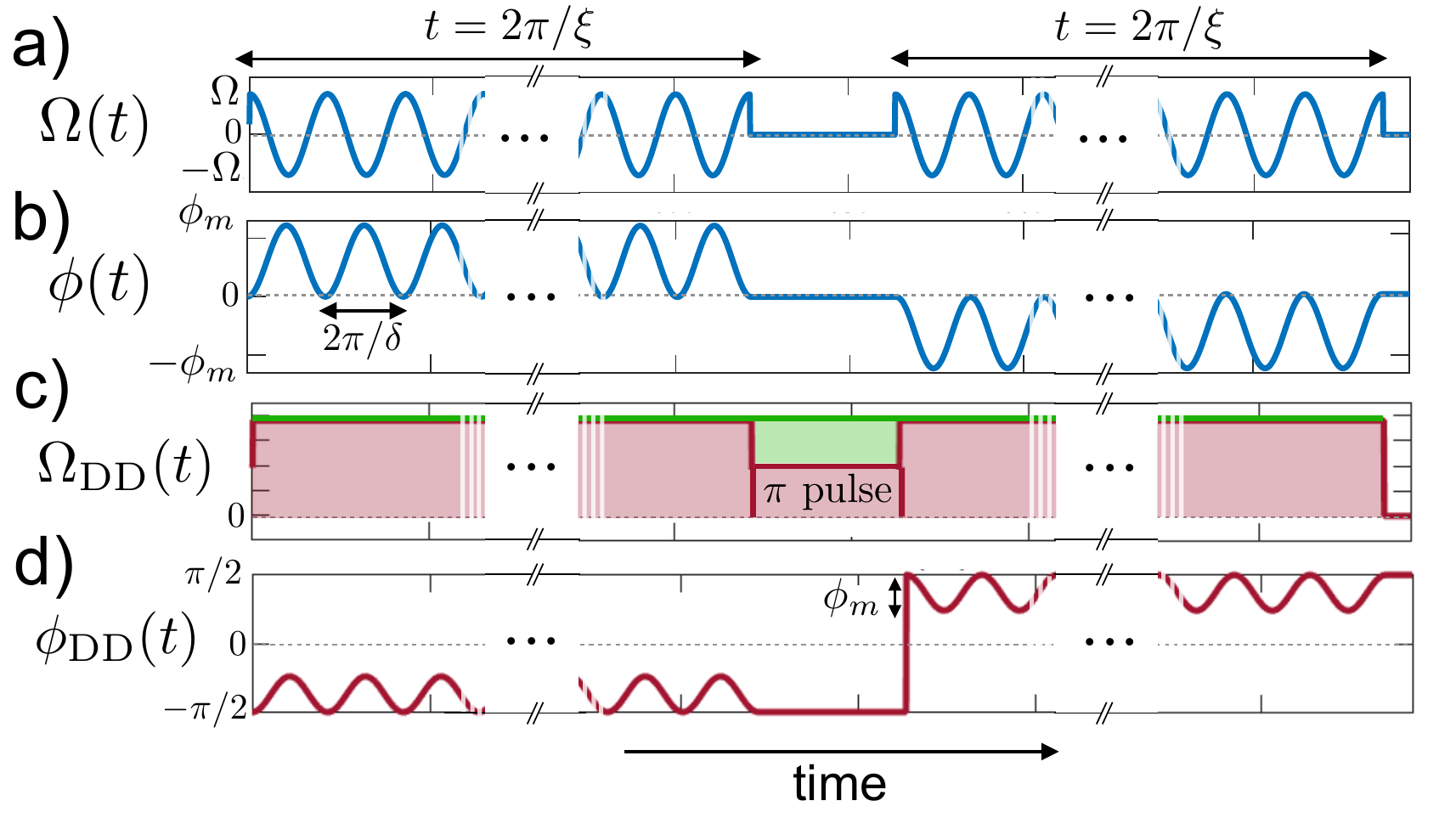}
\caption{Scheme of the control parameters. (a) Bichromatic driving $\Omega(t)=\Omega\cos{(\delta t)}$. (b) Phase modulation of the bichromatic driving. The change of sign during the second half of the evolution is due to the phase flip of the carrier driving. Here, $\phi_m=\max{(|\phi(t)|)}$ (c)  Carrier driving acting equivalently on both ions except in the middle, where each ion undergoes a $180^{\circ}$ (red) and $360^{\circ}$ (green) rotation respectively. (d) Phase modulation of the DD field. During the second part of the evolution, the phase is flipped from $-\pi/2$ to $\pi/2$. } \label{fig:MWScheme}
\end{figure}

Our method can be further improved since the first term in Eq.~(\ref{HMSIDDI}), i.e. $\eta\nu (be^{-i\nu t}+\textrm{H.c.}) J_0S_z$, leads to undesired accumulative effects that we can correct. The latter can be calculated in a rotating frame with respect to $\tilde{\Omega}_\textrm{DD}S_y/2$ and computing the second-order Hamiltonian, which reads
\begin{equation}\label{HMSDDIIPrecise}
\tilde{H}\approx H_{\textrm G} - g_{\tilde{\Omega}}(2b^\dagger b +1)S_y - \frac{g_{\nu}}{2}(S_x^2+S_z^2) ,
\end{equation}
where, $g_{\tilde{\Omega}}=\frac{\tilde{\Omega}_\textrm{DD}\eta^2J_0^2}{1-{\tilde{\Omega}^2}_\textrm{DD}/\nu^2}$, $g_\nu=\frac{\nu\eta^2J_0^2}{1-{\tilde{\Omega}^2}_\textrm{DD}/\nu^2}$, see appendix~\ref{app:SecondOrder}. Although small, the terms $ g_{\tilde{\Omega}}(2b^\dagger b +1)S_y $ and $ \frac{g_{\nu}}{2}(S_x^2+S_z^2)$ spoil a superior gate performance. Hence, we will eliminate them by introducing refocusing techniques.  In particular, to nearly remove the $g_{\tilde{\Omega}}(2b^\dagger b +1)S_y$ term, we divide the evolution in two parts, and flip the phase of the carrier driving during the second part of the evolution. In this respect, a scheme of the control parameters can be seen in Fig.~\ref{fig:MWScheme}. This phase flip causes a change in the sign of $\Omega_\textrm{DD}$, (i.e. $\Omega_\textrm {DD} \rightarrow -\Omega_\textrm{DD}$) which acts as a refocusing of unwanted shifts in $S_y$. This strategy is also valid to minimise the errors due to constant shifts in $\Omega_\textrm{DD}$ as it can be seen in Fig.~\ref{fig:StaticErrors}(b). Note that the phase flip of the carrier forces us to also change the sign of $\phi(t)$, since Eq.~(\ref{TP}) should hold during the implementation of the gate. As we will see later, performing a large number of phase flips ($n_\textrm{PF}$) will further suppress fluctuations on the carrier driving, while it also limits the possible values for $\Omega_\textrm{DD}$, check appendix~\ref{app:InteractionPic} for additional details.

A partial refocusing of the term $\frac{g_{\nu}}{2}(S_x^2+S_z^2)$ in Eq.~(\ref{HMSDDIIPrecise}) is also possible by rotating one of the qubits in the middle and at the end of the gate via $\pi$ pulses. In particular, if these $\pi$ pulses are performed along the $y$ axis, i.e., each $\pi$ pulse equals $\exp{(i\pi/2\sigma_1^y)}$, the $S_x^2$ and $S_z^2$ operators change their sign simultaneously, while $S_y^2$ remains unchanged. The combined action of phase flips and $\pi$ pulses allows us to approximate Eq.~(\ref{HMSDDIIPrecise}) as  $\tilde{H}\approx H_{\textrm G}$. It is noteworthy to mention that off-resonant vibrational modes would contribute with accumulative factors similar to the last term in Eq.~(\ref{HMSDDIIPrecise}). Then, these are refocused by the two $\pi$ pulses as we will show in our numerical simulations. As our method removes undesired effects due to additional vibrational modes, it is directly applicable to produce entangling gates between any two ions in a large chain. In addition, one can always consider to concatenate sequences of two-qubit operations leading to multi-qubit gates~\cite{Nielsen10}.
For completeness in our analysis, in Fig.~\ref{fig:StaticErrors}(c) we plot the Bell state fidelity with respect to constant shifts in $\Omega(t)$. Note that our scheme also shows robustness with respect to this kind of errors, since a shift in  $\Omega(t)$ rotates with frequency $\delta$ which diminishes its effect.

\subsection{Optimal refocusing and results} 
\begin{figure}[t!]
\centering
\includegraphics[width=0.7\textwidth]{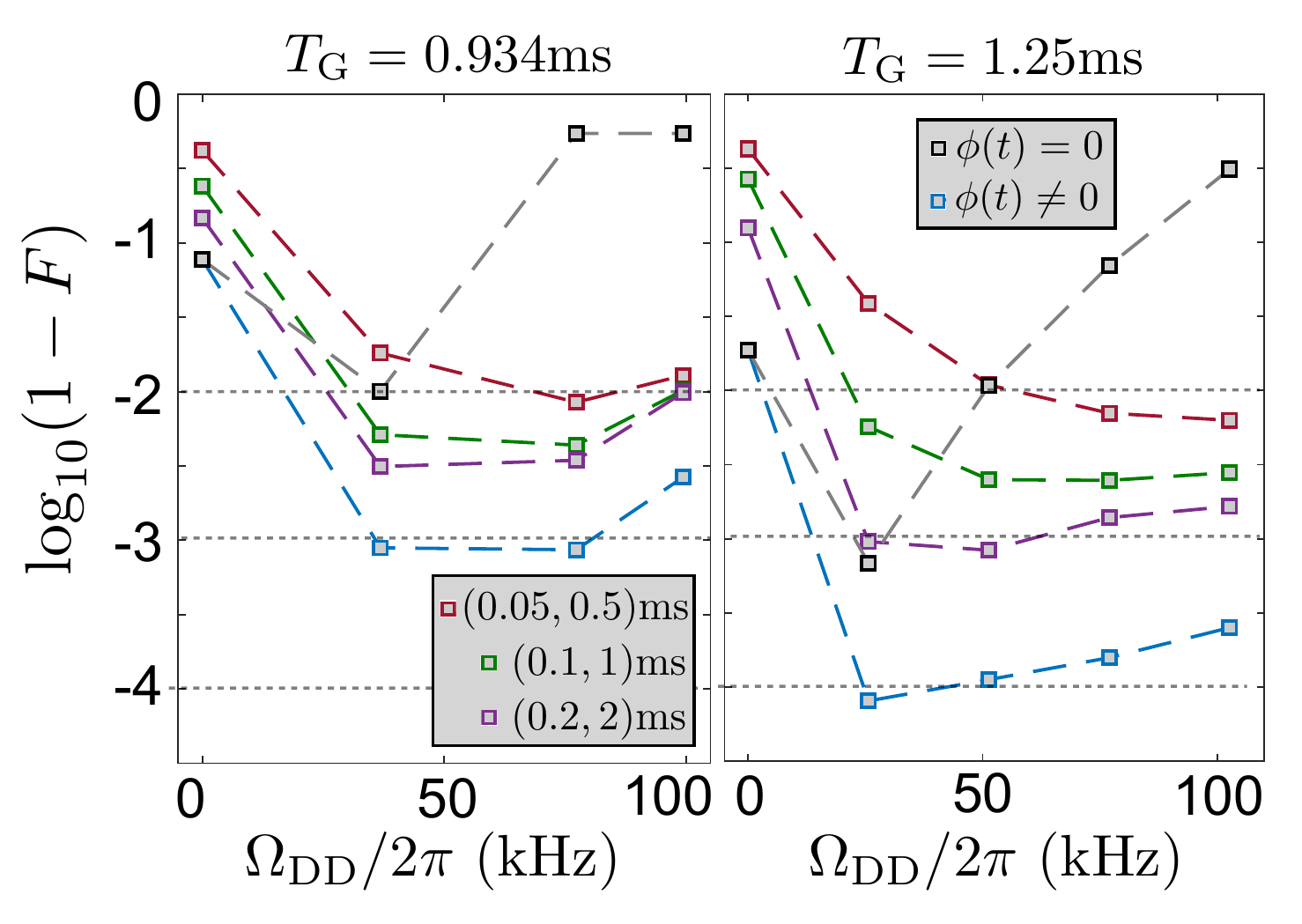}
\caption{Logarithm of the Bell state infidelity as a function of $\Omega_\textrm{DD}$, for $g_{B}=20.9$ T/m, $\nu=(2\pi)\times138$~kHz and $\Omega= (2\pi)\times37$~kHz  (left panel), and  $g_{B}=38.5$ T/m, $\nu=(2\pi)\times207$~kHz and $\Omega= (2\pi)\times26.6$~kHz  (right panel). Blue and black squares take into account crosstalk effects and the presence of the off-resonant vibrational mode, with and without phase modulation, respectively. Other curves include motional heating of the center-of-mass mode, and fluctuations of the magnetic field as well as the driving fields. The red, green and purple squares correspond to different error parameters $(\tau,T_2)=(0.05,0.5)$, $(0.1,1)$, and $(0.2,2)$ ms that characterise  magnetic field fluctuations.}\label{fig:FluctuatingErrors}
\end{figure}

To demonstrate the performance of our method in realistic experimental scenarios, we calculate the Bell state fidelity with fluctuating errors in both magnetic and driving fields, as well as in the presence of motional heating. Furthermore, our simulations include crosstalk terms, and the off-resonant breathing mode (the initial state of both motional modes is a thermal state with an average number of phonons $\bar{n}=1$). The results are shown in Fig.~\ref{fig:FluctuatingErrors} for two different parameter regimes. These are $g_{B}=20.9$ T/m, $\nu=(2\pi)\times138$~kHz and $\Omega= (2\pi)\times37$~kHz on the left figure, and $g_{B}=38.5$ T/m, $\nu=(2\pi)\times207$~kHz and $\Omega= (2\pi)\times26.6$~kHz on the right figure (note both regimes have a LD parameter $\eta=0.011$).

Blue squares indicate Bell state infidelities obtained with our method. Here, a total number of 31 phase flips were employed. Notice that, for ${\Omega}_\textrm{DD}=0$ fidelities are below $99\%$ even without fluctuating errors. We identify that this is due to the crosstalk of the MW driving fields at Rabi frequency $\Omega$, since these induce frequency shifts in the off-resonant qubits. If $\Omega_\textrm{DD} \neq 0$, these energy shifts are cancelled and we find fidelities ranging from $99,9\%$ to $99,99\%$. Note that these values are obtained using moderate MW radiation power, rather than with Rabi frequencies on the order of MHz as in section~\ref{sect:1_PDD}. The parameters in the right panel are more favourable for several reasons: first, the Rabi frequency is smaller than for the left case and the magnetic field gradient is larger; both lower crosstalk effects. Second, a larger trap frequency lowers the effect of the off-resonant mode. Finally, a smaller $\Omega/\delta$ ratio is also preferable to avoid any effect of higher-order Bessel functions. In this respect, note we always truncate the Jacobi-Anger expansion to the first order, see Eq.~(\ref{HMSIDDI}). Black squares indicate the infidelities obtained without phase modulation. As expected, phase modulation is crucial to remove energy shifts induced by ${\Omega}_\textrm{DD}$.

Other curves take into account the heating of the center-of-mass mode and fluctuating errors in the magnetic field and MW drivings. The effect of the former is introduced in our model with a dissipative term as the one described in Eq.~(\ref{MotionalHeating}) for two modes, whereas in this case $T=300$K was chosen. In the left panel, we consider a heating rate of $\dot{\bar{n}}\approx300$ ph/s~\cite{Weidt15,Weidt16,Brownnutt15}. For the right panel, we consider a more favourable scenario with $\dot{\bar{n}}\approx200$ ph/s. 

Magnetic and MW fluctuations are introduced via an Ornstein-Uhlenbeck (OU) stochastic process~\cite{Gillespie96}. Each point in Fig.~\ref{fig:FluctuatingErrors} corresponds to 100 realisations. The OU process is characterised by the correlation time $\tau_{B}$ and coherence time $T_2$ for the magnetic field fluctuations, while $\tau_\Omega$ and the relative amplitude error $\delta_{\Omega}$ are used for the MW driving field fluctuations. For the driving fields, we choose a correlation time of $\tau_B=500~\mu$s and a relative amplitude error of $\delta_{\Omega}=0.5\%$ in the left panel, and a correlation time of $\tau_\Omega=1$~ms and $\delta_{\Omega}=0.25\%$ in the right panel~\cite{Cohen17}. Different strengths for the magnetic field fluctuations are given by the red, green and purple squares, with parameters $(\tau,T_2)=(0.05,0.5)$, $(0.1,1)$, and $(0.2,2)$~ms, respectively.

Our numerical simulations predict fidelities above $99\%$ even for the worst case corresponding to $(\tau,T_2)=(0.05,0.5)$ ms and with current heating rates (left panel). In a more optimistic experimental scenario with $(\tau,T_2)=(0.2,2)$ ms, our protocol leads to fidelities larger than $99,9\%$ for distinct values of $\Omega_\textrm{DD}$ (right panel). 

In this section, we have proposed a method for the generation of  entangling gates that combines phase-modulated continuous MW  drivings with phase flips and refocusing $\pi$ pulses to produce  entangling gates with high fidelity.  Numerical simulations including the main sources of decoherence show that fidelities on Bell-state  preparation exceeding 99\% are possible within current experimental limitations, while  fidelities  larger  than  99.9\% are achievable  with further experimental improvement.

\chapter{Quantum Simulation of Light-Matter Interactions}
\label{chapter:chapter_2}
\thispagestyle{chapter}

Understanding the interactions that emerge among two-level atoms (qubits) and bosonic field modes is of major importance for the development of quantum technologies. The qubit-boson interaction governs the dynamics of distinct quantum platforms such as cavity QED~\cite{Haroche89,Raimond01}, trapped ions~\cite{Leibfried03} or superconducting circuit~\cite{Clarke08}, that can achieve the SC regime. Here, the qubit-boson Rabi coupling $g$ is usually much smaller than the field frequency, but larger than the coupling to the environment. In these conditions, the JC model~\cite{Jaynes63} that appears after applying the RWA provides an excellent description of the system. In the last decade, experiments in circuit QED have achieved couplings well above the USC regime ($g/\omega\gtrsim0.1$)~\cite{Yoshihara17}, hindering a perturbative treatment of the QRM. In the DSC regime, ($g/\omega\gtrsim1$) the full QRM has to be considered, and the associated physics is different from the one described by JC model~\cite{Casanova10}. 

In this chapter, we study two different extensions of the QRM, namely the Rabi-Stark model and the nonlinear QRM. In the former, a Stark coupling term is added to the QRM, which leads to multi-photon selective interactions in the SC and USC regimes. The latter is natural to the implementation of the QRM in trapped ions, when moving beyond the LD regime. 

\section{Selective interactions in the Rabi-Stark model}

The QRM with a Stark coupling term, named the Rabi-Stark model~\cite{Eckle17} (in the following we will also use that denomination) was first considered by Grimsmo and Parkins~\cite{Grimsmo13,Grimsmo14}. The study of its energy spectrum~\cite{Eckle17,Maciejewski14,Xie19} has revealed some interesting features such as a spectral collapse or a first-order phase transition~\cite{Xie19}, which connects it with the two-photon QRM~\cite{Travenec12,Maciejewski15,Travenec15,Felicetti18_1,Felicetti18_2,Cong19,Xie17} or with the anisotropic QRM~\cite{Xie14}. On the other hand, dynamical features of the JC model with a Stark coupling term have been studied in the past~\cite{Pellizzari94,Solano00,Franca01,Solano05,Franca05,Prado13}. The Stark coupling is useful to restrict the resonance condition and the Rabi oscillations to a preselected JC doublet, leaving the other doublets in a dispersive regime. This selectivity has found applications for state preparation and reconstruction of the bosonic modes in cavity QED~\cite{Pellizzari94,Franca01} or trapped ions~\cite{Solano00,Solano05,Franca05}. In light of the above, the dynamical study of the full QRM with a Stark coupling term in the SC and USC regimes is well justified.

In this section, we study the dynamical behaviour of the QRM with a Stark term, i.e. the Rabi-Stark model, and show that the interplay between the Stark and Rabi couplings gives rise to selective $k$-photon interactions in the SC and USC regimes. Note that, previously, $k$-photon (or multiphoton) resonances have been investigated in the linear QRM~\cite{Ma15,Garziano15}, driven linear qubit-boson couplings~\cite{Nha00,Chough00,Klimov04,Casanova18QRM,Puebla19_1} or nonlinear couplings~\cite{Shore93,Vogel95}, and recently have found applications for quantum information science~\cite{Macri18,Boas19}. In our case, $k$-photon transitions appear as higher-order processes of the linear QRM, while the Stark coupling is responsible for the selective nature of these interactions. This section is organised as follows: In section~\ref{subsect:OnePhoton} the Rabi-Stark model is introduced and we review the selective one-photon interactions that appear. In section~\ref{subsect:MultiPhoton} we use time-dependent perturbation theory to characterise the emergent $k$-photon interactions whose strength scales as $(g/\omega)^k$. Finally, in section~\ref{subsect:QRSTI} we introduce a method to simulate the Rabi-Stark model in a wide parameter regime using a single trapped ion. Moreover, we validate our proposal with numerical simulations which show an excellent agreement between the dynamics of the Rabi-Stark model and the one achieved by the trapped-ion simulator.

\subsection{Selectivity in one-photon interactions}\label{subsect:OnePhoton}

The Hamiltonian of the Rabi-Stark model is
\begin{equation}\label{QRS1}
H=\frac{\omega_0}{2}\sigma_z +\omega a^\dag a + \gamma a^\dag a \sigma_z + g(\sigma_+ +\sigma_-)(a+a^\dag)
\end{equation}
where $\omega_0$ is the frequency of the qubit or two-level system, $\omega$ is the frequency of the bosonic field, and $\gamma$ and $g$ are the couplings of the Stark and Rabi terms, respectively. Note that the Stark term is diagonal in the bare basis $\big\{|{\textrm e}\rangle,|{\textrm g}\rangle\big\}\otimes|n\rangle$ (where $\sigma_z|{\textrm e}\rangle=|{\textrm e}\rangle$, $\sigma_z|{\textrm g}\rangle=-|{\textrm g}\rangle$ and $a^\dagger a|n\rangle=n|n\rangle$), and it can be interpreted as a qubit energy shift that depends on the bosonic state. If we move to an interaction picture with respect to the first three terms in Eq.~(\ref{QRS1}), the system Hamiltonian reads (see appendix~\ref{app:QRSDyson} for additional details)
\begin{equation}\label{QRSIntPic}
H_I(t)=\sum_{n=0}^{\infty}\Omega_n(\sigma_+ e^{i\delta^+_{n} t}+\sigma_- e^{i\delta^-_{n} t})|n\!+\!1 \rangle\langle n| + \textrm{H.c.}
\end{equation}
where $\Omega_n=g\sqrt{n+1}$, $\delta_n^+=\omega+\omega^0_n$ and $\delta_n^-=\omega-\omega_n^0$, with $\omega_n^0=\omega_0+\gamma(2 n +1)$. If $\gamma=0$, these detunings are independent of the state $n$, and, for $|\delta^+| \gg \Omega_n$ and $\delta^-=\omega-\omega_0=0$ ($|\delta^-| \gg \Omega_n$ and $\delta^+=\omega+\omega_0=0$), a resonant JC (anti-JC) Hamiltonian is recovered when fast rotating terms are averaged out by invoking the RWA. In these conditions, the dynamics leads to Rabi oscillations between the states $|{\textrm e},n\rangle \leftrightarrow |{\textrm g},n+1\rangle$ ($|{\textrm g},n\rangle \leftrightarrow |{\textrm e},n+1\rangle$) for every $n$, and at a rate proportional to $\Omega_n$. These interactions are not selective as they apply to all Fock states in the same manner. 
\pagebreak

\begin{figure}
\centering
\includegraphics[width=0.8\textwidth]{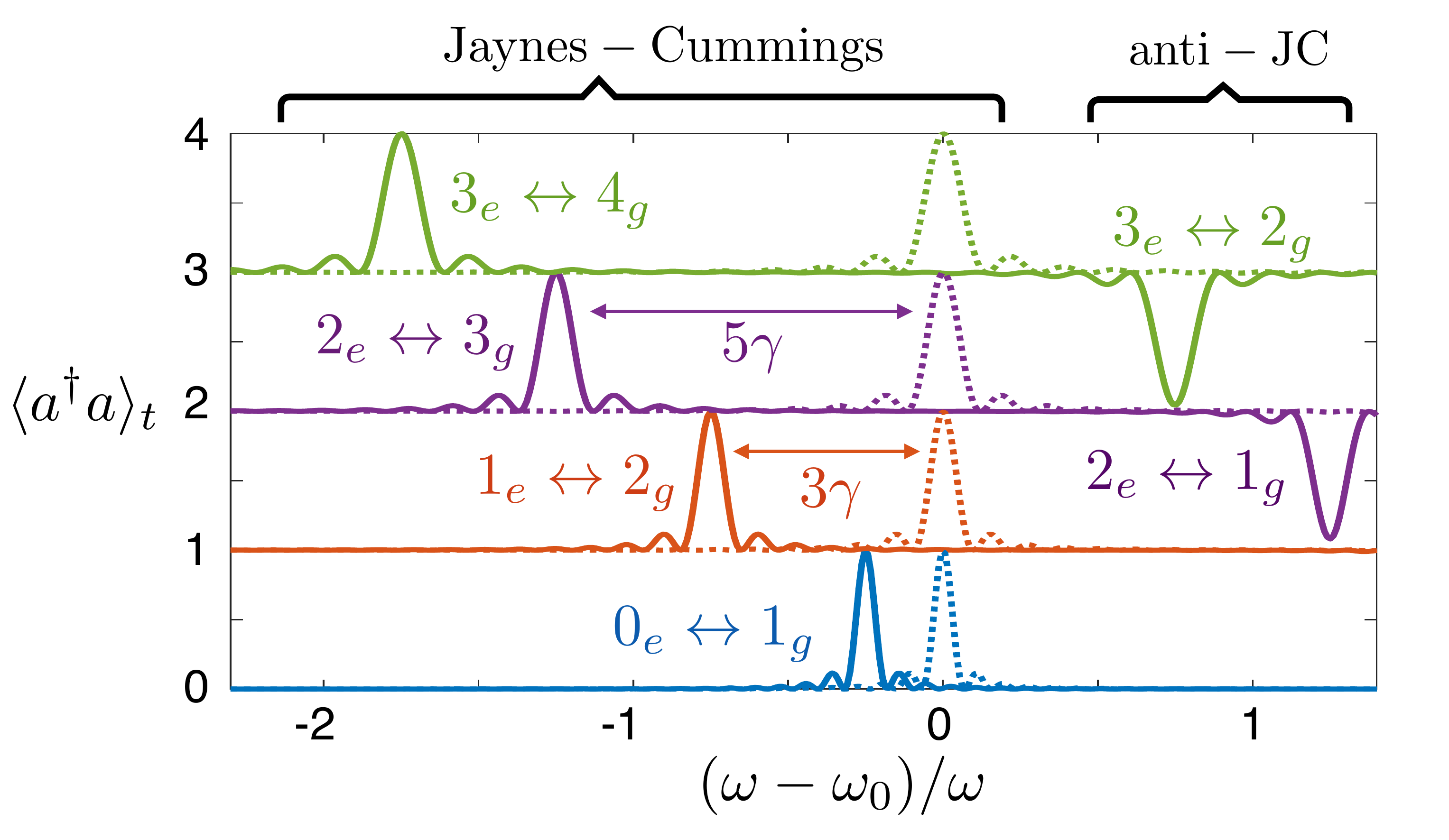}
\caption{One-photon selective interactions of the Rabi-Stark model. Hamiltonian (\ref{QRS1}) acts during a time $t=\pi/2\Omega_n$ and we calculate $\langle a^\dagger a\rangle$ for different ratios of $\omega_0/\omega$ and initial states $|\textrm{e},0\rangle$ (blue), $|\textrm{e},1\rangle$ (orange), $|\textrm{e},2\rangle$ (purple) and $|\textrm{e},3\rangle$ (green) with fixed couplings $\gamma/\omega=-0.25$ and $g/\omega=0.02$ (solid lines). If $\gamma=0$, all JC peaks would be at $\omega-\omega_0=0$ (dashed lines).}\label{fig:QRSOnePhoton}
\end{figure}


The presence of a nonzero Stark coupling $\gamma$ makes these detunings dependent on $n$, allowing to identify a resonance condition for a selected Fock state $n=N_0$, while the rest of Fock states stay out of resonance. From Eq.~(\ref{QRSIntPic}) we note that if $\delta^{-}_{N_0}=0$ ($\delta^{+}_{N_0}=0$) and $|\delta^{-}_{n\neq N_0}|\gg \Omega_{n\neq N_0}$ ($|\delta^{+}_{n\neq N_0}|\gg \Omega_{n\neq N_0}$), the dynamics of Hamiltonian~(\ref{QRS1}) will produce a resonant one-photon JC (anti-JC) interaction only in the subspace $\{|{\textrm e}\rangle,|{\textrm g}\rangle\}\otimes\{|N_0\rangle,|N_0+1\rangle\}$.  This is observed in Fig.~\ref{fig:QRSOnePhoton}, where resonance peaks appear for initial states $|{\textrm e},n\rangle$ with different number $n$. Here, a one-photon Rabi oscillation occurs if $\omega-\omega_0=\gamma(2n+1)$ i.e. $\delta^{-}_{n}=0$. In Fig.~\ref{fig:QRSOnePhoton}, we vary $(\omega-\omega_0)/\omega$ in the $x$ axis for fixed $\gamma/\omega=-0.25$ and $g/\omega=0.02$, and meet this resonance condition for $n=0,1,2,3$ that correspond to the four peaks on the left side (solid lines). The other two peaks on the right correspond to $\delta^{+}_{n}=0$ resonances leading to one-photon anti-JC interactions for $n=1,2$.

\subsection{Multi-photon interactions}\label{subsect:MultiPhoton}

As revealed previously, besides one-photon transitions, the Rabi-Stark Hamiltonian produces selective $k$-photon interactions. Unlike the selective one-photon interactions, which appear due to the interplay between the Stark term and the rotating or counter-rotating terms, these selective multi-photon interactions are a direct consequence of the interplay between the Stark term and both the rotating and counter-rotating terms. Calculating the Dyson series for Eq.~(\ref{QRSIntPic}), we obtain that the second order Hamiltonian is 
\begin{equation}\label{QRSSecondOrder1}
H_I^{(2)}=\sum_{n=0}^{\infty}\big(\Delta_n^{\textrm e}\sigma_+\sigma_- +\Delta_n^{\textrm g}\sigma_-\sigma_+ \big)|n\rangle\langle n|
\end{equation}
where $\Delta_n^{\textrm e}=\Omega^2_{n-1}/\delta^+_{n-1}-\Omega^2_{n}/\delta_n^-$ and $\Delta_n^{\textrm g}=\Omega^2_{n-1}/\delta^-_{n-1}-\Omega^2_n/\delta_n^+$, plus a time-dependent part oscillating with frequencies $\delta_{n+1}^++\delta_n^-=2\omega+2\gamma$, $\delta_{n+1}^-+\delta_n^+=2\omega-2\gamma$, and $\delta_{n}^\pm,\delta_{n+1}^\pm$ that is averaged out due to the RWA (see appendix~\ref{subapp:QRSSecondOrder} for the derivation of Eq.~(\ref{QRSSecondOrder1})).

The third order Hamiltonian leads to three-photon transitions described by the following Hamiltonian (see appendix~\ref{subapp:QRSThirdOrder} for the derivation)
\begin{equation}\label{QRSThirdOrder1}
H_I^{(3)}(t)=\sum_{n=0}^{\infty}\big(\Omega^{(3)}_{n+}e^{i\delta^{(3)}_{n+}t}\sigma_++\Omega^{(3)}_{n-}e^{i\delta^{(3)}_{n-}t}\sigma_-\big)|n\!+\!3\rangle \langle n| + \textrm{H.c.},
\end{equation}
where $\Omega^{(3)}_{n\pm}=g^3\sqrt{(n+3)!/n!}/2\delta^\pm_n(\omega\mp\gamma)$ and $\delta^{(3)}_{n\pm}=\delta^\pm_{n+2}+\delta^\mp_{n+1}+\delta_n^\pm=2\omega+\delta_{n+1}^\pm$. According to this, a JC type three-photon process occurs for $|{\textrm e},N_0\rangle$ if $\delta^{(3)}_{N_0-}=0$ producing population exchange between the states $|{\textrm e}, N_0\rangle\leftrightarrow|{\textrm g}, N_0+3\rangle$. For the state $|{\textrm g}, N_0\rangle$, anti-JC-type transitions to the state $|{\textrm e}, N_0+3\rangle$ occur when $\delta^{(3)}_{N_0+}=0$. In the following we will check the validity of these effective Hamiltonians by numerically calculating the dynamics of Hamiltonian~(\ref{QRS1}).

\begin{figure}
\centering
\includegraphics[width=1.0\textwidth]{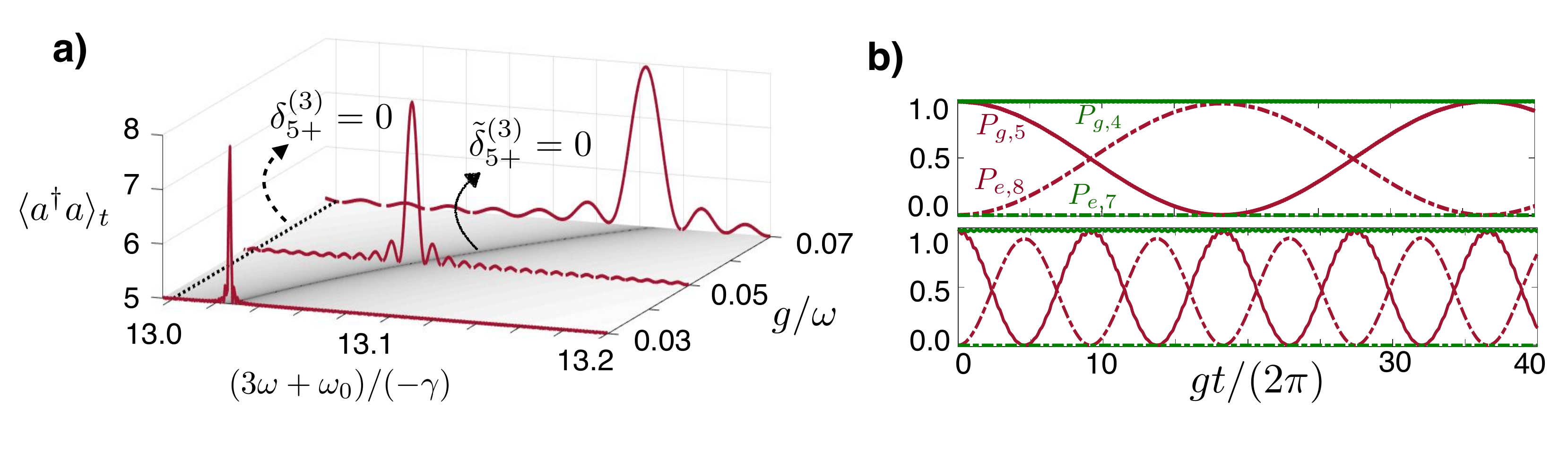}
\caption{Three-photon selective interactions of the Rabi-Stark model. (a) Resonance spectrum of anti-JC-like three-photon process for state $|{\textrm g},5\rangle$. After a time $t=\pi/2\Omega^{(3)}_{5+}$, $\langle a^\dagger a\rangle$ is shown for $\gamma/\omega=-0.4$. The peaks appear shifted from $\delta_{5+}^{(3)}=0$ (dashed line) at $\tilde{\delta}_{5+}^{(3)}=0$ which corresponds to the dark curve in the $xy$ plane representing the lower values of $\log_{10}{|\delta^{(3)}_{5+}|}$. (b) Time evolution of populations $P_{{\textrm g},4}$ (solid) and $P_{{\textrm e},7}$ (dashed) for initial state $|{\textrm g},4\rangle$ (green) and populations $P_{{\textrm g},5}$ and $P_{{\textrm e},8}$ for initial state $|{\textrm g},5\rangle$ (red) for $g/\omega=0.05$ (up) and $g/\omega=0.1$ (down).}
\label{fig:QRSThreePhoton}
\end{figure}

In Fig.~\ref{fig:QRSThreePhoton}(a) we let the system evolve for a time $t=\pi/2\Omega^{(3)}_{5+}$ for a fixed value of $\gamma/\omega=-0.4$ and calculate the average number of photons $\langle a^\dagger a\rangle$ for different values of $\omega_0/\omega$ and couplings $g/\omega$. We do this for the initial state $|{\textrm g},N_0=5\rangle$, near the resonance point $\delta^{(3)}_{5+}=3\omega+\omega_0+13\gamma=0$. We observe that  resonances do not appear when $\delta^{(3)}_{5+}=0$, see dashed line on the left, owing to a resonance frequency shift that depends on the value of $g$. To explain this we go to an interaction picture with respect to Hamiltonian~(\ref{QRSSecondOrder1}), then, the oscillation frequencies in Eq.~(\ref{QRSThirdOrder1}) will be shifted to $\tilde{\delta}^{(3)}_{n+}=\delta^{(3)}_{n+}+\Delta^\textrm{e}_{n+3}-\Delta_n^\textrm{g}$ and $\tilde{\delta}^{(3)}_{n-}=\delta^{(3)}_{n-}+\Delta^\textrm{g}_{n+3}-\Delta_n^\textrm{e}$. In the $xy$ plane of Fig.~\ref{QRSThirdOrder1}(a) we make a grayscale colour plot of $\log_{10}{|\tilde{\delta}^{(3)}_{5+}|}$ as a function of $\omega_0$ and $g$ and see that the minima of $\tilde{\delta}^{(3)}_{5+}$ (dark line) is in very good agreement with the point in which the three-photon resonance appears (the logarithm scale is used to better distinguish the zeros of $\tilde{\delta}^{(3)}_{5+}$). 

To show that the three-photon interaction applies only to the preselected subspace, in Fig.~\ref{fig:QRSThreePhoton}(b) we plot the evolution of initial states $|{\textrm g},4\rangle$ and $|{\textrm g},5\rangle$. As expected, the latter exchanges population with the state $|{\textrm e},8\rangle$ while the former remains constant. Besides, for $g/\omega=0.05$ (upper figure), the transition is slower but most of the population is transferred to $|{\textrm e},8\rangle$ at time $t=\pi/2\Omega^{(3)}_{5+}$. For $g/\omega=0.1$ (lower figure) the exchange rate is much faster but the transfer is not so efficient. 

In this context, higher-order selective interactions will be produced by the Rabi-Stark model and could in principle be tracked by the calculation of higher-order Hamiltonians. However, being a high-order process, its strength decreases with order $k$ since $\Omega^{(k)}/\omega\propto (g/\omega)^k$. Then, high-order processes require longer times to be observed which may exceed the coherence times of the system. In any case, we find interesting to study the case for a higher $k$. Following the same procedure as for calculating Eqs.~(\ref{QRSSecondOrder1}) and (\ref{QRSThirdOrder1}), we conclude that for even $k$, the $k$-th order Hamiltonian will not produce selective interactions as they will average out as a consequence of the RWA. For odd $k$, the $k$-th order Hamiltonian predicts a $k$-photon transition of the form
\begin{equation}\label{QRSkthOrder1}
H_I^{(k)}(t)=\sum_{n=0}^{\infty}\big(\Omega^{(k)}_{n+}e^{i\delta^{(k)}_{n+}t}\sigma_++\Omega^{(k)}_{n-}e^{i\delta^{(k)}_{n-}t}\sigma_-\big)|n\!+\!k\rangle\langle n| + \textrm{H.c.},
\end{equation}
where 
\begin{equation}\label{QRSdet}
\delta^{(k)}_{n\pm}=\sum_{s=0}^{k-1} \delta_{n+s}^\pm+\delta_{n+s+1}^\mp+\delta^\pm_{k}=(k-1)\omega+\delta^{\pm}_{n+(k-1)/2}
\end{equation}
and 
\begin{equation}\label{QRSRabiFreq}
\Omega^{(k)}_{n\pm}=\frac{g^k}{(k-1)!!(\omega\mp\gamma)^{\frac{k-1}{2}}}\sqrt{\frac{(n+k)!}{n!}}\prod^{k-2}_{s=1,3...}\frac{1}{\delta^{(s)}_{n\pm}}.
\end{equation}
Using Eqs.~(\ref{QRSdet}) and (\ref{QRSRabiFreq}) and with the help of numerical simulations, it is easy to find $k$-photon processes to validate the effective Hamiltonian (\ref{QRSkthOrder1}). Here, numerical simulations are required as the analytic calculation of the exact resonance frequencies of higher-order processes rapidly becomes challenging. For example, for tracking a JC-type five-photon interaction for $N_0$, we use the condition $\delta^{(5)}_{N_0-}=0$ to retrieve an approximate value for the qubit frequency of $\omega^c_0= 5\omega-\gamma(2N_0+5)$. Then, we calculate the time evolution governed by Hamiltonian~(\ref{QRS1}) for a time $t=\pi/2\Omega^{(5)}_{N_0-}$ and plot $\langle\sigma_+\sigma_-\rangle$ for different values of $\omega_0$ close to $\omega_0^c$ until we find a peak corresponding to the resonant five-photon interaction.

\begin{figure}[t!]
\centering
\includegraphics[width=0.78\linewidth]{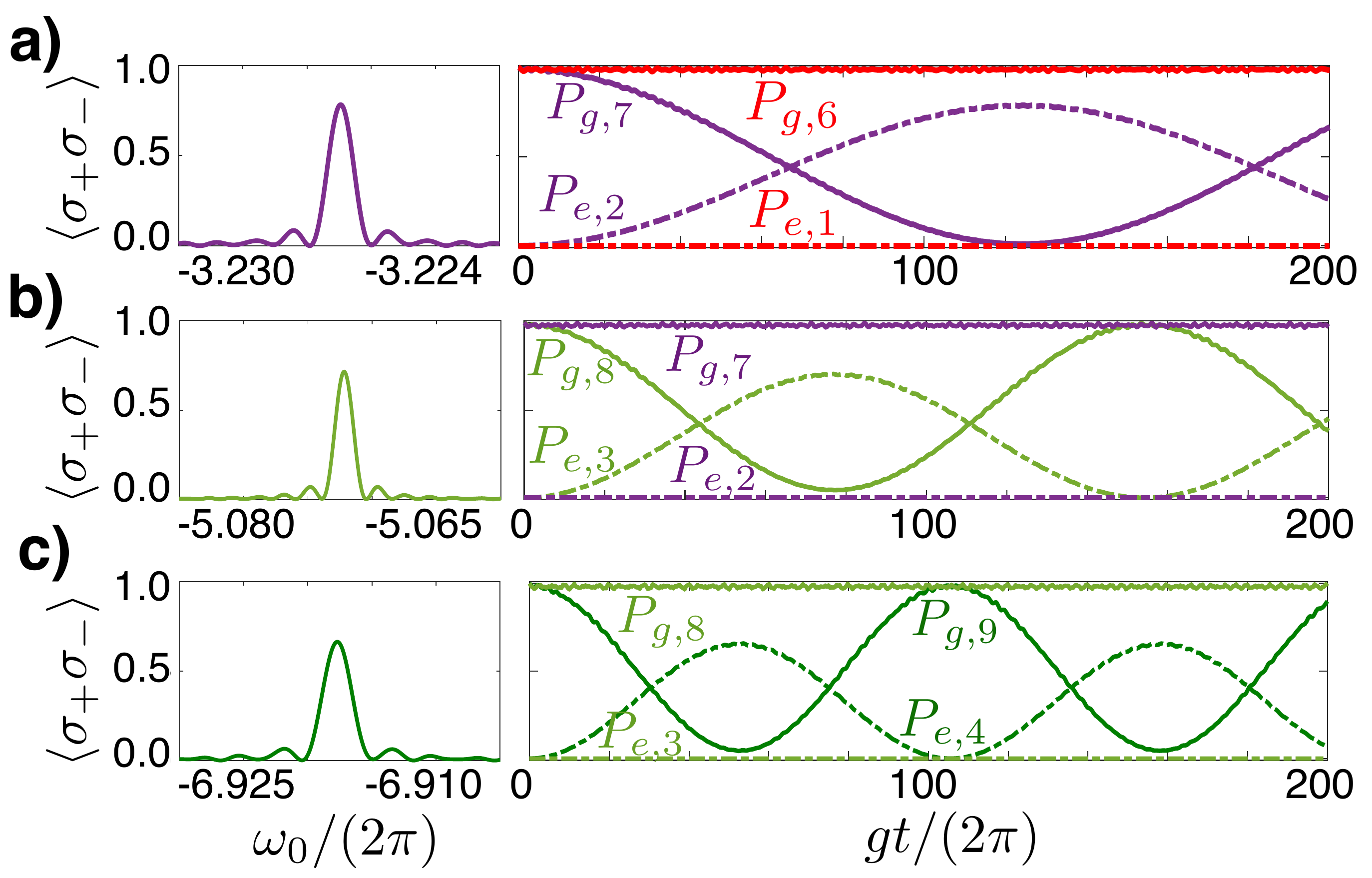}
\caption{Five-photon selective interactions of the Rabi-Stark model. a) On the left, $\langle\sigma_+\sigma_-\rangle$ is shown after a time $t=\pi/2\Omega^{(5)}_{2-}$, for different values of $\omega_0/\omega$ around $\omega_0^c=5\omega-\gamma(2\times2+5)$ and initial state $|{\textrm g},7\rangle$. Here, $g/\omega=0.1$ and $\gamma/\omega=0.9$. On the right, time evolution of populations $P_{{\textrm e},2}$ and  $P_{{\textrm g},7}$ for initial state $|{\textrm g},7\rangle$ and populations  $P_{{\textrm e},1}$ and  $P_{{\textrm g},6}$ for initial state $|{\textrm g},8\rangle$, for $\omega_0/\omega=-3.227$. b) and c) The same procedure with initial state $|{\textrm g},8\rangle$ and $|{\textrm g},9\rangle$, where the peaks appear for $\omega_0/\omega=-5.072$ and $\omega_0/\omega=-6.918$. } 
\label{fig:QRSFivePhoton}
\end{figure}

As an example, in Fig.~\ref{fig:QRSFivePhoton} we show these resonances for $N_0=2$, $3$ and $4$, with $g/\omega=0.1$ and $\gamma/\omega=0.9$. We find resonance peaks for $\omega_0/\omega=-3.227$,$-5.072$ and $-6.918$ which are close to the ones obtained with the approximate formula $\omega_0^c/\omega=-3.1$,$-4.9$ and $-6.7$. In comparison with the three-photon processes, five-photon transitions are slower, and the population transfer to the preselected state is partial for $g/\omega=0.1$. It is interesting to note that the revival of the initial state as well as the selectivity condition are maintained at the beginning of the USC regime. Note that for $\omega_0/\omega=-3.227$, an exchange between states $|{\textrm g},7\rangle \leftrightarrow |{\textrm e},2\rangle$ occurs while the neighbouring states $ |{\textrm g},6\rangle$ and  $|{\textrm e},1\rangle$ are completely out of resonance. In this respect, with larger coupling constants such as $g/\omega\approx0.3$ one would still get signatures of selectivity, but the interaction will not longer be a JC (or anti-JC) type $k$-photon interaction as it would involve states out of the selected JC (or anti-JC) doublet. In Fig.~\ref{fig:QRSFivePhoton} the population transfer from $|{\textrm g},N_0+5\rangle$ to $|{\textrm e},N_0\rangle$ is already partial, and interestingly, the remaining population goes to states $|{\textrm g},N_0+1\rangle$ and $|{\textrm g},N_0-1\rangle$.

To experimentally verify our predictions regarding the selective $k$-photon interactions of the Rabi-Stark model, in the next section we propose an experimental implementation of the model.

\subsection{Implementation with trapped ions}\label{subsect:QRSTI}

Trapped ions are excellent quantum simulators~\cite{Leibfried03,Blatt12}, with experiments implementing the one-photon QRM~\cite{Pedernales15,Puebla16,Lv18} and proposals for the two-photon QRM~\cite{Felicetti15,Puebla17}. In the following, we propose a route to simulate the Rabi-Stark model using a single trapped ion. 

The Hamiltonian of a single trapped ion interacting with co-propagating laser beams labeled with $j$ can be written, in an interaction picture with respect to the free energy Hamiltonian $H_0=\frac{\omega_I}{2}\sigma_z+\nu a^\dagger a$, as~\cite{Leibfried03}  
\begin{equation}\label{TIHamil}
H=\sum_{j}\frac{\Omega_j}{2} \sigma^+e^{i\eta(ae^{-i\nu t}+a^\dag e^{i\nu t})}e^{-i(\omega_j-\omega_I)t}e^{i\phi_j} +\textrm {H.c.}.
\end{equation}
Here $\Omega_j$ is the Rabi frequency, $\eta$ and $a^\dagger (a)$ are the LD parameter and the creation (annihilation) operator acting on vibrational phonons, $\nu$ is the trap frequency, $\omega_j-\omega_I$ is the detuning of the laser frequency $\omega_j$ with respect to the carrier frequency $\omega_I$, and $\phi_j$ accounts for the phase of the laser. 

\begin{figure}[t!]
\centering
\includegraphics[width=1\linewidth]{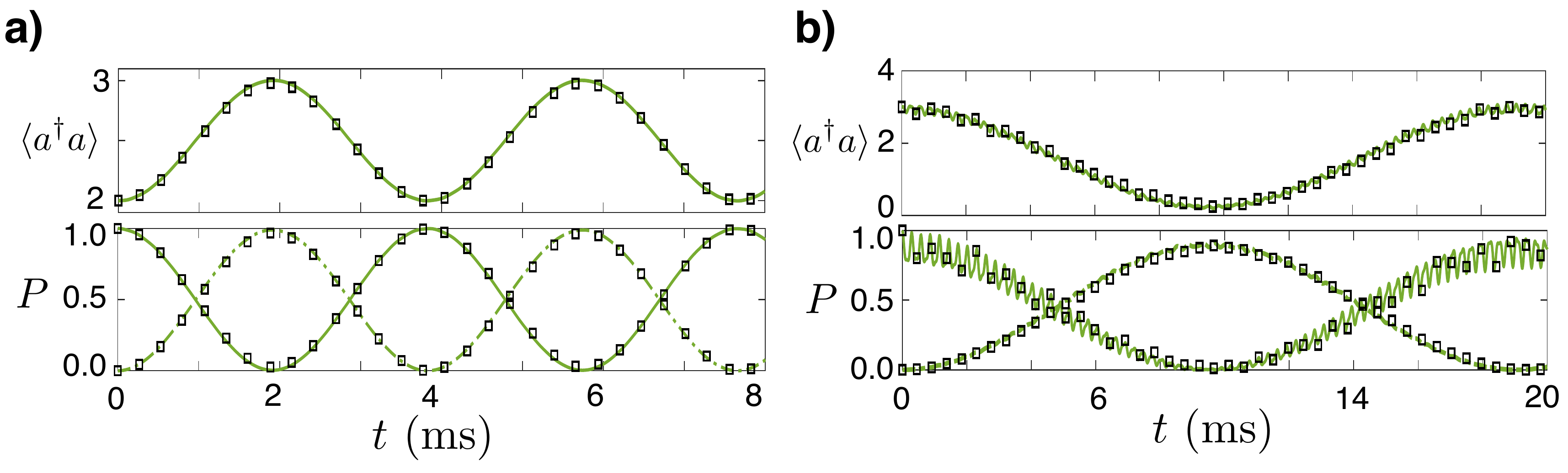}
\caption{Selective one-photon and three-photon interactions with a trapped ion. a) Time evolution of the mean number of phonons and populations $P_{+,2}$ (solid) and $P_{-,3}$ (dashed) starting from state $|+,2\rangle$ for $g/\omega^{\textrm R}=0.05$, $\gamma/\omega^{\textrm R}=-0.4$ and $\omega_0^{\textrm R}/\omega^{\textrm R}=3$. b) Time evolution of the mean number of phonons and populations $P_{+,3}$ (solid) and $P_{-,0}$ (dashed) starting from state $|+,3\rangle$ for $g/\omega^{\textrm R}=0.3$, $\gamma/\omega^{\textrm R}=-0.1$ and $\omega_0^{\textrm R}/\omega^{\textrm R}=-2.4385$. Solid green lines evolve according to Eq.~(\ref{TIQRS}) while black squares evolve according to Eq.~(\ref{TIHamil}).} 
\label{fig:QRSTI}
\end{figure}

As a possible implementation of the Rabi-Stark model we consider two drivings acting near the first red and blue sidebands, and a third one on resonance with the carrier interaction $\omega_{\textrm S}=\omega_I$. The Hamiltonian in the LD regime, $\eta \sqrt{\langle a^\dagger a\rangle}\ll 1$, and after the vibrational RWA approximation, reads
\begin{equation}\label{Scheme2}
H_\textrm{LD}=-ig_ra \sigma^+ e^{-i\delta_rt}  -ig_b a^\dagger \sigma^+ e^{-i\delta_bt} -  \hat{g}_{\textrm S}\sigma^++\textrm{H.c.}
\end{equation}
where $\omega_{r,b}=\omega_I\mp \nu +\delta_{r,b}$,  $g_{r,b}=\eta\Omega_{r,b}/2$, $\phi_{r,b,{\textrm S}}=-\pi$ and $\hat{g}_{\textrm S}=\frac{\Omega_{\textrm S}}{2}(1-\eta^2/2)-\frac{\Omega_{\textrm S}}{2}\eta^2a^\dag a=\frac{\Omega_0}{2}-\gamma^{\textrm R} a^\dagger a$. Dependence of the carrier interaction on the phonon number appears when considering the expansion of $e^{i\eta(a+a^\dag)}$ up to the second order in $\eta$. At this point, if $\delta_r=-\delta_b=\omega^{\textrm R}$, Eq.~(\ref{Scheme2}) can already be mapped to a Rabi-Stark model in a frame rotated by $-\omega^{\textrm R}a^\dagger a$. However, the engineered Hamiltonian cannot explore all regimes of the model, as $\Omega_0$ and $\gamma^{\textrm R}$ cannot be independently tuned, thus restricting the Hamiltonian to regimes where $\gamma^{\textrm R}\ll\Omega_0$. This issue can be solved by moving to an interaction picture with respect to $\frac{\Omega_\textrm{DD}}{2}\sigma_x - \omega^{\textrm R}a^\dagger a$, where $\Omega_\textrm{DD}=-(\Omega_0+\omega_0^{\textrm R})$, and by shifting the detunings by $\delta_{r,b}=\Omega_\textrm{DD}\pm\omega^{\textrm R}$. The resulting Hamiltonian after neglecting terms oscillating at $\Omega_\textrm{DD}$ is
\begin{equation}\label{TIQRS}
H^{II}_\textrm{LD}=\frac{\omega_0^{\textrm R}}{2}\sigma_x+\omega^{\textrm R} a^\dagger a + g^{\textrm R}\sigma_y(a+a^\dagger)+\gamma^{\textrm R} a^\dagger a\sigma_x.
\end{equation}
Here $g^{\textrm R}=(\eta\Omega_r/4)(1-\epsilon_{\textrm S})$ if $\Omega_b=\Omega_r(1-\epsilon_{\textrm S})/(1+\epsilon_{\textrm S})$ with $\epsilon_{\textrm S}=\Omega_{\textrm S}/\nu$. See appendix~\ref{app:QRSTI} for a detailed derivation of Eq.~(\ref{TIQRS}). Notice that Eqs.~(\ref{QRS1}) and (\ref{TIQRS}) are equivalent by simply changing the qubit basis. For the latter, the diagonal basis is given by $\{|+\rangle,|-\rangle\}\otimes|n\rangle$, where $\sigma_x|\pm\rangle=\pm|\pm\rangle$. The parameters of the model are now $\omega^{\textrm R}_0=-(\Omega_0+\Omega_\textrm{DD})$, $\omega^{\textrm R}=(\delta_r-\delta_b)/2$, and $\gamma^{\textrm R}=\eta^2\Omega_{\textrm S}/2$. Regimes where $\gamma^{\textrm R}<0$ can be also reached by taking $\phi_{\textrm S}=0$, however, the frequency of the rotating frame changes to $\Omega_\textrm{DD}=\Omega_0-\omega_0^{\textrm R}$. Moreover, in this case $g^{\textrm R}=(\eta\Omega_r/4)(1+\epsilon_{\textrm S})$ if $\Omega_b=\Omega_r(1+\epsilon_{\textrm S})/(1-\epsilon_{\textrm S})$.

In the following, we verify the feasibility of the proposal by comparing the dynamics generated by the Hamiltonian~(\ref{TIHamil}) with the one of the Rabi-Stark model at Eq.~(\ref{TIQRS}). The results are shown in Figs.~\ref{fig:QRSTI}(a) and~\ref{fig:QRSTI}(b) for one-photon and three-photon oscillations respectively. The experimental parameters we use in Fig.~\ref{fig:QRSTI}(a) are $\nu=(2\pi)\times4.98$ MHz for the trapping frequency, $\eta=0.1$ for the LD parameter and $\Omega_{\textrm S}=(2\pi)\times120$ kHz for the carrier driving, leading to a Stark coupling of $|\gamma^{\textrm R}|=(2\pi)\times0.6$ kHz. We consider a Stark coupling of $\gamma^{\textrm R}/\omega^{\textrm R}=-0.4$, a Rabi coupling of $g^{\textrm R}/\omega^{\textrm R}=0.05$, and $\omega_0^{\textrm R}=\omega^{\textrm R}-\gamma^{\textrm R}(2N_0+1)$ with $N_0=2$. To achieve this regime the experimental parameters are $\Omega_r=(2\pi)\times2.94$ kHz, $\Omega_b=(2\pi)\times3.08$ kHz, and $\Omega_\textrm{DD}=(2\pi)\times114.86$ kHz. We observe that with an initial state $|+,2\rangle$, there is an exchange of population with the state $|-,3\rangle$. 
In Fig.~\ref{fig:QRSTI}(b) we show that selective three-photon oscillations of the Rabi-Stark model can be observed in some milliseconds. Starting from $|+,3\rangle$, we can observe coherent population exchange with state $|-,0\rangle$. Here, the LD parameter is $\eta=0.05$ and the parameters of the model are $\gamma^{\textrm R}/\omega^{\textrm R}=-0.1$, $g^{\textrm R}/\omega^{\textrm R}=0.3$ and $\omega_0^{\textrm R}/\omega^{\textrm R}=-2.4385$ for which we require $\Omega_r=(2\pi)\times35.2$ kHz, $\Omega_b=(2\pi)\times36.9$ kHz, and $\Omega_\textrm{DD}=(2\pi)\times123.5$ kHz. Although in the previous case we have focused on the Rabi-Stark model in the SC and USC regimes, it is noteworthy to mention that our method is still valid for larger ratios of $g/\omega$. Thus, our method represents a simple and versatile route to simulate the Rabi-Stark model in all important parameter regimes.

In this section, we studied the dynamics of the Rabi-Stark model QRM in the SC and USC regimes and characterise the novel $k$-photon interactions that appear by using time-dependent perturbation theory. Due to the Stark-coupling term, these $k$-photon interactions are selective, thus their resonance frequency depends on the state of the bosonic mode. Finally, and with the support of detailed numerical simulations, we proposed an implementation of the Rabi-Stark model with a single trapped ion. The numerical simulations show an excellent agreement between the dynamics of the trapped-ion system and the Rabi-Stark model.

\section{Nonlinear quantum Rabi model in trapped ions}

The study of the nonlinear QRM covered in this section is also a study of the nonlinear behaviour of a single trapped ion when it is far away from the LD regime. While in the past, research beyond the LD regime was mainly focussed on the nonlinear JC model~\cite{Vogel95,MatosFilho96_1,MatosFilho96_2,Stevens98}, its implications in laser cooling~\cite{Morigi97,Morigi99,Foster09} or for possible applications to simulate Frack-Condon physics~\cite{Hu11} has also been investigated. To set up the stage for a subsequent analysis, in section~\ref{subsect:JCmodels} we first briefly review the JC model and take this as a reference to show the difference with the nonlinear JC model. The appearance of nonlinear terms in the Hamiltonian suppresses the collapses and revivals for a coherent-state evolution typical from linear cases. In section~\ref{subsect:antiJCmodel}, we investigate how the nonlinear anti-JC model, which appears as the counterpart of nonlinear JC model, can be combined with controlled depolarising noise, to generate arbitrary $n$-phonon Fock states. Moreover, the latter could in principle be done without a precise control of pulse duration or shape, and without the requirement of a previous high-fidelity preparation of the motional ground state. Finally, in section~\ref{subsect:NQRMTI}, we propose the quantum simulation of the nonlinear quantum Rabi model by simultaneous off-resonant nonlinear JC and anti-JC interactions. 

\subsection{JC models in trapped ions}\label{subsect:JCmodels}

The Hamiltonian describing a laser-cooled two-level ion trapped in a harmonic potential and driven by a monochromatic laser field can be expressed as 
\begin{equation}\label{IonHamil}
H=\frac{\omega_I}{2}\sigma_z+\nu a^{\dag}a+\frac{\Omega}{2}\sigma^x[e^{i(\eta(a+a^\dag)-\omega_\textrm{ L} t+\phi)}+\textrm{H.c.}],
\end{equation}
where $\omega_0$ is the two-level transition frequency, $\sigma_z,\sigma^x$ are  Pauli matrices associated to this two-level system, $\Omega$ is the Rabi frequency, $\omega_\textrm{ L}$ is the driving laser frequency, and $\phi$ is the phase of the laser field.

In the LD regime, moving to an interaction picture with respect to $H_0=\frac{\omega_I}{2}\sigma_z+\nu a^{\dag}a$, and after the application of the so-called optical RWA, the Hamiltonian in Eq.~(\ref{IonHamil}) can be written as~\cite{Leibfried03}
\begin{equation}\label{NQRMLDregime}
H_\textrm{int}^\textrm{LD}=\frac{\Omega}{2}\sigma^+[1+i\eta(ae^{-i\nu t}+a^\dag e^{i\nu t})]e^{i(\phi-\delta t)}+\textrm{H.c.},
\end{equation}
where $\delta=\omega_\textrm{ L}-\omega_I$ is the laser detuning and the condition $\eta \ll1$ allows to keep only the zero and first order terms in the expansion of $\exp{[i\eta(a+a^\dag)]}$. When $\delta=-\nu$ and $\Omega\ll \nu$, after applying the vibrational RWA, the dynamics of such a system is described by the JC Hamiltonian, $H_\textrm{JC}=i g_r (\sigma^+ a - \sigma^-a^{\dag})$, where $g_r=\eta\Omega/2$ and $\phi=0$. This JC model is analytically solvable and generates population exchange between states $|\textrm{g},n\rangle \leftrightarrow |\textrm{e},n\!-\!1\rangle$ with rate $\Omega_{n,n-1}=\eta\Omega\sqrt{n}$. On the other hand, when the detuning is chosen to be $\delta=\nu$, the effective model is instead described by the anti-JC model $H_\textrm{aJC}=i g_b (\sigma^+ a^\dag - \sigma^-a)$, where $g_b=\eta\Omega/2$, which generates population transfer between states $|\textrm{g},n\rangle \leftrightarrow |\textrm{e},n\!+\!1\rangle$ with rate $\Omega_{n,n+1}=\eta\Omega\sqrt{n+1}$.

\begin{figure}[t!]
\centering
{\includegraphics[width=0.9 \textwidth]{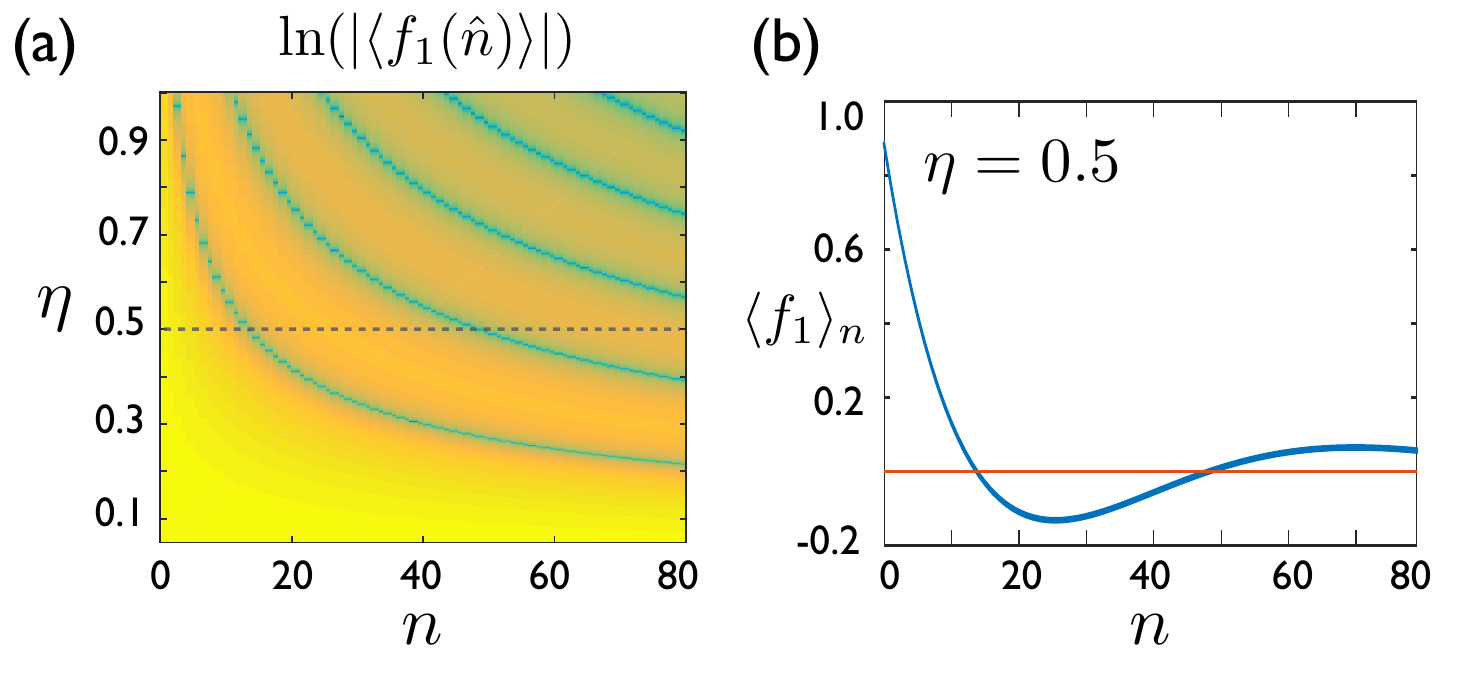}}
\caption{(a) Logarithm of the absolute value of the operator $f_1(\hat{n})$ evaluated for different Fock states $|n\rangle$ and LD parameters $\eta$. Dark (blue) regions represent cases where $f_1(\hat{n})|n\rangle\approx 0$. (b) Nonlinear function $f_1(n)$ for a fixed value of the LD parameter $\eta=0.5$ (oscillating blue curve). Zero value (horizontal orange line)\label{fig:NLZeros}}
\end{figure}

When the trapped-ion system is beyond the LD regime, the simplification of the exponential term described above is not justified and Eq.~(\ref{NQRMLDregime}) reads
\begin{eqnarray}\label{BLDregime}
H_\textrm{int}=\frac{\Omega}{2}\sigma^+ e^{i\eta(a^+ e^{i\nu t}+a e^{-i\nu t})-i(\delta t-\phi)}+\textrm{H.c.}\label{IntHam}.
\end{eqnarray}
When $\delta=-\nu$ and $\Omega \ll \nu$, after applying the vibrational RWA, the effective Hamiltonian describing the system is given by the nonlinear JC model~\cite{Vogel95}, which can be expressed as
\begin{eqnarray}
H_\textrm{nJC}=ig_r[\sigma^+ f_1(\hat{n}) a - \sigma^- a^\dag f_1(\hat{n})],
\end{eqnarray}
where the nonlinear function $f_1$~\cite{Vogel95} is given by
\begin{equation}\label{NLfunc}
f_1(\hat{n})=e^{-\eta^2/2}\sum_{l=0}^{\infty}\frac{(-\eta^2)^l}{l!(l+1)!}a^{\dag l} a^l,
\end{equation}
with $a^{\dag l} a^l=\hat{n}!/(\hat{n}-l)!$. The dynamics of this model can also be solved analytically, and as the linear JC model, yields to population exchange between states $|\textrm{g},n\rangle \leftrightarrow |\textrm{e},n\!-\!1\rangle$. However, in this case the Rabi frequencies are $\tilde{\Omega}_{n,n-1}= |f_1(n-1)|\Omega_{n,n-1}=\eta\Omega\sqrt{n} |f_1(n-1)|$, where $f_1(n)$ corresponds to the value of the diagonal operator $f_1$ evaluated on the Fock state $|n\rangle$, i.e. $f_1(n)\equiv\langle f_1(\hat{n})\rangle_n$.
If the detuning in Eq.~(\ref{BLDregime}) is chosen to be $\delta=\nu$, and $\Omega \ll \nu$, then the application of the vibrational RWA yields the nonlinear anti-JC model,
\begin{eqnarray}
H_\textrm{naJC}=ig_b[\sigma^+a^\dag f_1(\hat{n}) - \sigma^- f_1(\hat{n})a ],
\end{eqnarray}
which, as the linear anti-JC model, generates population exchange between states $|\textrm{g},n\rangle \leftrightarrow |\textrm{e},n\!+\!1\rangle$ with rate $\tilde{\Omega}_{n,n+1}=|f_1(n)|\Omega_{n,n+1}=\eta\Omega\sqrt{n+1} |f_1(n)|$. The nonlinear function $f_1$ depends on the LD parameter $\eta$ and on the Fock state $| n \rangle$ on which it is acting. The LD regime is then recovered when $\eta\sqrt{\langle(a+a^\dagger)^2\rangle}\ll 1$. In this regime, $|f_1(n)|\approx1$, and thus the dynamics is the one corresponding to the linear models.

Beyond the LD regime, the nonlinear function $f_1$, which has an oscillatory behaviour both in $n\in\mathbb{N}$ and $\eta\in\mathbb{R}$, needs to be taken into account.  In Fig.~\ref{fig:NLZeros}(a), we plot the logarithm of the absolute value of $f_1(n,\eta)$ for different values of $n$ and $\eta$, where the green regions represent lower values of $\log{(|f_1(n,\eta)|)}$, i.e, values for which $f_1\approx 0$. This oscillatory behaviour can also be seen in Fig.~\ref{fig:NLZeros}(b) where we plot the value of $f_1$ as a function of the Fock state number $n$ for $\eta=0.5$. For this specific case, we can see that the function is close to zero around $n=14$ and $n=48$, meaning that for $\eta=0.5$, the rate of population exchange between states $|\textrm{g},15\rangle \leftrightarrow |\textrm{e},14\rangle$ and $|\textrm{g},49\rangle \leftrightarrow |\textrm{e},48\rangle$ will vanish for the nonlinear JC model. The same will happen to the exchange rate between states $|\textrm{g},14\rangle \leftrightarrow |\textrm{e},15\rangle$ and $|\textrm{g},48\rangle \leftrightarrow |\textrm{e},49\rangle$ for the nonlinear anti-JC model.

\begin{figure}[t!]
\centering
{\includegraphics[width=0.9 \textwidth]{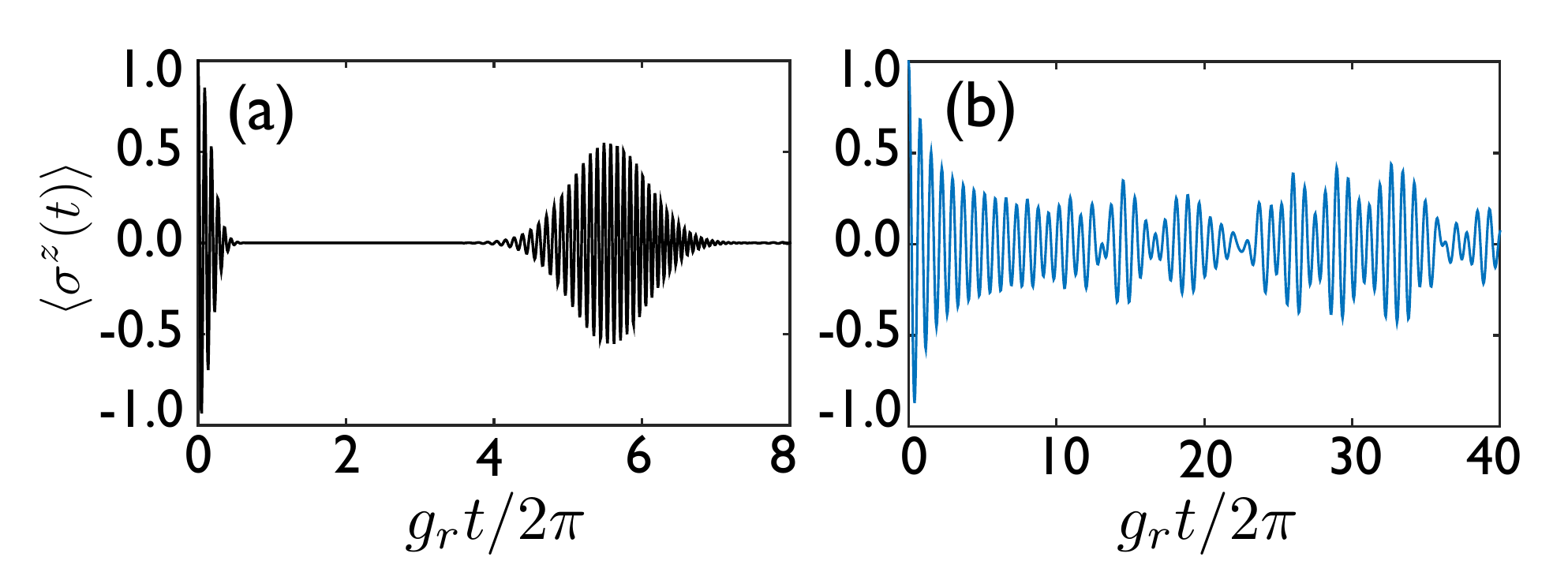}}
\caption{Average value of $\sigma_z$ operator versus time for a coherent initial state $|\alpha=\sqrt{30}\rangle$ after (a) linear JC and (b) nonlinear JC evolution, both with the same coupling strength $g_r$ and $\eta=0.5$ for the nonlinear case. As shown in (a), there exists an approximate collapse and subsequent revival in the JC model dynamics, while for the nonlinear JC model this is not the case. \label{JaynesCollapse}}
\end{figure}

We observe approximate collapses and revivals for an initial coherent state\footnote{A coherent state is defined as $|\alpha\rangle=e^{-|\alpha|^2/2}\sum_{n=0}^\infty\frac{\alpha^n}{\sqrt{n!}}|n\rangle$} with an average number of photons of $|\alpha|^2=30$ by evolving with the JC model, as shown in Ref.~\cite{Gerry04}, see Fig.~\ref{JaynesCollapse}(a). Here, we plot $\langle\sigma^z(t)\rangle=\langle \psi(t)|\sigma^z|\psi(t)\rangle$ for a state that evolves according to the JC model. Comparing the same case for the nonlinear JC model with $\eta=0.5$, as depicted in Fig.~\ref{JaynesCollapse}(b), we appreciate that in the latter case the collapses and revivals vanish, and the dynamics is more irregular. This can seem natural given that the phenomenon of revival takes place whenever the most significant components of the quantum state, after some evolution time, turn out to oscillate in phase again, which may be more unlikely if the dynamics is nonlinear. Notice that we let the case of the nonlinear JC model evolve for a longer time, since the nonlinear function $f_1$ effectively slows down the evolution.

\subsection{Fock-state generation with a dissipative nonlinear anti-JC model}\label{subsect:antiJCmodel}

In this section we study the possibility of using the dynamics of the nonlinear anti-JC model introduced in the previous section to, along with depolarising noise, generate high-number Fock states in a dissipative manner. In particular, the depolarising noise that we consider corresponds to the spontaneous relaxation of the internal two-level system of the ion. Such a dissipative process, combined with the dynamics of the JC model in the LD regime (linear JC model), is routinely exploited in trapped-ion setups for the implementation of sideband cooling. It is noteworthy to mention that the effect of nonlinearities on sideband cooling protocols, which arise when outside the LD regime, have also been a matter of study~\cite{Morigi97,Morigi99}.

\begin{figure}[t!]
\centering
{\includegraphics[width=1 \linewidth]{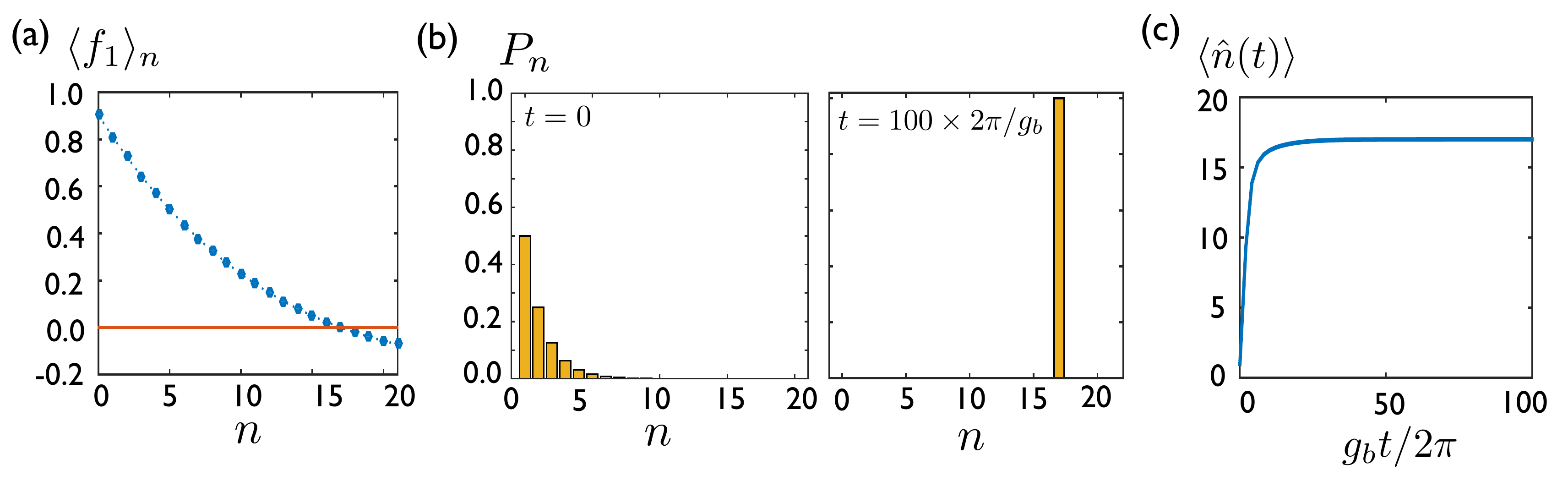}}
\caption{(a) The nonlinear function $f_1$ evaluated at different Fock states $n$, for the case of $\eta=0.4518$ (decreasing blue curve). Zero value (horizontal orange line). For this value of the LD parameter, $f_1|17\rangle=0$. (b) Phonon statistics of the initial thermal state with $\langle n\rangle=1$ (c) Time evolution of the average value of the number operator $\hat{n}$ starting from the state in (b) and following the evolution for the preparation of Fock state $| 17 \rangle$, that is during a nonlinear anti-JC model with spontaneous decay of the two-level system. (d) Phonon statistics at the end of the protocol, $t=100\times2\pi/g_b$, with all the population concentrated in Fock state $|17\rangle$. \label{fig:FockStatePrep}}
\end{figure}

Our method works as follows: we start in the ground state of both the motional and the internal degrees of freedom $|\textrm{g},0\rangle$ (as we will show later, our protocol works as well when we are outside the motional ground state, as long as the population of Fock states higher than the target Fock state is negligible). Acting with the nonlinear anti-JC Hamiltonian we induce a population transfer from state $|\textrm{g},0\rangle$ to state $|\textrm{e},1\rangle$, while at the same time, the depolarising noise transfers population from $|\textrm{e}, 1\rangle$ to $|\textrm{g}, 1\rangle$. The simultaneous action of both processes will ``heat" the motional state, progressively transferring the population of the system from one Fock state to the next one. Eventually, all the population will be accumulated in state $|\textrm{g},n\rangle$, where a blockade of the propagation of population through the chain of Fock states occurs, if $f_1(n)=0$, as the transfer rate between states $|\textrm{g},n\rangle$ and $|\textrm{e},n+1\rangle$ vanishes, $\tilde{\Omega}_{n,n+1}=0$. We point out that the condition $f_1(n)=0$ can always be achieved by tuning the LD parameter to a suitable value, i.e. for every Fock state $|n\rangle$, where $n>0$ there exists a value of the LD parameter $\eta$ for which $f_1(n,\eta)=0$. As an example, we choose the LD parameter $\eta=0.4518$, for which $f_1(17)=0$, and simulate our protocol using the master equation
\begin{eqnarray}\label{LabMasterEq}
\dot{\rho}=-i[H_\textrm{naJC},\rho]+\frac{\Gamma}{2} (2\sigma^-\rho \sigma^+ - \sigma^+ \sigma^-\rho-\rho \sigma^+ \sigma^-),
\end{eqnarray}
where  $\Gamma=2g_{b}$ is the decay rate of the internal state.

In Fig.~\ref{fig:FockStatePrep} we numerically show how our protocol is able to generate the motional Fock state $ |17 \rangle$, starting from a thermal state with $\langle n\rangle=1$. In other words, one can obtain large final Fock states starting from an imperfectly cooled motional state, by a suitable tuning of the LD parameter. As an advantage of our method compared to previous approaches~\cite{Meekhof96}, we do not need a fine control over the Rabi frequencies or pulse durations, given that the whole wave-function, for an arbitrary initial state with motional components smaller than $n$, will converge to the target Fock state $|n\rangle$. We want to point out that this protocol relies only on the precision to which the LD parameter can be set, which in turn depends on the precision to which the wave number $k$ and the trap frequency $\nu$ can be controlled. These parameters enjoy a great stability in trapped-ion setups~\cite{Johnson16}, and therefore we deem the generation of high-number Fock states as a promising application of the nonlinear anti-JC model dynamics.

\subsection{Nonlinear quantum Rabi model}\label{subsect:NQRMTI}

Here we propose to implement the nonlinear quantum Rabi model (NQRM) in all its parameter regimes via the use of the Hamiltonian in Eq.~(\ref{IntHam}). We consider off-resonant first-order red and blue sideband drivings with the same coupling $\Omega$ and corresponding detunings $\delta_r$, $\delta_b$. The interaction Hamiltonian after the optical RWA reads~\cite{Leibfried03,Pedernales15},
\begin{eqnarray}
H_\textrm{int} = \sum\limits_{j=r,b}\frac{\Omega_j}{2}\sigma^+e^{i\eta(a^{\dag}e^{i \nu t}+a e^{-i \nu t})}e^{-i(\delta_j t-\phi_j)}+\textrm{H.c.},
\end{eqnarray}
where $\omega_r=\omega_I-\nu+\delta_r$ and $\omega_b=\omega_I+\nu+\delta_b$, with $\delta_r,\delta_b\ll \nu \ll \omega_I$ and $\Omega_r=\Omega_b \ll \nu$. We consider the system beyond the LD regime and set the laser-field phases to $\phi_{r,b}=0$. If we invoke the vibrational RWA, i.e. neglect terms that rotate with frequencies in the order of $\nu$, the remaining terms read
\begin{equation}
H_\textrm{int}=ig^\textrm{ R}\sigma^+\big(f_1ae^{-i \delta_r t}+a^{\dag}f_1e^{-i \delta_b t}\big)+\textrm{H.c.},
\end{equation}
where  $g^\textrm{ R}=\eta\Omega_r/2$ and  $f_1\equiv f_1(\hat{n})$ was introduced in Eq.~(\ref{NLfunc}). The latter corresponds to an interaction picture Hamiltonian of the NQRM with respect to the free Hamiltonian $H_0=\frac{1}{4}(\delta_b+\delta_r)\sigma_z +\frac{1}{2}(\delta_b-\delta_r)a^\dag a$. Therefore, undoing the interaction picture transformation, we have
\begin{equation}\label{NQRM}
H_\textrm{nQRM}=\frac{\omega_0^{\textrm R}}{2}\sigma_z+\omega^{\textrm R} a^{\dag}a+i g^\textrm{ R} (\sigma^+ - \sigma^-)(f_1a+a^{\dag}f_1),
\end{equation}
where $\omega_0^{\textrm R}=-\frac{1}{2}(\delta_r+\delta_b)$ and $\omega^{\textrm R}=\frac{1}{2}(\delta_r-\delta_b)$.  Equation~(\ref{NQRM}) represents the general form of the NQRM, where $\omega_0^{\textrm R}$ is the level splitting of the simulated two level system, $\omega^{\textrm R}$ is the frequency of the simulated bosonic mode and $g$ is the coupling strength between them, which in turn will be modulated by the nonlinear function $f_1(\hat{n},\eta)$. The different regimes of the NQRM will be characterised by the relation among these four parameters. First, in the LD regime or $\eta\sqrt{\langle (a+a^{\dag})^2 \rangle}\ll 1$, Eq.~(\ref{NQRM}) can be approximated to the linear QRM~\cite{Pedernales15}. Beyond the LD regime, in a parameter regime where $|\omega^{\textrm R}-\omega_0^{\textrm R}|\ll g^\textrm{ R}\ll|\omega^{\textrm R}+\omega_0^{\textrm R}|$, the RWA can be applied. This would imply neglecting terms that rotate at frequency $\omega^{\textrm R}+\omega_0^{\textrm R}$ in an interaction picture with respect to $H_0$, leading to the nonlinear JC model studied in section~\ref{subsect:JCmodels}. On the other hand, the nonlinear anti-JC model would be recovered in a regime where $|\omega^{\textrm R}+\omega_0^{\textrm R}|\ll g^\textrm{ R}\ll|\omega^{\textrm R}-\omega_0^{\textrm R}|$. It is worth mentioning that the latter is only possible if the frequency of the two-level system and the frequency of the mode have opposite sign. The USC and DSC regimes are defined as $0.1\lesssim g^\textrm{ R}/\omega^{\textrm R} \lesssim 1$ and $g^\textrm{ R}/\omega^{\textrm R}\gtrsim1$ respectively, and in these regimes the RWA does not hold anymore.

\begin{figure}[]
{\includegraphics[width=1 \linewidth]{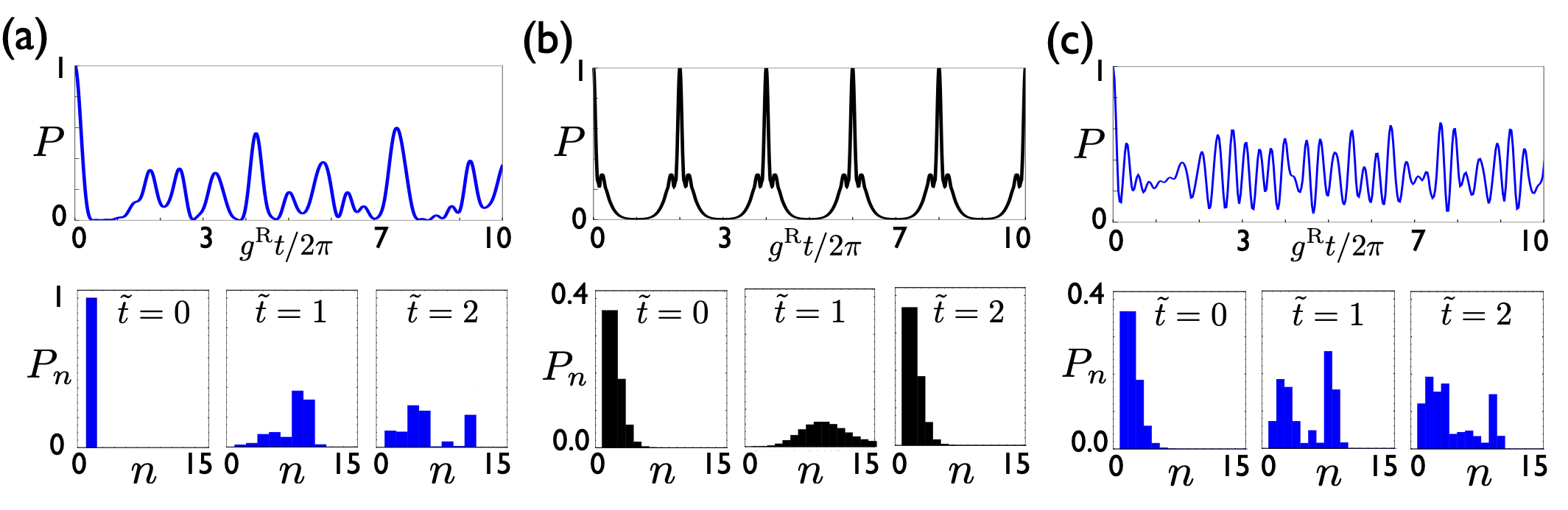}}
\caption{Fidelity with respect to the initial state $P(t)=|\langle\psi_0|\psi(t)\rangle|^2$ versus time is shown in the upper figures. In the lower figures, phonon statistics is shown at different times, where $\tilde{t}=g^\textrm{ R}t/2\pi$. In (a), $|0,\textrm{g}\rangle$ is chosen as initial state, while in (b) and (c) the initial state is $|\alpha\!=\!1,\textrm{g}\rangle$. In (a), the evolution occurs under the NQRM with LD parameter $\eta=0.67898$, where $f_1|7\rangle=0$, $g^\textrm{ R}/\omega^\textrm{R}=4$ and $\omega_0^\textrm{R}=0$. Observing the phonon statistics we see how Fock states where $n>7$ never get populated. In (b), the state evolve under the linear QRM. In (c), the state evolve under the NQRM with LD parameter $\eta=0.57838$, where $f_1|10\rangle=0$, $g^\textrm{ R}/\omega^\textrm{R}=3.7$ and $\omega_0^\textrm{R}=0$. \label{fig:NQRMEvolution}}
\end{figure}

As an example, here we investigate the NQRM in the DSC regime with initial Fock state $|0, \textrm{g}\rangle$, where $|0\rangle$ is the ground-state of the bosonic mode, and $|\textrm{g} \rangle$ stands for the ground state of the two-level system. In Fig.~\ref{fig:NQRMEvolution}(a), we study the case for $\eta=0.67898$, where $f_1|7\rangle=0$, $g^\textrm{ R}/\omega^\textrm{R}=4$ and $\omega_0^\textrm{R}=0$. More specifically, a quantum simulation of the model in this regime can be achieved with the following detunings and Rabi frequency: $\delta_r=2\pi\times11.31$ kHz, $\delta_b=-2\pi\times11.31$ kHz, $g^\textrm{ R}=2\pi\times45.24$ kHz and $\Omega_r=2\pi\times 133.26$ kHz. In Ref.~\cite{Casanova10}, it was shown that the linear QRM shows collapses and revivals and a round trip of the phonon-number wave-packet along the chain of Fock states, when in the DSC regime. Here, we observe that in the nonlinear case, Fig.~\ref{fig:NQRMEvolution}(a), collapses and revivals do not present the same clear structure, having a more irregular evolution. Most interestingly, the system dynamics never surpasses Fock state $|n\rangle$, for which $f_1(n)=0$.  Regarding the simulated regime of the nonlinear QRM, we point out that the nonlinear term also contributes to the coupling strength. Therefore, to keep the NQRM in the DSC regime, the ratio $g^\textrm{ R}/\omega^\textrm{R}$ should be larger than that for the linear QRM since $f_1(n)< 1$ always. Summarising, our result illustrates that the Hilbert space is effectively divided into two subspaces by the NQRM, namely those spanned by Fock states below and above Fock state $| n \rangle$.  We denote the Fock number $n$, where $f_1|n\rangle=0$, as ``the barrier'' of the NQRM.

To benchmark the effect of the barrier, we also provide simulations starting from an initial coherent state with $\alpha=1$ whose average phonon number is $\langle n \rangle=|\alpha|^2=1$, and make the comparison between the QRM and the NQRM in the DSC regime. For the parameter regime $g^\textrm{ R}/\omega^\textrm{R}=2$ and $\omega_0^\textrm{R}=0$, the fidelity with respect to the initial coherent state in the linear QRM performs periodic collapses and full revivals as it can be seen in Fig.~\ref{fig:NQRMEvolution}(b). In the lower figures of Fig.~\ref{fig:NQRMEvolution}(b), we observe a round trip of the phonon-number wave packet, similarly to what was shown in Ref.~\cite{Casanova10} for the case of the linear QRM starting from a Fock state. The NQRM, on the other hand, has an associated dynamics that is aperiodic and more irregular, as shown in Fig.~\ref{fig:NQRMEvolution}(c), and never crosses the motional barrier produced by the corresponding $f_1(n)=0$. This suggests that the NQRM could be employed as a motional filter. This filter is determined by the location of the barrier with respect to the initial state distribution. Here, by filter we mean that the population of Fock states above a given threshold can be prevented. For the simulation we choose the LD parameter $\eta=0.57838$ for which $f_1|10\rangle=0$, which is far from the centre of the distribution of the initial coherent state, as well as most of its width. The simulated parameter regime corresponds to the DSC regime with $g^\textrm{ R}/\omega^\textrm{R}=3.7$ and $\omega_0^\textrm{R}=0$. This case could also be simulated with trapped ions with detunings of $\delta_r=2\pi\times11.31$kHz and $\delta_b=-2\pi\times11.31$kHz, and a Rabi frequency of $\Omega_r=2\pi\times 133.26$kHz. As for the case corresponding to initial Fock state $|0,\textrm{g}\rangle$, the evolution of the NQRM in the coherent state case, depicted in Fig.~\ref{fig:NQRMEvolution}(c), never exceeds the barrier.

In summary, in this section we have proposed the implementation of the nonlinear QRM in arbitrary coupling regimes with trapped-ion quantum simulators. The nonlinear term that appears in our model is characteristic of the region beyond the LD regime. This nonlinear term causes the blockade of motional propagation at $|n\rangle$ whenever $f_1(\hat{n})|n\rangle=0$. In order to compare our model with the linear QRM, we have plotted the evolution of the population of the internal degrees of freedom of the ion evolving under the linear JC and nonlinear JC models and observe that for the latter the collapses and revivals disappear. Also, we have proposed a method for generating large Fock states in a dissipative manner, making use of the nonlinear anti-JC model and the spontaneous decay of the two-level system. Finally, we have studied the dynamics of the linear and nonlinear full QRM on the DSC regime and notice that the nonlinear case can act as a motional filter.

\chapter{Boson Sampling with Ultracold Atoms}
\label{chapter:chapter_3}
\thispagestyle{chapter}

The realization of quantum supremacy is a milestone on the path towards fault-tolerant quantum computing. On the one hand, it demonstrates that quantum speedup is indeed possible, which shall serve to discard the existence of some unknown and fundamental physical principle capable of hindering the realization of a quantum computer~\cite{Bassi13}. On the other hand, it constitutes the first violation of the so-called extended Church-Turing thesis~\cite{Elthan97}, which states that any physically realizable computational model can be efficiently simulated by a classical Turing machine. The latter can only be claimed if the computational complexity of the task performed by the quantum machine is known. For example, the outperformance of classical computers by quantum annealers or quantum simulators would not serve as a violation of the extended Church-Turing thesis, as the computational complexity of the problem being solved is not well defined. In this sense, a rigorous test of quantum supremacy has to be based on solid assumptions about the computational complexity of the accomplished task~\cite{Aaronson17}. 

Quantum sampling problems enjoy the fact that their computational complexity can be precisely assessed~\cite{Terhal04}. These problems consist on generating samples according to a probability distribution associated to the output of a quantum circuit. Moreover, these models can be implemented by quantum computers without using error correcting codes~\cite{Lund17}, dramatically reducing their experimental cost in comparison to other quantum algorithms. Particularly, the so-called boson-sampling problem~\cite{Aaronson11,Gard15,Lund17} has received a lot of attention because non-interacting particles, for example photons, are enough for its realization. 

\begin{figure}[t]
\centering
	\includegraphics*[width=1\columnwidth]{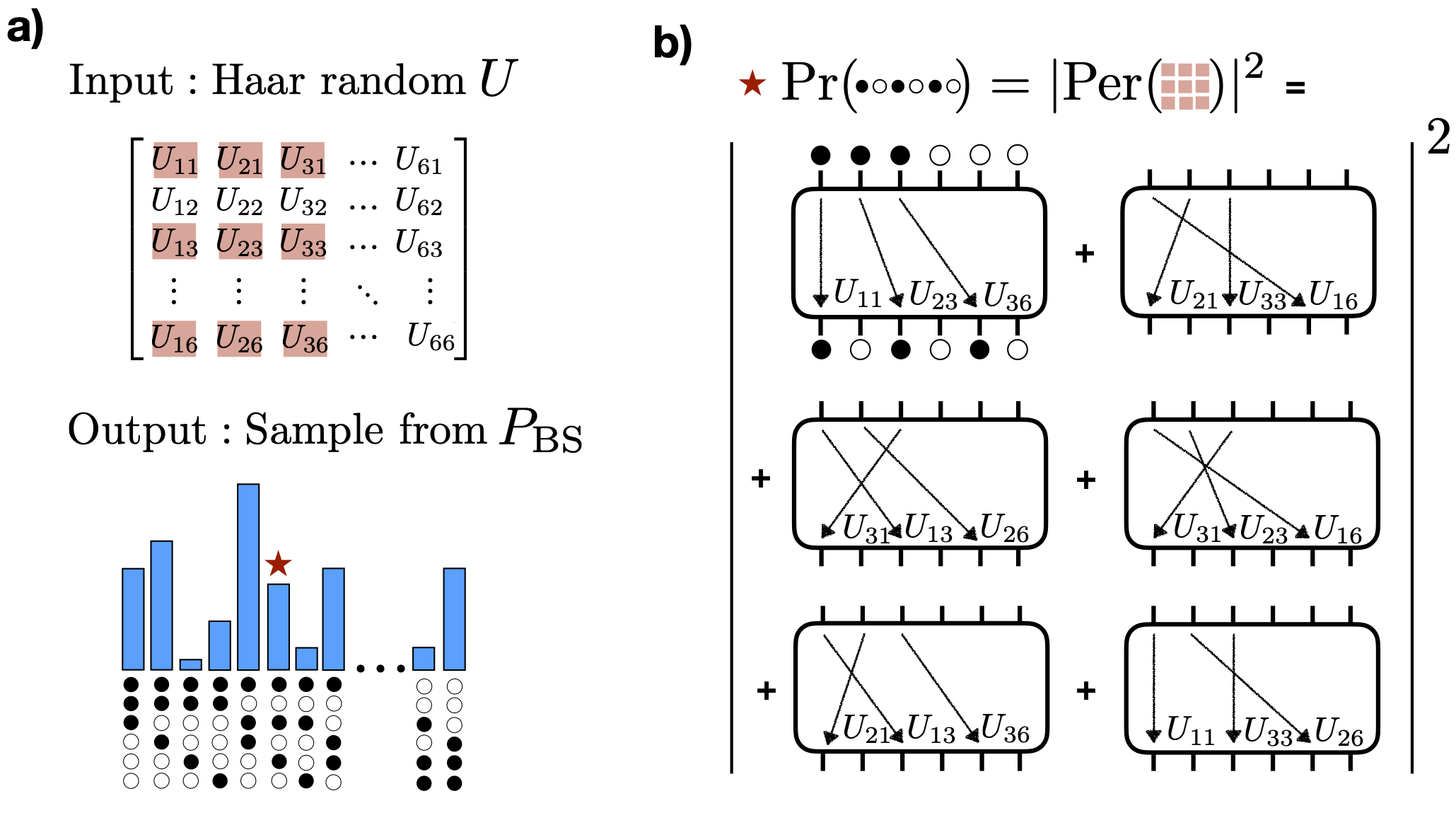}
	\caption{a) Given an $M\times M$ Haar random unitary matrix $U$, a probability distribution can be defined by the modulus squared permanents of $N\times N$ submatrices. Boson sampling consists in sampling from this probability distribution $P_{BS}$. b) In an experimental boson sampler, $P_{BS}$ contains the probabilities of all possible outcomes of the machine, given an $N$-boson initial state. For example, the $3\times 3$ submatrix $A_S$ formed by the 9 elements in red squares in a) relates the initial state $a_1^\dagger a_2^\dagger a_3^\dagger|0\rangle$ with the output state $a_1^\dagger a_3^\dagger a_6^\dagger|0\rangle$. As a result of quantum interference among $N!$ physical paths, the probability of this output state is given by the modulus squared permanent of $A_S$. \label{BSProblem}}
\end{figure}

Boson sampling consists in the problem of sampling from the probability distribution of the outcomes generated when you introduce $N$ bosons in a linear interferometer with $M$ modes.   Both initial and final states are written in the Fock basis, and the linear interferometer is described by an $M\times M$ unitary matrix. The probability amplitude of each outcome state is proportional to the permanent of an $N\times N$ submatrix, where the rows and columns relate the initial state with that particular outcoming state. The permanent appears due to all $N!$ different physical paths contributing to the same outcome, see Fig.~\ref{BSProblem}(b). Even if it looks similar to the determinant, whose computation is efficient by classical means, calculating the permanent of a complex-valued matrix is a $\#$P-hard computational problem~\cite{Valiant79}, meaning that the simulation of the boson sampling problem is believed to be extremely inefficient for a classical computer. It has being conjectured that sampling, even approximately, from the probability distribution of a boson sampler is already a $\#$P-hard problem~\cite{Aaronson11,Aaronson11b}. This, along with boson sampling being robust against experimental imperfections~\cite{Rohde12a,Rohde12b,Arkhipov15,Drummond16,Dittel18}, has motivated the rapid progress of boson sampling machines.

Up to now, quantum boson sampling has been realized using photonic quantum circuits~\cite{Broome13,Tillmann13,Spring13,Crespi13,Carolan14,Spagnolo14,Tillmann15,Carolan15,Loredo17,Wang17c,Wang18}, with the current record being 20 input and 14 output photons in 60 modes~\cite{Wang19_BS}. Although significant conceptual and technological advances have been made scaling-up these photonic devices to a larger number of photons~\cite{Motes14,Lund14,Bentivegna15,Aaronson16b,He17,Hamilton17,Wang18}, the simultaneous control of more than twenty photons remains, so far, inaccessible experimentally. In addition, the progress of classical methods to simulate boson sampling has been significant in the last few years~\cite{Neville17,Wu18,Lundow19}. These two facts complicate a near-term quantum-supremacy test by a photonic boson-sampling machine. On the other hand, alternative methods for doing boson sampling with trapped ions~\cite{Shen14}, superconducting circuits~\cite{Peropadre16} or optical lattices~\cite{Deshpande18,Muraleedharan19} have been proposed. However, up to now only a proof-of-principle experiment using coupled vibrational modes of a single ion has been realized~\cite{Shen17}.

In this chapter, we present a scalable method to implement boson sampling using ultracold atoms in state-dependent optical lattices. In our scheme, atoms cooled into their vibrational ground state play the role of indistinguishable bosons, while both the lattice sites and the internal state of the atoms serve as the bosonic modes. Polarization-synthetized optical lattices can be used to realize state-dependent lattice-shift operations~\cite{Robens16pol,Robens17}, which allows bringing together spatially separated modes. In addition, pairwise interactions among these modes, analogous to the beamsplitters used in photonic devices, can be achieved via the combination of MW radiation with site-resolved optical pulses. The latter is the basic building block of our proposal, and has already been demonstrated in Ref.~\cite{RobensThesis}, where the Hong-Ou-Mandel interference between two optically trapped atoms is reported. 

The chapter is organized as follows: In section~\ref{PSOLforQC} we describe the method to realize quantum circuits with ultracold atoms in polarization-synthesized optical lattices. In section~\ref{ScalingAndErrors} we discuss the case of the boson-sampling quantum circuit and the scaling of such a device to tens of atoms. We study how the two-body collisions affect to the rate in which valid samples are generated, and also how they change the form of the final probability distribution. In section~\ref{subsec:additional}, we also consider other sources of error like dephasing or imperfect ground state cooling.

\section{Quantum circuits with spin-dependent optical lattices}\label{PSOLforQC}

In this section we present ultracold atoms in state-dependent optical lattices as a scalable architecture to realize discrete-time quantum circuits, and, in particular, boson sampling. The idea is based on the fact that any $M\times M$ unitary matrix $U$ describes the evolution of a particular $M$-mode linear interferometer, allowing only nearest-neighbour coupling among the modes~\cite{Reck94,Clements16}. Linear interferometers of noninteracting atoms are an interesting alternative to photonic interferometers because of the ability of controlling a large number of particles in atomic systems. Moreover, these systems can exploit controlled coherent collisions~\cite{Chin10} among particles at the same site. These nonlinear processes could increase the amount of quantum correlations in the output states~\cite{Brunner18}, possibly making the classical simulation of such a sampler even harder than for the linear case.

\subsubsection{Preparing an array of identical atoms}

Interference between two optical laser beams is routinely used to create arrays of optical micropotentials called optical lattices. Atoms can be trapped in those lattice sites and repositioned one by one using state-dependent moving potentials~ \cite{Kim16,Barredo16,Endres16,Robens17,Kumar18}. As a result, these atoms can be prepared in states were the position of each atom in the lattice is predefined. Currently, the record in the number of assorted atoms is in $N=111$~\cite{OhldeMello19}, well above the $N=20$ photons achieved with photonic devices~\cite{Wang19_BS}. Furthermore, ideas to increase this number up to $1000$ exist for the case of two-dimensional state-dependent optical lattices~\cite{Robens17}. After being rearranged, the atoms can be cooled down to their vibrational ground state by means of sideband cooling. Using this, a purity or ground-state probability of up to $90\%$ has being demonstrated~\cite{Kumar18}, the main limitation being the small trap frequency along at least one of the confining directions. In this respect, probabilities close to unity are expected in deep three-dimensional optical lattices. 
Alternative techniques to create low-entropy atom ensembles exist, which involve the preparation of a Mott insulator state and the subsequent selection of atoms at predefined lattice sites. With this method, a purity of $99\%$ per site has being experimentally achieved for $12$ atoms~\cite{Lukin19,Rispoli19}.

\begin{figure*}[t]
	\includegraphics*[width=\linewidth]{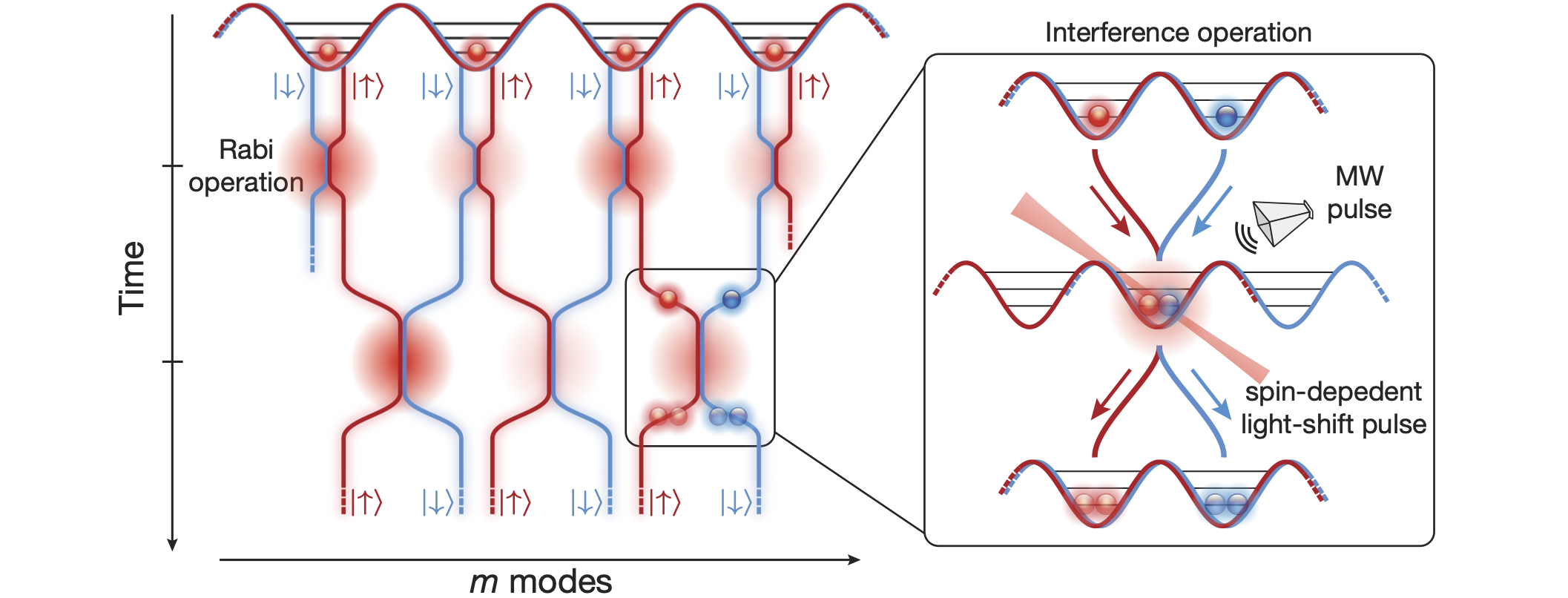}
	\caption{\label{BSwithPSOLScheme}
	 Illustration of the quantum-circuit scheme based on ultracold atoms in polarization-synthesized optical lattices. Each lattice site hosts two modes of the quantum circuit, represented by two atomic internal states, $|{\uparrow}\rangle$ and $|{\downarrow}\rangle$.  A representative initial state is shown in the figure, where every second mode is occupied. Atoms are displaced by state-dependent shift operations, while their internal states are coupled using MW radiation and light shifts. Inset:  A combination of local addressing pulses and MW pulses realize the equivalent of a photonic beamsplitter. }
\end{figure*}

\subsubsection{Wiring the quantum circuit with polarization-synthesized optical lattices}

For the realization of discrete-time quantum circuits, we consider polarization-synthesized optical lattices \cite{Robens17}. These consist of two independent optical potentials that are polarization selective, trapping atoms depending on the polarization of the transition associated with an internal state. For that, two states of the 6s$^2S_{1/2}$ hyperfine manifold of the $^{133}$Cs atom can be used, e.g., $|\!\!\uparrow\rangle=|F=4,m_F=4\rangle$ and $|\!\!\downarrow\rangle=|F=3,m_F=3\rangle$. The lattice wavelength is set at $\lambda_L=870$ nm. Then, due to different polarizability~\cite{Deutsch98,Grimm00,LeKien13}, atoms in $|\!\!\uparrow\rangle$ or $|\!\!\downarrow\rangle$ feel only either one of the two periodic potentials:
\begin{equation}
V_{\uparrow,\downarrow}(x) = V^0_{\uparrow,\downarrow}\cos^2\{2\pi/\lambda_L[x-x_{\uparrow,\downarrow}(t)]\},
\end{equation} 
where the position of both lattice sites $x_{\uparrow,\downarrow}(t)$ can be independently controlled with subnanometer precision by a fast polarization synthesizer~\cite{Robens16pol}. Also, the depth of the micropotentials $V^0_{\uparrow,\downarrow}$ is sufficiently large to suppress the tunnelling of atoms to neighbouring sites.

In Fig.~\ref{BSwithPSOLScheme} we illustrate how different lattice sites can be ``wired'' by using fast, state-dependent shifts of the optical lattice, generating discrete-time quantum circuits with ultracold atoms. In our scheme, the modes of the quantum circuit are represented by both the different lattice sites and the two internal states. Thus, with our method, $M/2$ lattice sites are sufficient to represent $M$ modes. The shift operation preserves the coherence between the two internal states~\cite{Robens15}, and its duration can be as short as the trapping period, which is about $3$ femtoseconds. The geometry of the circuit depicted in Fig.~\ref{BSwithPSOLScheme} is equivalent to those realized with photonic systems, except that in our case a single spatial dimension is enough, in contrast to the two spatial dimensions used in photonic circuits~\cite{Carolan15}. As in photonic circuits, the evolution of an $M$-mode interferometer coupled pairwise at discrete time steps is represented by an $M{\times}M$ unitary matrix $U$~\cite{Reck94,Clements16}. The matrix representation of a general pairwise interaction acting at time $t$ on lattice site $s$ is  
\begin{equation}
	\label{eq:basiccoupling}
	T(t,s)=\left(\begin{array}{cc} e^{-i\phi}\cos(\theta/2)&-\sin(\theta/2)\\e^{-i\phi}\sin(\theta/2)&\cos(\theta/2) \end{array}\right),
\end{equation}
where $\phi$ is a phase imprinted onto only the $|\!\!\uparrow\rangle$ mode, and $\theta$ is the angle by which the pseudospin (i.e., the two coupled modes, $|\!\!\uparrow\rangle$ and $|\!\!\downarrow\rangle$) is rotated around the $y$ axis of the Bloch sphere. Interestingly, a general $M{\times}M$ unitary matrix $U$ can be represented by the product of $M(M{-}1)/2$ independent operations $T(t,s)$~\cite{Reck94,Clements16}. Notice that, as each operation is characterized by two parameters, the whole protocol is then defined by $M(M{-}1)$ independent parameters, corresponding to the number of free parameters characterizing a generic $M\times M$ unitary matrix, up to $M$ phase shifts applied to the $M$ output modes, which are irrelevant for most applications. In principle, all these operations can be implemented in $M$ time steps, resulting in an $M$-step circuit depth~\cite{Clements16}.

\subsubsection{Arbitrary mode coupling by light pulses}\label{sec:mode_coupling}

Two modes in the same lattice site can be coupled by combining site-resolved light pulses with global MW pulses, realizing $T(t,s)$. More specifically, site-resolved light pulses are used to imprint a local differential phase shift between the two hyperfine states, $|\!\!\uparrow\rangle$ and $|\!\!\downarrow\rangle$, while global MW radiation results in a global Hadamard operation. The latter can be implemented by a $\pi/2$ MW pulse, rotating the pseudospin by $90$ degrees around the $x$ axis of the Bloch sphere. This is represented by a Hadamard-like transformation $H_{2\times 2} = \exp(-i \sigma_x \pi/4)$. Differential light shifts can be realized with cesium atoms~\cite{Deutsch98,Grimm00,LeKien13}. In addition, these pulses can be focused so that they act only onto the target sites~\cite{Weitenberg11,Preiss15,Robens16mbg}, allowing the realization of independent rotations around the $z$ axis, $A(\varphi_s)=\exp[-i \sigma_z \varphi/2]$, where the angle $\varphi_s$ is controlled by the product of the laser intensity and the pulse duration. The duration of the global MW pulse is about $150\mu$s, however, can be shortened to tens of microseconds just by increasing the MW power. On the other hand, the local laser pulses can be realized in about $10\mu$s using approximately $1\mu$W of laser power per addressed lattice site. Moreover, in the case of the latter, the probability of error by scattering of light is of the order of $10^{-5}$. As the main advantage, this method avoids using site-resolved Raman pulses in order to couple the two modes. 

The described quantum gates are sufficient to build a generic $T(t,s)$ operation, as shown by the following formula:
\begin{equation}\label{eq:decomposition}
T'(t,s)=	e^{i\phi/2}\,T(t,s)= H_{2\times 2}^\dagger A(\theta) H_{2\times 2} A(\phi).
\end{equation}
Notice that the $e^{i\phi/2}$ is global only to the pseudospin subspace and, thus, we should account for it when building the $M\times M$ unitary. However, it can be shown that the algorithm in Clements \emph{et al.}~\cite{Clements16} can be easily adapted to employ $T'(t,s)$ as the basic building block of the quantum circuit, instead of $T(t,s)$. Thus, the control of this global phase is not necessary for the purpose at hand. In the inset of Fig.~\ref{BSwithPSOLScheme} an illustration of the application of a $T(t,s)$ operation between a $|\!\!\uparrow\rangle$ mode in site $s$ and a $|\!\!\downarrow\rangle$ mode in site $s+1$ is shown. First, we use a spin-dependent displacement operation to bring the $|\!\!\downarrow\rangle$ mode in site $s+1$ to the site $s$. Then, the quantum gates described above can be applied, and the operation is finished by shifting back the $|\!\!\downarrow\rangle$ mode to site $s+1$.

\subsubsection{Site- and state-resolved detection of individual atoms}

Being able to measure the output state is a necessary condition for any useful quantum computation. A fluorescence image is captured to measure the final state~\cite{Alberti15,Robens16mbg}. Using a high-resolution objective lens, the position of the atoms in the optical lattices can be reconstructed with high-fidelity, exceeding $99\%$ \cite{Robens16b}.
This fluorescence technique provides information about the occupation of the lattice sites, however, to discriminate between the two modes in the same lattice sites, a spin-sensitive detection scheme is needed. For that, as demonstrated in Ref.~\cite{Robens16b}, a long-distance state-dependent shift can be performed, turning apart atoms in different states, $|\!\!\uparrow\rangle$ and $|\!\!\downarrow\rangle$, that were initially on the same lattice site. 

Ideally, one should also be able to detect how many atoms are in each site. Unfortunately, standard fluorescence imaging produces pairwise atom losses, allowing only the parity of the occupation number to be measured. To solve this, one could try to spread the atoms along the perpendicular lattice sites before imaging, for example, through a ballistic expansion. Similar methods are employed for number-resolving photodetection~\cite{Hadfield09,Carolan14}, and have been recently adapted to optical lattices~\cite{Omran15,Lukin19}. Alternatively, one could exploit interaction blockade to induce occupation-dependent tunneling to distinct sites of a multilayer optical lattice \cite{Preiss15b}.

\section{Scaling of atomic boson sampling}\label{ScalingAndErrors}

Quantum circuits with ultracold atoms, described in section~\ref{PSOLforQC}, provide a way to implement boson sampling, as an alternative to photonic devices. Currently, atomic systems allow controlling a hundred of particles distributed in hundreds of lattice sites, which makes these systems highly attractive to scale up boson sampling and achieve quantum supremacy. The initial state could be given by $N$ atoms uniformly distributed along the first $N$ sites $|\psi_0\rangle =\sum_{s=1}^{N}\hat{a}^{\dagger}_{2s-1} |0\rangle$, where $|0\rangle$ is the vacuum of all $M$ modes. If the generated unitary $U$ is chosen randomly according to the Haar measure, then to sample from the output probability distribution $P(n_1,n_2,..., n_{M})=|\langle n_1,n_2,...,n_{M}|\hat{U}|\psi_0\rangle|^2$ is believed to be a hard task for classical computers~\cite{Aaronson11}. Although the formal mathematical proof of hardness requires $N\leq M^{1/6}$, typically, a more feasible condition, i.e. $N\leq M^{1/2}$, is assumed sufficient.

In order to propose a quantum supremacy demonstration experiment by solving the boson sampling problem, it is necessary to take into account experimental imperfections. In this section we study how particle loss may, on the one hand, decrease the rate in which valid samples can be generated, and, on the other hand, change the complexity of the probability distribution we are sampling from.

\subsection{A simple model for the sampling rate}

The repetition rate of the experiment is directly related to the time necessary to prepare the initial state, make all the interference operations, and measure the final state. The initial state can be prepared efficiently in about $100$ ms up to $100$ atoms~\cite{Robens17}, that is why we assume a fixed time $t_\textrm{in}$ for carrying out this initial step. The time required to make all interference operations, however, directly depends on the number of modes in the system, which grows as $M=N^2$ as a requirement of the problem itself. As we already mentioned, any unitary transformation of dimension $M\times M$ can be done by $M(M-1)/2$ two-dimensional unitary operations~\cite{Reck94,Clements16} from which we consider that, in average, $(M-1)/2$ can be done in parallel, i.e., at the same time (see Fig.~\ref{BSwithPSOLScheme}). The final state measurement is done in a single operation lasting about $t_\textrm{det}=50$ ms, which detects the position and spin of the atoms by fluorescence imaging.
All this suggests that the processing time scales as $t_\textrm{pr}\equiv R_\textrm{pr}^{-1} \approx N^2 t_\textrm{op}+t_\textrm{in} +t_\textrm{det}$ with respect to the number of particles $N$. Here, $t_\textrm{op}$ is the time required to make an interference operation, while $t_\textrm{in}$ and $t_\textrm{det}$ are the initialization and measurement times respectively. In addition, we are mainly interested in detecting the output states that have at most one atom in each mode, since the probability of those events is the one predicted to be the hardest for classical algorithms. Thus, the states belonging to this subspace, called collision-free-subspace, are post-selected from all the output states. With a quadratic scaling of the number of modes, approximately a fraction $e^{-1}$ of the final states correspond to these kind~\cite{Arkhipov12}.

The above analysis describes the ideal situation where all atoms are prepared in the ground state of motion, none of them is lost in the experiment, and the measurement is carried out with 100\% efficiency. However, in a realistic scenario, one should account for experimental errors that affect the rate at which post-selected samples are generated. Following a similar approach as the one presented in Refs.~\cite{Wang17c,Neville17}, we estimate the sampling rate as
\begin{equation}\label{atomrate}
R=\frac{1}{e}R_\textrm{pr}\eta_\textrm{d}^{N}P_\textrm{surv},
\end{equation}
where $P_\textrm{surv}$ is the survival probability of $N$ atoms and $\eta_\textrm{d}$ is the detection efficiency per atom. The latter is controlled significantly well with the best reported detection efficiency of $99\%$ \cite{Robens16b}. The survival probability of the $N$ atoms during the complete evolution will depend on the processing time $t_\textrm{pr}$. In principle, an atom may escape from the trap due to background vapor collisions. The survival probability that accounts for this effect for single atoms is given by the exponential formula $\exp{(-t_\textrm{pr}/\tau_\textrm{bg})}$, where $\tau_\textrm{bg}$ is the mean lifetime of a single atom before it is lost by background collisions. 

Also, the presence of more than one atom in the same lattice site may cause the loss of one of the atoms, or even the loss of both of them due to inelastic collisions (e.g., spin exchanging collisions). For this type of collisions, the survival probability of a pair of atoms can be written as a $(1+(t_\textrm{ pr}-t_\textrm{ in})/\tau_\textrm{tb})^{-1}$ function of time \cite{Roberts00}, where $\tau_\textrm{tb}$ is the mean lifetime of two-body collisions. Then, one has to consider the probability of having a particle pair on the same lattice site and per each step of the process. If we start from a configuration that has all the atoms placed on different lattice sites, this probability can be considered zero until $t_\textrm{in}$. During the process of interference operations and measurement, this probability cannot be ignored. To approximate the value of this probability, we consider that the bosons are at all times uniformly distributed. We can then write the survival probability of $N$ atoms per time step $\tau$ as
\begin{equation}\label{SurvProbStep}
P_\textrm{step}(\tau)=(e^{-N\frac{\tau}{\tau_\textrm{bg}}})\sum_{k=0}^{N/2} P_\textrm{pair}(k)\Big[1+\frac{\tau}{\tau_\textrm{tb}}\Big]^{-k}
\end{equation}
where $P_\textrm{pair}(k)$ is the probability of finding $k$ pairs distributed in different sites. For example, $k=0$ corresponds to the case where all atoms are on distinct lattice sites. $P_\textrm{pair}(k)$ can be analytically calculated, leading to $(3/2)^k/(e^{3/2}k!)$ for large $N$, see appendix~\ref{pair_distribution} for details. Equation~(\ref{SurvProbStep}) does not take into account states with more than two particles in a lattice site. For a more accurate description, $P_\textrm{pair}(k)$ can be substituted by $P(k_2,k_3,k_4,k_5)$, meaning the probability of having $k_2$ pairs, $k_3$ trios, $k_4$ quartets, and so on. In Fig.~\ref{BSEstimations}(a), different configurations are shown for $N=4$ and $M=12$. However, and as proved in appendix~\ref{pair_distribution}, this generalized probability tends to the same Poissonian distribution $(3/2)^{k_2}/(e^{3/2}k_2!)$ at large $N$. With this model, the total survival provability can be written as
\begin{equation}\label{SurvProb}
P_\textrm{surv}(N)=P_\textrm{step}(t_\textrm{ op})^MP_\textrm{step}(t_\textrm{ det}). 
\end{equation}
Later, we will benchmark this model with exact numerical simulations for a small number of particles $N$. Up to then, we can assume this model to be correct and compare it with other models that describe the scaling of the boson sampling problem in photonic circuits or classical computers.

\begin{table}
\centering
\caption{Experimental parameters}
\label{table2}
\begin{tabular}{{ c c c c c c}}
\hline
\hline
 & Conservative & State of the art \\
\hline
Spin addresing + Lattice shift time ($t_\textrm{ op}$) & $150\mu$s +$20\mu$s    &$30\mu$s +$3\mu$s  \\
\hline
Initialization time ($t_\textrm{ in}$) & $0.75$s    &$0.1$s    \\
\hline
Detection time ($t_\textrm{ det}$) & $60$ms    &$30$ms    \\
\hline
Detection efficiency ($\eta_\textrm{ d}$) & 0.99    & 0.999   \\
\hline
One-body loss lifetime ($\tau_\textrm{ bg}$)& $60$s    &$360$s    \\
\hline
Two-body loss lifetime ($\tau_\textrm{ tb}$) & $40$ms    &$400$ms    \\
\hline
\hline
\end{tabular}
\end{table}

Conservative and state of the art values for the experimental parameters can be found in Table~\ref{table2}. For the two-body losses, we take $\tau_\textrm{tb}=400$ms as the state-of-art value, which is in principle achievable using Feshbach resonances~\cite{Chin04}. These two cases are represented by the two solid blue curves in Fig.~\ref{BSEstimations}(b), were the lower and upper curves corresponding to current and best cases, respectively. We have also included an additional line (dotted blue) that represents the case with the best reported parameters except for the mean lifetime of two-body collisions, which is assumed to remain $\tau_\textrm{tb}=40$ms. Notice that the upper blue lines meet the solid gray line (representing the classical computing rate) around $N\gtrsim30$, thus, predicting quantum advantage for $N\gtrsim30$ particles.

\begin{figure}[t]
\centering
	\includegraphics*[width=\columnwidth]{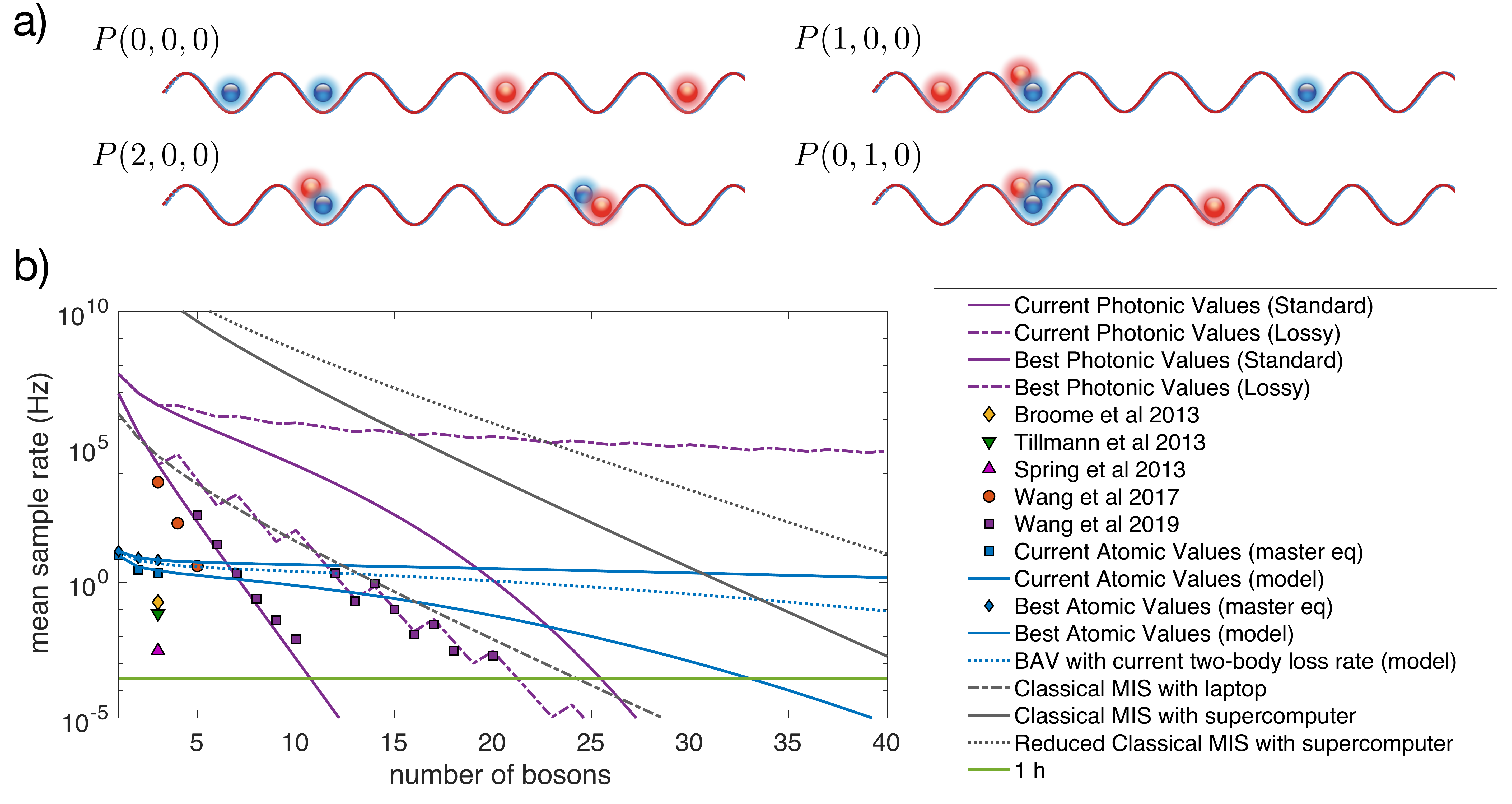}
	\caption{ a) Examples of states with $M=12$ and $N=4$. $P(0,0,0)$ is the ratio between the number of states with zero pairs, trios and quartets, and the total number of states. $P(1,0,0)$ corresponds to the proportion of states with a single particle pair, and zero trios and quartets. $P(2,0,0)$ and $P(0,1,0)$ for states with two particle pairs, zero trios and zero quartets, and a single particle trio, zero pars and zero quartets, respectively. b) Scaling of boson sampling machines. The sampling rate versus the number of particles is shown for atomic (blue), photonic (purple), and classical (grey) boson sampling devices. If experimental parameters are improved, both photonic and atomic devices may surpass the classical algorithms, at $N\approx 23$ and $N\approx 30$, respectively. In the case of the photonic device, for the dashed curve we allow $\approx 30\%$ of lost photons. \label{BSEstimations}}
\end{figure}

Photonic boson sampling devices also suffer from particle loss, which is, actually, the main limitation when scaling up. Although the rate in which indistinguishable photons are created is of $R_0=76$MHz, the rate in which valid experimental samples are generated in current setups drops down to $R=295$Hz for 5-photon boson sampling~\cite{Wang19_BS}. If the number $M$ of modes scale quadratically with $N$, the sampling rate of a photonic experiment can be characterized by
\begin{equation}
R=\frac{1}{e}\frac{R_0}{N}\eta^N
\end{equation}
where $R_0/N$ is the rate in which an $N$-photon initial state can be prepared and $\eta=\eta_\textrm{f} \eta_\textrm{c}^d$ is the survival probability of a single photon during the experiment. The latter is a product of a fixed survival probability $\eta_\textrm{f}$ that accounts for errors that do not increase with the number of particles $N$ and the transmission probability of the photon through the circuit $\eta_\textrm{c}^d$, which depends on the transmission probability per unit of length $\eta_\textrm{c}$ and the optical depth $d$, that increases quadratically ($d=M=N^2$ for a square circuit~\cite{Clements16}) with $N$. If photon loss is allowed in the model, the sampling rate becomes
\begin{equation}
R=\frac{1}{e}\frac{R_0}{N}\binom{N}{N-k_l}\eta^{N-k_l}(1-\eta)^{k_l}, 
\end{equation}
where $k_l$ is the number of lost photons. In boson sampling with lost photons, if the loss of $2$ photons is allowed $R$ increases by a factor of $N(N-1)/2$ with respect to the case without losses.

From the $R=295$Hz count rate reported in Ref.~\cite{Wang19_BS} for the 5-photon boson sampling, and taking into account that the transmission rate of the $60\times60$ optical circuit is $98,7\%$, and, thus, $\eta_\textrm{c}=0.987^{1/60}$, we can extract a fixed survival probability of $\eta_\textrm{f}=0.14$. According to Ref.~\cite{Wang18},  the single-photon source, interferometer and detection efficiencies can be increased up to $0.8$, $0.9$, and $0.9$ respectively, obtaining $\eta_\textrm{f}=0.65$. In Fig.~\ref{BSEstimations}(b), lower purple curves (both solid and dashed) assume $\eta_\textrm{f}=0.14$ and a circuit transmission rate of $\eta_\textrm{c}=0.987^{1/60}$, while the upper curves assume $\eta_\textrm{f}=0.65$ and a perfect transmission rate. With the best parameters, quantum advantage could be achieved around $N\approx23$ with a $30\%$ photon loss.

Finally, the time required by the Metropolised independence sampling algorithm introduced in Ref.~\cite{Neville17} to produce a valid sample scales as 
\begin{equation}\label{computerrate}
1/R=100\tilde{a}N^22^N
\end{equation}
where $\tilde{a}$ relates to the speed of the classical computer. This value has been reported to be $\tilde{a}=3\times10^{-15}$ s in the case of the Tianhe 2 supercomputer~\cite{Wu18}. For the case of a regular computer we choose this value to be $\tilde{a}=3\times10^{-9}$ s. If comparing with the photonic boson sampler with photon loss, we should substitute $N$ by $N-k_l$ in Eq.~(\ref{computerrate}). The latter corresponds to the dotted grey line in Fig.~\ref{BSEstimations}(b), which crosses the photonic sampling rate at around $N\approx 23$.

Since both $t_\textrm{pr}$ and $d$ scale quadratically with the number of particles, the atomic and photonic sampling-rate formulas scale worse than the classical algorithm. However, and as it is shown in Fig.~\ref{BSEstimations}(b), at meaningful time scales (i.e., for all practical purposes) both experimental machines are expected to sample much faster than classical supercomputers. In the case of the atomic device, we show that sampling faster than a classical supercomputer is possible around $N\gtrsim 30$, both when $\tau_\textrm{tb}=40$ms and $\tau_\textrm{tb}=400$ms. 

One thing that has to be considered is that two-body losses may change the form the probability distribution we are sampling from. If that is the case, we need to quantify how far are we from the probability distribution that corresponds to the boson sampling problem. For that, in the next section we present a Hamiltonian model that will serve us to quantify this distance between both probability distributions in terms of the state fidelity.

\subsection{Hamiltonian model for particle loss}\label{subsec:HamilModel}
In this section, we use the second quantization formalism of quantum mechanics to build a Hamiltonian model that will represent the boson sampling problem with particle loss. The system will contain $N$ particles in $M$ bosonic modes, and will be subjected to incoherent one-body and two-body losses. Those losses can be represented, as we will see, by an anti-Hermitian part in the system Hamiltonian. 

Using second quantization, the free-energy Hamiltonian corresponding to a one-dimensional optical lattice of $M/2$ sites (for simplicity, we will assume that the number of modes is even) where each site can hold atoms in two different atomic states is written as
\begin{equation}
H_0=\sum_{s=1}^{M/2} \omega_\uparrow a^\dagger_{2s-1,\uparrow}a_{2s-1,\uparrow} + \omega_\downarrow a^\dagger_{2s,\downarrow}a_{2s,\downarrow} 
\end{equation}
where $a^\dagger_j (a_j)$ is the creation (annihilation) operator associated to the $j$-th mode. From now on and for simplicity, $a_{j,\uparrow}\equiv a_{j}$ and $a_{j,\downarrow}\equiv a_{j}$, odd or even $j$ numbers being associated to $|\!\!\uparrow\rangle$ or $|\!\!\downarrow\rangle$ atomic states, respectively. 

\begin{figure}[t]
\centering
	\includegraphics*[width=\columnwidth]{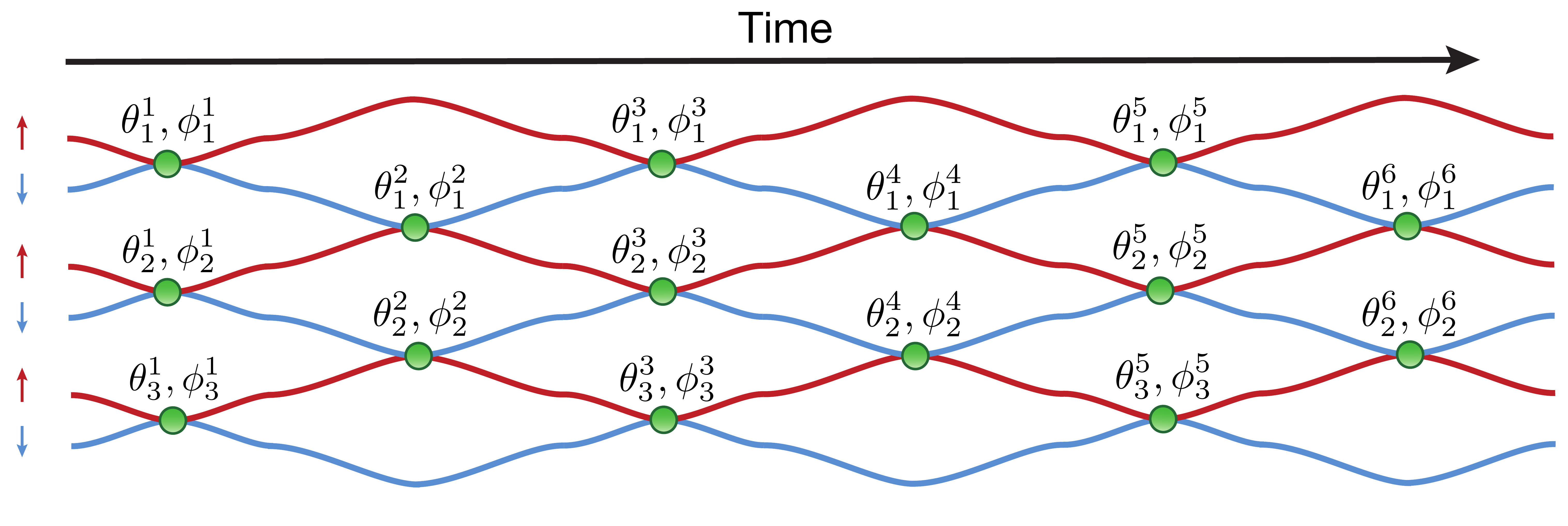}
\caption{ Implementation of a $6\times6$ unitary operation in $6$ time steps. At each time step $t$, a $2\times 2$ unitary transformation is applied on each site $s$, characterized by $\theta_s^t$ and $\phi_s^t$. Red and blue lines represent bosonic modes associated with state $|\!\!\uparrow\rangle$ and $|\!\!\downarrow\rangle$. Green circles represent the light shift generated by a focused laser beam, which allows to perform an independent operations on each lattice site. \label{CircuitFigure}}
\end{figure}

Coherent population exchange between $|\!\!\uparrow\rangle$ and $|\!\!\downarrow\rangle$ states can be achieved using a MW field with frequency $\omega_\uparrow-\omega_\downarrow$, giving rise to Rabi oscillations. The effective Hamiltonian associated to this process is, in an interaction picture with respect $H_0$,
\begin{equation}\label{BSHamil}
H_{x}=\sum_{s=1}^{M/2} \frac{\Omega_0}{2} (a^\dagger_{2s-1}a_{2s}e^{i\varphi_0} + a_{2s-1}a^\dagger_{2s}e^{-i\varphi_0} )
\end{equation}
where $\Omega_0$ and $\varphi_0$ are the Rabi frequency and the phase associated to the MW driving. As anticipated in section \ref{PSOLforQC}, local differential light shifts can be produced by focused laser beams, which allow to shift the energies defined by $H_0$. This can be represented by the following Hamiltonian, 
\begin{equation}\label{shiftHamil}
H_{z}=\sum_{s=1}^{M/2} \frac{\Omega_s}{2} (a^\dagger_{2s-1}a_{2s-1} - a^\dagger_{2s}a_{2s}),
\end{equation}
where the value of $\Omega_s$ can be different for each site $s$. Combining the evolution of Hamiltonians (\ref{BSHamil}) and (\ref{shiftHamil}), any $2\times2$ unitary tranformation between on-site modes can be represented. The time evolution operator corresponding to this generic operation would be
\begin{equation}\label{OneTimeEvol}
\hat{U}_{t}=e^{-iH^{(\varphi=\pi)}_{x}\tau/4} e^{-iH^{\vec{\theta}}_{z}\tau/4} e^{-iH^{(\varphi=0)}_{x}\tau/4} e^{-iH^{\vec{\phi}}_{z}\tau/4},
\end{equation}
where $\tau=2\pi/\Omega_0$, and $\Omega_{s}\tau/4=\theta_s$ at $\exp{[-iH^{\vec{\theta}}_{z}\tau/4]} $ and $\Omega_{s}\tau/4=\phi_s$ at $\exp{[-iH^{\vec{\phi}}_{z}\tau/4]}$. The operation described in Eq.~(\ref{OneTimeEvol}) applies simultaneously in all $M/2$ lattice sites, and it is characterized by $M$ independent parameters $\vec{\theta}_t=(\theta_1,\theta_2,...,\theta_{M/2})$ and $\vec{\phi}_t=(\phi_1,\phi_2,...,\phi_{M/2})$. Those parameters describe $M/2$ arbitrary unitary operations, each as defined in Eq.~(\ref{eq:decomposition}), between all on-site modes. An example is shown in Fig.~\ref{CircuitFigure} for the case $M=6$. As explained previously, the ability to shift the optical potential associated to one of the atomic states allows to connect neighbouring modes of the circuit. Thus, the following Hamiltonians, can also be implemented: 
\begin{equation}\label{BSHamilp}
H'_{x}=\sum_{s=1}^{M/2-1} \frac{\Omega_0}{2} (a^\dagger_{2s}a_{2s+1}e^{i\varphi_0} + a_{2s}a^\dagger_{2s+1}e^{-i\varphi_0} ),
\end{equation}
and
\begin{equation}\label{shiftHamilp}
H'_{z}=\sum_{s=1}^{M/2-1} \frac{\Omega_s}{2} (a^\dagger_{2s}a_{2s} - a^\dagger_{2s+1}a_{2s+1}).
\end{equation}

Up to now, we have assumed the number of modes in the circuit $M$ to be even. In the case $M$ is odd, the summatories in Eqs.~(\ref{BSHamil}), (\ref{shiftHamil}), (\ref{BSHamilp}) and (\ref{shiftHamilp}) have to sum up to $(M-1)/2$. Doing the same as in Eq.~(\ref{OneTimeEvol}) with Hamiltonians (\ref{BSHamilp}) and (\ref{shiftHamilp}), we are able to implement $M/2$ two-dimensional unitary operations among neighbouring modes. Concatenating these operations as
\begin{equation}\label{AllTimeEvol}
\hat{U}=\hat{U}_M \hat{U}_{M-1}...\hat{U}_{2}\hat{U}_1,
\end{equation}
we are able to transform the $M$ modes of the system according to a Haar random unitary matrix $U$. Unitaries $U$ and $\hat{U}$ should not be mixed: while $U$ is an $M\times M$ unitary matrix, $\hat{U}$ is an infinite-dimentional unitary operator acting on the Hilbert space of quantum states. Both can be related by the following formula
\begin{equation}\label{ModeTransformation}
\hat{U}^\dagger a^\dagger_{j} \hat{U} =\sum_{i=1}^MU_{ji}a^\dagger_i, 
\end{equation}
which describes how the $j$-th mode transforms after all operations in Eq.~(\ref{AllTimeEvol}).

To model the effect of particle loss we can use a Lindblad master equation of the following form
\begin{equation}\label{Lindblad}
\dot{\rho}=-i[H,\rho] +\sum_b \Gamma_b \mathcal{L}_b(\rho),
\end{equation}
where
\begin{equation}\label{Lindblad2}
\mathcal{L}_b(\rho)=F_b \rho F_b^\dagger - \frac{1}{2}\{F^\dagger_b F_b,\rho\},
\end{equation}
and $F_b$ represents a jump operator acting on the system and producing the loss. The jump operator corresponding to one-particle loss due to collisions with background gas would be an annihilation operator acting independently on each mode $a_m$. Thus, the superoperator corresponding to this process would be 
\begin{equation}\label{background}
\mathcal{L}_\textrm{ bg}(\rho)=\sum^M_{m=1} a_m \rho a_m^\dagger - \frac{1}{2}\{\hat{n}_m,\rho\},
\end{equation}
with a loss rate given by $\Gamma_\textrm{ bg}=1/\tau_\textrm{ bg}$. For two-body collisions, we need to consider two scenarios: on the one hand, two particles on the same mode can collide. The jump operator would be $a^2_m/\sqrt{2}$ and the superoperator 
\begin{equation}\label{twobody1}
\mathcal{L}_\textrm{ tb1}(\rho)=\frac{1}{2}\sum^M_{m=1} a^2_m \rho (a_m^\dagger)^2 - \frac{1}{2}\{\hat{n}_m(\hat{n}_m-1),\rho\},
\end{equation}
with loss rate $\Gamma_\textrm{ tb}=1/\tau_\textrm{ tb}$. On the other hand, two particles on neighbouring modes on the same site could collide, in which case the jump operator would be $a_{2s-1}a_{2s}$, where $s$ is the site index. This jump operator would change to $a_{2s}a_{2s+1}$ when the lattice configuration is shifted to produce the interactions described in Eqs.~(\ref{BSHamilp}) and (\ref{shiftHamilp}). Thus, we would have two different superoperators acting depending on the lattice configuration. Those would be 
\begin{equation}\label{twobody2}
\mathcal{L}_\textrm{ tb2}(\rho)=\sum^{M/2}_{s=1} a_{2s-1}a_{2s}  \rho a^\dagger_{2s-1}a^\dagger_{2s} - \frac{1}{2}\{\hat{n}_{2s-1}\hat{n}_{2s},\rho\},
\end{equation}
or
\begin{equation}\label{twobody2p}
\mathcal{L}'_\textrm{ tb2}(\rho)=\sum^{M/2}_{s=1} a_{2s}a_{2s+1}  \rho a^\dagger_{2s}a^\dagger_{2s+1} - \frac{1}{2}\{\hat{n}_{2s}\hat{n}_{2s+1},\rho\},
\end{equation}
both with decay rate $\Gamma_\textrm{ tb}=1/\tau_\textrm{ tb}$.

In the boson sampling problem the number of particles is conserved. In other words, the Hamiltonian commutes with $\hat{N}=\sum_m\hat{n}_m$, the operator corresponding to the total number of particles. In the master equation, the elements that describe population transfer from the $N$ particle subspace to another subspace with a lower amount of particles are the first terms in Eqs.~(\ref{background}), (\ref{twobody1}), (\ref{twobody2}), and (\ref{twobody2p}). If we are only interested in the $N$ particle subspace, we can safely ignore these terms, as they do not affect this subspace. Doing so, we loose the unitarity of the master equation, which will now describe loss of probability, $\textrm{ tr}(\rho)\leq1$. In exchange, the description of the problem becomes simpler and the dynamics of the lossy system is given by the non-Hermitian Hamiltonian 
\begin{equation}
\tilde{H}=H-iV -i\frac{\Gamma_\textrm{ bg}}{2}\hat{N},
\end{equation}
where $V$ changes from
\begin{equation}
V=\frac{\Gamma_\textrm{ tb}}{4}\sum_{m=1}^M\hat{n}_m(\hat{n}_m-1) 
+\frac{\Gamma_\textrm{ tb}}{2}\sum_{s=1}^{M/2}\hat{n}_{2s-1}\hat{n}_{2s} 
\end{equation}
to
\begin{equation}
V'=\frac{\Gamma_\textrm{ tb}}{4}\sum_{m=1}^M\hat{n}_m(\hat{n}_m-1) 
+\frac{\Gamma_\textrm{ tb}}{2}\sum_{s=1}^{M/2-1}\hat{n}_{2s}\hat{n}_{2s+1} 
\end{equation}
depending on the lattice configuration. It can be proven that, at each step, the anti-Hermitian part commutes with the Hermitian part, i.e. $[H,V]=[H',V']=0$, and, in addition, we have that $[\hat{N},V]=[\hat{N},V']=[\hat{N},H]=[\hat{N},H']=0$, thus, the time evolution operator corresponding to the time step $t$ can be written as $\hat{U}_t e^{-V\tau} e^{-\Gamma_\textrm{ bg}\tau\hat{N}/2}$. At the end, the time evolution operator can be written as
\begin{equation}\label{LossyTimeEvol}
\hat{U} e^{-M\tau\Gamma_\textrm{ bg}\hat{N}/2}e^{-V_{M}\tau}e^{-V_{M-1}\tau}...e^{-V_{2}\tau} e^{-V_{1}\tau}
\end{equation}
where $V_t$ is given by $\hat{U}^\dagger_{t-1,1}V\hat{U}_{t-1,1}$ when $j$ odd, and $\hat{U}^\dagger_{t-1,1}V'\hat{U}_{t-1,1}$ when $j$ even, with $\hat{U}_{t,1}=\prod_{j=1}^{t}\hat{U}_j$. The probability of having an $N$-particle state at the output is then given by 
\begin{equation}
p=e^{-M\tau\Gamma_\textrm{ bg}N}\langle\psi_0|\big(\prod_{t=1}^M e^{-V_{t}\tau}\big)^\dagger \prod_{t=1}^M e^{-V_{t}\tau} |\psi_0\rangle,
\end{equation}
where $|\psi_0\rangle$ is the initial state of the system, and $\hat{N}|\psi_0\rangle=N|\psi_0\rangle$. Notice that the exponential decay produced by uncorrelated particle loss $\exp{(-N\Gamma_\textrm{ bg}M\tau)}$ has the same form as in Eq.~(\ref{SurvProbStep}). Using Eq.~(\ref{LossyTimeEvol}), we can also write the fidelity between the final states produced by the cases with and without losses. This is given by\footnote{Typically, the fidelity between two pure states can be written as $|\langle\psi_1|\psi_2\rangle|^2$. However, if these are not normalized, the correct expression is $|\langle\psi_1|\psi_2\rangle|^2/(\langle\psi_1|\psi_1\rangle\langle\psi_2|\psi_2\rangle)$}
\begin{equation}\label{generalfidelity}
F=|\langle\psi_0| \prod_{t=1}^M e^{-V_{t}\tau} |\psi_0\rangle|^2 \bigg/\langle\psi_0|\big(\prod_{t=1}^M e^{-V_{t}\tau}\big)^\dagger \prod_{t=1}^M e^{-V_{t}\tau} |\psi_0\rangle.
\end{equation}
From Eq.~(\ref{generalfidelity}), we learn that uncorrelated particle loss does not affect the state fidelity nor the form of the final probability distribution. In consequence, one can claim that homogeneous one-body losses do not change the complexity associated to the probability distribution of the boson sampling problem. Unfortunately, one can not claim the same thing for two-body losses, which is why we need to know how both the fidelity and the success probability scale with the number of particles $N$. These quantities can be numerically calculated for a small number of particles, however, their calculation becomes intractable for more than five particles because the Hilbert space dimension increases exponentially. Thus, it is convenient to have analytical expressions for $F$ and $p$. For that, we need to do some approximations.

\subsubsection{Weak two-body losses}
\begin{figure}[t]
\centering
	\includegraphics*[width=1\columnwidth]{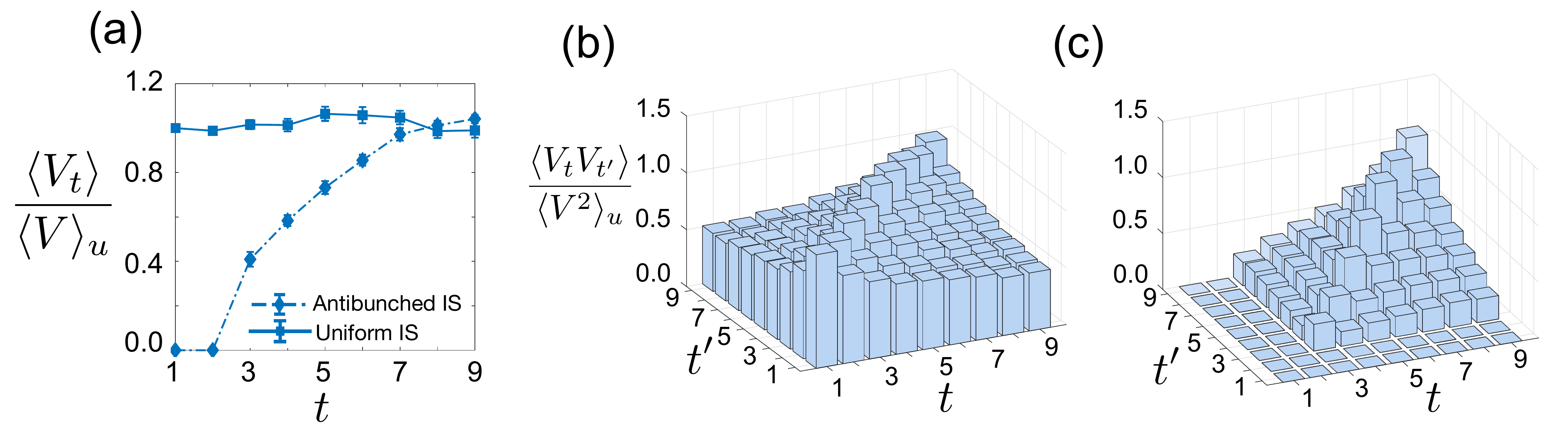}
	\caption{Numerically calculated $\langle V_t\rangle$ and $\langle V_t V_{t'}\rangle$ for $N=3$ and $M=9$. (a) Mean value of $\langle V_t\rangle$ for 30 different random unitaries. Solid and dashed lines correspond to uniform and anti-bunched initial states, respectively. (b) and (c) show the mean value of $\langle V_t V_{t'}\rangle$ for 30 unitaries and for the cases of uniform and anti-bunched initial states, respectively. \label{Fig:AvVs} }
\end{figure}

If the two-body loss rate is small, meaning $\Gamma_\textrm{ tb} M\tau\ll1$, we can expand all exponentials to the second order, and we get that the success probability is  
\begin{equation}
p\approx e^{-M\Gamma_\textrm{ bg}\tau N}\Big\{1-2\tau\sum_{t}^M\langle V_t \rangle +\tau^2[\sum_{t}^M\langle V_t^2\rangle +\sum_{t, t'}^M\langle V_t V_{t'}\rangle]\Big\}  +O(\tau^3)
\end{equation}
and the fidelity is 
\begin{equation}
F\approx 1-\tau^2\Big\{\sum_{t, t'}^M\langle V_t V_{t'}\rangle -\langle V_t \rangle\langle V_{t'}\rangle\Big\} +O(\tau^3).
\end{equation}
In Fig.~\ref{Fig:AvVs}, we can see the form of $\langle V_t\rangle$ and $\langle V_t V_{t'}\rangle$ for two different initial states and for $N=3$. For a uniform initial state $|\psi_u\rangle$, where all configurations have the same probability amplitude, $\langle V_t\rangle$ and $\langle V_t V_{t'}\rangle$ maintain, in average, the same value throughout the process. With an anti-bunched initial state, where the average distance among atoms is maximal, $\langle V_t\rangle$ and $\langle V_t V_{t'}\rangle$ are zero at the beginning and they increase their value with each step $t$. When increasing the number of particles, this behaviour is not expected to change, as long as the number of modes scales quadratically with $N$. Thus, one can assume the following lower bounds for $p$ and $F$
\begin{equation}\label{lowerprob}
p \gtrsim e^{-M\Gamma_\textrm{ bg}\tau N}\Big\{1-2M \tau \langle V \rangle_u +M\tau^2(1 +M)\langle V^2\rangle_u\Big\}  +O(\tau^3)
\end{equation}
and 
\begin{equation}\label{lowerfidelity}
F \gtrsim 1-M^2\tau^2\Big\{\langle V^2\rangle_u -\langle V \rangle^2_u\Big\} +O(\tau^3),
\end{equation}
where we assume that $\langle V_t\rangle\lesssim \langle V \rangle_u$ and $\langle V_t V_{t'}\rangle \lesssim\langle V^2\rangle_u$.

Now, one can analytically calculate the values of $\langle V\rangle_u$ and $\langle V^2\rangle_u$, which, for large $N$, give $\langle V\rangle_u=3\Gamma_\textrm{ tb}/4$ and $\langle V^2\rangle_u=(3/2 + 9/4)\Gamma^2_\textrm{ tb}/4$, see appendix \ref{App:AvVs} for more details. Using these results, Eqs.~(\ref{lowerprob}) and (\ref{lowerfidelity}) give
\begin{equation}\label{lowerprob2}
p \gtrsim e^{-M\Gamma_\textrm{ bg}\tau N}\Big\{1-3M \Gamma_\textrm{ tb}\tau/2 +15 M^2\Gamma^2_\textrm{ tb}\tau^2(1 +1/M)/36\Big\}  +O(\tau^3)
\end{equation}
and 
\begin{equation}\label{lowerfidelity2}
F \gtrsim 1-3M^2\Gamma^2_\textrm{ tb}\tau^2/8 +O(\tau^3).
\end{equation}
From the above equations, we can infer that when increasing the number of particles, $\Gamma_\textrm{ tb}\tau$ has to scale as $1/M$, either by decreasing $\tau$ or $\Gamma_\textrm{ tb}$, if we want to maintain a constant value for $F$ and $p$. Notice that maintaining a constant value for the fidelity would imply that the complexity of the generated sample is kept constant~\cite{Aaronson16b}. 

According to Eq.~(\ref{lowerfidelity2}), achieving quantum advantage at $N\approx30$ with $99\%$ state fidelity would require $\tau_\textrm{ tb}$ to be around five thousand times the operation time $\tau$. For example, if $\tau \sim 100\mu$s, then $\tau_\textrm{ tb}\sim 500$ms. This value is, in principle, achievable using Feshbach resonances~\cite{Chin04}. However, it may well be that high fidelities are not necessary to achieve quantum supremacy. Advantage with respect classical computers may be possible in a regime where $M\Gamma_\textrm{ tb}\tau \gtrsim1$, as long as sampling from the generated probability distribution, different from the one generated by standard boson sampling, is a hard task. This is why it is interesting to try to extend the previous analysis to the strong-loss case.

\subsubsection{Strong two-body losses}

\begin{figure}[t]
\centering\includegraphics*[width=1\columnwidth]{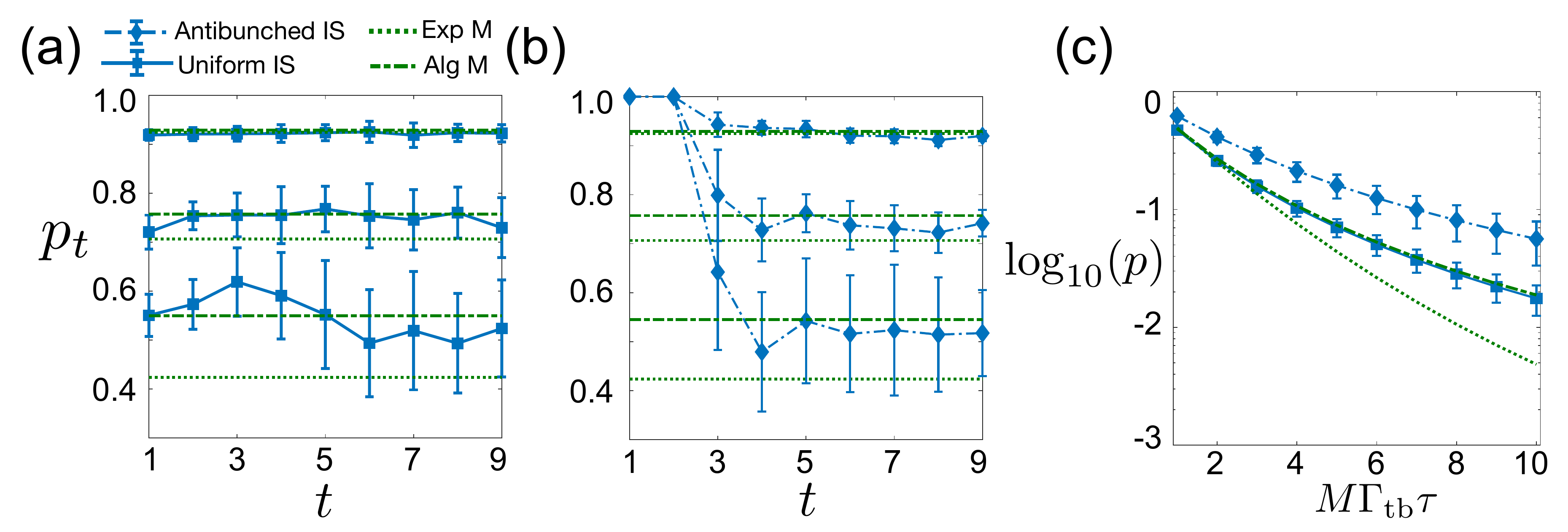}
	\caption{(a) and (b) Success probability per time step $p_t$, where $p=\prod_{t}p_t$, for different decay rates $M\Gamma_\textrm{ tb}\tau=1$ (upper lines), $M\Gamma_\textrm{ tb}\tau=5$ (middle lines), and $M\Gamma_\textrm{ tb}\tau=20$ (lower lines), $N=3$, and for a uniform and antibunched initial state, respectively. Blue lines represent the exact calculation, where each point is calculated after 30 different random unitaries. Dashed and dotted green lines represent the results of the models with algebraic and exponential decay, respectively. (c) Success probability versus $M\Gamma_\textrm{ tb}\tau$. One can see how the algebraic model (dashed green) is in very good agreement with the exact case with uniform initial state (solid blue). \label{Fig:Probs} }
\end{figure}

In the following, we try to estimate how the success probability would scale in a scenario where losses are strong and $M\Gamma_\textrm{ tb} \tau\ll1$ does not hold true. Exact numerical simulations can be used to calculate the behaviour of the success probability throughout the process. Similar to the case with weak losses, probability loss per time step is more or less constant for uniform initial states, see Fig.~\ref{Fig:Probs}(a). To estimate this value, we assume we have a uniform state at each time step, then, according to the model presented in the previous section, probability loss per time step would be given by\footnote{The loss rate of 3 particles (a particle trio) on the same site is given by $3\Gamma_\textrm{ tb}$. Notice that there are 3 different ways to form a pair, thus, the probability of a two-body collision is 3 times higher.}
\begin{equation}\label{simplemodel2}
P_\textrm{ step}=e^{-N\Gamma_\textrm{ bg}\tau}\sum_{k_3=0}^{N/3}\sum_{k_2=0}^{N/2-3k_3} P(k_2,k_3)\Big[1+3\Gamma_\textrm{ tb}\tau\Big]^{-k_3}\Big[1+\Gamma_\textrm{ tb}\tau\Big]^{-k_2},
\end{equation}
where $P(k_2,k_3)$ gives the probability having $k_2$ pair and $k_3$ trios. We ignore the cases where $4$ or more particles are in a lattice site, since these probabilities are zero for $N=3$. Notice that Eq.~(\ref{simplemodel2}) does not correspond to applying the loss Hamiltonian evolution operator to the uniform state for a time $\tau$, i.e. $e^{-N\Gamma_\textrm{ bg}\tau}\langle e^{-V\tau}\rangle_u$, which would end up with a similar result with exponential decays $\exp{(-\Gamma_\textrm{ tb}\tau)}$ instead of algebraic $(1+\Gamma_\textrm{ tb}\tau)^{-1}$. Interestingly, the formula with the algebraic decay approaches better to the exact result, as it can be seen in Fig.~\ref{Fig:Probs}(c). This tells us that the average state in the lossy case is not really uniform, as it is in the case without any loss. On the other hand, the excellent agreement between our model and the exact results justifies our model and its use in Fig.~\ref{BSEstimations}(b).

According to this model, advantage with respect classical computers can not be reached within the current experimental values. However, sampling faster than a classical supercomputer would be possible around $N\approx34$ with $\tau_\textrm{ tb}\approx 40$ms and $\tau\approx 33\mu$s. In this regime, $M\Gamma_\textrm{ tb}\tau\approx 1$, and most likely the generated probability distribution is different from the one required by standard boson sampling. Unfortunately, to determine wether it is still hard to sample from that probability distribution is beyond the scope of this work.

\subsection{Additional considerations}\label{subsec:additional}

Particle loss is the main experimental imperfection that affects to the sampling rate. However, there are other sources of error that, without affecting the sampling rate, can change the form the final probability distribution, i.e. the final state. Undesired differential light shifts may appear due to fluctuations of the magnetic field or imperfect light polarization of the optical lattice~\cite{Alberti14}. These light shifts will induce errors in the application of the $2\times2$ unitary transformations, which could deviate the total $M\times M$ unitary transformation from the ideal Haar random unitary. In this regard, Leverrier \emph{et al.}~\cite{Leverrier15} proved that the hardness of the final probability distribution is guaranteed, as long as the error introduced by each operation scales as $1/N^2$. A uniform light shift can be represented in our Hamiltonian model, just by including a term like $H_\textrm{ shift}=\epsilon\Omega_0\sum_{s=1}^{M/2}\hat{n}_{2s}$ through all the process. For $N=3$, a light shift with $\epsilon=8\times10^{-3}$ yields $F\approx 99\%$. Maintaining this fidelity at $N\approx30$ would then require $\epsilon \sim 8 \times10^{-5}$, which, for $\Omega_0\approx(2\pi)\times3$ MHz, would equal to an uncontrolled light shift of $(2\pi)\times 240$ Hz.

\pagebreak

Another relevant issue is the degree of distinguishability among the atoms. The boson sampling problem assumes that the particles are indistinguishable. In our case, two atoms on the same mode will be indistinguishable only if they share the same motional state. For that, ground-state cooling of all vibrational degrees of freedom is required. Moreover, the transport of atoms could change their motional state by inducing unwanted excitations. In this respect, Refs.~\cite{Rohde12a,Shchesnovich15} prove that quantum advantage is also possible with partially distinguishable particles. More precisely, Shchesnovich~\cite{Shchesnovich15} shows that, while increasing $N$, the final state fidelity (which measures how close we are to the boson sampling probability distribution) would remain constant, as long as the fidelity between the vibrational states of two atoms scales as $1-O(1/N)$.

Summarizing, in this chapter we have proposed a method to realize boson sampling with ultracold atoms in optical lattices. Using simple error models, we compare the scaling of atomic, photonic and classical boson sampling machines up to tens of particles, and conclude that, in the near future, atomic and photonic machines could achieve quantum advantage with $N\approx30$ and $N\approx23$ particles, respectively. We benchmark the atomic error model using exact numerical simulations of a Hamiltonian model that includes one-body and two-body losses. This two models are in good agreement for $N=3$ with weak and strong two-body losses\footnote{Numerical simulations of the Hamiltonian model become too demanding for $N>4$.}. Finally, with strong two-body losses one may still sample faster than a classical supercomputer, however, the probability distribution one is sampling from does not correspond to the boson sampling probability distribution, and further work is needed to prove wether it is still hard to sample from that probability distribution.

\chapter{Quantum Sensing with NV Centers}
\label{chapter:chapter_4}
\thispagestyle{chapter}

The ability to manipulate a small quantum system using electromagnetic radiation, and initialize and read-out its quantum state, can be used to gain control over nearby systems. It is the case of nanoscale NMR, whose goal is that of detecting and controlling magnetic field emitters (as nuclear spins) with high frequency and spatial resolution~\cite{Rondin14,Schmitt17,Boss17,Zopes18a,Glenn18, Zopes18b}. The NV center is a prominent quantum sensor candidate owing to its long coherence times working on room temperature environments. Its electronic spin is controlled with MW radiation, while initialization and read-out is done with optical fields~\cite{Schirhagl14, Wu16}. At room temperature, the decay time $T_1$ of the NV center is of the order of milliseconds~\cite{Degen17}. Coherence times due to dephasing $T_2$ are typically shorter, due to the interaction among $^{13}$C nuclei in the diamond and the NV~\cite{Maze08}. Here, DD techniques have played an important role, taking the coherence time $T_2$ to the decay time~$T_1$~\cite{Souza12}. On the other hand, DD techniques are also employed to couple the NV to a target signal, e.g. classical electromagnetic radiation~\cite{Taylor08} or the hyperfine fields emitted by nuclear spins~\cite{Abobeih18,Muller14b, Wang16a}. In particular, DD techniques generate filters that allow the passage of signals with only specific frequencies~\cite{Haase18}.  It is the accuracy of this filter what determines the fidelity in the detection and control on the target signal.  Continuous or pulsed (or stroboscopic) DD schemes are typically considered. While the former requires to fulfill the Hartmann-Hahn condition~\cite{Hartmann62, Casanova19}, the latter uses the time spacing among $\pi$ pulses to induce a rotation frequency in the NV, matching that of the target signal~\cite{Taminiau12}. Pulsed DD schemes have advantages over continuous methods, such as the achievement of enhanced frequency selectivity by using large harmonics of the generated modulation function~\cite{Taminiau12, Casanova15}. Another advantage is the demonstrated robustness against control errors of certain pulse sequences such as those of the XY family~\cite{Maudsley86, Gullion90, Souza11, Wang17a, Arrazola18}. However, the use of large harmonics makes DD sequences sensitive to environmental noise, and leads to signal overlaps which hinders  spectral readout~\cite{Casanova15}. As we will show, these issues can be minimised by applying a large static magnetic field $B_z$.  Unfortunately, the performance of pulsed DD techniques under large $B_z$ gets spoiled unless $\pi$ pulses are fast, i.e. highly energetic, compared with nuclear Larmor frequencies (note these are proportional to $B_z$). This represents a serious disadvantage, especially when DD sequences act over biological samples, since fast $\pi$ pulses require high MW power causing damage as a result of the induced heating~\cite{Cao20}.

In this chapter, we propose a design of amplitude modulated decoupling pulses that solves these problems and achieves tuneable, hence highly selective, NV-nuclei interactions. This can be done without fast $\pi$ pulses, i.e. with low MW power, and involving large magnetic fields. We use an NV center in diamond to illustrate our method, although this is general thus applicable to arbitrary hybrid spin systems. Furthermore, our protocol can be incorporated to standard pulsed DD sequences such as the widely used XY8 sequence, demonstrating its flexibility. We note that a different approach based on a specific continuous DD method~\cite{Casanova19} has been proposed to operate with NV centers under large $B_z$ fields.

\section{Model: NV center for nanoscale NMR}
We consider an NV center coupled to nearby nuclear spins. This is described by 
\begin{equation}\label{original}
H = D S_z^2 - \gamma_e B_z S_z -\sum_j \omega_\textrm{ L} I_j^z + S_z \sum_j \vec{A}_j \cdot \vec{I}_j,
\end{equation}
where $D = (2\pi)\times 2.87$ GHz is the so-called zero-field splitting, $\gamma_e = -(2\pi)\times 28.024$ GHz/T is the electronic gyromagnetic ratio of the NV center, and $B_z$ is the intensity of the magnetic field applied in the NV axis (the $z$ axis). The nuclear Larmor frequency is $\omega_\textrm{ L} = \gamma_n B_z$ with $\gamma_n$ the nuclear gyromagnetic ratio. $S_z =|1\rangle\langle 1|-|-\!1\rangle\langle -1|$ is the spin-$1$ operator representing the NV center, $|1\rangle$ and $|\!-\!\!1\rangle$ being two of the three hyperfine levels of the NV. The nuclear spin-$1/2$ operators are $I_j^\alpha = 1/2 \ \sigma_j^{\alpha}$ ($\alpha = x,y,z$) and $\vec{A}_j$ is the hyperfine vector mediating the NV-nucleus coupling. The latter is given by 
\begin{equation}\label{hypervec}
\vec{A}_j=\frac{\mu_0\gamma_e\gamma_n\hbar}{4\pi r_j^3}\bigg[\hat{z}-\frac{3(\hat{z}\cdot\vec{r}_j)\vec{r}_j}{r^2_j} \bigg],
\end{equation}
where $\vec{r}_j$ is the position vector of nucleus $j$ with respect to the origin (the vacancy site), and $\mu_0=4\pi\times10^{-7} \ \textrm{ T}\cdot \textrm{m/A}$ is the vacuum permeability.

The internal state of the NV is manipulated with an external MW field polarized in the $x$ axis. The Hamiltonian corresponding to such field is
\begin{equation}\label{controlHamil}
 H_\textrm{ c} = \sqrt{2} \Omega(t) S_x \cos{[\omega t -\phi]},
 \end{equation}
where $S_x = \frac{1}{\sqrt 2} (|1\rangle\langle 0| +|\!-\!1\rangle\langle 0| + \textrm{ H.c.})$, $\phi$ is the pulse phase, and $\omega$ is the MW driving frequency. If this frequency equals the natural frequency of the $|1\rangle \leftrightarrow |0\rangle$ transition ($D+ \gamma_e B_z/2$), Eq.~(\ref{original}) is, in the rotating frame defined by $D S_z^2 - \gamma_e B_z S_z$,
\begin{equation}\label{simulations}
H = \sum_j \omega_j \ \hat{\omega}_j\cdot \vec{I}_j + \frac{\sigma_z}{2}\sum_j \vec{A}_j\cdot \vec{I}_j + \frac{\Omega(t)}{2} (|1\rangle\langle 0| e^{i\phi} + \textrm{ H.c.}),
\end{equation}
where $\sigma_z=|1\rangle\langle1|-|0\rangle\langle0|$, and $\hat{\omega}_j = \vec{\omega}_j / |\vec{\omega}_j|$ with $\vec{\omega}_j = \omega_\textrm{ L} \hat{z} - \frac{1}{2} \vec{A}_j$ (note that $|\vec{\omega}_j| = \omega_j$). It is worth mentioning that the use of the $|1\rangle \leftrightarrow |0\rangle$ transition translates into a position-dependent shift of the Larmor frequency of the nuclei, $\omega_j \approx \omega_\textrm{ L}- \frac{1}{2}A_j^z$. For some applications, it may be convenient to avoid this shift using the $|1\rangle \leftrightarrow |-1\rangle$ transition instead\footnote{For an example check article 14 from the list of publications.}.

In pulsed DD methods resonance is achieved by the stroboscopic application of the MW driving leading to periodic $\pi$ rotations in the NV electronic spin. In an interaction picture with respect to $\frac{\Omega(t)}{2} (|1\rangle\langle 0| e^{i\phi} + \textrm{ H.c.})$, the stroboscopic application of these MW pulses is described as
\begin{equation}
H =  -\sum_j \omega_j \ \hat{\omega}_j\cdot \vec{I}_j +F(t) \frac{\sigma_z}{2}\sum_j \vec{A}_j\cdot \vec{I}_j,
\end{equation}
with the modulation function $F(t)$ taking periodically the values $+1$ or $-1$, depending on the number of $\pi$ pulses on the NV. Notice that a similar thing has been done in section~\ref{sect:1_PDD} in the context of trapped ions.

A common assumption of standard DD techniques is that  $\pi$ pulses are nearly instantaneous, thus highly energetic. However, in real cases we deal with finite-width pulses such that a time $t_{\pi} = \frac{\pi}{\Omega}$ is needed to produce a $\pi$ pulse (if $\Omega$ constant during the pulse). This has adverse consequences on the NV-nuclei dynamics such as the appearance of spurious resonances~\cite{Lorentz15, Haase16, Lang17}, or the drastic reduction of the NMR sensitivity at large $B_z$~\cite{Casanova18MW}.  In Ref.~\cite{Casanova18MW} a strategy to signal recovery is presented. Here, we extend this approach to achieve selective nuclear interactions. In the following, we demonstrate that the introduction of extended pulses with tailored $\Omega$ leads to tunable NV-nuclei interactions with low-power MW radiation. 

\section{DD with instantaneous MW pulses}
\begin{figure}[t]
\centering
\includegraphics[width=0.7\columnwidth]{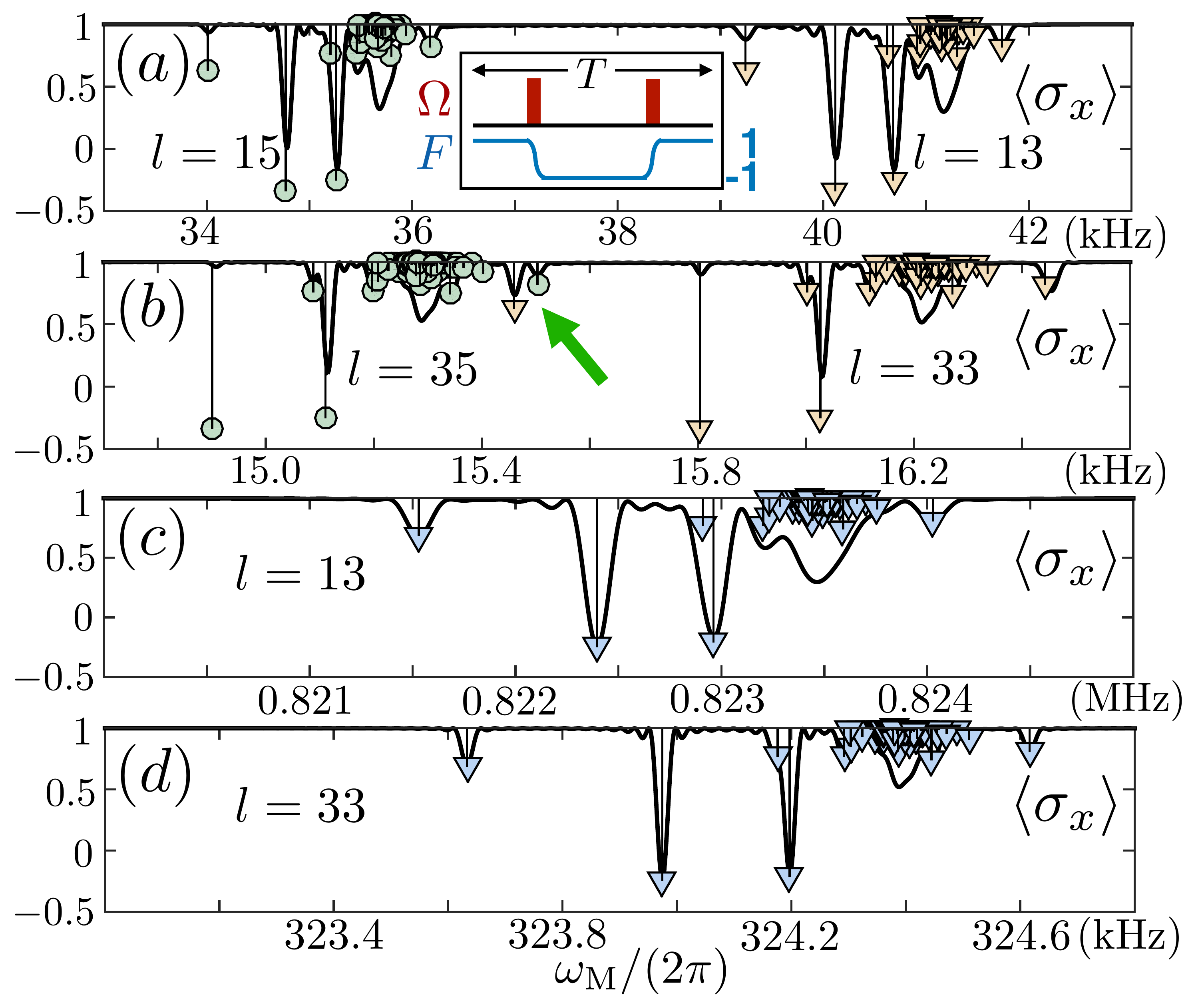}
\caption{Signal  (black-solid)  harvested with instantaneous $\pi$ pulses, $ B_z = 500$ G in (a) (b) and $B_z = 1$ T in (c) (d). Circles and triangles are the theoretically expected values for $\langle \sigma_x \rangle$. In (a) we select $l=13, 15$ and their signals are clearly separated. In (b) we use $l=33, 35$ and observe a spectral overlap (green arrow). In (c) (d) the spectral overlap is removed owing to a large $B_z$, while the signal (black-solid) matches the theoretically expected values. Final sequence time for (a) (c) is $\approx 0.5$ ms, and $\approx 1.2$ ms for (b) (d).}
\label{instantaneous}
\end{figure}

We consider the widely used XY8 = XYXYYXYX scheme, X (Y) being a $\pi$ angle rotation around the $x$ ($y$) axis of the Bloch sphere corresponding to states $|0\rangle$ and $|1\rangle$. The sequential application of XY8 on the NV leads to a periodic, even, $F(t)$ that  expands in harmonic functions as $F(t)=\sum_n f_n \cos{(n\omega_\textrm{ M} t)}$, where $f_n = 2/T\int_0^T F(s)  \cos{(n \omega_\textrm{ M} s)} ds$, and $\omega_\textrm{ M} = \frac{2\pi}{T}$  with $T$ the period of $F(t)$. See an example of $F(t)$ in the inset of Fig.~\ref{instantaneous}~(a). In an interaction picture with respect to $-\sum_j \omega_j \ \hat{\omega}_j\cdot \vec{I}_j $, Eq.~(\ref{simulations}) is\footnote{To prove it, one can use the following: $e^{i a\left(\hat{n} \cdot \vec{\sigma}\right)} ~ \vec{\sigma}~  e^{-i a\left(\hat{n} \cdot \vec{\sigma}\right)} = \vec{\sigma}  \cos (2a) + \hat{n} \times \vec{\sigma} ~\sin (2a)+ \hat{n} ~ (\hat{n} \cdot \vec{\sigma}) ~ (1 - \cos (2a))~$.}
\begin{equation}\label{modulated}
H = \sum_{n,j} \frac{f_n \cos{(n \omega_\textrm{ M} t)} \sigma_z}{2} \bigg[A_{j}^{x} I^x_j   \cos{(\omega_j t)} + A_{j}^{y} I^y_j   \sin{(\omega_j t)} + A_{j}^{z} I^z_j\bigg],
\end{equation}
where $A_{j}^{x,y,z} = |\vec{A}_{j}^{x,y,z}|$ with $\vec{A}_{j}^{x} = \vec{A}_{j} -  (\vec{A}_{j}\cdot \hat{\omega}_j) \ \hat{\omega}_j$, $\vec{A}_{j}^{y} =   \hat{\omega}_j\times \vec{A}_{j}$, $\vec{A}_{j}^{z} =  (\vec{A}_{j}\cdot \hat{\omega}_j) \ \hat{\omega}_j$, and $I_{x}^j = \vec{I}_j \cdot \hat x_j$,  $I_{y}^j = \vec{I}_j \cdot \hat{y}_j$, $I_{z}^j = \vec{I}_j \cdot \hat{z}_j$ with $\hat{x}_j = \vec{A}_{j}^{x}/ A_{j}^{x}$, $\hat{y}_j = \vec{A}_{j}^{y}/ A_{j}^{y}$ and $\hat{z}_j = \vec{A}_{j}^{z}/ A_{j}^{z}$. Notice that the Cartesian basis vectors have been redefined for each nucleus $j$.

Now, one selects a harmonic in the expansion of $F(t)$ and the period $T$, to create a resonant interaction of the NV with a target nucleus (namely the $k$th nucleus). To this end, in Eq.~(\ref{modulated}) we set $n=l$, and $T$ such that $l \omega_\textrm{ M} \approx \omega_k$. After eliminating fast rotating terms we get 
\begin{eqnarray}\label{big}
H &\approx& \frac{f_l A_k^x}{4}\sigma_z [I_k^{-} e^{i(\omega_k - l\omega_\textrm{ M}) t} +  \textrm{ H.c.}]\nonumber\\
&+&\sum_{j\neq k}  \frac{f_l A_j^x}{4}\sigma_z [I_j^{-} e^{i(\omega_j - l\omega_\textrm{ M}) t} +  \textrm{ H.c.}]\nonumber\\
&+&\sum_{n\neq l} \sum_j \frac{f_n A_j^x}{4}\sigma_z  [I_j^{-}e^{i(\omega_j - n\omega_\textrm{ M})t} + \textrm{ H.c.}].
\end{eqnarray}

By inspecting the first line of (\ref{big}), one finds that nuclear spin addressing at the $l$th harmonic is achieved when
\begin{equation}\label{resonancecond}
l \omega_\textrm{ M} = l  \frac{2\pi}{T}= \omega_k.
\end{equation}
With this resonance condition, the first line in (\ref{big}) is the resonant term $f_l A_k^x/4\sigma_z I_k^x$,  while detuned contributions (those in second and third lines) would average out by the RWA. More specifically, with Eq.~(\ref{resonancecond}) at hand we can remove the second line in Eq.~(\ref{big}) if  
\begin{equation}\label{cond1}
|\omega_j - \omega_k| \gg  f_l A_j^x/4. 
\end{equation}
Detuned contributions corresponding to harmonics with $n\neq l$ are in the third line of~(\ref{big}). These can be neglected if
 \begin{equation}\label{cond2}
 |\omega_j - n/l \omega_k| \approx \omega_\textrm{ L} (l-n)/l \gg f_n A_j^x/4  \ \ \forall n. 
 \end{equation}
To strengthen condition~(\ref{cond1}), one can reduce the value of $f_l$ by selecting a large harmonic (see later), while condition~(\ref{cond2}) applies better for large values of $B_z$ since $\omega_\textrm{ L}\propto B_z$. After neglecting the off resonant terms in Eq.~(\ref{big}), the Hamiltonian is 
\begin{equation}
H= \frac{f_lA_k^x}{4}\sigma_z I_k^x.
\end{equation}
For the above Hamiltonian the dynamics can be exactly solved, and the evolution of $\langle \sigma_x \rangle$ (when the initial state is $\rho=|+\rangle\langle +| \otimes \frac{1}{2} \mathbb{I}$, i.e. we consider the nucleus in a thermal state) reads
\begin{equation}
\langle \sigma_x \rangle = \cos{\bigg(\frac{f_lA_k^{x}}{4}t\bigg)}.
\end{equation} 
This is the ideal signal retrieved by perfect nuclear addressing when $\omega_\textrm{ M}=\omega_k/l$, and its represented by the depth of each panel (circles or triangles) in Fig.~\ref{instantaneous}.

Assuming instantaneous $\pi$ pulses, standard DD sequences with  constant $\Omega$~\cite{Maudsley86, Gullion90, Souza11} lead to $|f_l| = \frac{4}{\pi l}, 0$ for $l$ odd, even. Thus, large harmonics (i.e. with large $l$) reinforce condition~(\ref{cond1}) as they lead to a smaller value for $f_l$. In Fig.~\ref{instantaneous} we compute the signal corresponding to the NV observable $\langle\sigma_x\rangle$ in a sample that contains 150 $^{13}$C nuclei ($\gamma_{^{13}\textrm{ C}} = (2\pi)\times 10.708$ MHz/T). To obtain sufficient spectral resolution we use large harmonics. Figure~\ref{instantaneous} (a) shows the signal for $l=13, 15$ and the theoretically expected values for $\langle\sigma_x\rangle$ (triangles for $l=13$ and circles for $l=15$) that would appear if perfect single nuclear addressing is considered. We observe that the computed signal does not match with the theoretically expected values. In addition to a flawed accomplishment of conditions~(\ref{cond1}, \ref{cond2}),  this is also a consequence of using large harmonics since, for large $l$, the period $T=2\pi l/\omega_k$ and the spacing between $\pi$ pulses grows, see  inset in Fig.~\ref{instantaneous}(a), spoiling the efficient elimination of the $\sigma_z A_{j}^{z} I^z_j$ terms in Eq.~(\ref{modulated}) by the RWA. In the inset of Fig.~\ref{instantaneous} (a) there is a sketch of the pulse structure we repeatedly apply ($20$ times in (a) and (b), while in (c) and (d) that structure is used 400 times) to get the signals in Fig.~\ref{instantaneous}, red blocks are instantaneous $\pi$ pulses, while their associated $F(t)$ is in blue. Working with even larger harmonics introduces the problem of spectral overlaps. These appear when the signal associated to a certain harmonic contains resonance peaks corresponding to other harmonics. In Fig.~\ref{instantaneous} (b) one can see (green arrow) how a peak of $l=35$ (green circle) is mixed with the signal of $l=33$ (orange triangle). This is an additional disadvantage since the interpretation of the spectrum gets challenging.

Condition~(\ref{cond2}) is strengthened using a large $B_z$. This also implies a larger resonance frequency (namely $\omega_k$) for each nucleus. Addressing large $\omega_k$ is beneficial since the period $T$ (note that, in resonance $T=2\pi l/\omega_k$) and the interpulse spacing get shorter turning into a better cancellation of $\sigma_z A_{j}^{z} I^z_j$ terms. In Fig.~\ref{instantaneous} (c) (d), we use a large $B_z=1$ T and the spectral overlap is removed, while the computed signal matches the theoretically expected values (blue triangles). 

Unfortunately, to consider $\pi$ pulses as instantaneous in situations with large $B_z$ is not correct, since nuclei have time to evolve during $\pi$ pulse execution leading to signal drop~\cite{Casanova18MW}.  Hence, if one cannot deliver enough MW power to the sample, the results in Fig.~\ref{instantaneous} (c) and (d) are not achievable.

\section{A solution with extended pulses}
In realistic situations $\pi$ pulses are finite, thus the value of $f_l=2/T\int_0^T F(s)  \cos{(l \omega_\textrm{ M} s)} ds$ has to be computed by considering the intrapulse contribution. This is (for a generic $m$th pulse)  $2/T  \int_{t_m}^{t_m+t_{\pi}} F(s) \cos{(l\omega_\textrm{ M} s)} \ ds$, with $t_\pi$ being the $\pi$ pulse time and $t_m$ the instant we start applying MW radiation, see Fig.~\ref{whole} (a). In addition, the  $F(t)$ function must hold the following conditions: Outside  the $\pi$ pulse region $F(t)=\pm 1$, while $F(t)$ is bounded as $-1 \leq F(t) \leq 1$ $\forall t$, Fig.~\ref{whole} (a). 

\begin{figure}[t]
\centering
\includegraphics[width=0.8\columnwidth]{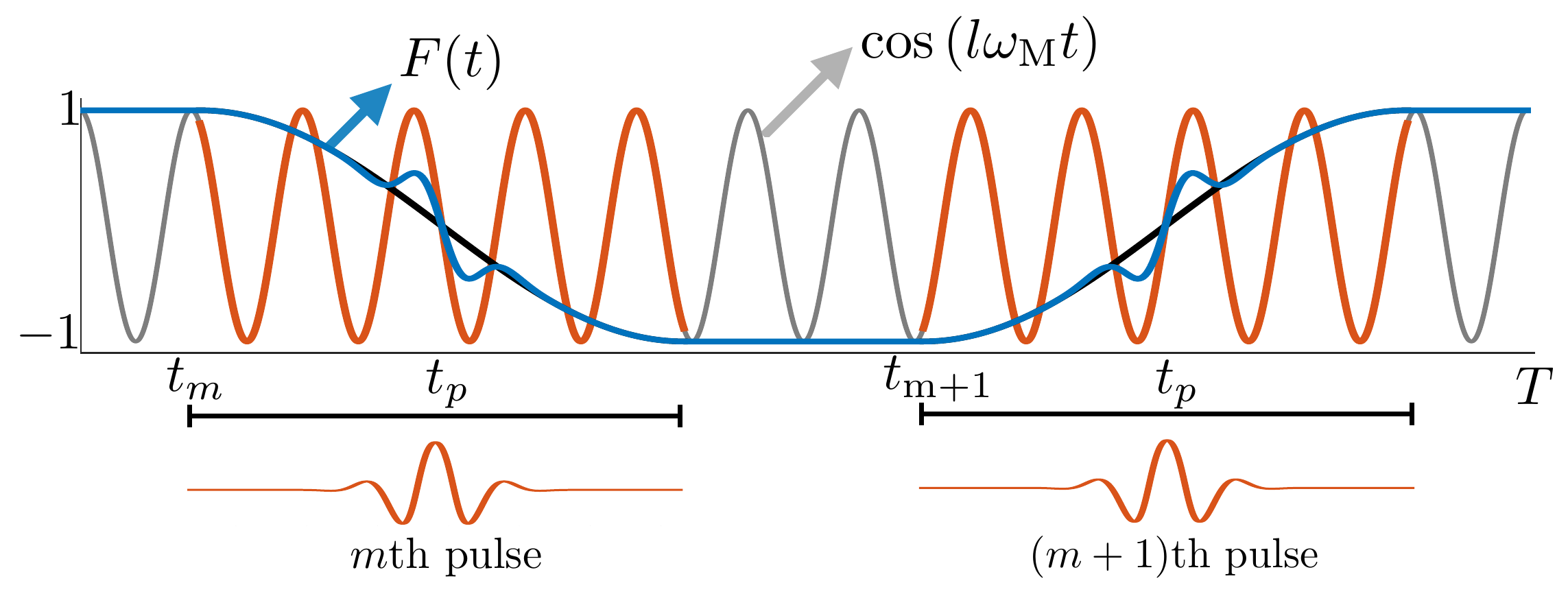}
\caption{Upper panel, one period of $F(t)$ (solid-blue) including the intrapulse behavior, and the $\cos{(l\omega_\textrm{ M}t)}$ function. Extended $\pi$ pulses span during $t_{\pi}$ (intrapulse regions appear marked in red). In this example, $l=13$ and $t_{\pi}\approx4.5\times(T/l)$. Solid-black, behavior of $F(t)$ in case standard top-hat pulses are applied. Bottom panel, train of modulated $\Omega(t)$ leading to $F(t)$. }
\label{fig:modulatedfun}
\end{figure}

\begin{figure}[t]
\centering
\includegraphics[width=0.8\columnwidth]{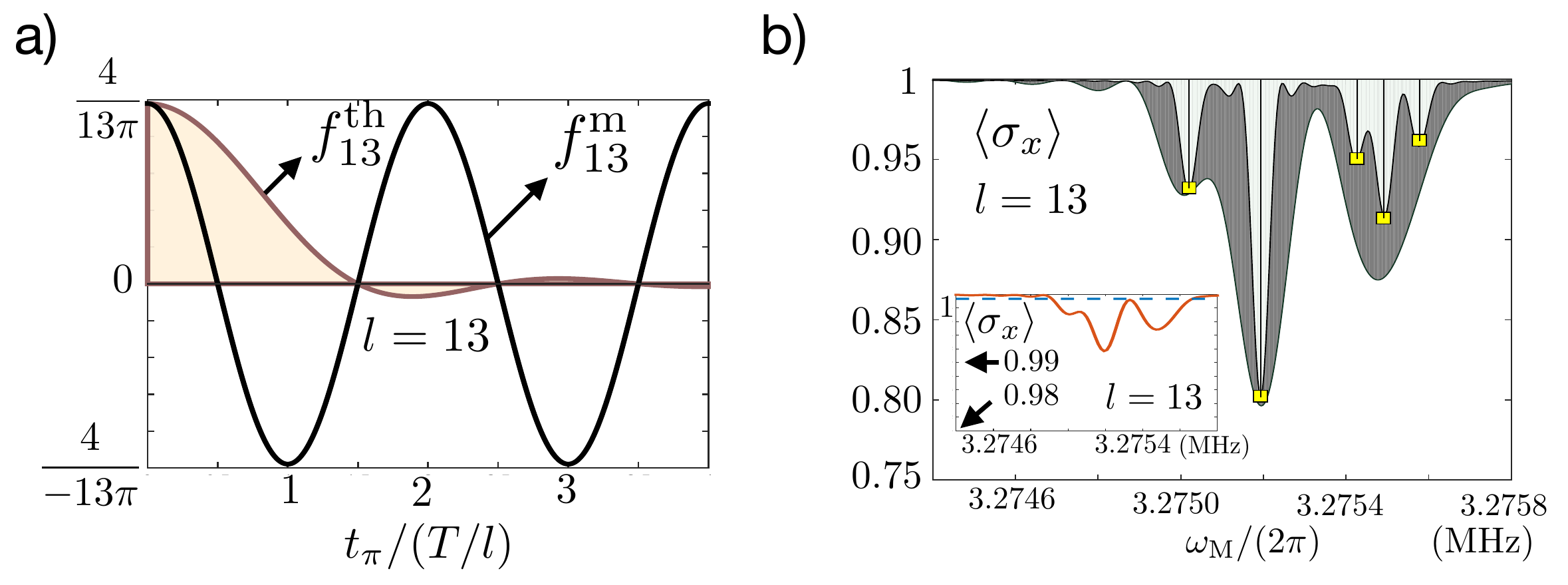}
\caption{ (a) $f_l^\textrm{ m}$ (black-solid) and $f_l^\textrm{ th}$ (curve on the yellow area) as a function of the ratio $t_{\pi}/(T/l)$ for $l=13$. (b) $\langle \sigma_x\rangle$ (curves over dark and clear areas) for the conditions discussed in the main text. Inset, $\langle \sigma_x\rangle$ computed with top-hat pulses. For all numerical simulations in (b) we assume a $1\%$ of error in  $\Omega(t)$~\cite{Cai12}.}
\label{whole}
\end{figure}

Now, we present a design for $F(t)$ that satisfies the above conditions, cancels intrapulse contributions, and leads to tunable NV-nuclei interactions. In particular, for the $m$th pulse
\begin{equation}\label{modulatedF}
F(t) = \cos{\big[\pi(t - t_{m})/t_\pi \big]} + \sum_{q}\alpha_{q}(t) \sin{\big[ q l \omega_\textrm{ M}(t - t_{p}) \big]}.
\end{equation} 
Here, $\alpha_{q}(s)$ are functions to be adjusted (see later) and $t_p=t_m+t_{\pi}/2$ is the central point of the $m$th pulse, Fig.~\ref{fig:modulatedfun}. We modulate $F(s)$ in the intrapulse region such that  (for the $m$th pulse) $\int_{t_m}^{t_m+t_{\pi}} F(s) \cos{(l\omega_\textrm{ M} s)} \ ds =0$ and $F(t)$ cancels the intrapulse contribution. Once we have $F(t)$, we find that the associated Rabi frequency is 
\begin{equation}\label{modulatedOmega}
\Omega(t) = \frac{\partial}{\partial t} \arccos[F(t)]=-\frac{1}{\sqrt{1-F(t)^2}},
\end{equation} 
if $F(t)$ differentiable. See appendix~\ref{FindOmega} for additional details. Now, the value of the $f_l$ coefficient obtained with the \textit{ modulated} $F(t)$ in Eq.~(\ref{modulatedF}) (from now on denoted $f_l^\textrm{ m}$) depends only on the integral out of $\pi$ pulse regions. This can be calculated leading to
\begin{equation}\label{modulatedf}
f_l^\textrm{ m} = \frac{4}{\pi l }\cos{\bigg(\pi  \frac{t_{\pi}}{T/l}\bigg)}\sin{(\pi l /2)},
\end{equation} 
which is our main result. For the derivation, see appendix~\ref{calcext}. By modifying the ratio between $t_\pi$ (the extended $\pi$ pulse length) and $T/l$ we can select a value for $f^\textrm{ m}_l$ and achieve tunable NV-nuclei interactions. According to Eq.~(\ref{modulatedf}), $f_l^\textrm{ m}$ can be taken to any amount between $-\frac{4}{l\pi}$ and $\frac{4}{l\pi}$, see solid-black curve in Fig.~\ref{whole}(a). In addition, owing to the periodic character of Eq.~(\ref{modulatedf}), one can get an arbitrary value (between $-\frac{4}{l\pi}$ and $\frac{4}{l\pi}$) for $f_l^\textrm{ m}$ even with large  $t_{\pi}$.  This implies highly extended $\pi$ pulses, thus a low delivered MW power. On the contrary, for standard $\pi$ pulses in the form of \textit{ top-hat} functions (i.e. generated with constant $\Omega$) one finds (see appendix~\ref{calcth} for the calculation)
\begin{equation}\label{tophatcoeff}
f_l^\textrm{ th} = \frac{4\sin{(\pi l /2)}\cos{(\pi l t_{\pi}/T)}}{\pi l (1-4l^2t_{\pi}^2/T^2 )}.
\end{equation}
Unlike $f^\textrm{ m}_l$, the expression for $f_l^\textrm{ th}$ shows a decreasing fashion for growing  $t_{\pi}$. Note that $ |f_l^\textrm{ th}| \propto [t_{\pi}/(T/l)]^{-2}$. This behaviour can be observed in Fig.~\ref{whole}(a), curve over the yellow area. Hence, standard top-hat pulses cannot operate with a large $t_{\pi}$, as this leads to a strong decrease of $f_l^\textrm{ th}$, thus to signal loss. 

To show the performance of our theory, we select, a Gaussian form for $\alpha_{1}(t) = a_1 e^{-(t-t_p)^2/2c_1^2}$  and set $\alpha_{q}(t)=0$, $\forall q>1$. See one example of a modulated $F(t)$ in Fig.~\ref{fig:modulatedfun} (solid-blue) as well as the behavior of $F(t)$ if common top-hat $\pi$ pulses are used (solid-black). Once we choose the $t_{\pi}$, $l$, and $c$ parameters that will define the shape of $F(t)$, we select the remaining constant $a_1$ such that it  cancels the intrapulse contribution, i.e. $\int_{t_m}^{t_m+t_{\pi}} F(s) \cos{(l\omega_\textrm{ M} s)} \ ds =0$. By inspecting Eq.~(\ref{modulatedF}) one easily finds that a natural fashion for $a_1$ is given by
\begin{equation}\label{a1ratio}
a_1=-\frac{\int_{t_m}^{t_m+t_\pi} \cos{\big[\pi(s - t_{m})/t_\pi \big]}  \cos{(l\omega_\textrm{ M} s)} \ ds}{\int_{t_m}^{t_m+t_\pi} e^{-\frac{(s-t_p)^2}{2c^2}}\sin{\big[ l \omega_\textrm{ M}(s - t_{p}) \big]}  \cos{(l\omega_\textrm{ M} s)} \ ds}.
\end{equation}

In Fig.~\ref{whole}(b) we simulated a sample containing 5 protons\footnote{ Note that the presence of extended pulses disables the use of numerical techniques, as those in \cite{Maze08}, to simulate large nuclear samples. Hence, because of machine restrictions, here we focus in a sample that includes five nuclei and the NV} at an average distance from the NV of $\approx 2.46$ nm.  Numerical simulations have been performed starting from Eq.~(\ref{simulations}) without doing further assumptions. The 5-H target cluster has the hyperfine vectors (note $\gamma_\textrm{ H} = (2\pi)\times 42.577$ MHz/T) $\vec{A}_1 = (2\pi)\times[-1.84, -3.19, -11.02]$, $\vec{A}_2 = (2\pi)\times[2.38, 5.04, -8.78]$, $\vec{A}_3 = (2\pi)\times[8.09, 2.66, -1.02]$,  $\vec{A}_4 = (2\pi)\times[4.26, 2.46, 3.48]$, and $\vec{A}_5 = (2\pi)\times[4.07, 1.00, -7.09]$ kHz. We simulate two different sequences, leading to two signals, using our extended $\pi$ pulses  under a large magnetic field $B_z=1$ T. Vertical panels with yellow squares mark the theoretically expected resonance positions and signal contrast. For the first computed signal, curve over dark area in Fig.~\ref{whole}(b), we display a XY8 sequence where the phase of each X (Y) extended pulse is $\phi=0$ ($\phi=\pi/2$). The modulated Rabi frequency $\Omega(t)$ is selected such that it leads to $f_{13}^\textrm{ m}=4\pi/13 = 0.0979$ for $l=13$ (note this corresponds to the maximum value for $f_{13}^\textrm{ m}$) with  a pulse length $t_{\pi} = 6 \times(T/l)$. In addition, we take the width of the Gaussian function $\alpha_{1}(t)$ as $c_1 = 0.07 t_{\pi}$. The scanning frequency $\omega_\textrm{ M}$ spans around $\gamma_\textrm{ H} B_{z}/l$ for $l=13$, see horizontal axis in Fig.~\ref{whole}(b).  After repeating the XY8 sequence  $400$ times, i.e. 3200 extended  $\pi$ pulses have been applied leading to a final sequence time of $t_{f} \approx 0.488$ ms, we get the signal over the dark area.  As we observe in Fig.~\ref{whole}(b), this sequence does not resolve all nuclear resonances of the 5-H cluster. 
\pagebreak

To overcome this situation, we make use of the tunability of our method, and simulate a second sequence with extended $\pi$ pulses  leading to the signal over the clear area in Fig.~\ref{whole}~(b). This has been computed with a smaller value for $f_{13}^\textrm{ m} = 0.0979/3 = 0.0326$ which is achieved with $t_{\pi} \approx 6.4\times(T/l)$, i.e. a slightly longer $\pi$ pulse than those in the preceding situation, and $c_1=0.07 t_{\pi}$.  As the $f_{13}^\textrm{ m}$ coefficient is now smaller, we have repeated the XY8 sequence $400\times 3$ times (i.e. 9600 pulses) to get the same contrast than in the previous case. The final time of the sequence is $t_{f} \approx 1.5$ ms. As we observe in Fig.~\ref{whole}(b), our method faithfully resolves all resonances in the 5-H cluster, and reproduces the theoretically expected signal contrast. It is noteworthy to comment that the tunability offered by our method will be of help for different quantum algorithms with NV centers~\cite{Ajoy15, Perlin18, Casanova16, Casanova17}.

\section{MW power and nuclear signal comparison}
In the inset of Fig.~\ref{whole}(b) we plot the signals one would get using standard top-hat pulses with the same average power than our extended pulses in  Fig.~\ref{whole} (b).  We use that the energy of each top-hat and extended $\pi$ pulse, $E^\textrm{ th}(t_\pi)$ and $E^\textrm{ ext}(t_\pi)$, is $\propto \int \Omega^2(s) ds$ where the integral extends during the $\pi$ pulse duration (top-hat or extended). For an explicit  derivation of the energy relations see appendix~\ref{energydelivery}. The solid-orange signal in the inset has been computed with a XY8 sequence containing  3200 top-hat $\pi$ pulses with a constant $\Omega \approx (2\pi)\times 18.2$ MHz. For this value of $\Omega$, a top-hat $\pi$ pulse contains the same average power than each extended $\pi$ pulse used to compute the signal over dark area in  Fig.~\ref{whole}(b), i.e. $E^\textrm{ th}(t_\pi)= E^\textrm{ ext}(t_\pi)$. Unlike our method, the sequence with standard top-hat $\pi$ pulses produces a signal with almost no-contrast. Note that the vertical axis of inset in Fig.~\ref{whole}(b) has a maximum depth value of 0.98, and the highest contrast achieved with top-hat pulses falls below 0.99. The dashed signal in the  inset has been obtained with top-hat $\pi$ pulses with $\Omega\approx(2\pi)\times 4.68$ MHz. Again, this is done to assure we use the same average power than the sequence leading to the curve over the clear area in Fig.~\ref{whole}(b). In this last case, we observe that the signal harvested with standard top-hat $\pi$ pulses does not show any appreciable contrast. These results indicate that our method using pulses with modulated amplitude is able to achieve tunable electron nuclear interactions, while regular top-hat pulses with equivalent MW power fail to resolve these interactions. 

In summary, in this chapter we presented a general method to design extended $\pi$ pulses which are energetically efficient, and incorporable to stroboscopic DD techniques such as the widely used XY8 sequence. Our method leads to tunable interactions, hence selective, among an NV quantum sensor and nuclear spins at large static magnetic fields which represents optimal conditions for nanoscale NMR.

\chapter{Conclusions}
\label{chap_conclusions}
\thispagestyle{chapter}

In this Thesis we have designed protocols that tailor light-matter interactions for specific applications in quantum platforms such as trapped ions, ultracold atoms in optical lattices, and NV centers. In short, we have used DD techniques to design quantum operations that are robust against errors in environmental and control fields, achieving high-fidelity quantum logic in trapped ions and energy-efficient NMR at the nanoscale with NV centers in diamond. We have also studied generalised models of light-matter interaction, leading to the discovery of selective $k$-photon interactions in the Rabi-Stark model and a proposal for preparing non-classical quantum states using the NQRM. Moreover, we have shown how the appropriate tailoring of interactions among ultracold atoms in optical lattices could lead to solve the boson sampling problem faster than the best supercomputers, thus demonstrating quantum supremacy. More specifically:

In chapter~\ref{chapter:chapter_1}, we proposed two different methods to generate robust high-fidelity two-qubit gates with MW-driven trapped ions. On the one hand, we have demonstrated that pulsed DD schemes are efficient generators of fast and robust two-qubit gates. Particularly, our MW sequence induces all motional modes to cooperate, which results in a faster gate. In addition, our sequence is specifically designed to be robust against fluctuations in the magnetic and MW fields. Then, it achieves fidelities larger than $99.9\%$ including these realistic sources of decoherence. On the other hand, we proposed a different method that uses phase-modulated continuous DD combined with phase flips and refocusing $\pi$ pulses, to produce entangling gates with high fidelity. Contrary to the previous case, here we considered low-power MW radiation which matches current experimental scenarios. In particular, we demonstrated that fidelities on the preparation of maximally entangled Bell states exceeding $99\%$ are possible within current experimental limitations. Moreover, we also showed that fidelities larger than $99.9\%$ are reachable with minimal experimental improvements. Summarising, with the help of DD methods, we have developed two gate schemes that can arguably improve the fidelities of current entangling operations up to the threshold required for the application of quantum error correction techniques. Finally, in the same direction, one could study whether amplitude modulated pulses, like the ones used in chapter~\ref{chapter:chapter_4}, can bring extra robustness to our gates.

\pagebreak
In chapter~\ref{chapter:chapter_2}, we have studied two different models of light-matter interaction and propose means for their quantum simulation with a laser-driven trapped ion. First, we have considered the Rabi-Stark model, both in the SC and USC regimes. Here, we discovered $k$-photon interactions whose resonance frequency depends on the state of the bosonic mode. We have identified that this selective behaviour is caused by the Stark term, and developed an analytical framework to characterise these selective interactions. Second, we propose the NQRM, as a natural extension of the QRM in trapped-ion systems. The nonlinear term $f_1(\hat{n})$, which appears when moving outside the LD regime, causes the blockade of motional-state propagation at $|n\rangle$, whenever $f_1(\hat{n})|n\rangle=0$. In the SC regime, we compared the linear and nonlinear Jaynes-Cummings models, and observe that collapses and revivals of coherent states disappear. As an application of the model, we proposed a method to generate large-$n$ Fock states in a dissipative manner, making use of the nonlinear anti-JC model, and the spontaneous decay of the two-level system. Regarding this, we find interesting to study how the time needed to prepare the Fock state scales with the number $n$\footnote{We thank Prof. Jonathan P. Home for bringing up this suggestion.}. Finally, we showed how the Rabi-Stark and the NQRM can be implemented using a single trapped ion. Natural follow-ups of these works could involve proposals to realise these models in cavity or circuit QED systems or the extension to more qubits.

In chapter~\ref{chapter:chapter_3}, we have introduced a method to realise boson sampling with ultracold atoms in optical lattices. The group of Prof. Dieter Meschede and Dr. Andrea Alberti had previously demonstrated discrete-time random walks~\cite{Robens16b} and two-particle quantum interference~\cite{RobensThesis} using MW pulses combined with spin-dependent optical potentials. Boson sampling, equivalent to a multiparticle quantum walk, is hard to simulate classically, as proved by Aaronson and Arkhipov~\cite{Aaronson11}. On the one hand, we have studied the effect of particle loss in the boson sampling problem. In particular, we studied how correlated two-body losses hinders the rate in which valid experimental samples can be generated. With low particle loss, we proved that the samples are close enough to the boson sampling probability distribution (BSPD). Interestingly, we also observed that sampling faster than a classical supercomputer is possible for strong particle losses, however, the probability distribution we are sampling from does not resemble the BSPD. It is still an open question whether it is hard to sample from the probability distribution generated by a boson sampler with strong two-body losses. We also estimated how small other experimental errors have to be when increasing the number of bosons, in order to keep sampling from the BSPD. These experimental errors include fluctuations of the magnetic field or imperfect ground-state cooling. We note that, in principle, DD techniques could be used to relax the constrains on the fluctuating errors. As a final remark, we notice that one of the advantages of using atoms for boson sampling may come from the point of view of verification. This is, to verify that the results delivered by the boson sampler are correct, even in regimes where the results cannot be retrieved with classical supercomputers. In this regard, instead of bosonic atoms, a fermionic species could in principle be loaded in the optical lattice. Doing the same operations to the fermionic atoms would also result in a sampling problem (fermion sampling). However, in this case, the final probability distribution can be calculated in a time that scales polynomially with the number of particles. One could then verify that the machine samples correctly from the fermionic probability distribution, and expect that the same will happen for the bosonic case.

In chapter~\ref{chapter:chapter_4}, we presented a general method to design extended MW pulses that achieve tuneable interactions, hence selective, among an NV quantum sensor and nuclear spins at large static magnetic fields. The latter represent optimal conditions for nanoscale NMR, enhancing the precision with which signals from magnetic field emitters can be retrieved by the NV center. At large magnetic fields, the values of the Larmor frequencies increase and these may surpass the value of the Rabi frequency associated to the applied MW radiation. This induces a reduction in the contrast of the NMR spectra, something that is solved by our amplitude-modulated extended $\pi$ pulses. Our method avoids having to increase the value of the Rabi frequency to that of the Larmor frequencies, leading to an energetically efficient method. Furthermore, the method is general and can be incorporated to all stroboscopic DD techniques such as the widely used XY8 sequence. During the course of this thesis, we have also proposed the use of amplitude-modulated pulses for double quantum magnetometry\footnote{Check article 14 from the list of publications}. In that case, a more complex two-tone stroboscopic driving is required, achieving a larger spectral signal.

All in all, this thesis explores new avenues in the control of quantum systems through light-matter interactions shaped by specifically designed radiation patterns. We expect that the results presented here will boost the experimental generation of MW-driven two-qubit operations with fidelities well above the required threshold to apply quantum error correction techniques, better quantum sensors performing NMR at the nanoscale, and the first quantum supremacy experiment using trapped atoms. Moreover, our results will help in the development of quantum technologies which, besides leading to technological progress, will certainly be key assets to unveil the unsolved questions of nature.

\addcontentsline{toc}{chapter}{APPENDICES}

\cleardoublepage

\appendix 

\renewcommand{\chaptermark}[1]{\markboth{\textit{Appendix \thechapter.  #1}}{}}

\chapter{Further Considerations on the Pulsed DD Two-Qubit Gate}
\label{appendix:appendix_a}

\section{Initial approximations}\label{app:InitialApp}

\subsubsection{Two-Level Approximation}
In this section we numerically argue that the presence of the additional hyperfine levels of the $^{171}$Yb$^+$ ion, the fluctuations of the magnetic field, and the effect of fast rotating terms does not threaten the gate fidelities claimed in section~\ref{sect:1_PDD}.  For numerical simplicity we have considered a single four level system and looked for the fidelity of the propagator after a  sequence of 20 $\pi$-pulses is applied. We find that the error (infidelity) is on the order of $10^{-5}$, hence being one order of magnitude below the gate errors reported in Table~\ref{table1}. Therefore, we conclude that the presence of the additional levels, counter-rotating terms, and the fluctuations of the magnetic field have a negligible effect on the final fidelity of the gate to the order claimed in section \ref{subsect:Tailored}. We detail now the parameters and conditions in our numerical simulations. 

In the hyperfine ground state of the $^{171}$Yb$^+$ ion, transitions can be selected with the appropriate polarisation of the control fields. However, experimental imperfections might generate unwanted leakage of population from the qubit-states to other states. On the other hand, the presence of fluctuations of the magnetic field may also result in imperfect $\pi$-pulses which may also damage the performance of the gate. To account for these experimental imperfections we simulate the following 4-level Hamiltonian
\begin{eqnarray}
H_{4l}&=&E_0 |0\rangle \langle 0 | + E_1|1\rangle \langle 1 | + E_2 |2\rangle \langle 2 | + E_3 |3\rangle \langle 3 | 
+ X(t) |1\rangle \langle 1 | -X(t) |3\rangle \langle 3 | \nonumber \\
&+& \Omega(t) ( | 0 \rangle \langle 1| + \epsilon_\perp | 0 \rangle \langle 2 | + \epsilon_\perp | 0 \rangle \langle 3 | +\textrm{H.c.} ) \cos{[\omega t + \phi(t)]},
\end{eqnarray}
where the energies of the hyperfine levels, $E_i$, are those corresponding to a $^{171}$Yb$^+$ ion in a magnetic field of $100$ G, and the qubit is codified in levels $ \{|0 \rangle, | 1 \rangle \}$. Function $X(t)$ represents a fluctuating magnetic field, which shifts the magnetically sensitive levels $|1\rangle $ and $|3\rangle$ in opposite directions. Numerically we have constructed this function as an OU process~\cite{Gillespie96}, where the parameters have been chosen such that the qubit-levels, in the absence of any pulses, show a coherence decaying exponentially with a $T_2$ coherence-time of $3$ms, as experimentally observed~\cite{Wineland98}. Particularly, this corresponds to values $\tau_B=50\mu$s for the correlation time, and $c_d=2/(\tau_B*T_2)$ for the diffusion constant of the OU process. $\Omega(t)$ is a step function taking exclusively values $\Omega$ and $0$, and $\epsilon_\perp$ is a small number representing the leaking of the qubit population through unwanted transitions. For the numerical analysis we have used unfavourable values for this set of parameters. More specifically, the Rabi frequency was assigned a value of $\Omega=(2\pi)\times20$MHz, which is already twice the maximum value used in all the other simulations throughout the analysis, having therefore a larger probability of exciting other, undesired, hyperfine transitions. Moreover, the simulations were performed for the longest sequence discussed in section \ref{subsect:Tailored}, which lasts $80\mu$s. 

\begin{figure}[t] 
\centering
\includegraphics[width=1\linewidth]{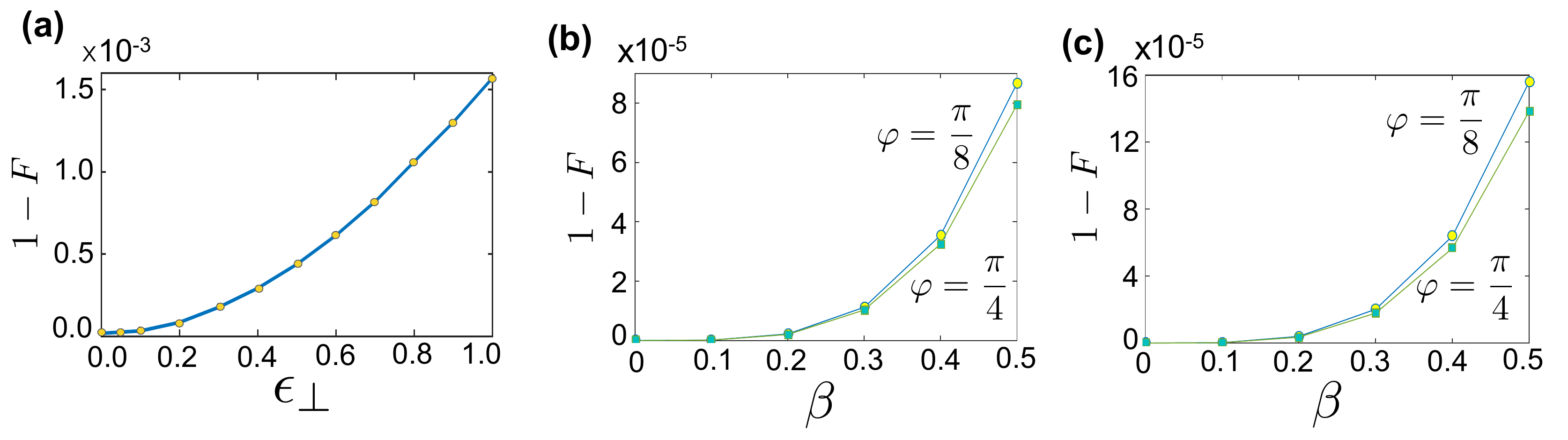}
\caption{a) State infidelity after a AXY-4 sequence, consisting of 20 $\pi$-pulses and a total time of $80\mu$s, vs the strength of the leakage of population to other spectator levels. Each point is the average of the infidelities of 100 runs of the sequence in the presence of stochastic fluctuations of the magnetic field. b) State infidelity as a consequence of coupling to a radial mode of the kind in Eq.~(\ref{radial}). We observe a growing infidelity for larger values of $\beta$ with $\beta=\Delta_r/\eta_1\nu_1$. The case of $g_B=150$T/m for two qubit gate phases $\varphi=\pi/4$ squares, and $\varphi=\pi/8$ circles. In  c) we use $g_B= 300$T/m and, again, we use squares for $\varphi=\pi/4$, and circles for $\varphi=\pi/8$. For both plots we used $\psi_{4}=|{\textrm e}\rangle \otimes (|{\textrm g}\rangle - i |{\textrm e}\rangle) + |{\textrm g}\rangle \otimes |{\textrm g}\rangle$ (up to normalisation) as the initial state.}
\label{fig:Infidelity}
\end{figure}
We compare the propagator resulting from our simulations to the identity, which is what one would expect after an even number of $\pi$-pulses, 20 in our case, and we compute a value for the fidelity according to the definition
\begin{equation}
F_{A,B}=\frac{|\textrm{Tr}(AB^\dag)|}{\sqrt{\textrm{Tr}(AA^{\dag}) \textrm{Tr} (BB^\dag)}},
\end{equation}
where $F_{A,B}$ is the fidelity between operators $A$ and $B$. To account for the stochastic effects of the OU process that models the fluctuations of the magnetic field, we have averaged the resulting fidelities over $100$ runs of our numerical simulator. In Fig.~\ref{fig:Infidelity} we show the value of the infidelity, $1-F$, for a number of values of $\epsilon$. We can see that the error grows non-linearly with the strength of the leakage $\epsilon$ due to polarisation errors in the control fields. However, for alignment errors below $20\%$ ($\epsilon_\perp=0.2$) we obtain that the infidelity is smaller than $10^{-4}$. Hence, for polarisation errors below $20\%$, the effect of additional hyperfine levels, magnetic field fluctuations, and fast counter rotating terms should only be detectable in the fifth significant order of the gate fidelity, and not alter the $99.9\%$ fidelity claimed in section~\ref{subsect:Tailored}.

\subsubsection{Coupling with radial modes}

In this section we study the influence of the motional radial modes of the ion in our proposal. To account for the effect of a given radial mode $d$, Hamiltonian~(\ref{Hamiltonianbare}) needs to be complemented with a term of the form
\begin{eqnarray}\label{radial}
 \nu_r d^{\dag}d + \Delta_r (d  + d^\dag) [ \sigma_1^z  +  \sigma_2^z].
\end{eqnarray}
Because of computational restrictions, in this analysis we will only consider one radial mode, and assume no motional decoherence. Term~(\ref{radial}) is justified because of the unavoidable presence of some remanent magnetic field gradient in the radial direction, which leads to the coupling $\Delta_r$ that we model as a fraction of the coupling $\eta_1\nu_1$ in Hamiltonian~(\ref{Hamiltonianbare}) of the main text, i.e. $\Delta_r= \beta\eta_1\nu_1$. We have compared the states evolved under Hamiltonian~(\ref{Hamiltonianbare}) in the main text including and not the coupling term in Eq.~(\ref{radial}), and computed the infidelity between them. In Fig.~\ref{fig:Infidelity} we show the results for different values of $\beta$. The value of $\nu_r =(2\pi)\times2.5$ MHz was used in the simulations, and the initial state for the qubits was chosen to be $\psi_4$ in Table~\ref{table1}, while a thermal state with 2 phonons was used as the initial state of the radial mode. We observe that even for large values of $\beta$, the impact of the radial mode is negligible, on the order of $10^{-5}$ for values of $\beta$ up to 0.4, which are experimentally unexpected.

\section{Two hyperfine ions under a magnetic field gradient}\label{app:twoions}

The Hamiltonian of the relevant hyperfine levels of the two-qubit system (composed, in our case, of two $^{171}$Yb$^+$ ions) under a $z$ dependent magnetic field can be expressed as
\begin{eqnarray}\label{model}
H =  \nu_1 a^\dag a + \nu_2 c^\dag c   &+& [\omega_{\textrm{e}} + \gamma_e B(z_1)/2] |{\textrm{e}}\rangle \langle {\textrm{e}}|_1 +  \omega_{\textrm{g}} |{\textrm{g}} \rangle \langle {\textrm{g}}|_1  \nonumber\\
      &+&   [\omega_{\textrm{e}} + \gamma_e B(z_2)/2] |{\textrm{e}}\rangle \langle {\textrm{e}}|_2 +  \omega_{\textrm{g}} |{\textrm{g}} \rangle \langle {\textrm{g}}|_2.
\end{eqnarray}
If we assume that the ions, which interact through direct Coulomb force, perform only small oscillations around their equilibrium positions, $z_j=z_j^0+q_j$, and we expand $B(z_{j})$ to the first order in $q_j$, then $B(z_j)=B_{j} + g_{B}  \ q_{j}$, where $B_j\equiv B(z_j^0)$ and $g_B\equiv \partial B/{\partial z_j}\big|_{z_j=z_j^0}$. With this, and up to an energy displacement, the Hamiltonian~(\ref{model}) reads 
\begin{eqnarray}\label{Hamil2}
H &=& \frac{1}{2}[\omega_{\textrm{e}} + \gamma_e B_1/2 - \omega_{\textrm g}] \ \sigma_1^z + \frac{1}{2}[\omega_{\textrm{e}} + \gamma_e B_2/2 - \omega_{\textrm g}] \ \sigma_2^z \\
&+& \nu_1 a^{\dag}a +  \nu_2 c^{\dag}c +\frac{\gamma_e g_{B}}{4}(q_1+q_2) +\frac{\gamma_e g_{B}}{4}(q_1\sigma_1^z+q_2\sigma_2^z), \nonumber
\end{eqnarray}
where we have used the relations $|{\textrm{e}}\rangle \langle {\textrm{e}}|_j=\frac{1}{2}(\mathds{1}+\sigma_{\! j} ^z)$ and $|{\textrm{g}}\rangle \langle {\textrm{g}}|_j=\frac{1}{2}(\mathds{1}-\sigma_{\! j}^z)$. At this moment, it may be useful to recall that the displacement of the ions from their equilibrium positions, $q_1$ and $q_2$ can be expressed in terms of the collective normal modes, $Q_1$ and $Q_2$, as
\begin{eqnarray}\label{quantizedpositions}
q_1&=& \frac{Q_1 -  Q_2}{\sqrt{2}}= \sqrt{\frac{\hbar}{4M\nu_1}} \big(a+a^\dag\big) - \sqrt{\frac{\hbar}{4M\nu_2}} \big(c+c^\dag\big), \nonumber \\ 
q_2&=& \frac{Q_1 + Q_2}{\sqrt{2}}= \sqrt{\frac{\hbar}{4M\nu_1}} \big(a+a^\dag\big) + \sqrt{\frac{\hbar}{4M\nu_2}} \big(c+c^\dag\big),
\end{eqnarray}
$M$ being the mass of each ion. Using these relations, which follow the prescription in~\cite{James98}, Eq.~(\ref{Hamil2}) can be rewritten as
\begin{eqnarray}\label{Hamil3}
H &=& \frac{1}{2}[\omega_{\textrm{e}} + \gamma_e B_1/2 - \omega_{\textrm g}]  \sigma_1^z + \eta_1\nu_1(a + a^\dag)  \sigma_1^z - \eta_2\nu_2 (c + c^\dag) \sigma_1^z\nonumber\\
&+& \frac{1}{2}[\omega_{\textrm{e}} + \gamma_e B_2/2 - \omega_{\textrm g}] \sigma_2^z + \eta_1\nu_1(a + a^\dag) \sigma_2^z + \eta_2\nu_2(c + c^\dag) \sigma_2^z\nonumber\\
&+& \nu_1 a^{\dag}a +  \nu_2 c^{\dag}c + \frac{\gamma_e g_{B}}{4} \sqrt{\frac{\hbar}{M \nu_1}} (a + a^\dag),
\end{eqnarray}
where we have defined $\eta_{1,2} \equiv  \frac{\gamma_e g_{B}}{8\nu_{1,2}}\sqrt{\frac{\hbar}{M \nu_{1,2}}}$ as the coupling strength between the qubits and the normal modes. The last term in Eq.~(\ref{Hamil3}) can be absorbed by a redefined bosonic operator $b = a + 2\eta_1$, which results in Hamiltonian
\begin{eqnarray}\label{simplifiyedJorge}
 \nonumber H= \nu_1 b^\dag b + \nu_2 c^\dag c &+& \frac{\omega_1}{2}\sigma_1^z + \eta_1\nu_1 (b+b^\dag) \sigma_1^z - \eta_2\nu_2(c+c^\dag)\sigma_1^z\\
     &+& \frac{\omega_2}{2}\sigma_2^z + \eta_1\nu_1 (b+b^\dag) \sigma_2^z + \eta_2\nu_2(c+c^\dag)\sigma_2^z\end{eqnarray}
where $\omega_{1,2} \equiv \omega_{\textrm{e}} - \omega_{\textrm g}  - 4\eta_1^2\nu_1  + \gamma_e B_{1,2}/2$. Furthermore, we can easily compute the quantity $\omega_{2} - \omega_{1}=\gamma_e (B_{2} - B_{1})/2=\gamma_e g_B (z^0_2-z^0_1)/2 = \gamma_e g_B \Delta z/2$.

\section{The interaction Hamiltonian}\label{app:IntHamil}

A bichromatic MW field of frequencies $\omega_j$ and phase $\phi$ will be applied to the system described by Eq.~(\ref{simplifiyedJorge}). The action of such MW field on the ions is described by the following Hamiltonian
\begin{equation}\label{MWField}
H_c(t)=\sum_{j=1}^2\Omega_j(t) (\sigma_1^x + \sigma_2^x)\cos(\omega_j t - \phi)
\end{equation}
where $\Omega_j$ is the Rabi frequency associated to the intensity of the MW field with frequency $\omega_j$. If we add this term to the Hamiltonian (\ref{simplifiyedJorge}), and we move to an interaction picture with respect to $H_0=\frac{\omega_1}{2} \sigma_1^z + \frac{\omega_2}{2} \sigma_2^z + \nu_1 b^{\dag}b +  \nu_2 c^{\dag}c$, the complete Hamiltonian in the interaction picture will be given by

\begin{eqnarray}\label{MWField2}
H^{\textrm I}(t)&=&
\eta_1\nu_1b e^{-i \nu_1 t} \sigma_1^z - \eta_2\nu_2ce^{-i\nu_2 t} \sigma_1^z + \eta_1\nu_1b e^{-i \nu_1 t} \sigma_2^z +\eta_2\nu_2ce^{-i\nu_2 t} \sigma_2^z \nonumber \\
&+& \big[\sum_{j=1}^2\Omega_j(t)\cos(\omega_j t - \phi) \big](\sigma_1^+ e^{i \omega_1 t} + \sigma_2^+ e^{i \omega_2 t} ) +\textrm{H.c.}
\end{eqnarray}
where we have use the relations $e^{i\theta a^\dagger a} a e^{-i\theta a^\dagger a}=ae^{-i\theta}$ and $e^{i\theta \sigma^z} \sigma^+ e^{-i\theta \sigma^z}=\sigma^+e^{i2\theta}$. Rewriting the last term leads to 
\begin{eqnarray}\label{MWField3}
H^{\textrm I}(t)&=&\eta_1\nu_1b e^{-i \nu_1 t} \sigma_1^z - \eta_2\nu_2ce^{-i\nu_2 t} \sigma_1^z + \eta_1\nu_1b e^{-i \nu_1 t} \sigma_2^z +\eta_2\nu_2ce^{-i\nu_2 t} \sigma_2^z \nonumber \\
&+& \frac{\Omega_1(t)}{2}\Big[\sigma_1^+ e^{i \phi}(e^{i2(\omega_1 t -\phi)} + 1) + \sigma_2^+( e^{i(\omega_1+ \omega_2) t}e^{-i\phi} + e^{-i(\omega_1-\omega_2)t}e^{i\phi}) \Big] \nonumber\\
&+& \frac{\Omega_2(t)}{2}\Big[\sigma_1^+( e^{i(\omega_2+ \omega_1) t}e^{-i\phi} + e^{-i(\omega_2-\omega_1)t}e^{i\phi})  + \sigma_2^+  e^{i \phi}(e^{i2(\omega_2 t -\phi)} + 1)\Big] \nonumber\\
&+&\textrm{H.c.}
\end{eqnarray}

At this point we can safely neglect the terms that rotate with frequencies $\pm|2\omega_1|,\pm|2\omega_2|$ and $\pm|\omega_1+\omega_2|$ by invoking the RWA. Because $|\omega_1|,|\omega_2| \gg \Omega_1,\Omega_2$, these terms will have a negligible effect on the evolution of the system. On the other hand, terms that rotate with frequencies $\pm|\omega_2-\omega_1|$ would have a significant effect on the evolution of the system. However this will be suppressed at the end of the two-qubit gate, because of the design of the pulse sequence. How this elimination occurs is covered in section \ref{subsect:Tailored} and appendix \ref{app:PulseP}. Hence, we can assume that these terms do not have any effect on the system and we can neglect them, thus the Hamiltonian is 
\begin{eqnarray}\label{MWField4}
 H^{\textrm I}(t)&=& \eta_1\nu_1(b e^{-i \nu_1 t} + b^\dag e^{i\nu_1 t}) \sigma_1^z - \eta_2\nu_2(ce^{-i\nu_2 t} + c^\dag e^{i \nu_2 t}) \sigma_1^z \\ \nonumber &+& \eta_1\nu_1(b e^{-i \nu_1 t} + b^\dag e^{i\nu_1 t}) \sigma_2^z + \eta_2\nu_2(ce^{-i\nu_2 t} + c^\dag e^{i \nu_2 t}) \sigma_2^z\\
\nonumber &+& \frac{\Omega_1(t)}{2}(\sigma_1^+ e^{i \phi} + \sigma_1^- e^{-i\phi}) + \frac{\Omega_2(t)}{2}(\sigma_2^+ e^{i \phi} + \sigma_2^- e^{-i\phi}),
\end{eqnarray}
which corresponds to Eq.~(\ref{casi}) in the main text.

\section{The time-evolution operator}\label{app:TimeEvol}

An analytical expression for the time evolution operator exits for any Hamiltonian of the form
\begin{eqnarray}\label{Hdd2}
H^\textrm{II}(t) = \sum_{j=1}^N\sum_{m=1}^M f_{j}(t)\eta_{jm}\nu_m (a_m e^{-i\nu_mt} + a_m^\dag e^{i\nu_mt}) \sigma_j^z.
\end{eqnarray}
The Hamiltonian for our ion system, undergoing a sequence of instantaneous $\pi$-pulses, is given by
\begin{eqnarray}\label{Hdd}
H^{\textrm II}(t) &=&  f_{1}(t)\eta_1\nu_1 (b e^{-i\nu_1t} + b^\dag e^{i\nu_1t}) \ \sigma_1^z \nonumber\\
&-& f_{1}(t)\eta_2\nu_2 (c e^{-i\nu_2t} + c^\dag e^{i\nu_2t})\ \sigma_1^z\nonumber\\   &+& f_{2}(t)\eta_1\nu_1 (b e^{-i\nu_1t} + b^\dag e^{i\nu_1t}) \ \sigma_2^z \nonumber\\
 &+& f_{2}(t)\eta_2\nu_2 (c e^{-i\nu_2t} + c^\dag e^{i\nu_2t})\ \sigma_2^z,
\end{eqnarray}
and therefore belongs to this category with $a_1=b$, $a_2=c$ and $\eta_{11}=\eta_{21}=\eta_1$, $\eta_{12}=-\eta_{22}=-\eta_2$.

The time evolution operator of a time dependent Hamiltonian is given by the Dyson series or equivalently by the Magnus expansion:
\begin{eqnarray}\label{Magnus}
U(t) = \exp{\big\{\Omega^\textrm{M}_1(t) +\Omega^\textrm{M}_2(t)+\Omega^\textrm{M}_3(t)+...\big\}},
\end{eqnarray}
where (in general for $t_0\neq0$)
\begin{eqnarray}\label{MagnusTerms}
\Omega^\textrm{M}_1(t,t_0) &=& -i\int_{t_0}^t dt_1H(t_1)\nonumber\\
\Omega^\textrm{M}_2(t,t_0) &=& -\frac{1}{2}\int_{t_0}^{t}dt_1\int_{t_0}^{t_1} dt_2 [H(t_1),H(t_2)] \\
\Omega^\textrm{M}_3(t,t_0) &=& \frac{i}{6} \int_{t_0}^{t}dt_1\int_{t_0}^{t_1} dt_2  \int_{t_0}^{t2} dt_3 \Big([H(t_1),[H(t_2),H(t_3)]]\nonumber\\ &+& [H(t_3),[H(t_2),H(t_1)]] \Big),\nonumber
\end{eqnarray}
and so on. In our case, $\Omega^\textrm{M}_k$ terms for $k>2$ are zero because $[H(s),[H(s'),H(s'')]]=0$. The first term $\Omega^\textrm{M}_1$ can be written as 
\begin{eqnarray}\label{Magnus1}
\Omega^\textrm{M}_1(t,t_0) =-i \sum_{j,m}\eta_{jm}\big[a_m G_{jm}(t,t_0) +a_m^\dag G_{jm}^*(t,t_0)\big]\sigma_j^z
\end{eqnarray}
where 
\begin{eqnarray}\label{Gfunc}
G_{jm}(t,t_0) =\nu_m\int_{t_0}^t dt' f_j(t')e^{-i\nu_m t'}.
\end{eqnarray}
The second term can be calculated to be

\begin{eqnarray}\label{Magnus2}
\Omega^\textrm{M}_2(t,t_0) &=& -\frac{1}{2}\int_{t_0}^{t}dt_1\int_{t_0}^{t_1} dt_2 [H(t_1),H(t_2)] = -\frac{i}{2}\int_{t_0}^{t}dt_1 \big[H(t_1),\Omega^\textrm{M}_1(t_1,t_0)] \nonumber\\ 
&=& -\frac{i}{2}\int_{t_0}^{t}dt_1\sum_{jm}\sum_{j'm'}(-i)\eta_{jm}\eta_{j'm'}\nu_m\nonumber\\
&\times&[f_j(a_me^{-i\nu_mt_1}+a_{m}^\dag e^{i\nu_m t_1})\sigma_j^z,(a_{m'}G_{j'm'}+a_{m'}^\dag G^*_{j'm'} )\sigma_{j'}^z] \nonumber\\
&=& -\frac{i}{2}\int_{t_0}^{t}dt_1\sum_{jj'}\sum_{m}(-i)\eta_{jm}\eta_{j'm}\nu_{m}\nonumber\\ &\times&\big(f_jG^*_{j'm}e^{-i\nu_m t_1}[a_m,a_m^\dag]+ f_{j}G_{j'm}e^{i\nu_m t_1}[a_m^\dag,a_m]\big)\sigma_j^z\sigma^z_{j'},
\end{eqnarray}
which, using $[a_m,a_{m'}^\dag]=\delta_{m,m'}$ and $(\sigma_j^z)^2=\mathds{1}$ becomes
\begin{eqnarray}\label{Magnus22}
\Omega^\textrm{M}_2(t,t_0) &=& i\varphi(t,t_0)\sigma_1^z\sigma_2^z + K(t,t_0)\mathds{1},
\end{eqnarray}
where the phase $\varphi$ is a time dependent function given by 
\begin{eqnarray}\label{GenPhase2}
\varphi(t,t_0)=\sum_m  \Im{ \int_{t_0}^{t}dt'\eta_{1m}\eta_{2m}\nu_m\big\{f_1(t')G_{2m}(t',t_0)+f_2(t')G_{1m}(t',t_0)\big\}e^{i\nu_m t'}}.
\end{eqnarray}
where $\Im$ indicates the imaginary part. We have ignored the term $K(t)$, as it will only contribute with a global phase. Finally, we can easily check that the time evolution operator can be written as 
\begin{equation}\label{timevol}
U(t,t_0) =U_s(t,t_0)U_c(t,t_0) 
\end{equation}
where 
\begin{equation}\label{twogate}
U_s(t,t_0)= \exp \bigg\{-\! i \sum_{j,m}\eta_{jm} \left[ a_m G_{jm}(t,t_0) + a_m^\dag G_{jm}^*(t,t_0)\right]\sigma_j^z \bigg\},
\end{equation}
and
\begin{equation}
U_c(t,t_0)=\exp \left[i \varphi(t,t_0) \sigma_1^z \sigma_2^z\right].
\end{equation}

\section{Properties of the $G_{jm}(t)$ and $\varphi(t)$ functions}\label{app:Properties}

\begin{figure*}[t!]
\centering
\includegraphics[width=1\linewidth]{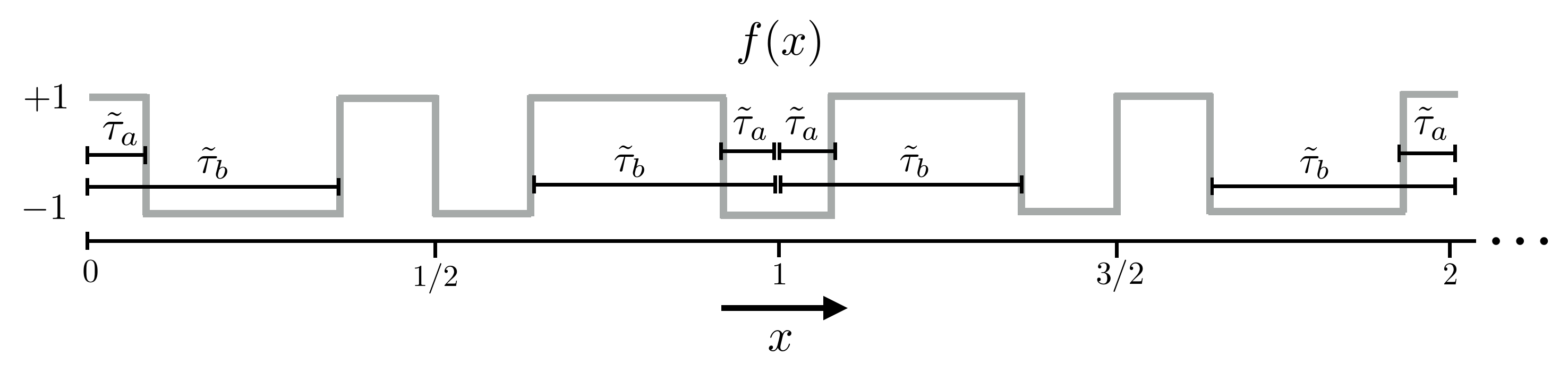}
\caption{ Modulation function $f(x)$ function that corresponds to the first two blocks of the AXY-4 pulse sequence. Time has been normalised by the characteristic time of the sequence $\tau$ ($x=t/\tau$), as well as $\tilde{\tau}_a=\tau_a/\tau$ and $\tilde{\tau}=\tau_b/\tau$.}\label{fig:AXY2}
\end{figure*}

Searching for all different sequences that fulfil the conditions $G_{jm}(T_\textrm{G})=0$ and $\varphi(T_\textrm{G})\neq0$ gets easy if we identify which are the indispensable variables that define the problem. The sequence function $f_1(t)=f_2(t)=f(t)$ is completely defined by the four parameters $\tau_a,\tau_b,\tau$, and $n_\textrm{B}$, for the case of an AXY-$n_\textrm{B}$ sequence. The duration of the sequence is of course only determined by two of them: $\tau$ and $n_\textrm{B}$ ($T_\textrm{G}=n_\textrm{B}\tau$). It is useful to rescale the domain on the $f$ function using $\tau$, the characteristic time of the sequence, as $t=x\tau$. Then, the property $f(t+\tau)=-f(t)$ becomes $f(x+1)=-f(x)$, and also $\tau_a,\tau_b$ may be redefined as $\tilde{\tau}_a=\tau_a/\tau$ and $\tilde{\tau}_b=\tau_b/\tau$. The AXY-4 sequence with this time rescaling is shown in Fig.~\ref{fig:AXY2}.  With this change of variable, the $G_{jm}$ functions at a time $T_\textrm{G}$ read
\begin{equation}\label{Gchange}
G_{jm}(T_\textrm{G})=\nu_m\int_{0}^{T_\textrm{G}}\!dt \ f(t)\ e^{-i\nu_m t}=\nu_m\tau\int_{0}^{M}\!dx \ f(x) \ e^{-i\nu_m \tau x}.
\end{equation}
Now, if we relate $\tau$ and $\nu$ as
\begin{equation}\label{nutau}
\nu \tau=2\pi r \ \ \mbox{with} \ \ r\in \mathbb{N}, 
\end{equation}
these functions become independent of the frequency $\nu$,
\begin{eqnarray}\label{Gtilde}
G_{j1}(T_\textrm{G})=\nu_1\tau \int_{0}^{M}\!dx \ f(x)\ e^{-i2\pi r x}=2\pi r\int_{0}^{M}\!dx \ f(x)\ e^{-i2\pi r x} \\
G_{j2}(T_\textrm{G})=\nu_2\tau \int_{0}^{M}\!dx \ f(x)\ e^{-i2\pi\sqrt{3} r x}=2\pi \sqrt{3}r\int_{0}^{M}\!dx \ f(x)\ e^{-i2\pi\sqrt{3} r x}.
\end{eqnarray}
This simplification works also for the $\varphi(t)$ function, that can be redefined in terms of $\tilde{\varphi}_1(t)$ and $\tilde{\varphi}_1(t)$ as 
\begin{eqnarray}\label{varphi}
\varphi(t)&=&\eta_1^2\tilde{\varphi}_1(t)-\eta_2^2\tilde{\varphi}_2(t)=\eta_1^2(\tilde{\varphi}_1(t)-\frac{1}{3\!\sqrt{3}}\tilde{\varphi}_2(t))=\eta_1^2\tilde{\varphi}(t),
\end{eqnarray}
where 
\begin{eqnarray}\label{phitilde}
\tilde{\varphi}_1(T_\textrm{G})=(2\pi r)^2 \ \Im{\int_{0}^{M}\!dx \int_{0}^{x}\!dy \ [f_1(x)f_2(y)+f_2(x)f_1(y)] \ e^{i2\pi r(x-y)}} \\
\tilde{\varphi}_2(T_\textrm{G})=(2\pi \sqrt{3}r)^2 \ \Im{ \int_{0}^{M}\! dx\int_{0}^{x}\!dy \ [f_1(x)f_2(y)+f_2(x)f_1(y)] \ e^{i2\pi \sqrt{3}r(x-y)}},
\end{eqnarray}
or
\begin{eqnarray}\label{phitilde}
\tilde{\varphi}(T_\textrm{G})=(2\pi r)^2 \ \Im{\int_{0}^{M}\!dx \int_{0}^{x}\!dy}  [f_1(x)f_2(y)+f_2(x)f_1(y)] \\ \times (e^{i2\pi r(x-y)}-\frac{1}{\sqrt{3}}e^{i2\pi \sqrt{3}r(x-y) }). 
\end{eqnarray}

Now, it is clear that the functions $\tilde{G}_{j1}(T_\textrm{G})$, $\tilde{G}_{j2}(T_\textrm{G})$, and $\tilde{\varphi}(T_\textrm{G})$, only depend on $\tilde{\tau}_a$, $\tilde{\tau}_b$, $n_\textrm{B}$ and $r$. Therefore, the functions plotted in Fig.~(\ref{Gplot}) (corresponding to cases $n_\textrm{B}=4$, $r=1,2,3$), do not depend on the frequencies $\nu_m$, but only on $\tilde{\tau}_a$ and $\tilde{\tau}_b$.

\section{Pulse propagator}\label{app:PulseP}

\begin{figure}[t]
\centering
\includegraphics[width=0.7\linewidth]{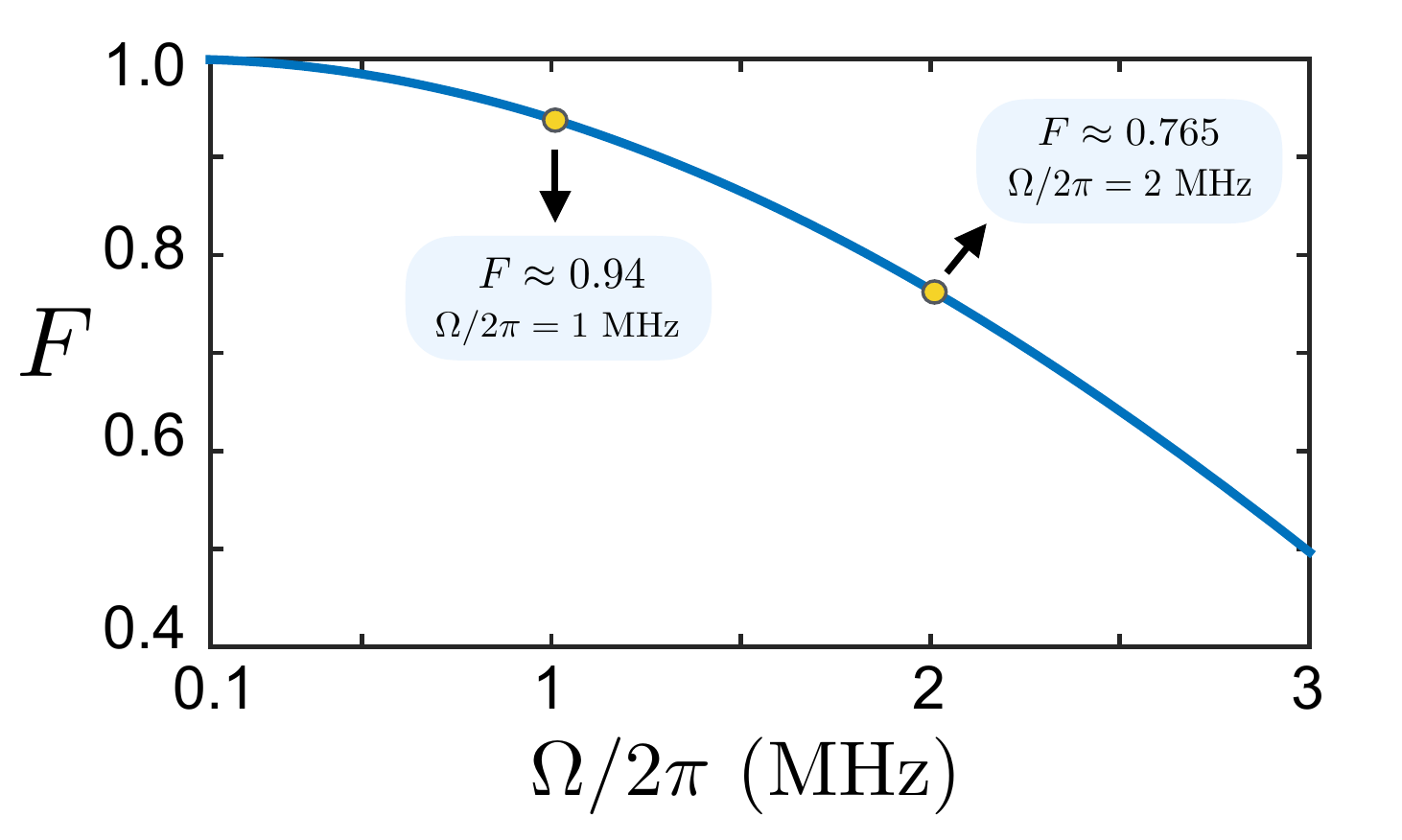}
\caption{Fidelity after 20 $\pi$-pulses between the propagators associated to the Hamiltonians $H =  \frac{\Omega_1}{2} \sigma_1^{\phi} +  \frac{\Omega_1}{2}[ \sigma_2^+  e^{i\phi} e^{i\delta_2t} + {\textrm H.c.}]$ and $H =  \frac{\Omega_1}{2} \sigma_1^{\phi}$ as a function of the Rabi frequency $\Omega$.  We can observe how the fidelity decays because of the fail of the RWA. For this plot we have taken $\delta_2\approx (2\pi)\times 45$ MHz, a value that is even larger than the ones we use in section \ref{subsect:Tailored}.}
\label{crosstalkfigure}
\end{figure}

Our strategy to produce fast $\pi$-pulses on a certain ion qubit and, at the same time, eliminate undesired effects on the off-resonant qubit consists in appropriately choosing the Rabi frequency of the driving. When a $\pi$-pulse is applied, say on the first qubit, we have the following Hamiltonian
\begin{equation}
H = \frac{\omega_1}{2} \sigma_1^z + \frac{\omega_2}{2} \sigma_2^z + \Omega_1 \cos(\omega_1 t - \phi) (\sigma_1^x + \sigma_2^x ).
\end{equation}
In a rotating frame with respect to $ \frac{\omega_1}{2} \sigma_1^z + \frac{\omega_2}{2} \sigma_2^z$ and after eliminating fast rotating terms, the above Hamiltonian reads
\begin{eqnarray}\label{ctH}
H =  \frac{\Omega_1}{2} \sigma_1^{\phi} +  \frac{\Omega_1}{2}[ \sigma_2^+  e^{i\phi} e^{i\delta_2t} + \textrm{H.c.}],
\end{eqnarray}
where $\sigma_1^{\phi} =\sigma_1^+ e^{i\phi} + \sigma_1^- e^{-i\phi}$ and $\delta_2 = \omega_2 - \omega_1$. At this level one can argue that, only when the Rabi frequency is small, the second term on the right-hand side of the above equation is negligible. This unavoidably limits the value of $\Omega_1$ and, consequently, how fast our decoupling pulses can be. In this respect, note that  for a value of $\omega_2 - \omega_1\approx (2\pi)\times45$ MHz, which is even larger than the ones used in section~\ref{sect:1_PDD}, we find that $\Omega_1$ should be significantly smaller than $(2\pi)\times1$ MHz if we want to eliminate the crosstalk between ions during 20 $\pi$-pulses, see Fig.~\ref{crosstalkfigure}. 

To eliminate this restriction, we can use the following expression 
\begin{eqnarray}\label{prop}
U_{[t:t_0]} &\equiv& \hat{T}e^{-i\int_{t_0}^t H(s) ds} = U_0 \tilde{U}_{[t:t_0]} \\ \nonumber
&\equiv& e^{-iH_\delta (t-t_0)} \hat{T}e^{-i\int_{t_0}^t  U^{\dag}_0 (H(s) - H_\delta) U_0ds} ,
\end{eqnarray}
where $\hat{T}$ is the time ordering operator, $H_\delta=-(\delta_2/2) \sigma_2^z$, and find the time evolution operator for Eq.~(\ref{ctH}) in a generic time interval $(t, t_0)$. This is 
\begin{equation}
U_{[t:t_0]} =  e^{-i \frac{\Omega_1}{2} \sigma_1^{\phi} (t-t_0) } e^{i\frac{\delta_2}{2} \sigma_2^{z} (t-t_0)} e^{-i  \gamma (t-t_0)  \hat{n}_0\cdot\vec{\sigma}_2} ,
\end{equation}
where $\gamma = \frac{1}{2}\sqrt{ \Omega_1^2 +  \delta_2^2 }$ and
 \begin{equation}\label{unitcross}
 \hat{n}_{0}=\frac{1}{2\gamma}\Big[\Omega\cos{(\phi-\varphi_0)},-\Omega\sin{(\phi+\varphi_0)}, \delta\Big],
 \end{equation}
with $\varphi_0=\delta t_0 /2$.

Note that the first and the second terms in Eq.~(\ref{ctH}) commute, which allows to apply the relation~(\ref{prop}) only to the part $\frac{\Omega_1}{2}[ \sigma_2^+  e^{i\phi} e^{i\delta_2t} + \textrm{H.c.} ]$. Finally, for the sake of realising a $\pi$-pulse we will set $(t-t_0) = t^{(1)}_{\pi} \equiv \frac{\pi}{\Omega_1}$ which gives rise to 
\begin{eqnarray}\label{piuno}
U_{t_\pi}^{(1)} =  e^{-i(\Omega_1/2) \sigma_1^{\phi} t_\pi } e^{i(\delta_2/2) \sigma_2^{z} t_\pi} e^{-i  \gamma t_\pi (\hat{n}_0\cdot\vec{\sigma}_2)}. 
\end{eqnarray}

In the same manner, for a $\pi$-pulse (with $ t^{(2)}_{\pi} \equiv \frac{\pi}{\Omega_2}$) in resonance with the second ion we would have 
\begin{eqnarray}\label{pidos}
U_{t_\pi}^{(2)} =  e^{-i (\Omega_2/2) \sigma_2^{\phi} t_\pi } e^{i(\delta_1/2) \sigma_1^{z} t_\pi} e^{-i  \gamma t_\pi(\hat{n}_0\cdot\vec{\sigma}_1)}. 
\end{eqnarray}
Equations~(\ref{piuno}) and~(\ref{pidos}) contain the terms corresponding to the $\pi$-pulses in which we are interested ($\exp{[-i \frac{\Omega_1}{2} \sigma_1^{\phi} t_\pi]}$ and $\exp{[-i \frac{\Omega_2}{2} \sigma_2^{\phi} t_\pi]}$) plus the crosstalk contributions we want to get rid off. To eliminate terms $\exp{[-i\gamma t_\pi (\hat{n}_0\cdot\vec{\sigma}_2)]}$ and $\exp{[-i  \gamma t_\pi (\hat{n}_0\cdot\vec{\sigma}_1)]}$, we will adjust the Rabi frequencies $\Omega_{1,2}$ such that 
\begin{equation}
\gamma t_\pi = \frac{1}{2} \sqrt{ (\Omega_{1,2})^2 +  (\delta_{2,1})^2 }\frac{\pi}{\Omega_{1,2}}   = \pi \times k, \mbox{with} \ k\in \mathbb{Z}. 
\end{equation}
In this case we have that $\exp{[-i  \gamma t_\pi(\hat{n}_0\cdot\vec{\sigma}_2)]} = \exp{[-i  \gamma t_\pi(\hat{n}_0\cdot\vec{\sigma}_1)}]] = \pm \mathds{1}$, i.e. the unwanted terms contribute as a global phase. Hence, only pure dephasing terms remain in both pulses, $\exp{[i\frac{\delta_2}{2} \sigma_2^{z} t_\pi]}$ and $\exp{[i\frac{\delta_1}{2} \sigma_1^{z} t_\pi]}$, which will be cancelled by our tailored MW sequences as explained in section \ref{subsect:Tailored}. 

\section{Heating rates estimation}\label{app:Heating}
To estimate the $\Gamma_{b,c}$ parameters, we will rely on the data provided in~\cite{Weidt16}, as well as on the scaling relations one can extract from~\cite{Brownnutt15}. More specifically we take as reference values (for the center-of-mass mode) $\dot{n}_\textrm{com}^\textrm{ref} = 41$ phonons/second for a frequency $\nu^\textrm{ref}_1/(2\pi) = 427$ kHz, and (for the breathing mode) $\dot{n}_\textrm{bre}^\textrm{ref} = 7$ phonons/second for a frequency $\nu^\textrm{ref}_2/(2\pi) = 459$ kHz,~\cite{Weidt16}. The latter values correspond to a configuration at room temperature ($T^\textrm{ref}=300$ K) with an ion-electrode distance of $d_\textrm{i-e}^\textrm{ref} \approx 310 \ \mu$m, that would give rise to a magnetic field gradient $g_B = 23.6$T/m. 

Our operating conditions require, for the first studied case, an ion-electrode distance of $d_\textrm{i-e}\approx 150 \ \mu$m, to generate a magnetic field gradient of $g_B = 150$T/m where $\nu_1=\nu$ and $\nu_2 = \sqrt{3} \nu$ with $\nu/(2\pi) = 150$ kHz, while we will consider low temperatures of $T=50$ K. In this situation one can derive new values for $\dot{n}_\textrm{com}$ and $\dot{n}_\textrm{bre}$ using scaling relations~\cite{Brownnutt15} which in our case are
\begin{eqnarray}\label{scaling1}
\dot{n}_\textrm{com}\approx \dot{n}_\textrm{com}^\textrm{ref} \ \bigg(\frac{\nu_1^\textrm{ref}}{\nu_1}\bigg)^2 \bigg(\frac{d_\textrm{i-e}^\textrm{ref}}{d_\textrm{i-e}}\bigg)^4 \bigg(\frac{T^\textrm{ref}}{T}\bigg)^{-2.13},
\end{eqnarray}
and
\begin{eqnarray}\label{scaling2}
\dot{n}_\textrm{bre}\approx \dot{n}_\textrm{bre}^\textrm{ref} \ \bigg(\frac{\nu_2^\textrm{ref}}{\nu_2}\bigg)^2 \bigg(\frac{d_\textrm{i-e}^\textrm{ref}}{d_\textrm{i-e}}\bigg)^4 \bigg(\frac{T^\textrm{ref}}{T}\bigg)^{-2.13}.
\end{eqnarray}
Then, one can use that, when close to the motional ground state, we have~\cite{Brownnutt15}  
\begin{equation}
\dot{n}_{\textrm{com,bre}} = \Gamma_{b,c} \ \bar{N}_{b, c},
\end{equation}
that together with the definition of $\bar{N}_{b, c} \equiv N^\textrm{thermal}_{b,c} =1/(e^{\hbar \nu_{1,2}/k_{\textrm B} T}-1)$, allows us to obtain the values for $\Gamma_{b,c}$. 

In the second studied case, a magnetic field gradient of $g_B = 300 \frac{\textrm T}{\textrm m}$ would require to locate the ions closer to the electrodes, which would induce more heating. We estimate a distance according to the relation $d_\textrm{i-e} = \sqrt{\frac{150}{300}} \ 150 \ \mu$m $\approx 106 \ \mu$m that assumes a dependence  $g_B \sim \frac{1}{d_\textrm{i-e}^2}$. This new distance can be used in Eqs.~(\ref{scaling1}) and~(\ref{scaling2}) to derive new values for the heating rates $\Gamma_{b,c}$.

\include{appendices/appendix_B}
\chapter{Further Considerations on the Selective Interactions of the QRM}
\label{appendix:appendix_c}

\section{Dyson series of the Rabi-Stark Hamiltonian}\label{app:QRSDyson}
The Rabi-Stark Hamiltonian as written in Eq.~(\ref{QRS1}) is
\begin{equation}\label{SQRS1}
H=\frac{\omega_0}{2}\sigma_z +\omega a^\dag a + \gamma a^\dag a \sigma_z + g(\sigma_+ +\sigma_-)(a+a^\dag)
\end{equation}
where $\sigma_z,\sigma_+,\sigma_-$ are operators of the two-level system and $a^\dagger$ and $a$ are infinite dimensional creation and annihilation operators of the bosonic field. Using the ket-bra notation, the two-level matrices are $\sigma_+=|e \rangle\langle g|$, $\sigma_-=|g \rangle\langle e|$ and $\sigma_z=|e \rangle\langle e|-|g \rangle\langle g|$ where $|e\rangle$ and $|g\rangle$ the excited and ground states of the two level system, respectively. On the other hand, the bosonic operators can be written as
\begin{eqnarray}\label{SBosOp}
a^\dagger=\sum_{n=0}^{\infty}\sqrt{n+1}|n+1\rangle\langle n|   \\
a=\sum_{n=0}^{\infty}\sqrt{n+1}|n\rangle\langle n+1| 
\end{eqnarray}
where $|n\rangle$ is the $n$-th Fock state. With this notation, the Hamiltonian in Eq.~(\ref{SQRS1}) can be rewritten as 
\begin{eqnarray}\label{SQRS2}
H=\sum_{n=0}^{\infty}\omega_n^e|e \rangle\langle e|\otimes |n \rangle\langle n|+\omega^g_{n}|g \rangle\langle g|\otimes |n \rangle\langle n|\nonumber\\+\Omega_n(|e \rangle\langle g|+\textrm{H.c.})\otimes(|n+1\rangle\langle n| +\textrm{H.c.})
\end{eqnarray}
where $\omega_n^e=(\omega  +\gamma )n+\omega_0/2$, $\omega_n^g=(\omega  -\gamma )n-\omega_0/2$ and $\Omega_n=g\sqrt{n+1}$. We can move to an interaction picture with respect the diagonal part of Eq.~(\ref{SQRS2}), and the non-diagonal elements will rotate as
\begin{eqnarray}\label{SNonDiag2}
|e \rangle\langle g|\otimes |n+1 \rangle\langle n| &\rightarrow& |e \rangle\langle g| \otimes |n+1 \rangle\langle n| e^{i(\omega^e_{n+1}-\omega_n^g)t} \\
|g \rangle\langle e|\otimes |n+1 \rangle\langle n|  &\rightarrow& |g \rangle\langle e| \otimes |n+1 \rangle\langle n| e^{-i(\omega^e_n-\omega_{n+1}^g)t} \\
|e \rangle\langle g|\otimes |n \rangle\langle n+1| &\rightarrow& |e \rangle\langle g| \otimes |n \rangle\langle n+1| e^{i(\omega^e_{n}-\omega_{n+1}^g)t} \\
|g \rangle\langle e|\otimes |n \rangle\langle n+1|  &\rightarrow& |g \rangle\langle e| \otimes |n \rangle\langle n+1| e^{-i(\omega^e_{n+1}-\omega_{n}^g)t} 
\end{eqnarray}
where $\delta^+_{n}=\omega_{n+1}^e-\omega_n^g=\omega+[\omega_0+\gamma(2n+1)]$ and $\delta^-_{n}=\omega_{n+1}^g-\omega_{n}^e=\omega-[\omega_0+\gamma(2n+1)]$. The Hamiltonian in the interaction picture can be then rewritten as
\begin{equation}\label{SQRSIntPic}
H_I(t)=\sum_{n=0}^{\infty}\Omega_n(\sigma_+ e^{i\delta^+_{n} t}+\sigma_- e^{i\delta^-_{n} t})\otimes |n+1\rangle\langle n| +\textrm{H.c.}
\end{equation}
which corresponds to Eq.~(\ref{QRSIntPic}). 

\subsection{Second-order Hamiltonian}\label{subapp:QRSSecondOrder}
The second order Hamiltonian that corresponds to Eq.~(\ref{SQRSIntPic}) is given by~\cite{Sakurai94}
\begin{equation}\label{S2Dyson}
H^{(2)}(t)=-i\int_0^t dt' H_I(t)H_I(t').
\end{equation}
We can write $H_{I}(t)$ as
\begin{equation}\label{SredHI}
H_I(t)=\sum_{n=0}^{\infty} \Omega_n \Big(S_{n}(t)|n+1\rangle\langle n| +S^\dagger_{n}(t)|n\rangle\langle n+1|\Big),
\end{equation}
and, then, 
\begin{eqnarray}\label{S2orderQRS}
H^{(2)}(t)=-i\sum_{n,n'}\Omega_{n}\Omega_{n'}\Big(S_{n}(t)|n+1\rangle\langle n| +S^\dagger_{n}(t)|n\rangle\langle n+1|\Big)\nonumber \\ \times \int_0^tdt'\Big(S_{n'}(t')|n'+1\rangle\langle n'| +S^\dagger_{n'}(t')|n'\rangle\langle n'+1|\Big),
\end{eqnarray}
which gives $H^{(2)}=H_A^{(2)}+H_B^{(2)}$, where
\begin{eqnarray}\label{S2orderQRS_2}
H^{(2)}_A(t)=-i\sum_{n}\Omega_{n}^2\Big(S_{n}(t)\int_0^tdt'S^\dagger_{n}(t')\Big)|n+1\rangle\langle n+1| \nonumber\\ +\Omega^2_{n}\Big(S^\dagger_{n}(t)\int_0^tdt'S_{n}(t')\Big)|n\rangle\langle n| 
\end{eqnarray}
gives diagonal elements and
\begin{eqnarray}\label{S2orderQRS_3}
H^{(2)}_B(t)=-i\sum_{n}\Omega_{n}\Omega_{n+1}\Big(S_{n+1}(t)\int_0^tdt'S_{n}(t')\Big)|n+2\rangle\langle n| \nonumber\\+\Omega_{n}\Omega_{n+1}\Big(S^\dagger_{n}(t)\int_0^tdt'S^\dagger_{n+1}(t')\Big)|n\rangle\langle n+2| 
\end{eqnarray}
is related with two-photon processes. Calculating the two-level operators we obtain 
\begin{eqnarray}\label{S2orderSpinOp}
S_{n}(t)\int_0^tdt'S^\dagger_{n}(t')&=&\frac{i}{\delta^+_n}\sigma_+\sigma_- + \frac{i}{\delta^-_n}\sigma_-\sigma_+ \nonumber\\ &-&\frac{i}{\delta^+_n}\sigma_+\sigma_-e^{i\delta^+_n t} - \frac{i}{\delta^-_n}\sigma_-\sigma_+e^{i\delta^-_n t} \\
S^\dagger_{n}(t)\int_0^tdt'S_{n}(t')&=&-\frac{i}{\delta^+_n}\sigma_-\sigma_+ - \frac{i}{\delta^-_n}\sigma_+\sigma_- \nonumber\\ &+& \frac{i}{\delta^+_n}\sigma_-\sigma_+e^{-i\delta^+_n t} + \frac{i}{\delta^-_n}\sigma_+\sigma_-e^{-i\delta^-_n t} \\
S_{n+1}(t)\int_0^tdt'S_{n}(t')&=& -\frac{i}{\delta^-_{n}}\sigma_+\sigma_-(e^{i(\delta^+_{n+1} +\delta^-_{n})t}-e^{i\delta^+_{n+1}t})  \nonumber\\&-& \frac{i}{\delta^+_{n}}\sigma_-\sigma_+(e^{i(\delta^-_{n+1} +\delta^+_{n})t} -e^{i\delta^-_{n+1}t})\\
S^\dagger_{n}(t)\int_0^tdt'S^\dagger_{n+1}(t')&=&  \frac{i}{\delta^-_{n+1}}\sigma_-\sigma_+(e^{-i(\delta^+_{n} +\delta^-_{n+1})t} -e^{-i\delta^+_{n}t}) \nonumber\\ &+& \frac{i}{\delta^+_{n+1}}\sigma_+\sigma_-(e^{-i(\delta^-_{n} +\delta^+_{n+1})t} -e^{i\delta^-_{n}t}).
\end{eqnarray}
We can ignore the terms oscillating with $\pm\delta^{\pm}_n$, as these frequencies correspond to resonances of the first-order Hamiltonian and one-photon processes. Keeping the other terms we have that
\begin{eqnarray}\label{S2orderQRS_4}
H^{(2)}_A\approx\sum_{n}\Omega_{n}^2\Big(\frac{\sigma_+\sigma_-}{\delta_n^+} +\frac{\sigma_-\sigma_+}{\delta_n^-}\Big)|n+1\rangle\langle n+1|  -\Omega^2_{n}\Big(\frac{\sigma_-\sigma_+}{\delta^+_n} + \frac{\sigma_+\sigma_-}{\delta^-_n}\Big)|n\rangle\langle n| 
\end{eqnarray}
and
\begin{eqnarray}\label{S2orderQRS_5}
H^{(2)}_B(t)\approx \sum_{n}-\Omega_{n}\Omega_{n+1}\Big(\frac{\sigma_+\sigma_-}{\delta^-_{n}}e^{i(\delta^+_{n+1} +\delta^-_{n})t}   + \frac{\sigma_-\sigma_+}{\delta^+_{n}}e^{i(\delta^-_{n+1} +\delta^+_{n})t} \Big)|n+2\rangle\langle n| \nonumber\\ 
+\Omega_{n}\Omega_{n+1}\Big(\frac{\sigma_-\sigma_+}{\delta^-_{n+1}}e^{-i(\delta^+_{n} +\delta^-_{n+1})t}   + \frac{\sigma_+\sigma_-}{\delta^+_{n+1}}e^{-i(\delta^-_{n} +\delta^+_{n+1})t} \Big)|n\rangle\langle n+2|. 
\end{eqnarray}
The two-photon transition terms in Eq.~(\ref{S2orderQRS_5}) oscillate with frequencies $\delta^+_n+\delta^-_{n+1}=2\omega-2\gamma$ and $\delta^+_{n+1}+\delta^-_{n}=2\omega+2\gamma$, which are zero only in the points of the spectral collapse. Thus, we do not expect to see two-photon transitions in the regime where the Hamiltonian is bounded from below. The terms in Eq.~(\ref{S2orderQRS_4}) will induce an additional Stark shift that can induce a shift in the resonance conditions of the higher order processes, as we will see later. The Hamiltonian can be simplified to
\begin{equation}\label{S2orderQRS_6}
H^{(2)}_A\approx\sum_{n}\Bigg\{\Big(\frac{\Omega_{n-1}^2}{\delta_{n-1}^+} -\frac{\Omega_{n}^2}{\delta_n^-}\Big)\sigma_+\sigma_- +\Big(\frac{\Omega_{n-1}^2}{\delta_{n-1}^-} -\frac{\Omega_{n}^2}{\delta_n^+}\Big)\sigma_-\sigma_+ \Bigg\}|n\rangle\langle n|.
\end{equation}

\subsection{Third-order Hamiltonian}\label{subapp:QRSThirdOrder}
The third order Hamiltonian is calculated by 
\begin{equation}\label{S3Dyson}
H^{(3)}(t)=(-i)^2\int_0^t dt' \int_0^{t'}dt'' H_I(t)H_I(t')H(t'').
\end{equation}
Following the same notation of the previous section, the third order Hamiltonian is 
\begin{eqnarray}\label{S3orderQRS}
H^{(3)}(t)=-\sum_{n, n', n''}\Omega_{n}\Omega_{n'}\Omega_{n''}\Big(S_{n}(t)|n+1\rangle\langle n| +\textrm{H.c.}\Big)\nonumber\\ \times\int_0^t dt' \Big(S_{n'}(t')|n'+1\rangle\langle n'| +\textrm{H.c.}\Big) \int_0^{t'}dt'' \Big(S_{n''}(t'')|n''+1\rangle\langle n''| +\textrm{H.c.}\Big).
\end{eqnarray}
If we focus on the three-photon resonances, the following Hamiltonian contains them
\begin{eqnarray}\label{S3orderQRS_A}
H_A^{(3)}(t)=-\sum_{n}\Omega_{n}\Omega_{n+1}\Omega_{n+2}S_{n+2}(t) \nonumber\\ \times\int_0^t dt' S_{n+1}(t') \Big(\int_0^{t'}dt'' S_{n}(t'')\Big)|n+3\rangle\langle n| +\textrm{H.c}.
\end{eqnarray}
The contribution of the two-level operators can be easily calculated by noticing that from
\begin{eqnarray}\label{S3order_TL1}
S_{n+2}(t) S_{n+1}(t') S_{n}(t'')&=&(\sigma_+ e^{i\delta^+_{n+2} t}+\sigma_- e^{i\delta^-_{n+2} t})\\&\times&(\sigma_+ e^{i\delta^+_{n+1} t'}+\sigma_- e^{i\delta^-_{n+1} t'}) (\sigma_+ e^{i\delta^+_{n} t''}+\sigma_- e^{i\delta^-_{n} t''}),\nonumber
\end{eqnarray}
only the following two terms are not zero (notice that $\sigma_\pm^2=0$)
\begin{equation}\label{S3order_TL2}
S_{n+2}(t) S_{n+1}(t') S_{n}(t'')=\sigma_+ e^{i\delta^+_{n+2} t} e^{i\delta^-_{n+1} t'}e^{i\delta^+_{n} t''} +\sigma_- e^{i\delta^-_{n+2} t} e^{i\delta^+_{n+1} t'}e^{i\delta^-_{n} t''}.
\end{equation}
After calculating the integral we obtain that the Hamiltonian is
\begin{eqnarray}\label{S3orderQRS_A}
H_A^{(3)}(t)=\sum_{n=0}^{\infty}\Omega_{n}\Omega_{n+1}\Omega_{n+2}\Bigg\{\frac{1}{\delta_n^+(\delta_{n+1}^-+\delta_n^+)}\Big(e^{i\delta^{(3)}_{+n}t} - e^{i\delta_{n+2}^+t}\Big) \nonumber\\+ \frac{1}{\delta^+_n\delta_n^-}\Big(e^{i2(\omega+\gamma)t} - e^{i\delta_{n+2}^+t}\Big)  \Bigg\}\sigma_+|n+3\rangle\langle n| +\textrm{H.c} \nonumber\\ 
+\Omega_{n}\Omega_{n+1}\Omega_{n+2}\Bigg\{\frac{1}{\delta_n^-(\delta_{n+1}^++\delta_n^-)}\Big(e^{i\delta^{(3)}_{-n}t} - e^{i\delta_{n+2}^-t}\Big) \nonumber
\\+ \frac{1}{\delta^-_n\delta_n^+}\Big(e^{i2(\omega-\gamma)t} - e^{i\delta_{n+2}^-t}\Big)  \Bigg\}\sigma_-|n\rangle\langle n+3| +\textrm{H.c}
\end{eqnarray}
where $\delta^{(3)}_{+n}=\delta_{n+2}^++\delta_{n+1}^{-}+\delta_n^+=2\omega+\delta^+_{n+1}$ and $\delta^{(3)}_{-n}=\delta_{n+2}^-+\delta_{n+1}^{+}+\delta_n^-=2\omega+\delta^-_{n+1}$. Ignoring the resonances $\delta^+_{n+2}$ or $\delta^-_{n+2}$ that correspond to the first-order processes, and $2(\omega\pm\gamma)$ which is only zero at the point of the spectral collapse we are leaved with 
\begin{eqnarray}\label{S3orderQRS_A2}
H_A^{(3)}(t)=\sum_{n=0}^{\infty}\Omega_{n}\Omega_{n+1}\Omega_{n+2}\Bigg\{\frac{1}{2\delta_n^+(\omega-\gamma)}\sigma_+e^{i\delta^{(3)}_{n+}t}  \nonumber\\+ \frac{1}{2\delta_n^-(\omega+\gamma)}\sigma_-e^{i\delta^{(3)}_{n-}t}  \Bigg\}|n+3\rangle\langle n| +\textrm{H.c},
\end{eqnarray}
where the 3-photon JC and anti-JC resonances are easily identified as $\delta_{n+}^{(3)}=0$ and $\delta_{n-}^{(3)}=0$ respectively. In a simplified way, Eq.~(\ref{S3orderQRS_A2}) is rewritten as
\begin{equation}\label{S3orderQRS_A3}
H_A^{(3)}(t)=\sum_{n=0}^{\infty}(\Omega^{(3)}_{n+}\sigma_+e^{i\delta^{(3)}_{n+}t}  + \Omega_{n-}^{(3)}\sigma_-e^{i\delta^{(3)}_{n-}t})|n+3\rangle\langle n| +\textrm{H.c},
\end{equation}
where $\Omega^{(3)}_{n+}=g^3\sqrt{(n+3)!/n!}/2\delta^+_n(\omega-\gamma)$ and $\Omega^{(3)}_{n-}=g^3\sqrt{(n+3)!/n!}/2\delta^-_n(\omega+\gamma)$, as shown in section~\ref{subsect:MultiPhoton}.

\section{ Derivation of the Rabi-Stark Hamiltonian in trapped ions}\label{app:QRSTI}
In this section we will explain in detail how to go from Eq.~(\ref{Scheme2}) to Eq.~(\ref{TIQRS}) in the main text. Equation~(\ref{Scheme2}) reads
\begin{equation}\label{SScheme2}
H_{A}(t)=-i\frac{\eta\Omega_r}{2} a \sigma_+ e^{-i\delta_rt}  -i\frac{\eta\Omega_b}{2}  a^\dagger \sigma_+ e^{-i\delta_bt} + e^{i\phi_\textrm{S}}\hat{g}_{\textrm S}\sigma_++\textrm{H.c.}
\end{equation}
At this point the vibrational RWA has been applied, however for a precise understanding of the effective dynamics we need to consider the following additional terms that have been neglected, so that the total Hamiltonian is $H=H_A+H_B$, where
\begin{eqnarray}\label{SSchemeRemaining}
H_{B}(t)&=&-\frac{\Omega_r}{2} \sigma_+ e^{-i(-\nu +\delta_r)t} -\frac{\Omega_b}{2} \sigma_+ e^{-i(\nu +\delta_b)t} \nonumber\\ &+& i\eta e^{i\phi_{\textrm S}}\frac{\Omega_{\textrm S}}{2}\sigma_+(ae^{-i\nu t}+a^\dagger e^{i\nu t}) +\textrm{H.c.}
\end{eqnarray}
The first two terms are the off-resonant carrier interactions of the red and blue drivings which are usually neglected given that $\Omega_{r,b}\ll \nu$. The last term represents the coupling to the motional mode of the carrier driving. The latter does not commute with the first and second terms, and moreover, they both oscillate at similar frequencies (as $\delta_r,\delta_b\ll\nu$). The consequence is that, at second order, these terms produce interactions that cannot be neglected as we will see in the following. The second order effective Hamiltonian is
\begin{eqnarray}\label{SSecondOrder}
H^{(2)}(t)&=&-i\int_{0}^{t}H(t)H(t')dt' \nonumber\\&=&-i\int_{0}^{t}\Big(H_A(t)+H_B(t)\Big)\Big(H_A(t')+H_B(t')\Big)dt'.
\end{eqnarray}
We are only interested in terms arising from $\int_{0}^{t}H_B(t)H_B(t')dt'$ whose oscillating frequency is $\delta_r$ or $\delta_b$. These are
\begin{eqnarray}\label{SSecondOrderList}
\frac{\Omega_r}{2}\sigma_+e^{-i(-\nu+\delta_r)t}\int_{0}^t \!\!dt' i\eta\frac{\Omega_{\textrm S}}{2}e^{-i\phi_{\textrm S}}\sigma_- ae^{-i\nu t'} =-\eta\frac{\Omega_{\textrm S}\Omega_r}{4\nu}\sigma_+\sigma_-e^{-i\phi_{\textrm S}}e^{-i\delta_r t} a \\
\frac{\Omega_b}{2}\sigma_+e^{-i(\nu+\delta_b)t}\int_{0}^t \!\!dt' i\eta\frac{\Omega_{\textrm S}}{2}e^{-i\phi_{\textrm S}}\sigma_- a^\dagger e^{i\nu t'} =\eta\frac{\Omega_{\textrm S}\Omega_b}{4\nu}\sigma_+\sigma_-e^{-i\phi_{\textrm S}}e^{-i\delta_b t} a^\dagger \\
-i\eta\frac{\Omega_{\textrm S}}{2}e^{i\phi_{\textrm S}}\sigma_+a^\dagger e^{i\nu t}\int_0^t \!\!dt' \frac{\Omega_r}{2}\sigma_-e^{i(-\nu+\delta_r)t'}=\eta\frac{\Omega_{\textrm S}\Omega_r}{4(\nu-\delta_r)}\sigma_+\sigma_-e^{i\phi_{\textrm S}}e^{i\delta_r t} a^\dagger  \\
-i\eta\frac{\Omega_{\textrm S}}{2}e^{i\phi_{\textrm S}}\sigma_+a e^{-i\nu t}\int_0^t \!\!dt'\frac{\Omega_b}{2}\sigma_-e^{i(\nu+\delta_b)t'}=-\eta\frac{\Omega_{\textrm S}\Omega_b}{4(\nu+\delta_b)}\sigma_+\sigma_-e^{i\phi_{\textrm S}}e^{i\delta_b t} a \\
-\frac{\Omega_r}{2}\sigma_-e^{i(-\nu+\delta_r)t}\int_{0}^t\!\!dt'i\eta\frac{\Omega_{\textrm S}}{2}e^{i\phi_{\textrm S}}\sigma_+ a^\dagger e^{i\nu t'} =-\eta\frac{\Omega_{\textrm S}\Omega_r}{4\nu}\sigma_-\sigma_+e^{i\phi_{\textrm S}}e^{i\delta_r t} a^\dagger \\
-\frac{\Omega_b}{2}\sigma_-e^{i(\nu+\delta_b)t}\int_{0}^t\!\!dt'i\eta\frac{\Omega_{\textrm S}}{2}e^{i\phi_{\textrm S}}\sigma_+ a e^{-i\nu t'} =\eta\frac{\Omega_{\textrm S}\Omega_b}{4\nu}\sigma_-\sigma_+e^{i\phi_{\textrm S}}e^{i\delta_b t} a \\
i\eta\frac{\Omega_{\textrm S}}{2}e^{-i\phi_{\textrm S}}\sigma_-a e^{-i\nu t}\int_0^t\!\!dt' \frac{\Omega_r}{2}\sigma_+e^{-i(-\nu+\delta_r)t'}=\eta\frac{\Omega_{\textrm S}\Omega_r}{4(\nu-\delta_r)}\sigma_-\sigma_+e^{-i\phi_{\textrm S}}e^{-i\delta_r t} a  \\
i\eta\frac{\Omega_{\textrm S}}{2}e^{-i\phi_{\textrm S}}\sigma_-a^\dagger e^{i\nu t}\int_0^t\!\!dt' \frac{\Omega_b}{2}\sigma_+e^{-i(\nu+\delta_b)t'}=-\eta\frac{\Omega_{\textrm S}\Omega_b}{4(\nu+\delta_b)}\sigma_-\sigma_+e^{-i\phi_{\textrm S}}e^{-i\delta_b t} a^\dagger.
\end{eqnarray}
If we assume that $1/(\nu\pm\delta_j)\sim1/\nu $ and reorganise all the terms we get that the second-order effective Hamiltonian is 
\begin{equation}\label{SSecondOrderEff}
H_B^{(2)}(t)\approx \eta\frac{\Omega_{\textrm S}\Omega_r}{4\nu} (ie^{-i\phi_{\textrm S}}e^{-i\delta_r t}a+\textrm{H.c.})\sigma_z- \eta\frac{\Omega_{\textrm S}\Omega_b}{4\nu} (ie^{-i\phi_{\textrm S}}e^{-i\delta_b t}a^\dagger+\textrm{H.c.})\sigma_z
\end{equation}
which can be incorporated to the first-order Hamiltonian in Eq.~(\ref{SScheme2}) giving 
\begin{eqnarray}\label{SEffectiveQRS}
H_\textrm{ eff}(t)&=&-i(2g^{(1)}_r \sigma_+ -g^{(2)}_{r}e^{-i\phi_\textrm{S}}\sigma_z)ae^{-i\delta_rt}  -i(2g_b^{(1)}\sigma_+ +g_{b}^{(2)}e^{-i\phi_\textrm{S}}\sigma_z) a^\dagger e^{-i\delta_bt} \nonumber\\ &+& \frac{\Omega_0}{2} \sigma_+e^{i\phi_{\textrm S}}  -\eta^2\frac{\Omega_\textrm{S}}{2}a^\dagger a\sigma_+e^{i\phi_\textrm{S}} +\textrm{H.c.}, 
\end{eqnarray}
where $g_{r,b}^{(1)}=\eta\Omega_{r,b}/4$ and $g_{r,b}^{(2)}=\eta\Omega_\textrm{S}\Omega_{r,b}/4\nu$ and $\Omega_0=\Omega_\textrm{S}(1-\eta^2/2)$. Now, if we assume $\phi_\textrm{S}=0$ or $\pi$ and move to a frame with respect $\frac{\Omega_\textrm{DD}}{2}\sigma_{x}$, we obtain (below $\Omega\equiv\Omega_\textrm{DD}$ for clarity)
\begin{eqnarray}\label{SEffectiveQRS2}
H_\textrm{eff}^{I}&=& \frac{\omega_0^{R}}{2}\sigma_+ \mp \eta^2\frac{\Omega_\textrm{S}}{2}a^\dagger a\sigma_+ -i\Big(g^{(1)}_r (\sigma_x+i\sigma_ye^{- i\Omega t\sigma_x}) \mp g^{(2)}_{r}\sigma_ze^{- i\Omega t\sigma_x}\Big)ae^{-i\delta_rt}  \nonumber\\
 &-&i\Big(g_b^{(1)}(\sigma_x+i\sigma_ye^{- i\Omega t\sigma_x}) \pm g_{b}^{(2)}\sigma_ze^{- i\Omega t\sigma_x}\Big) a^\dagger e^{-i\delta_bt} +\textrm{H.c.},
\end{eqnarray}
where $\Omega_\textrm{DD}=\pm\Omega_0-\omega_0^\textrm{ R}$ for $\phi_\textrm{S}=0$ and $\phi_\textrm{S}=\pi$ respectively. Using that
\begin{eqnarray}\label{SRotatingPaulis}
\sigma_ye^{- i\Omega t\sigma_x}=\cos{(\Omega_\textrm{DD}t)} \sigma_y-  \sin{(\Omega_\textrm{DD}t)}\sigma_z =\tilde{\sigma}_+e^{ i\Omega_\textrm{DD}t} + \tilde{\sigma}_-e^{- i\Omega_\textrm{DD}t}\\
\sigma_ze^{- i\Omega t\sigma_x}=\cos{(\Omega_\textrm{DD}t)} \sigma_z+  \sin{(\Omega_\textrm{DD}t)}\sigma_y = -i(\tilde{\sigma}_+e^{ i\Omega_\textrm{DD}t} - \tilde{\sigma}_-e^{- i\Omega_\textrm{DD}t}),
\end{eqnarray}
where $\tilde{\sigma}_{\pm}=(\sigma_y\pm i\sigma_z)/2$, and that the detunings are chosen to be $\delta_r=\Omega_\textrm{DD}+\omega^\textrm{ R}$ and $\delta_b=\Omega_\textrm{DD}-\omega^\textrm{ R}$, Eq.~(\ref{SEffectiveQRS}) is rewritten as
\begin{eqnarray}\label{SEffectiveQRS3}
H_\textrm{eff}^{I}&=& \frac{\omega_0^\textrm{ R}}{2}\sigma_+ \mp \eta^2\frac{\Omega_{\textrm S}}{2}a^\dagger a\sigma_+ \nonumber\\ &+& \Big(g^{(1)}_r (-i\sigma_x+[\tilde{\sigma}_+e^{i\Omega_\textrm{DD}t} +\textrm{H.c.}]) \pm g^{(2)}_{r}(\tilde{\sigma}_+e^{i\Omega_\textrm{DD}t} -\textrm{H.c.}\Big)ae^{- i\Omega_\textrm{DD} t} e^{- i\omega^\textrm{ R} t} \nonumber\\
&+& \Big(g_b^{(1)}(-i\sigma_x+[\tilde{\sigma}_+e^{ i\Omega_\textrm{DD}t} + \textrm{H.c.}]) \mp g_{b}^{(2)}(\tilde{\sigma}_+e^{ i\Omega_\textrm{DD}t} - \textrm{H.c.}\Big) a^\dagger e^{- i\Omega_\textrm{DD} t}e^{ i\omega^\textrm{ R} t}  \nonumber\\ &+&\textrm{H.c.}
\end{eqnarray}
where all terms rotating with $\pm\Omega_\textrm{DD}$ or higher can be neglected by the RWA. After the approximation we have
\begin{eqnarray}\label{SEffectiveQRS4}
H_\textrm{eff}^{I}= \frac{\omega_0^\textrm{ R}}{2}\sigma_x\mp \eta^2\frac{\Omega_\textrm{S}}{2}a^\dagger a\sigma_+ &+& (g^{(1)}_r  \pm g^{(2)}_{r})\tilde{\sigma}_+ a e^{- i\omega^\textrm{ R} t} \nonumber\\ &+&(g_b^{(1)} \mp g_{b}^{(2)})\tilde{\sigma}_+  a^\dagger e^{ i\omega^\textrm{ R} t} +\textrm{H.c.}
\end{eqnarray}
which in a rotating frame with respect $-\omega^\textrm{ R}a^\dagger a$ is
\begin{equation}\label{SEffectiveQRS5}
H_\textrm{eff}^{II}= \frac{\omega^\textrm{ R}_0}{2}\sigma_x +\omega^\textrm{ R} a^\dagger a + g_{\textrm{JC}}(\tilde{\sigma}_+ a+\tilde{\sigma}_- a^\dagger)   +g_{\textrm{aJC}}(\tilde{\sigma}_+  a^\dagger+\tilde{\sigma}_- a) \mp \eta^2\frac{\Omega_\textrm{S}}{2}a^\dagger a\sigma_x 
\end{equation} 
where $g_{\textrm{JC}}=\eta\Omega_r(1\pm\Omega_{\textrm S}/\nu)/4$ and $g_{\textrm{aJC}}=\eta\Omega_b(1\mp\Omega_\textrm{S}/\nu)/4$, depending on the choice of the phase $\phi_\textrm{S}=0$ or $\phi_\textrm{S}=\pi$.

\chapter{Further Considerations on Boson Sampling with Ultracold Atoms}
\label{appendix:appendix_d}

\section{Pair distribution in the optical lattice}\label{pair_distribution}

To estimate the rate of two-body collisions in section~\ref{ScalingAndErrors}, we need to know the probability of finding $k$ pairs of atoms occupying the same lattice sites, regardless of their spin states. To calculate this probability, we assume that the quantum circuit realises a Haar-random unitary matrix $U$ and, importantly, that the system is in a linear superposition of all possible states, where the probabilities of all states are the same. This second assumption, namely the uniform distribution (on average) over all possible bosonic configurations, has been proven for the output modes of the quantum circuit~\cite{Arkhipov12} and, for convenience, it is assumed valid in this work for the intermediate stages of the quantum circuit too. This will be justified by exact numerical simulations in section~\ref{subsec:HamilModel}.

We can calculate the probability of finding exactly $k_2$ pairs by simply counting the number of configurations with $k_2$ pairs of atoms in the same site, and dividing it by the overall number of possible bosonic configurations.
This latter is given by the multiset coefficient $\big(\hspace{-2.5pt}\binom{M}{N} \hspace{-2.5pt}\big)$. To determine the number of configurations containing $k_2$ pairs, we should first consider that there are $\binom{M/2}{k_2}$ different ways in which $k_2$ pairs of atoms can be arranged in $M/2$ sites. Since for each site doubly occupied, there are three possible spin configurations, $|2\rangle_{\downarrow}|0\rangle_{\uparrow}$, 
$|1\rangle_{\downarrow}|1\rangle_{\uparrow}$, and $|0\rangle_{\downarrow}|2\rangle_{\uparrow}$, the previous number of combinations should be multiplied by $3^{k_2}$ to obtain the total number of configurations. Secondly, we should consider that there are $\binom{M/2-k_2}{N-2k_2}$ combinations in which the remaining $N-2k_2$ atoms could be arranged in the remaining $M/2-k_2$ sites. Since for each site singly occupied, there are only two possible spin configurations, $|1\rangle_{\downarrow}|0\rangle_{\uparrow}$ and $|0\rangle_{\downarrow}|1\rangle_{\uparrow}$, the previous number of combinations should be multiplied by $2^{N-2k_2}$ to obtain the total number of configurations in which the $N-2k_2$ single atoms could be arranged. Thus, the probability of finding $k_2$ pairs of atoms in distinct sites is:
\begin{equation}\label{NPairProb1}
P_\text{pair}(N,M;k_2)=\frac{3^{k_2}\binom{M/2}{k_2}2^{N-2k_2}\binom{M/2-k_2}{N-2k_2}}{\big(\hspace{-2.5pt}\binom{M}{N} \hspace{-2.5pt}\big)}\Big.
\end{equation}
Equation~(\ref{NPairProb1}) only takes into account particle pairs, forgetting about states that have more than two particles in a site. Those can be taken into account defining a more general expression $P(k_2,k_3,k_4,...)$, corresponding to the probability of having $k_2$ pairs, $k_3$ trios, $k_4$ quartets and so on. For simplicity, we will consider the case for only pairs and trios. The number of configurations in which $k_3$ trios can be placed in $M/2$ sites is given by $\binom{M/2}{k_3}$. This number has to be multiplied by $4^{k_3}$, as $4$ different states can represent a trio in a site: $|3\rangle_{\downarrow}|0\rangle_{\uparrow}$, 
$|2\rangle_{\downarrow}|1\rangle_{\uparrow}$, $|1\rangle_{\downarrow}|2\rangle_{\uparrow}$ and $|0\rangle_{\downarrow}|3\rangle_{\uparrow}$.  Then, the number of configurations in which $k_2$ pairs can be placed in the remaining $M/2-k_3$ sites is given by $\binom{M/2-k_3}{k_2}$. Finally, the number of configurations in which the remaining $N-2k_2-3k_3$ particles can be placed in $N-2k_2-3k_3$ sites is given by $\binom{M/2-k_2-k_3}{N-2k_2-3k_3}$. The probability of having $k_2$ pairs and $k_3$ trios is then given by 
\begin{equation}\label{NPairProb2}
P(k_2,k_3)=4^{k_3}\binom{M/2}{k_3}3^{k_2}\binom{M/2-k_3}{k_2}2^{N-2k_2-3k_3}\binom{M/2-k_2-k_3}{N-2k_2-3k_3}\Big/\bigg(\hspace{-2.5pt}\binom{M}{N} \hspace{-2.5pt}\bigg).
\end{equation}
In the limit of large $N$, both Eqs.~(\ref{NPairProb1}) and (\ref{NPairProb2}) converge to a Poissonian distribution
\begin{equation}\label{eq:SitePoissonian}
\lim_{N\to\infty} P_\text{pair}(N,M;k)=\frac{\lambda^{k}}{e^{\lambda}k!},
\end{equation}
with average value $\lambda=3/2c$. Here, the constant factor $c$ denotes the ratio $c=M/N^2$, which in the main text has been simply assumed equal to 1. As an example, the probability associated to the collision-free subspace (the subspace of states where all atoms are at different sites), i.e. $P_\text{pair}(N,M;0)$, tends to $1/\exp(3/2c)$ for large $N$.

To prove how Eq.~(\ref{NPairProb2}) tends to Eq.~(\ref{eq:SitePoissonian}) for large $N$, it is useful to recall that 
\begin{equation}\label{MultisetProd}
\bigg(\hspace{-2.5pt}\binom{M}{N} \hspace{-2.5pt}\bigg)=\frac{M^N}{N!}\prod_{a=0}^{N-1}(1+a/M)\approx \exp{(\sum_{a=0}^{N-1}a/M)}\approx\exp{(N^2/2M)},
\end{equation}
and 
\begin{equation}\label{BinomialProd}
\binom{M}{N} =\frac{M^N}{N!}\prod_{a=0}^{N-1}(1-a/M)\approx \exp{(-\sum_{a=0}^{N-1}a/M)}\approx\exp{(-N^2/2M)},
\end{equation}
for $M\ll N$. Using these relations, Eq.~(\ref{NPairProb2}) can be rewritten as
\begin{equation}\label{AlmostLimit}
P(k_2,k_3)\approx \frac{1}{k_2!k_3!}\bigg(\frac{N^3}{2M^2}\bigg)^{k_3}\bigg(\frac{3N^2}{2M}\bigg)^{k_2}e^{-3N^2/2M}.
\end{equation}
For a $M=cN^2$ scaling, $N^3/2M^2$ tends to zero, which means that the probability of having trios tends to zero for large $N$. Of course, the same will happen with the probability of having four or more particles in a site. For $k_3=0$, the probability of having $k_2$ pairs can be written as
\begin{equation}\label{Poissonian2}
P_\textrm{ pair}(k_2)=P(k_2,0)\approx \frac{1}{k_2!}\bigg(\frac{3}{2c}\bigg)^{k_2}e^{-3/2c},
\end{equation}
which equals Eq.~(\ref{eq:SitePoissonian}).

\section{Derivation of average values of $V$ and $V^2$}\label{App:AvVs}

In the following we present a derivation of the average value of $V$ and $V^2$ evaluated in a uniform state, $|\psi\rangle_u={D}^{-1/2}\sum_{d=1}^D|d\rangle $, where $D=\big(\hspace{-2.5pt}\binom{M}{N} \hspace{-2.5pt}\big)$. It may be useful to recall the form of $V$, 
\begin{equation}
V=\frac{\Gamma_\textrm{ tb}}{4}\sum_{m=1}^M\hat{n}_m(\hat{n}_m-1) 
+\frac{\Gamma_\textrm{ tb}}{2}\sum_{s=1}^{M/2}\hat{n}_{2s-1}\hat{n}_{2s}. 
\end{equation}
Then, $\langle V\rangle_u$ is 
\begin{equation}
\langle V\rangle_u=\frac{\Gamma_\textrm{ tb}}{4}M\langle \hat{n}_m(\hat{n}_m-1)\rangle_u 
+\frac{\Gamma_\textrm{ tb}}{2}\frac{M}{2}\langle\hat{n}_{2s-1}\hat{n}_{2s}\rangle_u,
\end{equation}
where we used that, because of the symmetry of the uniform state, $\langle\sum_{m=1}^M \hat{A}_m\rangle_u=M\langle \hat{A}_m\rangle_u$.
To evaluate $\langle \hat{n}_m\rangle_u$, we can think on the number of states that can have $k$ bosons in mode $m$. This is just equivalent to the number of configurations in which you can put $N-k$ particles in $M-1$ modes, which is given by $\big(\hspace{-2.5pt}\binom{M-1}{N-k} \hspace{-2.5pt}\big)$. $\langle \hat{n}_m\rangle_u$ is then given by $\sum_{k=0}^Nk p_k$, where $p_k=\big(\hspace{-2.5pt}\binom{M-1}{N-k} \hspace{-2.5pt}\big)/\big(\hspace{-2.5pt}\binom{M}{N} \hspace{-2.5pt}\big)$. In the same way, $\langle \hat{n}_m^2\rangle_u$ is given by $\sum_{k=0}^Nk^2 p_k$. We call $p_k$, the probability of having $k$ particles in mode $m$. To evaluate $\langle \hat{n}_m\hat{n}_{m'}\rangle_u$, one needs to take into account the probability of having $k$ particles in mode $m$ while having $k'$ particles in mode $m'$, which is given by $p_{k,k'}=\big(\hspace{-2.5pt}\binom{M-2}{N-k-k'} \hspace{-2.5pt}\big)/\big(\hspace{-2.5pt}\binom{M}{N} \hspace{-2.5pt}\big)$. $\langle \hat{n}_m\hat{n}_{m'}\rangle_u$ is then given by $\sum_{k=0}^N\sum_{k'=0}^{N-k}kk' p_{k,k'}$. Using all this, 
\begin{equation}\label{avVsimple}
\langle V\rangle_u=\frac{\Gamma_\textrm{ tb}}{4}M \sum_{k=0}^Np_k k(k-1) 
+\frac{\Gamma_\textrm{ tb}}{2}\frac{M}{2} \sum_{k=0}^N\sum_{k'=0}^{N-k}p_{k,k'}kk' .
\end{equation}
To make the calculation of the summations in Eq.~(\ref{avVsimple}) easier, we can make use of 
\begin{equation}\label{BirthdayApprox}
\bigg(\hspace{-2.5pt}\binom{M}{N-k} \hspace{-2.5pt}\bigg)/\bigg(\hspace{-2.5pt}\binom{M}{N} \hspace{-2.5pt}\bigg)\sim \frac{N^k}{(M+N)^k}
\end{equation}
when $k\ll N,M$~\cite{Arkhipov12}. Then, $p_k\approx a_1\lambda_1^k$ and $p_{k,k'}\approx a_2\lambda_2^{(k+k')}$, where $a_j=\prod_{i=1}^j\frac{M-i}{M+N-i}$ and $\lambda_j=\frac{N}{M+N-j}$. The summations can be then reduced to geometric series, for example, 
\begin{equation}
\sum^N_{k=0} k(k-1) p_k=a_1\lambda_1^2\frac{\partial^2}{\partial k^2}\sum_{k=0}^N \lambda_1^k\approx a_1\lambda_1^2 \frac{\partial^2}{\partial k^2}\frac{1}{1-\lambda_1}=\frac{2a_1\lambda_1^2}{(1-\lambda_1)^3}.
\end{equation}
Now, assuming that $M=cN^2$, one can calculate the limit at large $N$ of $M2a_1\lambda_1^2/(1-\lambda_1)^3$, which gives $2/c$. The rest of the summations are done using Wolfram Mathematica, obtaining $\langle V\rangle_u=3\Gamma_\textrm{ tb}/4c$. 

To calculate $\langle V^2\rangle_u$, the same approach is followed. First, we write $V^2$,
\begin{eqnarray}
V^2&=&\frac{\Gamma_\textrm{ tb}^2}{16}\sum_{m=1}^M\sum_{m'=1}^M\hat{n}_m(\hat{n}_m-1)\hat{n}_{m'}(\hat{n}_{m'}-1) \nonumber\\
 &+&\frac{\Gamma_\textrm{ tb}^2}{4}\sum_{s=1}^{M/2}\sum_{s'=1}^{M/2}\hat{n}_{2s-1}\hat{n}_{2s}\hat{n}_{2s'-1}\hat{n}_{2s'} \\
&+&\frac{\Gamma_\textrm{ tb}^2}{4}\sum_{m=1}^M\sum_{s=1}^{M/2}\hat{n}_m(\hat{n}_m-1)\hat{n}_{2s-1}\hat{n}_{2s}.\nonumber
\end{eqnarray}
The first term can be written as
\begin{eqnarray}
&&\langle \sum_{m=1}^M\sum_{m'=1}^M\hat{n}_m(\hat{n}_m-1)\hat{n}_{m'}(\hat{n}_{m'}-1) \rangle_u \nonumber \\
&=&M\langle \hat{n}_m^2(\hat{n}_m-1)^2\rangle_u +M(M-1)\langle \hat{n}_m(\hat{n}_m-1) \hat{n}_{m'}(\hat{n}_{m'}-1)\rangle_u  \\
&=&M\sum_{k=0}^N p_k k^2 (k-1)^2 + M(M-1)\sum_{k=0}^N \sum_{k=0}^{N-k}p_{k,k'}k(k-1)k'(k'-1). \nonumber
\end{eqnarray}
Similar with the second term
\begin{eqnarray}
&&\langle \sum_{s=1}^{M/2}\sum_{s'=1}^{M/2}\hat{n}_{2s-1}\hat{n}_{2s}\hat{n}_{2s'-1}\hat{n}_{2s'} \rangle_u \nonumber \\
&=&\frac{M}{2}\langle \hat{n}_{2s-1}^2\hat{n}_{2s}^2 \rangle_u +\frac{M}{2}(\frac{M}{2}-1)\langle \hat{n}_{2s-1}\hat{n}_{2s}\hat{n}_{2s'-1}\hat{n}_{2s'} \rangle_u  \\
&=&\frac{M}{2}\sum_{k=0}^N\sum_{k'=0}^{N-k}p_{k,k'}k^2k'^2 +\frac{M}{2}(\frac{M}{2}-1)\sum_{k=0}^N\sum_{k'=0}^{N'}\sum_{k''=0}^{N''}\sum_{k'''=0}^{N'''}p_{k,k',k'',k'''}kk'k''k''',\nonumber
\end{eqnarray}
where $N'=N-k$, $N''=N-k-k'$, $N'''=N-k-k'-k''$, and $p_{k,k',k'',k'''}=\big(\hspace{-2.5pt}\binom{M-4}{N-k-k'-k''-k'''} \hspace{-2.5pt}\big)/\big(\hspace{-2.5pt}\binom{M}{N} \hspace{-2.5pt}\big)$, which can be approximated to $p_{k,k',k'',k'''}\approx a_4\lambda_4^{(k+k'+k''+k''')}$. The third term is:
\begin{eqnarray}
&&\langle \sum_{m=1}^M\sum_{s=1}^{M/2}\hat{n}_m(\hat{n}_m-1) \hat{n}_{2s-1}\hat{n}_{2s}\rangle_u \nonumber \\
&=&M\langle \hat{n}_m^2(\hat{n}_m-1)\hat{n}_{m'}\rangle_u +\frac{M}{2}(M-1)\langle \hat{n}_m(\hat{n}_m-1) \hat{n}_{m'}\hat{n}_{m''}\rangle_u  \\
&=&M\sum_{k=0}^N\sum_{k'=0}^{N'} p_{k,k'} k^2 (k-1)k' + \frac{M}{2}(M-1)\sum_{k=0}^N \sum_{k'=0}^{N'}\sum_{k''=0}^{N''} p_{k,k',k''}k(k-1)k'k'', \nonumber
\end{eqnarray}
where $p_{k,k',k''}=\big(\hspace{-2.5pt}\binom{M-3}{N-k-k'-k''} \hspace{-2.5pt}\big)/\big(\hspace{-2.5pt}\binom{M}{N} \hspace{-2.5pt}\big)$, which can be approximated to $p_{k,k',k''}\approx a_3\lambda_3^{(k+k'+k'')}$. Carrying out these summations using Wolfram Mathematica, assuming $M=cN^2$, and calculating the limit at large $N$ yields to $\langle V^2\rangle_u=(3/2c + 9/4c^2)\Gamma^2_\textrm{ tb}/4$.

\begin{figure}[t]
	\centering\includegraphics*[width=1\columnwidth]{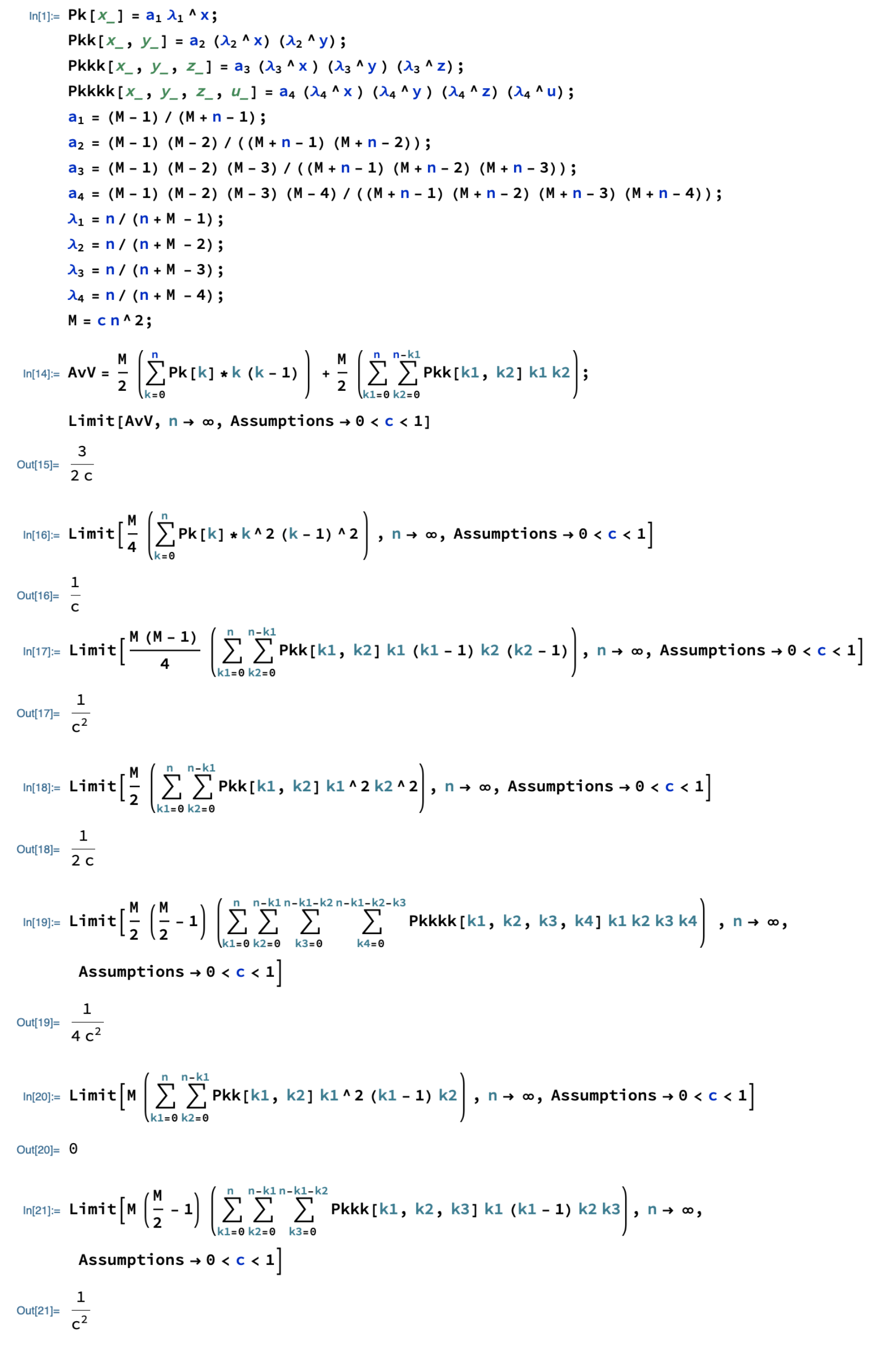}
	\caption{Wolfram Mathematica code to calculate the summatories.\label{mathematica_code}}
\end{figure}

\chapter{Further Considerations on Quantum Sensing with NV Centers}
\label{appendix:appendix_e}

\section{Finding $\Omega(t)$ from $F(t)$}\label{FindOmega}

The term representing the MW driving in Eq.~(\ref{simulations}) is $\frac{\Omega(t)}{2} (|1\rangle\langle 0| e^{i\phi} +\textrm{ H.c.})$, and the associated propagator for, e.g., the $m$th $\pi$-pulse is 
\begin{equation}
U_t = \exp{[-i\int_{t_m}^{t_m+t_{\pi}} \frac{\Omega(s)}{2} (|1\rangle\langle 0| e^{i\phi} +\textrm{ H.c.}) \ ds]}.
\end{equation}
During the $m$th $\pi$-pulse, i.e. in a certain time between $t_m$ and $t_m + t_{\pi}$, $U_t $ has the following effect on the electron spin $\sigma_z$ operator (in the following, $\sigma_\phi = |1\rangle\langle 0| e^{i\phi} +\textrm{ H.c.}$)
\begin{eqnarray}
e^{i\int_{t_m}^{t_m+t} \frac{\Omega(s)}{2} \sigma_\phi \ ds} \sigma_z e^{-i\int_{t_m}^{t_m+t} \frac{\Omega(s)}{2} \sigma_\phi \ ds} = e^{\big(i\int_{t_m}^{t_m+t} \Omega(s) \ ds \big)  \sigma_\phi }  \sigma_z \nonumber \\ =\cos{\bigg(\int_{t_m}^{t_m+t} \Omega(s) \ ds \bigg)} \sigma_z + i  \sin{\bigg(\int_{t_m}^{t_m+t} \Omega(s) \ ds \bigg)} \sigma_{\phi}\sigma_z.
\end{eqnarray}
In this manner, it is cleat that $F(t) = \cos{\big(\int_{t_m}^{t_m+t} \Omega(s) \ ds \big)}$. The other spin component, i.e. the one going with $ \sin{\big(\int_{t_m}^{t_m+t} \Omega(s) \ ds \big)}$, does not participate in the joint NV-nucleus dynamics for sequences with alternating pulses~\cite{Lang17} such as the XY8 $\equiv$ XYXYYXYX pulse sequence we are using in chapter~\ref{chapter:chapter_4}. A valid argument to neglect this term is that its period is twice the period of the pulse sequence, i.e. $2T$, and, thus, it is kept off resonance. Now, assuming that $F(t)$ and $\Omega(t)$ are differentiable, one can invert the expression $F(t) = \cos{\big(\int_{t_m}^{t_m+t} \Omega(s) \ ds \big)}$ and find
\begin{equation}
\Omega(t) = \frac{\partial}{\partial_t} \arccos[F(t)]=-\frac{1}{\sqrt{1-F(t)^2}}.
\end{equation}
The latter corresponds to Eq.~(\ref{modulatedOmega}).

\section{Calculation of $f_l$ coefficients}
\subsection{Coefficients for extended pulses}\label{calcext}

The analytical expression for the coefficients $f_l$ is given by 
\begin{equation}
f_{l}=\frac{2}{T}\int_{0}^{T} F(s) \cos{\Big(\frac{2 \pi l s}{T}\Big)}  \ ds,
\end{equation}
where $T=2 \pi /\omega_\textrm{ M}$. With a rescaling of the integrating variable given by $s=xT/2 $, this is rewritten as  
\begin{equation}\label{equation22}
f_{l}=\int_{0}^{2} F(x) \cos{(\pi l x)}  \ dx.
\end{equation}
The function inside the integral is symmetric or antisymmetric with respect to $x=1$, depending on $l$ been odd or even. This can be easily demonstrated by using $F(x+1)=-F(x)$ and $\cos{[\pi l (x+1)]}=\cos(\pi l)\cos{\pi l x}$. Thus, if $l$ is even and the function is symmetric with respect to $x=1$, the value of the integral will be zero. Anyway, one can work a general expression for Eq.~(\ref{equation22}). First, we can divide the integral in two parts,
\begin{equation}
f_{l}=\int_{0}^{1} F(x) \cos{(\pi l x)}  \ dx +\int_{1}^{2} F(x) \cos{(\pi l x)}  \ dx,
\end{equation}
and substitute $x$ for $x+1$ in the second integral. Using the symmetry properties specified above, the equation reduces to 
\begin{equation}
f_{l}=(1-\cos{(\pi l)})\int_{0}^{1} F(x) \cos{(\pi l x)}  \ dx.
\end{equation}
Now, from $x=0$ to $1$, $F(x)$ can be divided in three parts defined by $\tau_m\equiv 2 t_{m}/T$ and $1-\tau_m$, the times corresponding to the beginning and the end of the pulse,
\begin{equation}\label{integrals}
\int_{0}^{\tau_m} F(x) \cos{(\pi l x)}  \ dx +\int_{\tau_m}^{1-\tau_m} F(x) \cos{(\pi l x)}  \ dx+\int_{1-\tau_m}^{1} F(x) \cos{(\pi l x)}  \ dx,
\end{equation}
and the integral in the middle is zero for the extended pulses. This leaves us with the first and third integrals for which $F(x)$ is $1$ and $-1$ respectively, obtaining 
 \begin{equation}
f_{l}^\textrm{ m}=(1-\cos{(\pi l)})\Bigg\{\int_{0}^{\tau_m} \cos{(\pi l x)}  \ dx -\int_{1-\tau_m}^{1}  \cos{(\pi l x)}  \ dx\Bigg\},
\end{equation}
that leads to
 \begin{equation}
f_{l}^\textrm{ m}=\frac{1}{\pi l }(1-\cos{(\pi l)})\Bigg\{\sin{(\pi l \tau_m)} + \sin{(\pi l (1-\tau_m))} \Bigg\}.
\end{equation}
Using $\sin{[\pi l (1-\tau_m)]}=-\sin{(\pi l \tau_m)}\cos(\pi l)$ and $\sin^2{\theta}=(1-\cos{(2\theta)})/2$, the expression for $f_{l}^\textrm{ m}$ reduces to 
 \begin{equation}
f_{l}^\textrm{ m}=\frac{4}{\pi l }\sin^4{(\pi l/2)}\sin{(\pi l \tau_m)}.
\end{equation}
Now, by using the relation $T=4t_m+2t_\pi$,  $f_{l}^\textrm{ m}$ becomes
\begin{equation}\label{eqbat}
f_{l}^\textrm{ m}=\frac{4}{\pi l }\sin^4{(\pi l/2)}\sin{\Big(\pi l \Big(\frac{1}{2}+\frac{t_\pi}{T}\Big)\Big)},
\end{equation}
where $t_{\pi}$ is the duration of a $\pi$-pulse. Eq.~(\ref{eqbat}) is equivalent to Eq.~(\ref{modulatedf}) in the main text. To prove that, one may use the trigonometric identity $\sin(\theta+\pi l /2)=\sin{(\theta)}\cos{(\pi l /2)} +\cos{(\theta)}\sin{(\pi l /2)}$ which leads us to 
\begin{equation}\label{eqlast}
f_{l}^\textrm{ m}=\frac{4}{\pi l }\cos{\Bigg(\pi  \frac{t_{\pi}}{T/l}\Bigg)}\sin{(\pi l /2)},
\end{equation}
as $\sin^4{(\pi l /2)}\cos{(\pi l /2)}=0$ and $\sin^5{(\pi l /2)}=\sin{(\pi l /2)}$. 

\begin{figure}[t]
\includegraphics[width=0.9\columnwidth]{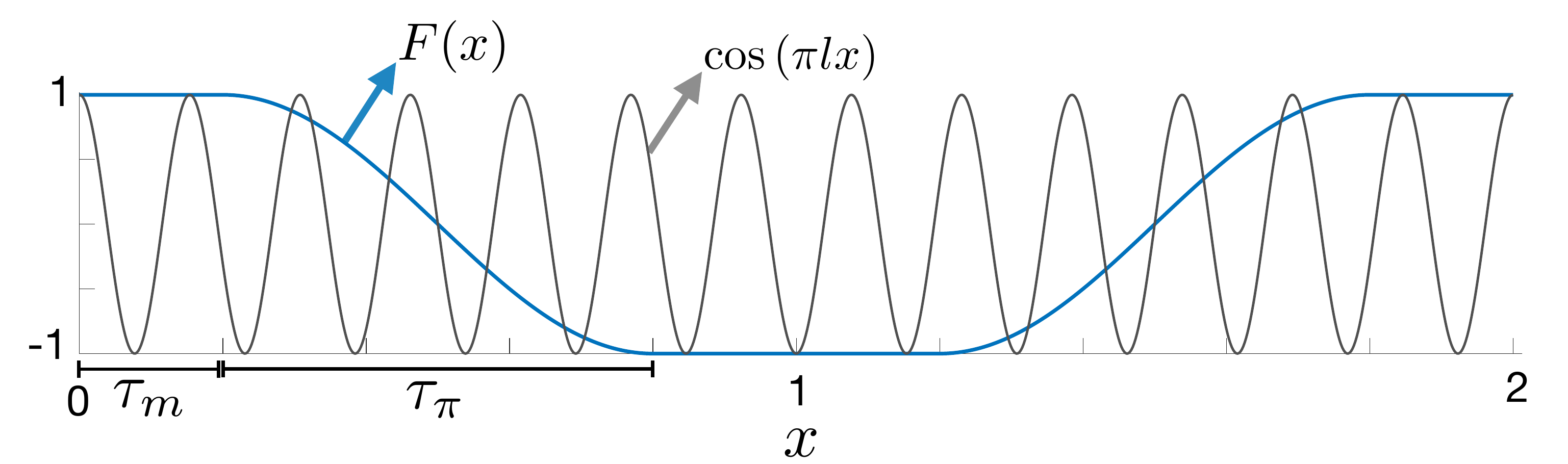}
\caption{Plot of $F(x)$ and $\cos{(\pi l x)}$ (where $l=13$) functions between $x=0$ and $x=2$, corresponding to $t=0$ and $t=T$ respectively. }
\label{Subfig}
\end{figure}

\subsection{Coefficients for top-hat pulses}\label{calcth}
For calculating the value of $f_l$ coefficients in the case of top-hat pulses, we just need to sum the contribution of the second integral on Eq.~(\ref{integrals}), which is not zero for top-hat pulses. The value of $F(s)$ during the pulse is $F(s)=\cos{[\pi(s-t_m)/t_\pi]}$, where $t_{p}=t_m+t_\pi/2$. With the rescaling of the integrating variable introduced in the previous section this is rewritten as
\begin{equation}
F(s)=\cos{[\pi(x-\tau_m)/\tau_\pi]},
\end{equation}
where $\tau_\pi=2 t_\pi /T $. So, we need to solve the following integral 
\begin{equation}
\int_{\tau_m}^{1-\tau_m} F(x) \cos{(\pi l x)}  \ dx=\int_{\tau_m}^{1-\tau_m}  \cos{[\pi(x-\tau_m)/\tau_\pi]}\cos{(\pi l x)}  \ dx
\end{equation}
which is not zero. To solve the integral, we can displace the reference frame by a factor of $\tau_p=1/2$, by the change of variable $x=y+\tau_m+\tau_\pi/2=y+1/2$. Now, the integral will be centered at zero and will look like 
\begin{eqnarray}
\int_{-\tau_{\pi}/2}^{\tau_{\pi}/2}  \cos{[\pi y/\tau_{\pi}+\pi/2]} \cos{[\pi l (y+1/2)]}  \ dy\nonumber\\=-\int_{-\tau_{\pi}/2}^{\tau_{\pi}/2}  \sin{(\pi y/\tau_{\pi})} \cos{[\pi l (y+1/2)]}  \ dy,
\end{eqnarray}
which using $\cos[\pi l (y+1/2)]=\cos{(\pi l y)}\cos{(\pi l /2)} - \sin{(\pi l y)}\sin{(\pi l /2)}$ becomes
\begin{eqnarray}
\sin{(\pi l /2)}\int_{-\tau_{\pi}/2}^{\tau_{\pi}/2}  \sin{(\pi y/\tau_{\pi})} \sin{(\pi l y)}  \ dy \nonumber\\- \cos{(\pi l /2)}\int_{-\tau_{\pi}/2}^{\tau_{\pi}/2}  \sin{(\pi y/\tau_{\pi})} \cos{(\pi l y)}  \ dy.
\end{eqnarray}
The second integral is zero owing to symmetry reasons, i. e. $\int_{-a}^{a}F(x)dx=0$ if $F(-x)=-F(x)$. Again, because of symmetry arguments, the first integral is
\begin{equation}
2\sin{(\pi l /2)}\int_{0}^{\tau_{\pi}/2}  \sin{(\pi y/\tau_{\pi})} \sin{(\pi l y)}  \ dy,
\end{equation}
which using trigonometric identities reads 
\begin{equation}
\sin{(\pi l /2)}\Big\{\int_{0}^{\tau_{\pi}/2}  \cos{(\pi y(l-1/\tau_{\pi})}   \ dy -\int_{0}^{\tau_{\pi}/2}  \cos{(\pi y(l+1/\tau_{\pi})}   \ dy.\Big\}
\end{equation}
Solving the integrals one gets
\begin{equation}
\frac{-1}{\pi}\sin{(\pi l /2)}\cos{(\pi l\tau_{\pi}/2)}\Big\{ \frac{1}{l-1/\tau_{\pi}}  +  \frac{1}{l+1/\tau_{\pi}} \Big\},
\end{equation}
which is simplified to 
\begin{equation}
\frac{2l\tau_\pi^2}{\pi(1-l^2\tau_{\pi}^2)}\sin{(\pi l /2)}\cos{(\pi l\tau_{\pi}/2)}.
\end{equation}
It is straightforward to prove that the sum of  the three integrals in Eq.~(\ref{integrals}) gives 
\begin{equation}
f_{l}^\textrm{ th}=\frac{4 \sin{(\pi l /2)}\cos{(\pi l t_{\pi}/T)}}{\pi l(1-4l^2t_{\pi}^2/T^2)},
\end{equation}
which correspond to the expression written in the main text.

\section{Energy delivery}\label{energydelivery}
The Poynting vector, that describes the energy flux for an electromagnetic wave, is given by 
\begin{equation}
\vec{P}=\frac{1}{\mu_{0}} \vec{E} \times \vec{B},
\end{equation}
where $\mu_{0}$ is the vacuum permeability, and $\vec{E}$ and $\vec{B}$ are the electric field and magnetic field vectors at the region of interest, i.e. the NV center. The latter, in the nanoscale, is sufficiently small compared with the wavelength of the MW radiation to assume a plane wave description of the radiation, so the magnetic field can be written as
\begin{equation}
\vec{B}=\vec{B}_{0}(t)\cos{(\vec{k}\cdot \vec{x}-\omega t +\phi)},
\end{equation}
where $\vec{k}$ is the wavevector and $\omega$ the frequency of the microwave field. We will also assume an extra time dependence $B_{0}(t)$ whose time scales will be several orders of magnitude larger than the period $2\pi/\omega$. From Maxwell equations in vacuum it is derived that, for such a magnetic field, $\vec{k}\cdot \vec{B}=0$, $\vec{k}\cdot \vec{E}=0$, and $\vec{E}\cdot \vec{B}=0$. From the equation $\vec{\nabla}\times\vec{B}=\frac{1}{c^2}\partial \vec{E}/\partial t$, it follows that
\begin{equation}\label{electric}
\vec{E}=c^2\int\!dt \ (\vec{\nabla}\times\vec{B})=-c^2\int\! dt \ (\vec{k}\times\vec{B_0}(t)) \sin{(\vec{k}\cdot \vec{x}-\omega t +\phi)}.
\end{equation}
We choose $\vec{B}$ to be perpendicular to the NV axis ($z$ axis), specifically, on the $x$ axis. The control Hamiltonian, is then 
\begin{equation}
H_{c}(t)=-\gamma_e\vec{B}\cdot \vec{S}=\gamma_eB_{x}(t)S_x \cos{(\omega t - \phi)},
\end{equation}
where $\vec{S}$ corresponds to the spin of the NV center, $\gamma_e$ is the gyromagnetic ratio of the electron and $\vec{x}=0$ the position of the NV. To recover Eq.~(\ref{controlHamil}) of the main text, we require that $\sqrt{2} \Omega(t)=\gamma_e B_x(t)$. The magnetic field vector at $\vec{x}=0$ is then
\begin{equation}
\vec{B}(t)=\frac{\sqrt{2}\Omega(t)}{\gamma_e} \cos{(\omega t -\phi)} \hat{x}\\
\end{equation}
and the electric field is, from Eq.~(\ref{electric}),
\begin{equation}
\vec{E}(t)=\frac{\sqrt{2}\omega c}{\gamma_e}\int dt \Omega(t) \sin{(\omega t -\phi)} \ \hat{k}\times \hat{x},
\end{equation}
which, using the wave equation $\partial^2 \vec{E}/\partial^2 t=c^2\nabla^2 \vec{E}=\omega^2\vec{E}$, converts into
\begin{equation}\label{elec2}
\vec{E}(t)=\frac{\sqrt{2}}{k\gamma_e} \frac{\partial}{\partial t}\Big[ \Omega(t) \sin{(\omega t -\phi)} \Big]\ \hat{x}\times \hat{k}.
\end{equation}

\subsection{The case of top-hat $\pi$ pulses}

For top-hat pulses we have that $\partial\Omega(t)/\partial t=0$ during the pulse, thus, the energy delivery per unit of area we obtain for top-hat pulses is
\begin{equation}
E^\textrm{ th}(t_{\pi})=\int_0^{t_\pi}dt |\vec{P}(t)|=\frac{c}{\mu_0}\frac{2}{\gamma_{e}^2}\int_0^{t_\pi}dt \ \Omega^2\cos^2(\omega t-\phi)
\end{equation}
which gives
\begin{equation}\label{eth}
E^\textrm{ th}(t_{\pi})=\frac{c}{\mu_0}\frac{\Omega^2}{\gamma_e^2} \Big\{t_\pi+\frac{1}{2\omega}\sin{(2\omega t_\pi -2\phi)}\Big\}.
\end{equation}
The second part of the formula is upper bounded by $(2\omega)^{-1}$, which , on the other hand, is several orders of magnitude smaller than $t_\pi$, thus negligible. As $t_{\pi}=\pi/\Omega$, Eq.~(\ref{eth}) can be rewritten as
\begin{equation}\label{Stophat}
E^\textrm{ th}(t_{\pi})\approx \frac{\pi c}{\mu_0}\frac{ \Omega}{\gamma_e^2},
\end{equation}
meaning that the energy increases linearly with the Rabi frequency.

\subsection{The case of extended $\pi$ pulses}
To study the case of an extended $\pi$ pulse, we need to calculate both terms on Eq.~(\ref{elec2}), which are non zero in general. The complete expression is given by 

\begin{equation}\label{Sextended}
E^\textrm{ ext}(t_{\pi})=\frac{c}{\mu_0}\frac{2}{\gamma_e^2}\int_{0}^{t_\pi}\! \!dt \ \Bigg[ \Omega^2(t)\cos^2{(\omega t -\phi)} +\frac{1}{\omega} \Omega(t)\frac{\partial\Omega(t)}{\partial t}\cos{(\omega t -\phi)}  \sin{(\omega t -\phi)} \Bigg] .
\end{equation}
As a final comment, for all cases simulated in the main text we find that the second term at the right hand side of Eq.~(\ref{Sextended}) is negligible, thus it can be written 
\begin{equation}
E^\textrm{ ext}(t_{\pi})\approx \frac{c}{\mu_0}\frac{2}{\gamma_e^2}\int_{0}^{t_\pi}\! \!dt \ \Bigg[ \Omega^2(t)\cos^2{(\omega t -\phi)}  \Bigg] .
\end{equation}

\subsection{Equivalent top-hat Rabi frequency}
To calculate the constant Rabi frequency leading to top-hat pulses with the same energy than extended pulses, one has to equal $E^\textrm{ top-hat}(t_{\pi}) = E^\textrm{ ext}(t_{\pi})$ and extract the value of the constant $\Omega$. With Eqs.~(\ref{Stophat}, \ref{Sextended}) one can easily find that 
\begin{equation}
\Omega=\frac{\mu_0 \gamma_e^2}{\pi c} E^\textrm{ ext}(t_{\pi}).
\end{equation}

\renewcommand\bibname{Bibliography}
\addcontentsline{toc}{chapter}{Bibliography}
\thispagestyle{chapter}
\bibliographystyle{sofia}
\bibliography{others/bibliography}

\cleardoublepage


\end{document}